\documentstyle[preprint,aps,harvard,epsf,floats]{revtex}

\newcommand{\beq}{\begin{equation}}
\newcommand{\eeq}{\end{equation}}
\citationstyle{dcu}
\def\ie{{\it i.e.}}
\def\eg{{\it e.g.}}

\def\th{\theta}

\def\np#1#2#3{Nucl. Phys. {\bf B#1} (#2) #3}
\def\pl#1#2#3{Phys. Lett. {\bf #1B} (#2) #3}

\def\prl#1#2#3{Phys. Rev. Lett. {\bf #1} (#2) #3}

\def\prd#1#2#3{Phys. Rev. {\bf D#1} (#2) #3}
\def\ap#1#2#3{Ann. Phys. {\bf #1} (#2) #3}

\def\DD{{\cal D}}

\def\FF{{\cal F}}
\def\GG{{\cal G}}

\def\LL{{\cal L}}
\def\MM{{\cal M}}

\def\nsp{{$NS5^\prime$}}
\def\nspp{{NS5^\prime}}
\def\gs{g_s}
\def\ddp{$Dp$}
\def\op{$Op$}
\def\beq{\begin{equation}}
\def\eeq{\end{equation}}
\def\bea{\begin{eqnarray}}
\def\eea{\end{eqnarray}}
\def\nf{N_f}
\def\nc{N_c}

\begin{document}
\preprint{hep-th/9802067, RI-2-98, EFI-98-06}
\title{Brane Dynamics and Gauge Theory}
\author{Amit Giveon}
\address{Racah Institute of Physics, The Hebrew University\\
         Jerusalem 91904, Israel}
\author{David Kutasov}
\address{EFI and Department of Physics, University of Chicago\\
         5640 South Ellis Avenue, Chicago, IL 60637, USA\\
         and\\
         Department of Particle Physics, Weizmann Institute of Science\\
         Rehovot 76100, Israel}
\maketitle
\begin{abstract}
We review some aspects of
the interplay between the dynamics of branes in string theory
and the classical and quantum physics of gauge theories with different
numbers of supersymmetries in various dimensions.
\end{abstract}
\newpage
\tableofcontents
\newpage
%\pagestyle{myheadings}
%\markright{Ekert 3/25/96 Revised}
\section{Introduction}
\label{ISO}
Non-abelian gauge theories are a cornerstone of
the standard model of elementary particle physics.
Such theories (for example QCD) are often strongly
coupled at long distances and, therefore, cannot be
studied by the standard perturbative methods of
weakly coupled field theory.
In the last few years important progress was made
in the study of the strongly coupled dynamics in a
class of gauge theories -- Supersymmetric Yang-Mills
(SYM) theories. New understanding
of the constraints due to supersymmetry, 
the importance of solitonic objects and
electric-magnetic, strong-weak coupling duality,
led to many exact results on the vacuum structure of
various supersymmetric field theories.

Despite the fact that supersymmetry (a symmetry
relating bosons and fermions) is not present in
the standard model, there are at least three reasons
to study supersymmetric gauge theories:

\begin{itemize}
\item It is widely believed that an $N=1$
supersymmetric extension of the standard model
describes physics at energies not far above
those of current accelerators, and is directly
relevant to the hierarchy problem and unification
of couplings.

\item Supersymmetric gauge theories provide
examples of many phenomena believed to occur
in non-supersymmetric theories in a more tractable
setting. Therefore, they serve as useful toy models
for the study of these phenomena.

\item The study of supersymmetric field theories
has many mathematical applications.

\end{itemize}

Non-abelian gauge theories also appear in low energy
approximations to string theory, where supersymmetry
plays an important role. String theory is a
theory of quantum gravity which, moreover,
unifies gravity and gauge fields in a consistent quantum
theory. Traditionally, the theory has been formulated
in an expansion in a (string) coupling,
however, many of the outstanding problems in the subject
have to do with physics outside the weak coupling
domain. String theory
has also been undergoing rapid progress in the last few
years, which was driven by similar ideas to those
mentioned in the gauge theory context above.

Some of the highlights of the progress in
gauge and string theory that are relevant for
this review are:

\medskip
\noindent
{\bf 1. Strong-Weak Coupling Duality}

The physics of asymptotically free gauge theory
depends on the energy scale at which the theory is studied.
At high energies the theory becomes weakly coupled and is
well described in terms of the fundamental fields in the
Lagrangian (such as quarks and gluons). At low energies
the theory is often strongly coupled and 
can exhibit several different behaviors (or phases):
confining, Higgs, Coulomb, free electric and free
magnetic phases. 

In the confining phase, the energy of a pair of test charges
separated by a large distance $R$ grows linearly with $R$. Thus,
such charges cannot be infinitely separated. In the Higgs phase, 
the gauge bosons are massive and the energy of a pair of test charges
goes to a constant at large $R$. The Coulomb phase is characterized
by potentials that go like $1/R$, while the free electric and magnetic
phases have logarithmic corrections to this behavior. 
The standard model of elementary particle physics realizes the 
confining, Higgs and free electric phases; other models that go
beyond the standard model use the other phases as well.

The determination of the phase structure of non-abelian 
gauge theories is an important problem that is in general complicated
because it involves understanding the physics of strongly coupled
gauge theory. In the last few years, this problem has been solved
for many supersymmetric gauge theories. One of the main advances
that led to this progress was the realization that electric-magnetic,
strong-weak coupling duality is quite generic in field theory.

In a typical realization of such a duality, one studies
an asympotitcally free gauge theory that becomes more and more
strongly coupled as one goes to lower and lower energies.
The extreme low energy behavior is then found to be governed
by a different theory which may be {\em weakly} coupled, \eg\ 
because it is not asymptotically free. 

In other interesting situations, the original theory 
depends on continuous parameters (exactly marginal deformations),
and the duality relates the theory at different values of these
parameters. An example of this is the maximally supersymmetric
four dimensional gauge theory, $N=4$ SYM. This theory depends
on a complex parameter $\tau$, whose imaginary part is proportional
to the square of the inverse gauge coupling; the real part of
$\tau$ is a certain $\theta$ angle. The theory becomes weakly 
coupled when ${\rm Im}\tau\to\infty$. It has been proposed that 
it is invariant under a strong-weak coupling duality
$\tau\to-1/\tau$ in addition to the semiclassically manifest
symmetry $\tau\to\tau+1$. This symmetry is a generalization 
of the well known symmetry of electrodynamics which takes
$\vec E\to \vec B$ and $\vec B\to -\vec E$ and at the same
time exchanges electric and magnetic charges. In the last few 
years convincing evidence has been found for the validity of 
this duality symmetry of $N=4$ SYM.

Many interesting generalizations to theories with less 
supersymmetry have been found. For example, certain ``finite'' 
supersymmetric gauge theories (\eg\ $N=2$ SYM with gauge group 
$SU(N_c)$ and $N_f=2N_c$ ``flavors'' of fundamental
hypermultiplets) also appear to have such symmetries.
Furthermore, it has been discovered that different $N=1$ 
supersymmetric gauge theories may flow to the
same infrared fixed point and thus exhibit the
same long distance behavior. As we change the
parameters defining the different theories, one of the
descriptions  might become more weakly coupled
in the infrared while another might become more strongly
coupled. In some cases, this equivalence relates a strongly 
coupled interacting gauge theory to an infrared free one.
Interesting phenomena have also been shown to occur in other
dimensions; in particular, a large class of previously
unsuspected non-trivial fixed points in five and six
dimensional field theory has been found.

String theory has been known for a long time to be invariant
under a large discrete symmetry group known as T-duality.
This duality relates weakly coupled string theories and 
is valid order by order in the string coupling expansion.
It relates {\em different} spacetime backgrounds in which 
the string propagates. A simple example of T-duality is the
equivalence of string propagation on a circle of radii $R$
and $1/R$. A perturbative fundamental string state that
carries momentum $n/R$ around the circle is mapped by T-duality
to a perturbative fundamental string state corresponding
to a string winding $n$ times around the dual circle of radius
$1/R$.

In the last few years it has been convincingly
argued  that the perturbative T-duality group
is enhanced in the full string theory to
a larger symmetry group, known as U-duality, which relates
perturbative string states to solitons, and connects
different string vacua that were previously thought
of as distinct theories.
In certain strong coupling limits string theory becomes
eleven dimensional and is replaced by an inherently
quantum ``M-theory.'' At low energies M-theory reduces
to eleven dimensional supergravity; the full structure
of the quantum theory is not well understood as of this writing.

\medskip
\noindent
{\bf 2. Solitonic Objects}

Gauge theories in the Higgs phase often have
solitonic solutions that carry 
magnetic charge. Such monopoles and their 
dyonic generalizations (which carry both
electric and magnetic charge) play an
important role in establishing duality in gauge
theory. In supersymmetric gauge theories their 
importance is partly due to the fact that they
preserve some  supersymmetries and,
therefore, belong to special representations
of the supersymmetry algebra known as
``short'' multiplets, which contain fewer states
than standard ``long'' multiplets of the superalgebra.
Particles that preserve part of the supersymmetry are
conventionally referred to as being ``BPS saturated.''
Because of the symmetries, some of the properties of these
solitons can be shown to be independent of the coupling constants,
and thus certain properties can be computed exactly by
weak coupling methods.  Often, at strong coupling, they become
the light degrees of freedom in terms of which the long distance 
physics should be formulated.

In string theory analogous objects were found.
These are BPS saturated $p$-branes, $p$ dimensional
objects (with $p+1$ dimensional worldvolumes)
which play an important role in establishing
U-duality. In various strong coupling regions
different branes can become light and/or weakly
coupled, and serve as the degrees of freedom in
terms of which the dynamics should
be formulated. The study of branes preserving
part of the supersymmetry in string theory led to
fascinating connections, some of which will be reviewed
below, between string (or brane) theory and gauge
theory.

\medskip
\noindent
{\bf 3. Quantum Moduli Spaces Of Vacua}

SYM theories and string theories often have massless scalar
fields with vanishing classical potential and, therefore,
a manifold of inequivalent classical vacua $\MM_{cl}$, which
is parametrized by constant expectation values of these scalar
fields. In the non-supersymmetric case quantum effects
generically lift the moduli space $\MM_{cl}$,
leaving behind a finite number of quantum vacua.
In supersymmetric theories the quantum lifting of the
classical moduli space is severely constrained by 
certain non-renormalization theorems. The quantum
corrections to the scalar potential can often be described by 
a dynamically generated non-perturbative superpotential~\footnote{There 
are cases where the lifting of a classical moduli
space cannot be described by an effective
superpotential for the moduli~\cite{ADS}. We thank
N. Seiberg for reminding us of that.},
which is severely restricted by holomorphicity,
global symmetries and large field behavior.
One often finds an unlifted quantum moduli space $\MM_q$.
In many gauge theories the quantum superpotentials
were analyzed and the moduli spaces $\MM_q$ have been
determined. Partial success was also achieved in the
analogous problem in string theory.

\bigskip
Branes have proven useful
in relating string dynamics to low energy phenomena.
In certain limits brane configurations in string
theory are well described as solitonic solutions of
low energy supergravity, in particular black holes.
Interactions between branes are then mainly due to
``bulk'' gravity. In other limits gravity decouples
and brane dynamics is well described by the light modes
living on the worldvolume of the branes. Often, these
light modes describe gauge theories in various dimensions
with different kinds of matter. Studying the brane description
in different limits sheds new light on the quantum mechanics
of black holes,
as well as quantum gauge theory dynamics.
Most strikingly, both subjects are seen to be different
aspects of a single problem: the dynamics of branes in string
theory.

The fact that embedding gauge theories in string theory
can help analyze  strongly coupled low energy gauge
dynamics is a priori surprising. Standard Renormalization
Group (RG) arguments would suggest that at low energies one
can integrate out all fluctuations of the string except
the gauge theory degrees of freedom, which are governed
by SYM dynamics (gravity also decouples in
the low energy limit). This would seem
to imply that string theory cannot in principle
teach us anything about low energy gauge dynamics.

Recent work
suggests that while most of the degrees
of freedom of string theory are indeed irrelevant
for understanding low energy physics, there is a
sector of the theory that is significantly larger
than the gauge theory in question that should
be kept to understand the low energy structure.
This sector involves degrees of freedom
living on branes and describing their internal
fluctuations and embedding in spacetime.

We will see that the reasons for the ``failure''
of the naive intuition here are rather standard
in the general theory of the RG:
\begin{enumerate}
\item In situations where the long distance
theory exhibits symmetries, it is advantageous
to study RG trajectories along which the symmetries
are manifest (if such trajectories exist). The
string embedding of SYM often provides such a trajectory.
Other RG trajectories (\eg\ the standard QFT definition
of SYM in our case) which describe the same long
distance physics may be less useful for studying
the consequences of these symmetries, since they are
either absent throughout the RG flow, arising as
accidental symmetries in the extreme IR limit, or
are hidden in the variables that are being used.
\item Embedding apparently unrelated low energy
theories in a larger high energy theory can reveal
continuous deformations of one into the other that
proceed through regions in parameter space where
both low energy descriptions fail.
\item The embedding in string theory allows
one to study a much wider class of long
distance behaviors than is possible in 
asymptotically free gauge theory.
\end{enumerate}

In brane theory, gauge theory arises
as an effective low energy
description that is useful in some region
in the moduli space of vacua. Different
descriptions are useful in different regions
of moduli space, and in some regions the
extreme IR behavior cannot be given a field
theory interpretation. The underlying dynamics
is always the same -- brane worldvolume
dynamics in string theory.
Via the magic of string theory, brane dynamics provides
a uniform and powerful geometrical picture of
a diverse set of gauge theory
phenomena and points to hidden relations between
them.

The purpose of this review is to provide
an overview of some
aspects of the rich interplay between brane
dynamics and supersymmetric gauge theory in
different dimensions.
We tried to make the presentation relatively
self contained,
but the reader should definitely
consult reviews (some of which are listed below)
on string theory, D-branes, string duality, and the recent
progress in supersymmetric gauge theory,
for general background and
more detailed discussions of aspects that are
only mentioned in passing below.

\subsection{General References}

In the last few years there was a lot of
work on subjects relevant to this review.
Below we list a few of the recent 
original papers
and reviews that can serve as a guide to
the literature.

We use the following ``conventions''
in labeling the references: 
for papers with up to three authors
we list the authors' last names in the
text; if there are more than three
authors, we refer to the paper as
``First author {\em et al.}''
For papers which first appeared
as e-prints, the year listed is
that of the e-print; papers before
the e-print era are labeled by the
publication year. In situations where
the above two conventions do not
lift the degeneracy we assign
labels ``a,b,c,...''

For introductions to SUSY field theory
see for example~\cite{GGRS,WB}.
Electric-magnetic strong-weak coupling
duality in four dimensional gauge theory
dates back to the work of~\cite{MO}.
Reviews of the exact duality in $N=4$
SYM and additional references to the
literature can be found
in~\cite{Olive,Har,DiVec}.
\cite{Har} also includes a pedagogical
introduction to magnetic monopoles
and other BPS states.

The recent progress in $N=2$ SYM started
with the work of~\cite{SW9407,SW9408}.
Reviews
include~\cite{Bilal,DiVec,Ler,AGH}.
The recent
progress in $N=1$ SUSY gauge theory
was led by Seiberg; two of the
important original papers
are~\cite{Sei94a,Sei94b}.
Some reviews
of the work on $N=1$ supersymmetric
theories
are~\cite{AKMRV,Sei95,IS95,Giv,Peskin,Shifman}.

The standard reference on
string theory 
is~\cite{GSW}; for a recent
review see \cite{KIR}.
Dirichlet branes are described  
in~\cite{Pol95,PCJ,Pol96}. 
Solitonic
branes are discussed in~\cite{CHS}.
A comprehensive review on
solitons in string theory
is~\cite{DKL}.

T-duality is reviewed
in~\cite{GPR}. The non-perturbative
dualities and M-theory are discussed
in~\cite{HT,W9503,Sch95,Sch96,Tow6,Vafa97,Tow}
and many additional papers. A recent
summary for non-experts is~\cite{Sch97}.
Finally, reviews on applications 
of branes to black hole physics 
can be found, for example, in~\cite{Mal,You,Peet}.

\subsection{Plan}
\label{plan}

The plan of the review is as follows. In section
\ref{BST} we introduce the cast of characters --
the different $1/2$ BPS saturated branes in string
theory.

We start, in section \ref{LES}, by describing the
field content of ten and eleven dimensional supergravity 
and, in particular, the $p$-form gauge fields to which
different branes couple. In section \ref{BWC} we describe
different branes at weak string coupling, where they appear
as heavy non-perturbative solitons charged under
various $p$-form gauge fields. This
includes Dirichlet branes (D-branes) which are
charged under Ramond sector gauge fields and
solitonic branes charged under Neveu-Schwarz
sector gauge fields. We also describe orientifolds,
which are non-dynamical objects (at least
at weak string coupling) that are very useful
for applications to gauge theory.

In section \ref{MI} we discuss the interpretation
of the different branes in M-theory, the
eleven dimensional theory that is believed
to underlie all string vacua as well as eleven
dimensional supergravity. We show
how different branes in string theory  descend
from the membrane and fivebrane of M-theory, and
discuss the corresponding superalgebras.

In section \ref{DP} we describe the transformation
of the various branes under U-duality, the
non-perturbative discrete symmetry of compactified
string (or M-) theory.
In section \ref{WB} we initiate the discussion
of branes preserving less than $1/2$ of the SUSY,
with particular emphasis on their worldvolume
dynamics. We introduce configurations of branes
ending on branes that are central to the gauge theory
applications, and discuss some of their properties.

Section \ref{D4N4} focuses on configurations
of parallel Dirichlet threebranes which realize four
dimensional $N=4$ SYM 
on their worldvolume.
We describe the limit in which the worldvolume
gauge theory decouples from all the complications
of string physics and explain two known features
of $N=4$ SYM using branes. The Montonen-Olive
electric-magnetic duality symmetry is seen to
be a low energy manifestation of the $SL(2,Z)$
self-duality of ten dimensional type IIB string
theory; Nahm's description of multi-monopole
moduli space is shown to follow from the
realization of monopoles as D-strings stretched
between $D3$-branes preserving $1/2$ of the SUSY.
We also describe the form of the metric on monopole
moduli space, and some properties of the generalization
to symplectic and orthogonal groups obtained by studying
threebranes near an orientifold threeplane.

In section \ref{D4N2} we move on to brane
configurations describing four dimensional
$N=2$ SYM. In particular, in section \ref{BSB}
we explain, using a construction of branes suspended
between branes, the observation by
Seiberg and Witten that the metric on the Coulomb
branch of such theories is given by the period matrix
of an auxiliary Riemann surface $\Sigma$. In the brane
picture this Riemann surface becomes physical,
and is interpreted as part of the worldvolume
of a fivebrane. $N=2$ SYM is obtained in brane theory
by studying
the worldvolume theory of the fivebrane wrapped
around $R^{3,1}\times \Sigma$.
We also discuss the geometrical realization of the
Higgs branch and various deformations of the theory.

Section \ref{D4N1} is devoted to four dimensional
theories with $N=1$ SUSY. We describe the classical
and quantum phase structure of such theories
as a function
of the parameters in the Lagrangian, and
explain Seiberg's duality between different
theories using branes. In the brane construction, 
the quantum moduli spaces of members
of a dual pair provide different parametrizations
of a single space -- the moduli space
of the corresponding brane configuration.
Each description is natural in a different
region in parameter space.
Seiberg's duality in brane theory is thus
reminiscent of the well known correspondence
between two dimensional sigma models on
Calabi-Yau hypersurfaces in weighted projective
spaces and Landau-Ginzburg models with $N=(2,2)$
SUSY~\cite{KMS,MAR,GVW}, 
where the relation between the two
descriptions can be established by embedding both
in the larger framework of the (non-conformal) gauged
linear sigma model~\cite{W93}.

In section \ref{D3} we study three dimensional
theories. In section
\ref{D3N4} we establish using brane
theory two results in $N=4$ SYM. One is that
the moduli space of many such theories is
identical as a hyper-K\"ahler manifold to the
moduli space of monopoles in a {\em different}
gauge theory. The other is ``mirror symmetry,''
\ie\ the statement that many $N=4$ SUSY gauge
theories have mirror
partners such that the Higgs branch
of one theory is the Coulomb branch of its
mirror partner and vice-versa.
In section \ref{D3N2} we study $N=2$ SUSY theories.
We describe the quantum moduli space of $N=2$
SQCD using branes and show that the two dualities
mentioned above, Seiberg's duality and mirror
symmetry, can be extended to this case and teach
us new things both about branes and about gauge
theories. We also discuss the phase structure
of four dimensional $N=1$ SUSY gauge theory
compactified to three dimensions on a circle
of radius $R$.

In section \ref{D2} we 
consider two dimensional theories.
We study $2d$ $N=(4,4)$ supersymmetric 
theories and compactifications of $N=4$
supersymmetric models
from three to two dimensions 
on a circle.
We also discuss $N=(2,2)$ SUSY
theories in two dimensions.
In section \ref{D5} we study 
some aspects of five
and six dimensional theories, 
as well as compactifications
from five to four dimensions 
on a circle. 
Finally, in section \ref{DISC} 
we summarize the discussion
and mention some open problems.

\subsection{Omissions}
\label{ommiss}

In the following we briefly 
discuss issues that will
not be reviewed 
extensively~\footnote{This    
subsection may be skipped
on a first reading.}:

\begin{itemize}

\item {\em Gauge Theories 
in Calabi-Yau Compactifications:}
An alternative (but related) way 
to study low energy gauge theory
is to compactify string theory
to $D$ dimensions
on a manifold preserving the required
amount of SUSY, and take 
$M_{Planck}\to\infty$ to
decouple gravity and massive
string modes. This leads to
a low energy gauge theory,
some of whose properties can be 
related to the geometry of the
internal space.

In particular, 
compacifications of the type II string
on singular Calabi-Yau (CY) threefolds --
fibrations of ALE spaces over $CP^1$ -- 
are useful
in the study of $N=2$ SYM
theories~\cite{KKLMV,KLMVW}; for reviews
see~\cite{Ler,Kle}.
BPS states are related to type IIB 
threebranes wrapped
around 3-cycles which 
are fibrations of vanishing
2-cycles in the ALE space.
On the base the threebrane 
is projected to a self-dual string on
a Riemann surface $\Sigma$, 
which is the Seiberg-Witten
curve. The string tension is 
related to the Seiberg-Witten
differential $\lambda$.
The existence of stable 
BPS states is reduced to a geodesic
problem on $\Sigma$ with metric $|\lambda|^2$.

Similarly, to study 
$N=1$ SYM theories in four dimensions
one compactifies F-theory~\cite{Vaf} 
on Calabi-Yau fourfolds.
This ``geometric engineering'' was initiated
in~\cite{KKV,KV} and is reviewed in~\cite{Kle}.

\item {\em Probing the Geometry of Branes with Branes:}
We shall briefly describe a few (related) examples
where the geometry near branes can be probed by
lighter objects. In particular, we shall describe
the metric felt by a fundamental string
propagating in the background of solitonic
fivebranes, and by threebranes 
near parallel sevenbranes and 
orientifold sevenplanes. 
In the latter case, the geometrical
data is translated into properties of the four
dimensional $N=2$ supersymetric gauge theory
on the threebranes.

The interplay between the gauge dynamics on
branes and the geometry corresponding to the presence
of other branes was studied  
in~\cite{Dou96,Sen9605,BDS}
and was generalized in many directions.

For instance, fourbranes can be used to
probe the geometry of parallel eightbranes and
orientifold eightplanes, leading to 
an interesting connection 
between five dimensional gauge theory and
geometry~\cite{Sei9608,MS,DKV}.
Similarly, $p$-branes (with $p<3$) 
can be used to probe the
geometry near parallel 
$(p+4)$-branes and orientifold $(p+4)$-planes,
leading to relations between 
low dimensional 
($D<4$) gauge theories
and geometry~\cite{Sei9606,SW9607,DS,BSS},
some of which will be discussed in this review.
Other brane configurations that were used to
study the interplay between geometry and
gauge theory appear 
in~\cite{AKS,Sen9611,ASYT,DLS,Sen9702}.

\item {\em Branes in Calabi-Yau Backgrounds:}
As should be clear from the last two items,
there is a close connection between brane
configurations and non-trivial string backgrounds.
In general one may consider 
branes propagating in non-trivial backgrounds,
such as CY compactifications.
The branes may live at points
in the internal space
or wrap non-trivial cycles of the manifold.

Such systems are studied for example 
in~\cite{BSV,DM,DL,BJPSV,VZ,Int,OV97,HO,AO,BI,Ahn9705,AT,AOTa}
and references therein.
In some limits, they
are related
by duality transformations
to the webs of branes in flat space
that are extensively discussed 
below~\cite{OV95,Kut,EGK,OV97}. For example, a
useful duality, which we shall review below,
is the one relating the $A$-type singularity
on $K3$ to a configuration 
of parallel solitonic fivebranes.

\item {\em Quantum Mechanics of Systems of
$D0$-Branes, D-Instantons, Matrix Theory:}
The QM of $D0$-branes in type IIA string
theory (in general in the 
presence of other branes and orientifolds) led
to fascinating developments which are outside 
the scope of this review~\cite{DKPS,BSS,BPa,PR,BGS}.
Matrix theory was introduced in~\cite{BFSS};
reviews and additional references are
in~\cite{Ban,BS}.
D-instantons were studied, for example, 
in~\cite{GG97,GV}
and references therein.

\item {\em Non-Supersymmetric
Theories:} It is easy to 
construct brane configurations
in string theory
that do not preserve 
any supersymmetry. So far, not
much was learned about 
non-supersymmetric gauge theories
by studying such configurations
(for reasons that we shall explain).
Some recent discussions appear
in~\cite{BSTYa,W9706,Gom97,ES,BPb}.
Dynamical supersymmetry breaking in the brane picture
was considered recently in~\cite{BHOO98}.

\end{itemize}

\section{Branes In String Theory}
\label{BST}

In addition to fundamental
strings, in terms of which it is
usually formulated, string theory contains
other extended $p$ dimensional objects, known
as $p$-branes, that play an important
role in the dynamics. These objects can be
divided into two broad classes according to their
properties for weak fundamental string
coupling $\gs$:
``solitonic'' or Neveu-Schwarz
(NS) branes, whose tension (energy per unit $p$-volume)
behaves like $1/\gs^2$,
and Dirichlet or D-branes, whose tension is proportional
to $1/g_s$ (and which are hence much lighter than
NS-branes in the $\gs\to0$ limit).

In this section we describe some properties
of the various branes. In supergravity, these $p$-branes 
are charged under certain massless $(p+1)$-form gauge 
fields. We start with a description of the low energy effective
theory corresponding to type II strings in ten dimensions as well 
as eleven dimensional supergravity, the low energy limit of M-theory. 
We then describe branes preserving half of the
SUSY in weakly coupled string theory: D-branes,
orientifold planes, and solitonic and Kaluza-Klein
fivebranes. We present the interpretation of the
different branes from the point of view of the full
quantum eleven dimensional M-theory, and
their transformation properties under U-duality.
We finish the section with a discussion of
webs of branes preserving less SUSY.

Our notations are as follows:
the $1+9$ dimensional spacetime of string theory 
is labeled by $(x^0, x^1,\cdots, x^9)$. 
The tenth spatial dimension of M-theory
is $x^{10}$. The corresponding
Dirac matrices are $\Gamma^M$,
$M=0,1,2,\cdots,10$.
Type IIA string
theory has $(1,1)$
spacetime supersymmetry (SUSY);
the spacetime supercharges generated by left and right
moving worldsheet degrees of freedom $Q_L$, $Q_R$ have
opposite chirality:
\bea
\Gamma^0\cdots\Gamma^9 Q_L=& +Q_L \nonumber\\
\Gamma^0\cdots\Gamma^9 Q_R=& -Q_R
\label{BST1}
\eea
Type IIB string theory has $(2,0)$ spacetime SUSY,
with both left and right moving supercharges having the same
chirality:
\bea
\Gamma^0\cdots\Gamma^9 Q_L=& Q_L \nonumber\\
\Gamma^0\cdots\Gamma^9 Q_R=& Q_R
\label{BST2}
\eea
Thus, IIA string theory is non-chiral, while the
IIB theory is chiral. We will mainly focus on type
II string theories, but $1+9$ dimensional theories
with $(1,0)$ SUSY can be similarly discussed.
Type I string theory can be thought of as type II
string theory with orientifolds and D-branes and
is, therefore, a special case of the discussion below.
Heterotic strings do not have D-branes, but do have
NS-branes similar to those described below.

\subsection{Low Energy Supergravity}
\label{LES}

The spectrum of string theory contains a finite
number of light particles and an infinite tower
of massive excitations with string scale or higher
masses. To make contact with low energy 
phenomenology it is convenient to focus on the dynamics of
the light modes. This can be achieved by integrating
out the infinite tower of massive fluctuations of the string
and defining a low energy effective action for the light
fields. If one thinks (formally) of string theory as a theory
describing an infinite number of fields $\phi$, some of which
are light $\phi_l$, and the rest are heavy $\phi_h$, governed by
the classical action $S(\phi_l,\phi_h)$, the low energy
effective action $S_{\rm eff}(\phi_l)$ can in principle be
obtained by integrating out the heavy fields:
\beq
e^{iS_{\rm eff}(\phi_l)}=\int D\phi_h e^{iS(\phi_l,\phi_h)}
\label{effact}
\eeq
In principle (\ref{effact}) is exact, but in practice
it is far from clear how to find the action $S$ and
how to integrate out the massive modes of the string. 
At the same time, the effective action is mainly of interest
at energies much lower than the masses of the fields $\phi_h$, 
where it makes sense to integrate them out. To find $S_{\rm eff}$
at low energies one can study the S-matrix of the string 
in the low energy approximation and
construct a classical action that reproduces it. The leading
terms in such an action are typically determined by the symmetries,
such as gauge and diffeomorphism invariance, and supersymmetry.

Following the above discussion for type II string theory
leads to the two $9+1$ dimensional type II supergravity
theories, type IIA and type IIB. Ten dimensional type IIA 
supergravity can be obtained by dimensional reduction of the  
unique eleven dimensional supergravity theory, which is of 
interest in its own right as the low energy limit of M-theory;
thus we start with this case. 

Eleven dimensional supergravity includes the bosonic 
(\ie\ commuting) fields $G_{MN}$, the eleven dimensional metric,
and $A_{MNP}$, a three index antisymmetric gauge field 
($M,N,P=0,1,\cdots, 10$). The only fermionic field is the
gravitino, $\psi^M_\alpha$ ($\alpha=1,\cdots, 32$).
The Lagrangian describing these fields
can be found in~\cite{GSW}. One can check that there are $128$
on-shell bosonic and fermionic degrees of freedom.   

The presence of the three index gauge field $A_{MNP}$
implies that eleven dimensional supergravity couples naturally
to membranes and to fivebranes. For a membrane with
worldvolume $X^M(\sigma_a)$, ($a=1,2,3$), the coupling is
(see~\cite{BSTTT} for a discussion of the full supermembrane
worldvolume action)
\beq
\int d^3\sigma\epsilon^{abc}A_{MNP}(X)
\partial_a X^M\partial_b X^N\partial_c X^P
\label{coupmem}
\eeq
Equation (\ref{coupmem}) implies that the membrane of eleven dimensional
supergravity is charged under the three-form gauge field
$A_{MNP}$. The coupling of eleven dimensional supergravity to fivebranes 
is similar, with $A_{MNP}$ replaced by its dual $\tilde A_{MNPQRS}$
defined by $*dA=d\tilde A$.

Type IIA supergravity is obtained by dimensional reduction of
eleven dimensional supergravity on a circle. Denoting the $1+9$
dimensional indices by $\mu,\nu,\lambda=0,1,\cdots, 9$, the fields of eleven 
dimensional supergravity reduce as follows in this limit. The metric
$G_{MN}$ gives rise to the metric $G_{\mu\nu}$, a gauge field
$A_\mu=G_{\mu, 10}$ and a scalar $\Phi=G_{10, 10}$. 
The antisymmetric tensor $A_{MNP}$ similarly gives rise
to the antisymmetric tensors $A_{\mu\nu\lambda}$ and
$B_{\mu\nu}=A_{\mu\nu, 10}$. In the standard Neveu-Schwarz-Ramond
quantization of superstrings~\cite{GSW}, the fields $G_{\mu\nu}$,
$B_{\mu\nu}$ and $\Phi$ originate in the same sector of the string
Hilbert space, the Neveu-Schwarz (or NS) sector, while the
gauge fields $A_\mu$ and $A_{\mu\nu\lambda}$ are Ramond-Ramond (RR)
sector fields. The scalar field $\Phi$ is the dilaton; its expectation
value determines the coupling constant of the string theory. 
Since the potential for $\Phi$ in type II string theory vanishes,
the theory can be made arbitrarily weakly coupled. 

Just like in eq. (\ref{coupmem}), the existence of the gauge fields
implies that type II string theory naturally couples to various $p$-branes.
The existence of $B_{\mu\nu}$ means that the theory naturally
couples to strings (electrically, as in (\ref{coupmem})) and fivebranes
(magnetically, via the six-form gauge field dual to $B_{\mu\nu}$). 
Since the gauge field to which these branes couple is an NS sector
field we refer to these branes as NS branes. The string charged
under $B_{\mu\nu}$ is simply the fundamental string that is used
to define type II string theory, while the fivebrane is the $NS5$-brane
studied by~\cite{CHS}. 

The gauge fields $A_\mu$ and $A_{\mu\nu\lambda}$ couple electrically
to zero-branes (particles) and membranes and magnetically to
sixbranes and fourbranes, respectively. Since the corresponding
gauge fields originate in the RR sector, these branes are sometimes
referred to as Ramond branes (or D-branes, see below). 

Type IIB supergravity has $(2,0)$ chiral supersymmetry.
The massless spectrum contains again the NS sector fields
$G_{\mu\nu}$, $B_{\mu\nu}$ and $\Phi$ and the
associated NS string and fivebrane.
The spectrum of RR $p$-form gauge fields is
different from the IIA case. There is an 
additional scalar $\chi$, which
combines with $\Phi$ into a complex coupling of type IIB
string theory. The antisymmetric tensors one finds have two
and four indices, $\tilde B_{\mu\nu}$, $A_{\mu\nu\lambda\rho}$.
The existence of the former implies that the theory can couple
to another set of strings and fivebranes, the D-string and $D5$-brane.
The four-form  $A$ is self dual $*dA=dA$; it couples to a three-brane.

In what follows we will discuss some properties of the various
branes mentioned above. We start with a description
of their construction and properties in weakly coupled 
string theory.

\subsection{Branes In Weakly Coupled String Theory}
\label{BWC}

\subsubsection{D-Branes}
\label{DB}

\begin{figure}
\centerline{\epsfxsize=140mm\epsfbox{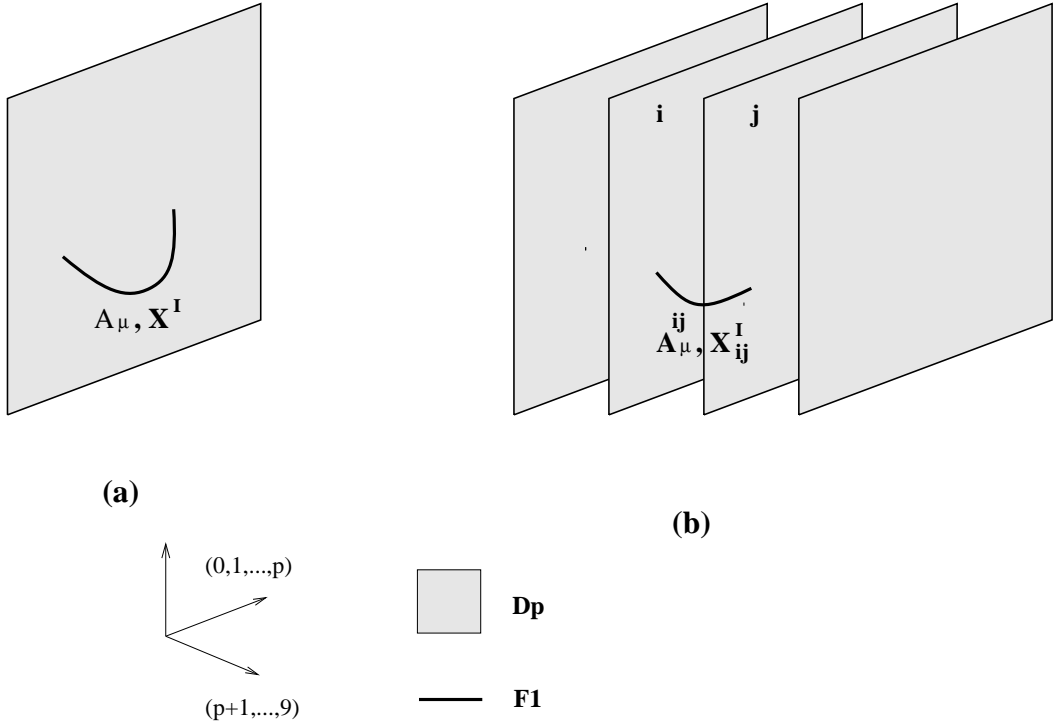}}
\vspace*{1cm}
\caption{Low lying states of fundamental strings
with both ends on D-branes describe a $U(1)$
gauge field and $9-p$ scalars living on a single
D-brane (a), or a $U(\nc)$ gauge field and $9-p$
adjoint scalars on a stack of $\nc$ D-branes (b).}
\label{one}
\end{figure}

In weakly coupled type II string theory, D-branes are
defined by the property that fundamental strings
can end on them~\cite{PCJ,Pol96}.
A Dirichlet $p$-brane (\ddp-brane)
stretched in the $(x^1,\cdots, x^p)$
hyperplane, located at a point in $(x^{p+1},\cdots,
x^9)$, is defined by including in the theory open strings
with Neumann boundary conditions for $(x^0,x^1,\cdots, x^p)$
and Dirichlet boundary conditions for $(x^{p+1},\cdots,
x^9)$ (see Fig.~\ref{one}).

\smallskip

The \ddp-brane is charged under a Ramond-Ramond (RR)
$(p+1)$-form potential of the type II string. As we saw,
in type IIA string theory there are such potentials
with even $p$ and, therefore, \ddp-branes with $p=0,2,4,6,8$.
Similarly, in type IIB string theory there are potentials
with odd $p$ and \ddp-branes
with $p=-1,1,3,5,7,9$. The $p=-1$ brane is the D-instanton,
while the $p=9$ brane is the $D9$-brane that fills the whole
$9+1$ dimensional spacetime and together with
the orientifold (to be described below) turns a type
IIB string into a type I one. The $D7$-brane is the 
''magnetic dual'' of the D-instanton; the $D8$-brane  
together with the orientifold turns a type IIA
string into a type I' one.

The tension of a \ddp-brane is
\beq
T_p={1\over \gs l_s^{p+1}}
\label{BST3}
\eeq
where $l_s$ is the fundamental string scale (the tension
of the fundamental string is $T=1/l_s^2$).
The \ddp-brane tension (\ref{BST3}) is equal to
its RR charge; D-branes are BPS saturated objects
preserving half of the thirty two
supercharges of type II string
theory. More precisely,
a \ddp-brane stretched along the $(x^1,\cdots, x^p)$
hyperplane preserves supercharges of the form $\epsilon_LQ_L
+\epsilon_RQ_R$ with
\beq
\epsilon_L=\Gamma^0\Gamma^1\cdots\Gamma^p\epsilon_R
\label{BST4}
\eeq
An anti \ddp-brane carries the opposite RR charge and
preserves the other half of the supercharges. Equation
(\ref{BST4}) can be thought of as arising from
the presence in the theory of open strings that
end on the branes. In the presence of such open
strings the left and right moving supercharges
$Q_L$, $Q_R$
(\ref{BST1},\ref{BST2}) are not independent;
eq. (\ref{BST4}) describes the reflection of
right to left movers at the boundary of the
worldsheet, which is confined to the brane.

The low energy worldvolume
theory on an infinite \ddp-brane
is a $p+1$ dimensional field theory invariant
under sixteen supercharges.
It describes the dynamics
of the ground states of
open strings both of whose
endpoints lie on the brane 
(Fig.~\ref{one}(a)).
The massless spectrum
includes a $p+1$ dimensional
$U(1)$ gauge field $A_\mu(x^\nu)$,
$9-p$ scalars $X^I(x^\mu)$
($I=p+1,\cdots, 9$, $\mu=0,\cdots, p$)
parametrizing fluctuations
of the \ddp-brane in the
transverse directions, and fermions
required for SUSY~\footnote{We will 
usually ignore the fermions
below. Their properties can be deduced
by imposing SUSY.}.
The low energy
dynamics can be obtained
by dimensional
reduction of $N=1$ SYM 
with gauge group $G=U(1)$ from $9+1$
to $p+1$ dimensions.
The bosonic part of 
the low energy worldvolume action is
\beq
S={1\over g_{SYM}^2}\int d^{p+1}x \left(
{1\over4}F_{\mu\nu}F^{\mu\nu}
+{1\over l_s^4}\partial_\mu X^I\partial^\mu X_I\right)
\label{BST5}
\eeq
The $U(1)$ gauge coupling on the brane $g_{SYM}$
is given by
\beq
g_{SYM}^2=\gs l_s^{p-3}
\label{BST6}
\eeq
The $\gs$ dependence in 
(\ref{BST3},\ref{BST6}) follows from the fact
that the kinetic term (\ref{BST5}) arises from
open string tree level (the disk), while the power of the
string length $l_s$ is fixed by dimensional analysis.

At high energies, the massless
degrees of freedom (\ref{BST5}) interact
with an infinite tower of ``open string'' states localized
on the brane, and with closed strings in the $9+1$ dimensional
bulk of spacetime.
To study SYM on the brane one needs to decouple the gauge
theory degrees of freedom from gravity and massive string modes.
To achieve that one can send $l_s\to0$ holding
$g_{SYM}$ fixed. This means (\ref{BST6})
$\gs\to 0$ for $p<3$, $\gs\to\infty$ for $p>3$. For
$p=3$ $g_{SYM}$ is independent of $l_s$ and the $l_s\to0$
limit describes $N=4$ SYM in $3+1$ dimensions. Note
that for $p\le3$ the above limit leads to a consistent
theory on the brane, whose UV behavior is just that
of $p+1$ dimensional SYM. For $p>3$ SYM provides
a good description in the infrared, but it must break
down at high energies.

Since D-branes are BPS 
saturated objects, parallel branes
do not exert forces on each other. The
low energy worldvolume dynamics on a
stack of $N_c$ nearby parallel \ddp-branes 
(Fig.~\ref{one}(b)) is a
SYM theory with gauge group $U(N_c)$ and
sixteen supercharges, arising from ground
states of open strings whose endpoints
lie on the branes~\cite{Pol94,W9510}.
The scalars $X^I$ (\ref{BST5})
turn into $N_c\times N_c$ matrices
transforming in the adjoint representation
of the $U(N_c)$ gauge group.
The $N_c$ photons in the Cartan subalgebra
of $U(N_c)$ and the diagonal
components of the matrices $X^I$ correspond
to open strings both of whose endpoints lie
on the same brane. The charged gauge bosons
and off-diagonal components of
$X^I$ correspond to strings whose endpoints
lie on different branes. Specifically, the
$(i,j)$, $(j,i)$ elements of $X^I$, $A_\mu$ arise
from the two orientations of
a fundamental string connecting
the $i$'th and $j$'th branes 
(see Fig.~\ref{one}(b)).

The generalization of (\ref{BST5}) to the case
of $\nc$ parallel \ddp-branes is described by
dimensional reduction of $N=1$ SYM with
gauge group $G=U(\nc)$ from $9+1$ to
$p+1$ dimensions. The bosonic part of the
$9+1$ dimensional low energy Lagrangian,
\bea
&\LL={1\over 4g_{SYM}^2} {\rm Tr}
F_{mn}F^{mn}; \;\;m,n=0,1,\cdots,9\nonumber\\
&F_{mn}=\partial_m A_n-
\partial_n A_m-i[A_m,A_n]
\label{BST55}
\eea
gives rise upon dimensional reduction to
kinetic terms for the $p+1$ dimensional
gauge field $A_\mu$ and adjoint scalars
$X^I$,
\beq
\LL_{\rm kin}={1\over g_{SYM}^2}{\rm Tr}\left(
{1\over4}F_{\mu\nu}F^{\mu\nu}
+{1\over l_s^4}\DD_\mu X^I\DD^\mu X_I\right)
\label{BST56}
\eeq
($\DD_\mu X^I=\partial_\mu X^I-i[A_\mu, X^I]$;
$F_{\mu\nu}=\partial_{[\mu}A_{\nu]}-i[A_\mu,A_\nu]$),
and to a potential for the adjoint scalars $X^I$,
\beq
V\sim {1\over l_s^8g_{SYM}^2}\sum_{I,J}{\rm Tr}
\;[X^I, X^J]^2
\label{BST7}
\eeq
Flat directions of the potential
(\ref{BST7}) corresponding to diagonal $X^I$
(up to  gauge transformations) parametrize
the Coulomb branch of the $U(N_c)$ gauge theory.
The moduli space of vacua is $(R^{9-p})^{\nc}/S_{\nc}$;
it is parametrized by the eigenvalues of $\vec X$,
\beq
\vec x_i=\langle\vec X_{ii}\rangle;\;\;\;i=1\cdots, \nc
\label{BST8}
\eeq
which label the transverse locations
of the $\nc$ branes.
The permutation group $S_{\nc}$ acts on
$\vec x_i$ as the Weyl group
of $SU(\nc)$.
For generic positions of the $N_c$ branes,
the off-diagonal components of $X^I$ as well as the
charged gauge bosons are massive (and the gauge symmetry
is broken, $U(N_c)\to U(1)^{N_c}$).
Their masses are read off (\ref{BST56}-\ref{BST8}):
\beq
m_{ij}={1\over l_s^2}|\vec x_i-\vec x_j|
\label{BST9}
\eeq
Geometrically (\ref{BST9}) can be thought of as
the minimal energy of a fundamental string
stretched between the $i$'th and $j$'th 
branes (Fig.~\ref{one}(b)).
When $n$ of the $N_c$ branes coincide, some of the
charged particles become massless (\ref{BST9})
and the gauge group is enhanced from $U(1)^{N_c}$
to $U(n)\times U(1)^{N_c-n}$.

\subsubsection{Orientifolds}
\label{OR}

An orientifold $p$-plane (\op-plane) is a generalization
of a $Z_2$ orbifold fixed plane to non-oriented
string theories~\cite{PCJ,Pol96}.
It can be thought of as the fixed plane
under a $Z_2$ symmetry which acts on the spacetime
coordinates and reverses the orientation of the string.
The fixed plane of the $Z_2$
transformation~\footnote{$z$, 
$\bar z$ parametrize the string
worldsheet; $z=\exp(\tau+i\sigma)$.}
\beq
x^I(z,\bar z)\leftrightarrow -x^I(\bar z, z);\;\;I=p+1,\cdots, 9
\label{BST10}
\eeq
is an \op-plane extending in the $(x^1,\cdots, x^p)$
directions and time.

\begin{figure}
\centerline{\epsfxsize=120mm\epsfbox{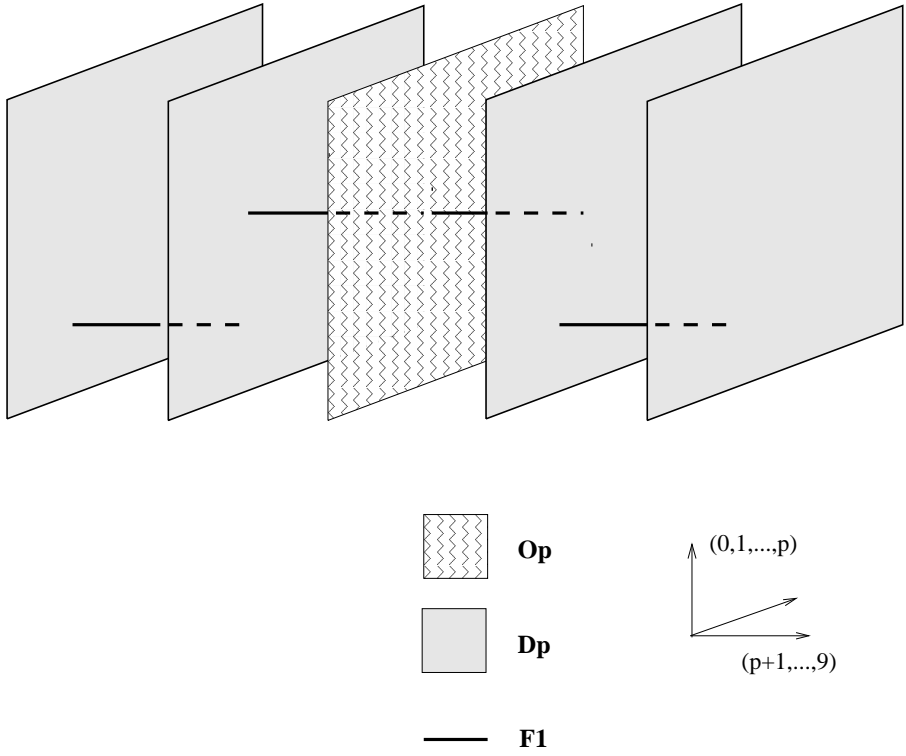}}
\vspace*{1cm}
\caption{An orientifold $p$-plane with two
adjacent parallel $Dp$-branes and their
mirror images. Fundamental strings stretched
between a D-brane and its image are projected
out for negative orientifold charge. Others
come in mirror pairs.}
\label{two}
\end{figure}

Like usual orbifold fixed planes,
the orientifold is not dynamical
(at least at weak string coupling).
It carries charge under
the same RR $(p+1)$-form
gauge potential, and breaks the
same half of the SUSY, as a parallel \ddp-brane.
In the presence of an \op-plane, the transverse
space $R^{9-p}$  is replaced by $R^{9-p}/Z_2$.
It is convenient to continue to describe the geometry as
$R^{9-p}$, add a $Z_2$ image for each
object lying outside the fixed plane and implement an
appropriate (anti-) symmetrization on the states.
Thus D-branes which are outside the orientifold
plane acquire mirror images (see Fig.~\ref{two}).
At the fixed plane one can sometimes
have a single D-brane which does not
have a $Z_2$ partner and hence cannot leave the singularity.
The RR charge of an \op-plane $Q_{Op}$ is equal (up to a sign)
to that of $2^{p-4}$ \ddp-branes (or
$2^{p-5}$ pairs of a \ddp-brane and its mirror).
Denoting the RR charge of a \ddp-plane by $Q_{Dp}$,
the orientifold charge is:
\beq
Q_{Op}=\pm 2\cdot 2^{p-5}Q_{Dp}
\label{BST11}
\eeq
(this will be further discussed later).
The (anti-) symmetric projection imposed on D-branes
by the presence of an orientifold plane leads to changes
in their low energy dynamics. On a stack of $N_c$
\ddp-branes parallel to an \op-plane one finds a
gauge theory with sixteen supercharges and the following
rank $[N_{c}/2]$ gauge group~\footnote{Our conventions
are $Sp(1)\simeq SU(2)$.} $G$:
\begin{itemize}
\item $Q_{Op}=+2\cdot 2^{p-5}Q_{Dp}$, $N_{c}$ even: $G=Sp(N_c/2)$.
\item $Q_{Op}=-2\cdot 2^{p-5}Q_{Dp}$: $G=SO(N_c)$.
\end{itemize}

\smallskip

The light matter consists of the ground states
of open strings stretched between different D-branes,
giving rise to a gauge
field for the group $G$ and $9-p$ scalars $X^I$ in the adjoint
of $G$. Positive orientifold charge gives rise to a symmetric
projection on the $N_c\times N_c$ matrices
$A_\mu$, $X^I$ and, therefore, a symplectic gauge group ($\nc$
must be even in that case; as is clear from (\ref{BST10}),
for the case of a symmetric projection it is
impossible to have a D-brane without an image
stuck at the orientifold),
while negative orientifold charge
leads to an antisymmetric projection
and to orthogonal gauge groups.

Geometrically,
$(N_c^2\pm N_c)/2$ of the $N_c^2$ oriented
strings stretched between the $N_c$ \ddp-branes
survive the (anti-) symmetric projection
due to the orientifold. The difference of $N_c$
between the symmetric and antisymmetric cases corresponds
to strings stretching between a \ddp-brane and its
mirror image. These strings transform to themselves
under the combined worldsheet and spacetime reflection
(\ref{BST10}); thus they are projected out in the
antisymmetric case, and give $2\times N_c/2$
massless modes in the symmetric case.

Since branes can only leave the orientifold
plane in pairs, there are $[\nc/2]$
``dynamical branes'' which are free to move.
Their locations in the transverse space $R^{9-p}$
parametrize the Coulomb branch of the theory.
The $[\nc/2]$ photons in the Cartan subalgebra
of $G$ and the scalars parametrizing the Coulomb
branch correspond to open strings both of whose
endpoints lie on the same brane.
When $n$ of the $[N_c/2]$ \ddp-branes coincide
outside the orientifold plane the gauge symmetry
is enhanced from $U(1)^{[N_c/2]}$ to $U(n)\times
U(1)^{[N_c/2]-n}$.
When $m$ of the $N_c$ branes coincide with the
orientifold plane the gauge group is enhanced
to $\left( SO(m)\; {\rm or}
\; Sp(m/2)\right)\times U(1)^{[(N_c-m)/2]}$.

For high dimensional orientifolds and D-branes
the discussion
above has to be slightly modified. In particular,
for $p\ge 7$ the rank of the gauge group $G$ 
is bounded since RR flux does not have
enough non-compact transverse directions to escape,
and therefore the total RR charge must vanish.
The case $p=9$ is further special,
since there are no transverse directions at all and
the reflection (\ref{BST10}) acts only on the worldsheet.
The requirement that the total RR charge vanish and
the orientifold charge $Q_{O9}=-32$ (see (\ref{BST11}))
are in this case directly related to the fact that the
gauge group of ten dimensional type I string theory
is $SO(32)$. The $p$ dependence in (\ref{BST11}) will
be discussed in section \ref{DP}.

\subsubsection{The Solitonic Fivebrane}
\label{SF}

The solitonic fivebrane~\cite{CHS}
which exists in weakly coupled type II
and heterotic string theory, is
a BPS saturated object which, like
the Dirichlet brane, preserves half
of the supersymmetry of the theory
and has tension
\beq
T_{NS}={1\over \gs^2 l_s^6}
\label{BST12}
\eeq
It couples magnetically to the
NS-NS sector $B_{\mu\nu}$ field
and can thus be thought of as
a magnetic dual of the fundamental
type II or heterotic string.
Since its tension is proportional
to $1/\gs^2$ it provides a non-trivial
background for a fundamental string
in leading order in $\gs$ (\ie\ on 
the sphere).
A fundamental string propagating in
the background
of $k$ parallel NS fivebranes
located at transverse positions
$\vec x_i$ is described by a
conformal field theory (CFT)
with non-trivial $G_{IJ}$, $B_{IJ}$,
$\Phi$ (metric, antisymmetric tensor and dilaton)
given by:
\bea
&e^{2(\Phi-\Phi_0)}=1+\sum_{j=1}^k{l_s^2\over
|\vec x-\vec x_j|^2}\nonumber\\
&G_{IJ}=e^{2(\Phi-\Phi_0)}\delta_{IJ};\;G_{\mu\nu}=
\eta_{\mu\nu}\nonumber\\
&H_{IJK}=-\epsilon_{IJKM}\partial^M\Phi
\label{BST13}
\eea
$I,J,K,M$ label the four directions transverse to
the fivebrane;
$\mu,\nu$ label the $5+1$ longitudinal directions.
$H$ is the field strength of $B$; $\Phi_0$ is the
value of the dilaton at infinity, related to the
string coupling at infinity $\gs=\exp\Phi_0$.
As is clear from (\ref{BST13}),
the effective string coupling $\exp\Phi$
depends on the distance from the fivebrane, diverging
at the core.

An NS fivebrane stretched in the
$(x^1,\cdots, x^5)$ hyperplane
preserves supercharges of the form
$\epsilon_LQ_L+\epsilon_RQ_R$, where for the type IIA
fivebrane

\bea
\epsilon_L=&\Gamma^0\Gamma^1\Gamma^2\Gamma^3
\Gamma^4\Gamma^5\epsilon_L\nonumber\\
\epsilon_R=&\Gamma^0\Gamma^1\Gamma^2\Gamma^3
\Gamma^4\Gamma^5\epsilon_R
\label{BST14}
\eea
while for the type IIB fivebrane
\bea
\epsilon_L=&\Gamma^0\Gamma^1\Gamma^2\Gamma^3
\Gamma^4\Gamma^5\epsilon_L\nonumber\\
\epsilon_R=&-\Gamma^0\Gamma^1\Gamma^2\Gamma^3
\Gamma^4\Gamma^5\epsilon_R
\label{BST15}
\eea
Thus, the non-chiral type IIA string theory gives rise
to a chiral fivebrane worldvolume theory with
(2,0) SUSY in six dimensions, while the chiral
type IIB theory gives rise to a non-chiral fivebrane
with (1,1) worldvolume SUSY. Equations
(\ref{BST14},\ref{BST15}) can be established
by a direct analysis of the supercharges
preserved by the background (\ref{BST13}).
As we will see later, string duality relates
them to the supercharges preserved by
D-branes (\ref{BST4}), and both have
a natural origin in eleven dimensions.

The light fields on the worldvolume of
a single type IIA NS fivebrane correspond 
to a tensor multiplet of six dimensional
$(2,0)$ SUSY,
consisting of a self-dual
$B_{\mu\nu}$ field and five scalars (and the
fermions needed for SUSY). On a single type
IIB fivebrane
one finds a vectormultiplet, \ie, a
six dimensional gauge field and four
scalars ($+$ fermions).

The four scalars in the vectormultiplet on the
type IIB fivebrane,
as well as four of the five scalars in the
tensor multiplet on the type IIA fivebrane,
describe fluctuations of the NS-brane in
the transverse directions. The fifth scalar
on a type IIA fivebrane lives on a circle
of radius $l_s$ and provides a hint
of a hidden eleventh dimension of quantum
type IIA string theory (more on this below).

The gauge coupling of the vector field
on the type IIB fivebrane is~\footnote{Since
the NS fivebrane is described by a CFT
on the sphere, one might have expected
the gauge coupling to go like 
$g_{SYM}^2\simeq \gs^2 l_s^2$ 
(in analogy to (\ref{BST6})). The 
form (\ref{BST16}) is obtained by 
taking into account the fact that the
worldvolume gauge field is a RR
field in the fivebrane CFT.}

\beq
g_{SYM}^2=l_s^2
\label{BST16}
\eeq

\noindent
Since NS-branes are BPS saturated objects,
parallel  branes do not exert forces on
each other. The low energy worldvolume
dynamics on a stack of $k$ parallel type IIB
$NS5$-branes is a $5+1$ dimensional (1,1) $U(k)$ SYM
theory (with sixteen supercharges), arising from
the ground states of D-strings stretched between
different NS-branes. It is described by
(\ref{BST56}, \ref{BST7}, \ref{BST16}) with $p=5$.
As for D-branes, the four scalars in the vector
multiplet are promoted to $k\times k$ matrices,
whose diagonal components parametrize the Coulomb
branch of the theory, $R^{4k}/S_k$.

The low energy theory describing a stack of
$k$ parallel type IIA $NS5$-branes is more
exotic. It can be thought of
as a non-abelian generalization
of the free theory of a tensor multiplet
on a single $NS5$-brane, and gives rise
to a non-trivial field theory
with $(2,0)$ SUSY in $5+1$ 
dimensions~\cite{W9507,Str,Sei9705117}.
It contains string-like low energy excitations
corresponding to Dirichlet membranes stretched
between the different $NS5$-branes. These strings
are charged under the self-dual $B_{\mu\nu}$
fields on the corresponding fivebranes and are
light when the fivebranes are close to each
other. The Coulomb
branch of the (2,0) theory, $(R^4\times S^1)^k
/S_k$, is parametrized by the expectation values
of the diagonal components of the five scalars
in the tensor multiplet. At the origin of the
Coulomb branch, the $(2,0)$ field theory corresponds
to a non-trivial superconformal field theory.

In the limit $\gs\to0$ the dynamics of the full type II
string theory simplifies and, in particular, all the
modes in the bulk of spacetime (including gravity)
decouple.
The dynamics of a type II string 
vacuum with $k$ NS fivebranes
remains non-trivial in the limit; 
in the type IIA  case
it is described at low energies by the
(2,0) field theory described above.
The theory of $k$ type IIB fivebranes has
(1,1) SUSY and reduces at low energies to the
(infrared free) $U(k)$ SYM theory; at finite
energies it is interacting. Providing
a useful description of the fivebrane theory
and, in particular, of the low energy
$(2,0)$ field theory of the IIA fivebranes
remains a major challenge as of this writing.

\subsubsection{The Kaluza-Klein Monopole}
\label{KKM}

Compactified type II string theory has additional
solitonic objects. One that will be particularly
useful later is the Kaluza-Klein (KK) monopole,
which is a fivebrane in ten 
dimensions~\cite{Tow,DKL}.
It is obtained when one of the ten directions,
call it $\rho$, is compactified on a circle of radius $R$.
The ten dimensional graviton gives rise in nine
dimensions to a gauge field $A_a=G_{a,\rho}$ ($a=0,
\cdots, 8$). The KK monopole carries magnetic charge
$R/l_s$ under this gauge field.
Like the monopole of $3+1$ dimensional gauge theory
it is localized
in three additional directions $\vec r$
and is extended in the remaining five.

The tension of the KK fivebrane is
\beq
T_{KK}={R^2\over \gs^2 l_s^8}
\label{KKK4}
\eeq
The factor of $1/\gs^2$ is due to the fact that,
like the NS fivebrane, the KK fivebrane ``gets its
tension'' from the sphere (\ie\ it is a ``conventional
soliton''). The other factors in
(\ref{KKK4}) are the square of the magnetic charge
and a $1/l_s^6$ due to the fact that this
is a fivebrane.

A fundamental string in the background of
$k$ parallel KK monopoles located at transverse positions
${\vec r}_i$ is described
by a CFT with the multi Taub-NUT metric ($B$=$\Phi$=const):
\bea
&ds^2=dx^{\mu}dx_{\mu}+ds^2_{\bot} \nonumber\\
&ds^2_{\bot}=Ud{\vec r}^2+U^{-1}(d\rho+{\vec \omega}\cdot
d{\vec r})^2
\label{KK1}
\eea
where $x^\mu$ label the $1+5$ longitudinal directions,
\beq
U=1+\sum_{j=1}^k {R\over 2|{\vec r}-{\vec r}_j|}
\label{KK2}
\eeq
and $\vec w$ is the multi Dirac monopole
vector potential
which satisfies
\beq
{\vec \nabla}\times {\vec \omega}={\vec \nabla}U
\label{KK3}
\eeq

In the limit $R\to\infty$ this background becomes an
ALE space with $A_{k-1}$ singularity.
On the other hand, in the $R\to 0$
limit the multi Taub-NUT background (\ref{KK1}-\ref{KK3})
is T-dual (in the $\rho$ direction and in an appropriate
sense~\cite{GHM})
to the multi NS fivebrane solution (\ref{BST13}) (more
on T-duality later).

\subsection{M-Theory Interpretation}
\label{MI}
All the different
ten dimensional string theories can be thought
of as asymptotic expansions around different
vacua of a single quantum theory.
This theory, known as ``M-theory,''
is in fact $1+10$ dimensional at
almost all points in its moduli space of vacua
(for a review see, for example~\cite{Sch96,Tow}
and references therein).

In the flat $1+10$ dimensional Minkowski vacuum the
theory reduces at low energies to eleven dimensional
supergravity. There is no adjustable dimensionless
coupling; the only parameter in the theory
is the eleven dimensional Planck scale $l_p$. Physics
is weakly coupled and well approximated by semiclassical
supergravity for length scales much larger than $l_p$.
It is strongly coupled at scales smaller than
$l_p$. The spectrum includes a three-form potential
$A_{MNP}$ ($M,N,P=0,1,\cdots, 10$)
whose electric and magnetic charges
appear as central extensions in the eleven dimensional
superalgebra,
\beq
\{Q_\alpha, Q_\beta\}=(\Gamma^MC)_{\alpha\beta}P_M
+{1\over2}(\Gamma_{MN}C)_{\alpha\beta}Z^{MN}+
{1\over 5!}(\Gamma_{MNPQR}C)_{\alpha\beta}
Y^{MNPQR}
\label{edsusy}
\eeq
where $\Gamma_{MN\cdots}$ are antisymmetrized
products of the $32\times 32$ Dirac matrices
in eleven dimensions, $C$ is the (real, antisymmetric)
charge conjugation matrix, $Z^{MN}$ is the electric
charge corresponding to $A_{MNP}$, and
$Y^{MNPQR}$ is the corresponding 
magnetic charge\footnote{In non-compact
space, only the charge per unit volume is
finite. Thus $Z$, $Y$ are best thought of
as providing charge densities.}.

A solitonic M-theory membrane/fivebrane ($M2/M5$)
carries electric/magnetic charge $Z/Y$
and breaks half
of the thirty two supercharges $Q$
(\ref{edsusy}). An $Mp$-brane ($p=2,5$)
stretched in the $(x^1,\cdots, x^p)$ directions preserves
the supercharges $\epsilon Q$ with
\beq
\Gamma^0\Gamma^1\cdots\Gamma^p\epsilon=\epsilon
\label{BST17}
\eeq
Its tension is fixed by SUSY to be $T_p=1/l_p^{p+1}$.
Large charge branes can be reliably described by
eleven dimensional supergravity. The metric
around a collection of $k$ $Mp$-branes located
at $\vec r=\vec r_j$ ($j=1,\cdots, k$;
$\vec r, \vec r_j$
are $10-p$ dimensional vectors) is given by:
\beq
ds^2=U^{-1/3}dx^\mu dx_\mu+U^{2/3} d\vec r\cdot
d\vec r
\label{mtf}
\eeq
where $x^\mu$ are the $p+1$ directions along the
brane, and
\beq
U=1+\sum_{j=1}^k{l_p^{8-p}\over|\vec r-\vec r_j|^{8-p}}
\label{UUU}
\eeq
and there is also a three index tensor field which we
do not specify.

The ten dimensional type IIA vacuum with string coupling $\gs$
can be thought of as a compactification of M-theory
on $R^{1,9}\times S^{1}$. Denoting the $1+9$
dimensional Minkowski space of type IIA string theory
by $(x^0,x^1,\cdots, x^9)$, and the compact direction
by $x^{10}$, the compactification radius $R_{10}$ and
$l_p$ are related to the type IIA parameters $\gs$,
$l_s$ by:
\beq
{R_{10}\over l_p^3}={1\over l_s^2}
\label{BST18}
\eeq
\beq
R_{10}=\gs l_s
\label{BST19}
\eeq
Thus, the strong coupling limit of type IIA string theory
$\gs\to\infty$ (or equivalently $R_{10}/l_p\to\infty$)
is described by the $1+10$ dimensional
Minkowski vacuum of M-theory.

Type IIA branes have a natural
interpretation in M-theory:
\begin{itemize}
\item A fundamental IIA string stretched (say) along $x^1$
can be thought of as an $M2$-brane wrapped around $x^{10}$
and $x^1$. It is charged under the gauge field
$B_{\mu 1}=A_{10\mu1}$.
Equation (\ref{BST18}) is the relation between the
wrapped membrane and string tensions.
\item The $D0$-brane corresponds to a Kaluza-Klein (KK)
mode of the graviton carrying momentum $1/R_{10}$
along the compact direction. It is electrically charged
under $G_{\mu,10}$. Equation (\ref{BST19})
relates the masses of the KK mode of the
graviton and $D0$-brane.
\item The $D2$-brane corresponds to a
``transverse'' $M2$-brane,
unwrapped around $x^{10}$.
It is charged under $A_{\mu\nu\lambda}$.
The tension of the $M2$-brane
$1/l_p^3$ reduces to (\ref{BST3}) using the relation
\beq
l_p^3=l_s^3\gs
\label{BST20}
\eeq
which follows from (\ref{BST18}, \ref{BST19}).
\item The $D4$-brane corresponds to an $M5$-brane
wrapped around $x^{10}$. It is charged under
the five-form gauge field $\tilde A_{10\mu_1\mu_2
\cdots\mu_5}$ dual to $A$ $(d\tilde A=* dA)$.
Its tension (\ref{BST3})
is equal to $R_{10}/l_p^6$ (\ref{BST20}).
\item The $NS5$-brane corresponds to a transverse
$M5$-brane, and is thus charged under $\tilde A_{\mu_1
\cdots\mu_6}$. Its tension (\ref{BST12}) is equal
to $1/l_p^6$.
\item The $D6$-brane is a KK monopole. It is
magnetically charged under the gauge field
$A_\mu=G_{\mu10}$.
\item The $D8$-brane is a mysterious object in
M-theory whose tension is known to be
$R_{10}^3/l_p^{12}$~\cite{EGKR}.
\end{itemize}
\noindent
All this can be summarized by decomposing
the representations of $SO(10,1)$ appearing
in (\ref{edsusy}) into representations
of $SO(9,1)$ and rewriting the supersymmetry
algebra (\ref{edsusy}) as
\bea
\{Q_\alpha, Q_\beta\}&=(C\Gamma^\mu)_{\alpha\beta}P_\mu
+(C\Gamma^{10})_{\alpha\beta}P_{10}
+(C\Gamma^\mu\Gamma^{10})_{\alpha\beta}Z_\mu
+{1\over2}(C\Gamma^{\mu\nu})_{\alpha\beta}Z_{\mu\nu}\nonumber\\
&+{1\over 4!}(C\Gamma^{\mu\nu\rho\sigma}\Gamma^{10})_{\alpha
\beta}Y_{\mu\nu\rho\sigma}
+{1\over5!}(C\Gamma^{\mu\nu\rho\sigma\lambda})_{\alpha\beta}
Y_{\mu\nu\rho\sigma\lambda}
\label{tdsusy}
\eea
where $9+1$ dimensional vector
indices are denoted by $\mu,\nu,\rho,\sigma,\cdots$.
The momentum in the eleventh direction
$P_{10}$ is reinterpreted in ten dimensions
as zero-brane charge; the spatial components
of $Z_\mu$ are carried by ``fundamental''
IIA strings. Similarly, $Z_{\mu\nu}$ is the
$D2$-brane charge, $Y_{\mu\nu\rho\sigma}$
is the $D4$-brane charge, and
$Y_{\mu\nu\rho\sigma\lambda}$ is carried by
$NS5$-branes.
The different preserved supersymmetries
(\ref{BST4}, \ref{BST14}) combine in eleven
dimensions into the single relation (\ref{BST17}).
Note that (\ref{tdsusy}) includes central charges
for $p$-branes with $p\le 5$. Higher branes
(\eg\ the $D6$-brane) are inherently tied
to compactification; therefore the corresponding
central charges have to be added to (\ref{tdsusy})
by hand. 

We mentioned above that the scalar
$X^{10}$ describing fluctuations of the
IIA fivebrane in $x^{10}$
lives on a circle of radius $l_s$.
{}From the point of view of compactified M-theory
it is clear that the scalar field $X^{10}$ lives
on a circle of radius proportional to $R_{10}$;
the proportionality constant is determined for
a canonically normalized $X^{10}$ by
dimensional analysis to be $1/l_p^3$ as scalars
in $5+1$ dimensions have scaling dimension
two. Using (\ref{BST18}) we arrive at the
conclusion that the radius of (canonically normalized)
$X^{10}$ is
$R_{10}/l_p^3=1/l_s^2$.
In the normalization used in
(\ref{BST5}), with $g_{SYM}=l_s$ (\ref{BST16}),
$X$ has dimensions of length and lives on a circle
of radius $l_s$.

The metric around an $M5$-brane
transverse to $x^{10}$ (\ref{mtf}, \ref{UUU})
goes over to that around the
$NS5$-brane (\ref{BST13}) as $R_{10}\to0$. To see
that, describe an $M5$-brane at $x^{10}=0$
on the circle
as an infinite stack of parallel fivebranes
located at $x^{10}=nR_{10}$ ($n=0, \pm1, \pm2,\cdots$).
The harmonic function $U$ (\ref{UUU}) is
\beq
U=1+\sum_n \left[{l_p^2\over|\vec x|^2+(nR_{10})^2}
\right]^{3\over2}
\label{tendl}
\eeq
As $R_{10}\to0$ one can replace the sum
by an integral and (\ref{tendl}) approaches
(using (\ref{BST18}))
\beq
U\simeq 1+l_s^2/|\vec x|^2
\label{BST2055}
\eeq
The component of the metric $G_{10,10}=U^{2/3}$
(\ref{mtf}) is related to the ten dimensional
dilaton via $G_{10,10}\equiv \exp(2\gamma)
=\exp(4\phi/3)$. The
string metric $\GG$ is  related to the eleven
dimensional metric $G$ by  a rescaling
$\GG=G\exp\gamma $. Performing the rescaling
leads to the ten dimensional form (\ref{BST13}).

Ten dimensional type IIB string theory
has a complex coupling
\beq
\tau=a+{i\over \gs},
\label{BST205}
\eeq
where $a$ is the expectation value
of the massless RR scalar.
In the eleven dimensional interpretation, the
ten dimensional type IIB
vacuum corresponds to M-theory compactified on  a
two-torus of complex structure $\tau$ and
vanishing area. Naively, the theory appears to be $1+8$
dimensional in this limit, but in fact as the size
of the torus goes to zero, the wrapping
modes of the $M2$-brane become light and
give rise to another non-compact
direction which we will label by $x^B$.

M-theory on a finite
two-torus corresponds to compactifying
$x^B$ on a circle of radius $R_B$.
In the special case $a=0$, the M-theory
two-torus is rectangular with sides $R_9, R_{10}$.
The mapping of the M-theory parameters $(R_9, R_{10},
l_p)$ to the type IIB ones $(R_B, \gs, l_s)$ is:
\beq
{R_{10}\over l_p^3}={1\over l_s^2}
\label{BST21}
\eeq
\beq
{R_9\over l_p^3}={1\over \gs l_s^2}
\label{BST22}
\eeq
\beq
{R_{9}R_{10}\over l_p^3}={1\over R_B}
\label{BST23}
\eeq
One way to establish (\ref{BST21}-\ref{BST23})
is to reinterpret the different type IIB branes
in M-theory:

\begin{itemize}
\item
A fundamental IIB string can be thought of
as an $M2$-brane wrapped around $x^{10}$. Equation
(\ref{BST21}) is the relation between the membrane
and string tensions.
\item
A $D1$-brane (D-string) that is not wrapped
around $x^B$ corresponds to an
$M2$-brane wrapped around $x^9$. Equation
(\ref{BST22}) is the relation between the membrane
and D-string tensions. A D-string wrapped around
$x^B$ corresponds to a KK mode of the eleven dimensional
supergraviton carrying momentum in the $x^{10}$ direction.
E.g., using (\ref{BST21}) and the relation
\beq
{1\over l_p^3}={R_B\over \gs l_s^4}
\label{BST24}
\eeq
which follows from (\ref{BST21}-\ref{BST23}), the masses
agree: $1/R_{10}=R_B/\gs l_s^2$.
\item
A KK mode of the supergraviton carrying
momentum in the $x^B$ direction in type IIB string
theory corresponds to
an $M2$-brane wrapped around $(x^9, x^{10})$;
eq. (\ref{BST23}) relates the masses of the two.
\item
A $D3$-brane unwrapped around $x^B$
corresponds to an $M5$-brane
wrapped on $(x^9, x^{10})$. The tension of the wrapped
$M5$-brane $R_9R_{10}/l_p^6$ reduces to (\ref{BST3})
using (\ref{BST21}, \ref{BST22}). A $D3$-brane wrapped
around $x^B$ corresponds to an $M2$-brane.
\item
A $D5$-brane wrapped around $x^B$
corresponds to an $M5$-brane
wrapped around $x^{10}$. The tension of the wrapped
$M5$-brane $R_{10}/l_p^6$ reduces to $R_B/\gs l_s^6$
using (\ref{BST24}).
A $D5$-brane unwrapped around $x^B$ corresponds
to a KK monopole charged under the gauge field
$G_{\mu,10}$ and wrapped around $x^9$.
\item
The $NS5$-brane wrapped around $x^B$
corresponds to an $M5$-brane wrapped on
$x^9$. Its tension $R_B/\gs^2 l_s^6$
is equal to that of the wrapped $M5$-brane $R_9/
l_p^6$.
An NS fivebrane unwrapped around $x^B$ corresponds
to a KK monopole charged under the gauge field
$G_{\mu,9}$ and wrapped around $x^{10}$.
\item
The $D7$-brane wrapped around $x^B$
corresponds to a KK monopole charged under
$G_{\mu, 10}$. A $D7$-brane unwrapped around
$x^B$ is related to the M-theory eightbrane
which reduces to the $D8$-brane of IIA string
theory.
\end{itemize}

Orientifolds correspond in M-theory to
fixed points of $Z_2$ transformations
acting both on space and on the supergravity
fields.

\subsection{Duality Properties}
\label{DP}

String (or M-) theory has a large moduli space of vacua
$\MM$ parametrized by the size and shape of the compact
manifold and the string coupling (as well as
the values of other background fields). At generic points
in $\MM$ the theory is eleven dimensional
and inherently quantum mechanical while at certain
degenerations it has different weakly coupled string
expansions.

The space of vacua $\MM$ is a non-trivial manifold;
in particular, it has an interesting global structure.
Some apparently distinct vacua are identified by the
action of a discrete group known as ``U-duality''~\cite{HT}.
Under this identification different
states of the theory are often mapped into each
other; an example is
the BPS branes discussed above. What looks like
a D-brane in one description may appear to be an NS-brane
in another, and may even correspond to an object of different
dimension.

An important  subgroup of U-duality is T-duality which takes
a weakly coupled vacuum to another weakly coupled vacuum
and is, therefore, manifest in string perturbation theory
(for a review see~\cite{GPR} and references therein).
Consider type IIA string theory in $1+8$ non-compact
dimensions with the $i$'th coordinate $x^i$ living on
a circle of radius $R_i$. At large $R_i$ the theory
becomes $1+9$ dimensional IIA string theory
while at small $R_i$ it
naively becomes $1+8$ dimensional. However, winding
type IIA strings with energy $nR_i/l_s^2$ become light
in the limit, producing a continuous Kaluza-Klein spectrum and
thus the theory becomes ten dimensional again.

{}From the discussion of the previous section it is clear what
the new $1+9$ dimensional theory is.
Weakly coupled type IIA string theory on a small
circle $R_i\to 0$ corresponds to M-theory on a vanishing
two-torus, which we saw before is just type IIB string
theory. How do different states in IIA string theory
map to their IIB counterparts?

The wrapped IIA string is a wrapped $M2$-brane
(see (\ref{BST18}) and subsequent discussion); the
modes becoming  light
in the $R_i\to0$ limit correspond to membranes wrapped
$n$ times around the shrinking two-torus labeled by $(x^i, x^{10})$.
Comparing their energy $nR_i R_{10}/l_p^3$ to
(\ref{BST23}) and using (\ref{BST18}-\ref{BST22})
we see that the IIB string one finds lives on a circle
of radius
\beq
R^{(B)}_i={l_s^2\over R^{(A)}_i}
\label{BST25}
\eeq
and has string coupling
\beq
\gs^{(B)}=\gs^{(A)}l_s/R^{(A)}_i
\label{BST26}
\eeq
We will refer to the transformation (\ref{BST25},
\ref{BST26}) as $T_i$ (T-duality in the $i$'th
direction).

The different branes of type IIA string theory
transform as follows under $T_i$:

\begin{itemize}
\item
As we just saw, a fundamental IIA string
wound $n$ times around
$x^i$ transforms into a fundamental IIB string
carrying momentum $n/R^{(B)}_i$.
An unwound fundamental IIA string carrying
momentum $m/R^{(A)}_i$
transforms under $T_i$ to a
fundamental IIB string wound $m$ times
around the $i$'th direction.
\item
A $D0$-brane corresponds in M-theory to
a KK graviton carrying momentum $1/R_{10}$.
As we saw earlier, in type IIB language this
is a D-string wrapped around the $i$'th direction.
\item
A $D2$-brane wrapped around $x^i$ corresponds in
M-theory to a transverse $M2$-brane wrapped around
$x^i$. We saw earlier that in type IIB language
this is a D-string unwrapped around $x^i$.
Similarly, a $D2$-brane unwrapped around $x^i$
was seen to correspond to an unwrapped $M2$-brane
and was interpreted in IIB language as a
$D3$-brane wrapped around $x^i$.
\item
At this point the pattern for Dirichlet branes should
be clear. A IIA Dirichlet $p$  brane wrapped around
$x^i$ is transformed under $T_i$ to an unwrapped
IIB Dirichlet
$p-1$ brane, while an unwrapped IIA Dirichlet $p$ brane
is transformed to a Dirichlet $p+1$ brane wrapped around
$x^i$:
\beq
T_i:\;
\text{\ddp\ wrapped on $x^i$}\longleftrightarrow
\text{$D(p-1)$ at a point on $x^i$}
\label{BST27}
\eeq
\item
Orientifold planes transform under $T_i$ in the same
way as D-branes (\ref{BST27}).
\item
A wrapped IIA NS fivebrane
transforms under $T_i$ to a wrapped
IIB NS fivebrane. An unwrapped IIA
NS fivebrane transforms into the KK
monopole carrying magnetic charge
under $G_{\mu, i}$:
\beq
T_i:\left\{
\begin{array}{ll}
\mbox{IIA $NS5$ wrapped on $x^i$}&\longleftrightarrow
\mbox{IIB $NS5$ wrapped on $x^i$}\\
\mbox{$NS5$ at a point on $x^i$}&\longleftrightarrow
\mbox{KK monopole charged under $G_{\mu, i}$}
\end{array}
\right.
\label{BST28}
\eeq
\end{itemize}
As a check, the tensions of the various (wrapped
and unwrapped)
Dirichlet and solitonic branes (\ref{BST3},
\ref{BST12}, \ref{KKK4})
transform under (\ref{BST25},
\ref{BST26}) consistently with
the above discussion.

The generalization to T-duality in more than one
direction $T_{i_1,i_2,\cdots, i_n}\equiv
T_{i_1}T_{i_2}\cdots T_{i_n}$ is straightforward:
\bea
T_{i_1,i_2,\cdots, i_n}:\; (R_{i_1},R_{i_2},\cdots, R_{i_n})
\longleftrightarrow &({l_s^2\over R_{i_1}}, {l_s^2\over R_{i_2}},
\cdots, {l_s^2\over R_{i_n}})\nonumber\\
\gs\longleftrightarrow
&\;\gs\prod_{\alpha=1}^n{l_s\over R_{i_\alpha}};\;
l_s\longleftrightarrow l_s
\label{TTT1}
\eea
For even $n$ it takes type IIA(B) to itself, while
for odd $n$ it exchanges the two.

The discussion above can be used to determine the
charge of the \op\ plane given in (\ref{BST11}).
Starting with the type I theory on $T^n$, which contains
a single $O9$-plane and thirty two $D9$-branes wrapped around
the $T^n$, and performing
T-duality, $T_{i_1,i_2,\cdots, i_n}$, 
we find a vacuum 
with $2^n$ orientifold $p$-planes, $p=9-n$,
one at each fixed
point on $T^n/Z_2$, as well as thirty two \ddp-branes.
The total RR $(p+1)$-form charge of the configuration is zero,
which leads to (\ref{BST11}).

Another interesting subgroup of U-duality is
S-duality of type IIB string theory in
$9+1$ dimensions~\cite{Sch95}, an $SL(2, Z)$ symmetry that acts
by fractional linear transformations with
integer coefficients on $\tau$ (\ref{BST205}).
In the M-theory interpretation of IIB
string theory, this $SL(2, Z)$ is the modular group
acting on the complex structure of the two-torus
(whose size goes to zero in the ten dimensional
limit). (For a review see~\cite{Sch96} and references therein).
We will focus on a $Z_2$ transformation
$S\in SL(2,Z)$ which acts as $\tau\to-1/\tau$;
we will furthermore restrict to the case of
vanishing RR scalar $a$ (namely a rectangular M-theory
two-torus),
in which case it acts on the coupling (\ref{BST205})
as strong-weak coupling duality: $\gs\to1/\gs$.
In the M-theory interpretation of IIB string theory
discussed in (\ref{BST21}-\ref{BST23})
$S$ acts geometrically by
interchanging $R_9\leftrightarrow R_{10}$.
Equations (\ref{BST21}, \ref{BST22}) imply that the
type IIB parameters $\gs, l_s$ transform as:
\beq
S:\; \gs\longleftrightarrow{1\over\gs};\;\;
l_s^2\longleftrightarrow l_s^2\gs
\label{BST30}
\eeq
Another way to arrive at (\ref{BST30}) is to require
that as the string coupling is inverted, the ten
dimensional Planck length $l_{10}^4=\gs l_s^4$ remains
fixed. {}From the discussion following eq. (\ref{BST23})
it is clear that the different IIB branes transform
under $S$ as follows:
\begin{itemize}
\item
The fundamental string is interchanged with the D-string.
\item
The $D3$-brane is invariant.
\item
The $NS5$-brane is interchanged with the $D5$-brane.
\item
The $D7$-brane transforms into a different
sevenbrane.
\end{itemize}
As a check, the tensions of the various branes (\ref{BST3},
\ref{BST12}) transform under (\ref{BST30}) consistently
with the above discussion. The transformations of
orientifold planes under $S$ are more intricate and will
be discussed in the context of particular applications
below.

The worldsheet dynamics on 
both the fundamental string and
D-string is that of a critical IIB string. At weak
string coupling the tension of the fundamental string
is much smaller than that of the D-string, and we can
think of the former as ``fundamental'' and of the latter
as a heavy soliton. At strong coupling, the D-string is
the lighter object and {\it it} should be used as the
basis for string perturbation theory. Since a
IIB string in its ground state
preserves half of the SUSY, it can be
followed from weak to strong coupling, and the above picture
is indeed reliable.

Under the full $SL(2,Z)$ S-duality group, the two different
kinds of strings are members of a multiplet of
$(p,q)$ strings, with the fundamental string corresponding
to $(p,q)= (1,0)$ and the D-string corresponding to
$(p,q)=(0,1)$. $p$ measures the charge carried by the string
under the NS-NS $B_{\mu\nu}$ field while $q$ measures the
charge under the RR $B_{\mu\nu}$ field. In M-theory the $(p,q)$
string corresponds to a membrane wrapped $p$ times around $x^{10}$
and $q$ times around $x^9$; it is stable when $p,q$ are
relatively prime.
A similar discussion applies to fivebranes that carry magnetic
charges under the two $B_{\mu\nu}$ fields and thus form a multiplet
of $(p,q)$ fivebranes.
There are also $(p,q)$ sevenbranes which carry magnetic
charge under the complex dilaton $\tau$.

In M-theory compactified on $T^d$, the $SL(2,Z)$ S-dualities
corresponding to different $T^2\subset T^d$ are subgroups of
the geometrical $SL(d, Z)$ symmetry group of $T^d$. Together
with T-duality (\ref{TTT1}) they generate the U-duality
group $E_{d(d)}(Z)$ of type II strings on $T^{d-1}$~\cite{EGKR}.

\subsection{Webs Of Branes}
\label{WB}

So far we discussed brane configurations which
preserve sixteen supercharges. In this
section we will describe some
configurations with lower supersymmetry.

We saw before that a stack of parallel D or NS-branes
preserves $1/2$ of the SUSY given by (\ref{BST4}) or
(\ref{BST14}, \ref{BST15}), respectively.
To find the SUSY preserved by a web of differently
oriented D and/or NS-branes one needs to impose
all the corresponding
conditions~\footnote{This analysis
is valid for widely separated branes and may miss
bound states.}
on the spinors
$\epsilon$.
The worldvolume dynamics on such a web of branes
is typically rather rich. We will next consider
it in a few examples.

\subsubsection{The \ddp\ -- $D(p+4)$ System}
\label{PP4}

\begin{figure}
\centerline{\epsfxsize=100mm\epsfbox{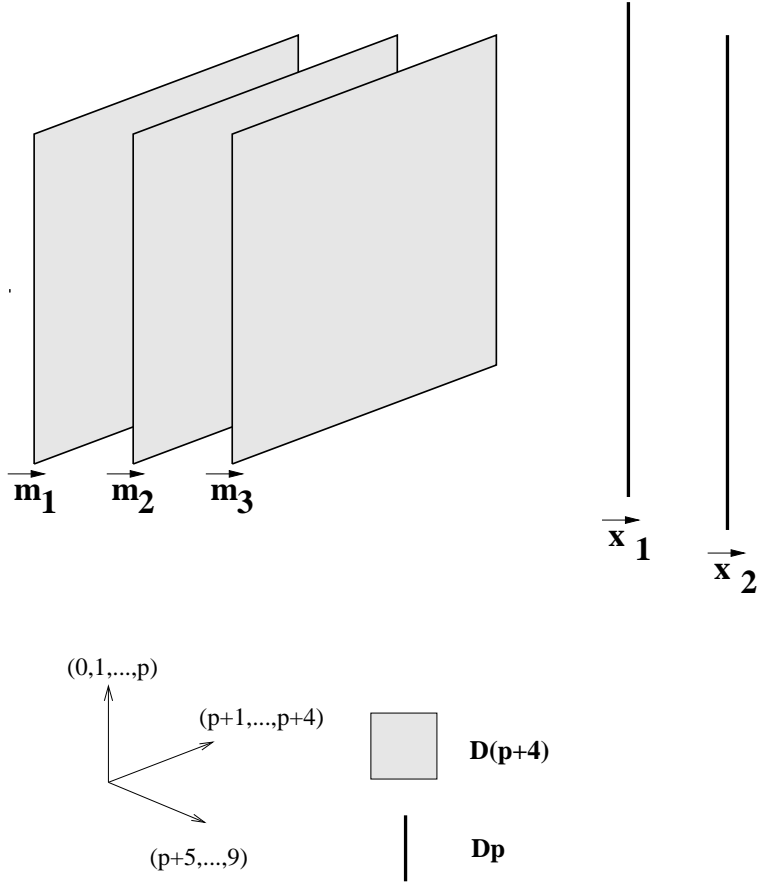}}
\vspace*{1cm}
\caption{The $Dp$ - $D(p+4)$ system consisting
of a stack of $\nc$ $Dp$-branes parallel to $\nf$
$D(p+4)$-branes. Locations in the transverse
space $(x^{p+5},\cdots, x^9)$ are labeled by
$\vec x_a$, $\vec m_i$, respectively.}
\label{three}
\end{figure}
\smallskip

Consider a stack of $N_c$ \ddp-branes
stretched in the $(x^1, \cdots, x^p)$
hyperplane ``parallel'' to a stack of
$\nf$ $D(p+4)$-branes stretched in
$(x^1, \cdots, x^{p+4})$ depicted in
Figure~\ref{three}.
Each stack preserves $1/2$ of the SUSY
and together they preserve $1/2\times 1/2=1/4$
of the thirty two supercharges of type II string theory.
The preserved supercharges are
those that satisfy (\ref{BST4})
\beq
\epsilon_L=\Gamma^0\Gamma^1\cdots\Gamma^p\epsilon_R=
\Gamma^0\Gamma^1\cdots\Gamma^{p+4}\epsilon_R
\label{BST36}
\eeq
The second equality in (\ref{BST36}) is a constraint
on $\epsilon_R$, $\epsilon_R=\Gamma\epsilon_R$ with
$\Gamma=\Gamma^{p+1}\Gamma^{p+2}\Gamma^{p+3}\Gamma^{p+4}$.
The matrix $\Gamma$ squares to the identity matrix, and
is traceless. Thus, half of its sixteen eigenvalues are $+1$
and half are $-1$. The constraint on $\epsilon_R$,
$\Gamma=1$, preserves eight
of the sixteen components of $\epsilon_R$. Given $\epsilon_R$,
the first equality in (\ref{BST36}) fixes $\epsilon_L$. Thus
the total number of independent supercharges preserved by
the configuration is eight.

The light degrees of freedom on each stack of
branes were discussed before. On the $N_c$ \ddp-branes
there is a $p+1$ dimensional $U(N_c)$ gauge theory
coupled to $9-p$ adjoint scalars and some fermions.
The adjoint scalars naturally split into $5-p$ fields
corresponding to fluctuations of the \ddp-branes
transverse to the $(p+4)$-branes which together
with the gauge field form the vectormultiplet of
a theory with eight supercharges, and the remaining
four fields, which form an adjoint hypermultiplet.

A similar theory with $N_c\to N_f$ and $p\to p+4$
lives on the $D(p+4)$-branes. Each of the two theories
has sixteen supercharges. The SUSY of the full theory
is broken down to eight supercharges by additional matter
corresponding to strings stretched between the two
stacks of branes. {}From the point of view of the \ddp-brane
this matter corresponds to $N_f$ flavors in the fundamental
representation of $U(N_c)$.
{}From the point of view of the $D(p+4)$-brane,
they are $\nc$ pointlike (in the transverse
directions) defects in the fundamental of
$U(\nf)$. When the
\ddp-branes are inside the $D(p+4)$-branes,
they can be thought of as small instantons~\cite{Dou95}.

It is important to emphasize that for an observer
that lives on the \ddp-brane, the degrees of freedom on the
$D(p+4)$-brane are non-dynamical background fields (at
least in infinite volume).
For example, the effective gauge coupling in $p+1$
dimensions $g_{p+1}$  of the $U(N_f)$ gauge field on
the $D(p+4)$-brane is given by
\beq
{1\over g_{p+1}^2}={V_{p+1,\cdots, p+4}
\over g_{p+5}^2}
\label{BST37}
\eeq
where $g_{p+5}$ is the $U(N_f)$ gauge coupling
in $p+5$ dimensions and
$V_{p+1,\cdots, p+4}$ is the volume of the
$D(p+4)$-brane worldvolume transverse to the
\ddp-brane. When this volume is infinite, the kinetic
energy of $U(N_f)$ excitations is infinite
as well and they are frozen at their classical values.
The same is true for other excitations on the
$D(p+4)$-brane. Thus, from the point of view
of the \ddp-brane, the $U(N_f)$ gauge symmetry
of the $D(p+4)$-brane is a global symmetry and the
only dynamical fields that appear due to the presence of
the $D(p+4)$-brane are the $N_f$ flavors corresponding
to strings stretched between the \ddp\ and $D(p+4)$-branes;
these modes are localized at the $Dp$-brane.

The relative locations in space of the various branes
correspond to moduli and couplings in the \ddp-brane
worldvolume theory.
Locations of the ``heavy'' $D(p+4)$-branes
correspond to couplings while locations of the
``light'' \ddp-branes are moduli:
\begin{itemize}
\item
The locations of the $D(p+4)$-branes in the transverse
space $(x^{p+5}, \cdots, x^9)$ $\vec m_i$ $(i=1,\cdots,
\nf)$ correspond to masses for the $\nf$ fundamentals.
\item
The locations of the \ddp-branes in $(x^{p+5}, \cdots, x^9)$
$\vec x_a$ $(a=1,\cdots, \nc)$ correspond to
expectation values of fields $\vec X$ in the adjoint
of $U(N_c)$ and parametrize the Coulomb
branch of the $U(N_c)$ gauge theory, as in (\ref{BST8}).
\item
The locations of the \ddp-branes parallel to the
$D(p+4)$-branes (in the $(x^{p+1}, \cdots, x^{p+4})$
directions) correspond to expectation values of an adjoint
hypermultiplet of $U(N_c)$.
\end{itemize}

One can think of the \ddp-branes as probing the
geometry near the $D(p+4)$-brane. For example, the
metric on the Coulomb branch of the $U(1)$ gauge theory
with $\nf$ flavors on a single \ddp-brane adjacent
to $\nf$ $D(p+4)$-branes is the
background metric of the $D(p+4)$-branes. This is
analogous (and in some cases U-dual)
to the situation described in section \ref{SF}
where we described the metric felt by a fundamental string
propagating in the background of solitonic fivebranes.

In general, some of the parameters that one can turn
on in the low energy field theory may be absent
in the brane configuration.
As an example, in the low energy $U(N_c)$
gauge theory with eight supercharges one can
add a mass term to the adjoint hypermultiplet and a
Fayet-Iliopoulos (FI) coupling, both which are absent in
the brane configuration. One way to understand this
is to note that theories with sixteen supercharges
do not have such couplings. The theory on  a stack
of isolated \ddp-branes has sixteen supercharges
and, while it is broken down to eight by the
presence of the $(p+4)$-branes, it inherits this
property from the theory with more SUSY.

Similarly, some of the moduli of the low energy
gauge theory may not correspond to geometrical
deformations in the brane description.
In the example above, the Higgs branch of the
$U(N_c)$ gauge theory, corresponding to non-zero
expectation values of the fundamentals, can be thought
of as the moduli space of instantons.
Each \ddp-brane embedded in the stack of
$N_f$ $(p+4)$-branes can be thought of as a
small (four dimensional) $U(N_f)$ instanton which
can grow and become a finite size instanton.
The moduli space of $N_c$ instantons in $U(N_f)$
is the full Higgs branch of the theory; it is not
realized geometrically. For a more detailed
discussion see~\cite{Dou96}. 

Clearly, the more of the couplings and moduli of
the gauge theory are represented geometrically,
the more useful the brane configuration is for
studying the gauge theory.

\subsubsection{More General Webs Of Branes}
\label{WBS}

The system described in the previous subsection
can be generalized in several directions:
applying U-duality transformations, rotating
some of the branes relative to others, adding
branes and/or orientifold planes, and considering
configurations of branes ending on branes. In this
and the next subsections
we will describe some of these possibilities:

\begin{itemize}
\item {\em Orientifolds:}
starting with the \ddp\ -- $D(p+4)$ system
we can add an \op-plane, an $O(p+4)$-plane,
or both, without breaking any further SUSY.
Adding an \op-plane leads to an $SO(\nc)$
or $Sp(\nc/2)$ gauge theory~\footnote{$\nf$ and
$\nc$ are even here.} on the \ddp-branes.
In gauge theory with eight supercharges and $\nf$
fundamentals the resulting global symmetry is
$Sp(\nf/2)$ or $SO(\nf)$, respectively.
Therefore, it is clear that an orthogonal
orientifold projection on the $p$-branes is
correlated with a symplectic projection on the
$(p+4)$-branes, and vice-versa.

A similar analysis
can be performed for the case of an $O(p+4)$-plane.
An example is type I theory, where
an orthogonal projection on ninebranes due to
an orientifold nineplane is correlated with a
symplectic projection on 
fivebranes~\cite{W9511,GPo}.
\item {\em The \ddp\ -- $D(p+2)$ System:}
compactifying the \ddp\ -- $D(p+4)$
system of section  \ref{PP4}
and considering different
limits gives rise to configurations with the
same amount of
SUSY in different dimensions. These can
be studied by using T-duality.
As an example, compactify $x^{p+1}$ on a circle,
T-dualize and then decompactify the resulting dual
circle. One finds a $D(p+1)$ -- $D(p+3)$ system;
a stack of
$\nc$ $D(p+1)$-branes whose worldvolume stretches
in $(x^0, x^1, \cdots, x^{p+1})$ and a stack of
$\nf$ $D(p+3)$-branes whose worldvolume lies in
$(x^0, x^1, \cdots, x^p, x^{p+2}, x^{p+3}, x^{p+4})$.
The two stacks of branes are now partially orthogonal,
with $p+1$ of their
$p+2$ and $p+4$ dimensional worldvolumes in common.

Formally, the degrees of freedom in the common dimensions
(which we will refer to as ``the intersection'')
are the same as before, however, one can no longer talk
about a $U(\nc)$ gauge theory on the intersection.
All matter in the adjoint of $U(\nc)$ is now classical,
as it lives on a ``heavy'' brane which has one infinite
direction ($x^{p+1}$) transverse to the intersection. The
only dynamical degrees of freedom on the $p+1$ dimensional
intersection region are the $\nf$ fundamentals of $U(\nc)$
which arise from $(p+1)-(p+3)$ strings.
Of course, re-compactifying $x^{p+1}$ restores the previous
physics, and we will usually implicitly consider this case
below.
\item {\em The \ddp\ -- $D(p+2)$ -- $D(p+2)^\prime$ System:}
to reduce the number of supersymmetries from eight to
four we can add to the previous system another stack of
differently oriented D-branes. A typical configuration consists
of a stack of $\nc$  \ddp-branes with worldvolume
$(x^0, x^1, \cdots, x^p)$,
$\nf$ $D(p+2)$-branes
$(x^0, x^1, \cdots, x^{p-1}, x^{p+1}, x^{p+2}, x^{p+3})$
and $\nf^\prime$ $D(p+2)^{\prime}$-branes
$(x^0, x^1, \cdots, x^{p-1}, x^{p+1}, x^{p+4}, x^{p+5})$.
The gauge group on the \ddp-branes is $U(\nc)$, with
the following matter:

1) $\nf$ fundamental hypermultiplets $Q$, $\tilde Q$
corresponding to strings stretched between the
\ddp\ and $D(p+2)$-branes, and $\nf^\prime$
fundamentals $Q^\prime$, $\tilde Q^\prime$
corresponding to strings stretched between
the \ddp\ and $D(p+2)^\prime$-branes.

2) $10-p$ adjoint fields whose expectation values
(\ref{BST8}) parametrize the locations of
the $p$-branes and the  Wilson line of the
worldvolume gauge field along the compact
$x^p$ direction. These can be split into:
a complex adjoint field $X$ describing
fluctuations of the \ddp-branes
in the $(x^{p+4}, x^{p+5})$ directions; a complex
adjoint field $X^\prime$ corresponding to
fluctuations in the $(x^{p+2}, x^{p+3})$
directions;  a complex adjoint $X^{\prime\prime}$
corresponding to fluctuations in the $x^{p+1}$ direction
as well as the gauge field $A_p$. 
$4-p$ adjoints parametrize
the Coulomb branch of the gauge theory.

$X$ couples to the $\nf$ flavors $Q$ and
$X^\prime$ couples to the $\nf^\prime$
flavors $Q^\prime$ via the superpotential
\beq
W=\tilde Q X Q + \tilde Q^\prime X^\prime Q^\prime
\label{BST375}
\eeq
Geometrically, the couplings (\ref{BST375})
are due to the fact that displacing
the \ddp-branes in the $(x^{p+4}, x^{p+5})$
directions stretches the $p-(p+2)$ strings
thus giving a mass to the quarks $Q$,
$\tilde Q$, etc.

More generally, the coupling matrix
of ($X$, $X^\prime$) and ($Q$, $Q^\prime$)
is governed by the relative angles between
the $D(p+2)$ and $D(p+2)^\prime$-branes.
Indeed, defining $v=x^{p+2}+ix^{p+3}$
and $w=x^{p+4}+ix^{p+5}$, one can check~\cite{BDL} that
arbitrary relative complex rotations
of the different $(p+2)$-branes in $v,w$
\beq
\left(
\begin{array}{c}
v\\
w\end{array}\right)
\longrightarrow
\left(\begin{array}{cc}
\cos\theta&\sin\theta\\
-\sin\theta&\cos\theta
\end{array}\right)
\left(\begin{array}{c}
v\\
w\end{array}\right)
\label{BST38}
\eeq
preserve four supercharges like the original
\ddp\ -- $D(p+2)$ -- $D(p+2)^\prime$ system.
When the relative angle between the
$D(p+2)$ and $D(p+2)^\prime$ branes goes to
zero, the SUSY is enhanced to eight supercharges
and one recovers the \ddp\  -- $D(p+2)$ system
described above.
\item {\em The NS -- \ddp\ System:}
starting with the $D3$ -- $D5$ system and performing
an $S$-duality transformation we find a system
consisting of $\nc$ $D3$-branes $(x^0, x^1, x^2, x^3)$,
and $\nf$ $NS5$-branes $(x^0, x^1, x^2, x^4, x^5, x^6)$
preserving eight supercharges.
T-duality (\ref{BST27}, \ref{BST28}) -- acting on 
any number of longitudinal directions of the NS-brane 
-- may be used to
turn this configuration into other configurations of
\ddp-branes and $NS5$-branes. Other T-dualities
(which act on one direction transverse to
the NS-brane)
map the system to configurations of \ddp-branes
wrapped around non-trivial cycles of ALE spaces.
Similarly to the D-brane case described above,
different NS-branes can be rotated with respect to
each other, by complex rotations of the form
(\ref{BST38}), which
preserve four of the eight supercharges.
\end{itemize}

\subsubsection{Branes Ending On Branes}
\label{BEB}
One of the important things branes can do is
end on other branes. D-branes are {\it defined}
by the property that fundamental strings can end on
them, and by a chain of dualities this can be
related to many other possibilities.

\begin{figure}
\centerline{\epsfxsize=140mm\epsfbox{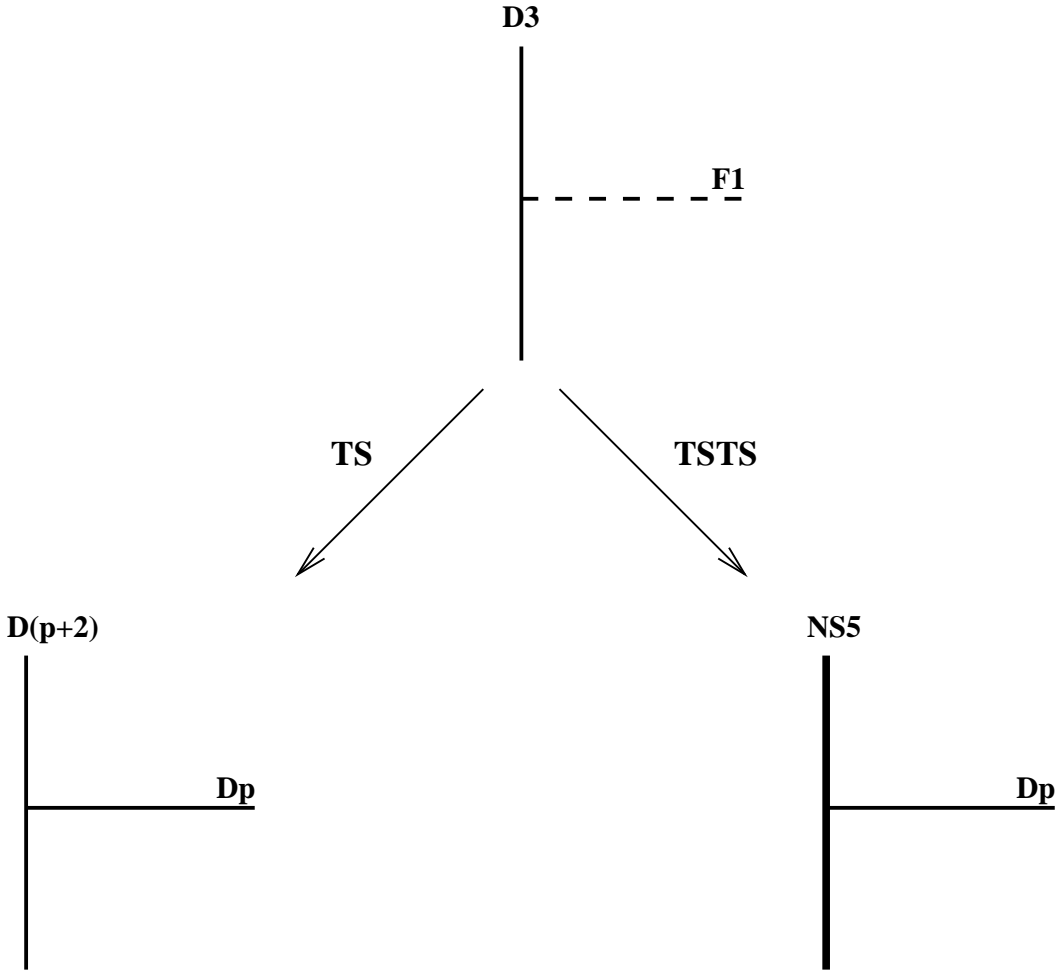}}
\vspace*{1cm}
\caption{U-duality relates a fundamental
string which ends on a $D3$-brane to
other supersymmetric configurations,
such as a $Dp$-brane which ends
on a $D(p+2)$-brane,
and a $Dp$-brane which ends
on an NS fivebrane.}
\label{four}
\end{figure}

Consider a fundamental string ending
on a $D3$-brane (Fig.~\ref{four}).
The $D3$-brane itself
preserves sixteen supercharges, and if we put the
open string ending on it in its ground state
it preserves $1/2$ of these, namely eight.
Performing $S$-duality we
reach a configuration of a D-string ending on the
$D3$-brane. By T-duality in $p-1$ directions
transverse to both branes we are led to a
configuration of a \ddp-brane ending on a
$D(p+2)$-brane with a $(p-1)+1$ dimensional
intersection.

\smallskip

For $p=3$, the configuration of a $D3$-brane
ending on a $D5$-brane can be mapped by applying
$S$-duality to a $D3$-brane ending on an $NS5$-brane.
Further T-duality along the fivebrane worldvolume
maps this to a configuration of a \ddp-brane
(with any $p\le6$)
ending on the $NS5$-brane.

In M-theory, many of the above configurations are related
to membranes ending on fivebranes. This is most
apparent for a $D2$-brane ending on an $NS5$-brane
in type IIA string theory as well as fundamental and D-strings
ending on the appropriate fivebranes. Others (\eg\
a $D4$-brane ending on an $NS5$-brane)
can be thought of as corresponding to a single
$M5$-brane with a convoluted worldvolume.

The worldvolume theory on a brane that ends
on another brane is a truncated version with
eight supercharges of the dynamics on an infinite
brane. The light fields are conveniently described
in terms of representations of $d=4$, $N=2$ SUSY
with spin $\le$ 1,
hypermultiplets and vectormultiplets:

\begin{itemize}
\item
For a \ddp-brane stretched in $(x^0, x^1,
\cdots, x^p)$, and ending (in the $x^p$
direction) on a $D(p+2)$-brane
stretched in $(x^0, x^1, \cdots, x^{p-1}, x^{p+1},
x^{p+2}, x^{p+3})$ and located at $x^p=0$,
the $p+1$ dimensional dynamics
now takes place on $R^{1,p-1}\times R^{+}$,
where the half line $R^{+}$ corresponds to $x^p\ge0$.
The three scalars corresponding to fluctuations
of the \ddp-brane along the $D(p+2)$-brane
$(X^{p+1}, X^{p+2}, X^{p+3})$ combine
with the $p$'th component of the \ddp-worldvolume
gauge field $A_p$ into a massless hypermultiplet
with free boundary conditions~\footnote{We will
soon see that the boundary conditions are
modified quantum mechanically.} at $x^p=0$.
The scalars describing fluctuations of the
\ddp-brane perpendicular to the $D(p+2)$-brane
$(X^{p+4}, \cdots, X^9)$ satisfy Dirichlet
boundary conditions $X^I(x^p=0)=0$ ($I=p+4,
\cdots, 9$).
These $6-p$ scalars
are paired by SUSY with the
gauge field  $A_\mu$, $\mu=0,1,\cdots, p-1$,
into a vectormultiplet. Thus, the
gauge field satisfies Dirichlet boundary conditions
as well.
\item
For a \ddp-brane stretched in $(x^0, x^1,
\cdots, x^{p-1}, x^6)$ and ending (in the $x^6$
direction) on an $NS5$-brane
stretched in $(x^0, x^1, \cdots, x^5)$,
the hypermultiplet contains the scalars
$(X^7, X^8, X^9)$ and the sixth component
of the \ddp-worldvolume gauge field $A_6$
and satisfies Dirichlet boundary conditions
at $x^6=0$.
The vectormultiplet consisting of the $6-p$
scalars $(X^p, \cdots, X^5)$ and the components
of the gauge field along $R^{1,p-1}$ is
(again, classically) free
at the boundary.
\end{itemize}

Quantum mechanically, we have to take into
account that
the end of a brane ending on another brane
looks like a charged object in the worldvolume
theory of the latter. Consider for example
the case of a fundamental string ending on
a \ddp-brane. It can be thought of as providing
a point-like source for the $p$-brane worldvolume
gauge field, leading to a Coulomb 
potential~\cite{CM,Gib}
\beq
A_0={Q\over r^{p-2}}
\label{BST39}
\eeq
where $Q$ is the worldvolume charge of the
fundamental string and $r$ the distance
from the charge on the $p$-brane.
To preserve SUSY it is clear from the form of
the action (\ref{BST5}) that in addition
to (\ref{BST39}) one of the $p$-brane worldvolume
scalar fields must be excited, say:
\beq
X^{p+1}={Ql_s^2\over r^{p-2}}
\label{BST40}
\eeq
The solution (\ref{BST39}, \ref{BST40})
preserves half of the sixteen worldvolume
supersymmetries and corresponds to a fundamental
string stretched along $x^{p+1}$ and ending on
the D-brane. We see that the string bends the
D-brane: the location of the brane becomes $r$
dependent (\ref{BST40}), approaching the
``classical'' value $x^{p+1}=0$ at large $r$
(for $p>2$). Standard charge quantization
implies that the quantum of charge in the
normalization (\ref{BST5}) is $Q=g_{SYM}^2$.
As $r\to0$,
$x^{p+1}\to\infty$; this corresponds to
a fundamental string ending on the $Dp$-brane.
Of course, a priori we only trust the solution
(\ref{BST39}, \ref{BST40}) for large $r$ where
the fields and their variations are small. As $r\to0$
higher order terms in the Lagrangian, that
were dropped in (\ref{BST5}), become important,
\eg\ one has to replace the Maxwell action by
the Born-Infeld action. A detailed discussion
of this and related issues appears  
in~\cite{CM,Gib,LPT,Tho,Has}.

A similar analysis can be performed in the
other cases mentioned above. The conclusion is
that when a brane ends on another
brane, the end of the first brane looks like
a charged object in the worldvolume theory
of the second brane. The latter is bent according
to (\ref{BST40}) with $p$ the codimension
of the intersection in the second brane, and $r$
the $p$ dimensional
distance to the end of the first brane on the
worldvolume of the second\footnote{This can be 
shown by U-dualizing to a fundamental string
ending on a $Dp$-brane.}.

\begin{figure}
\centerline{\epsfxsize=140mm\epsfbox{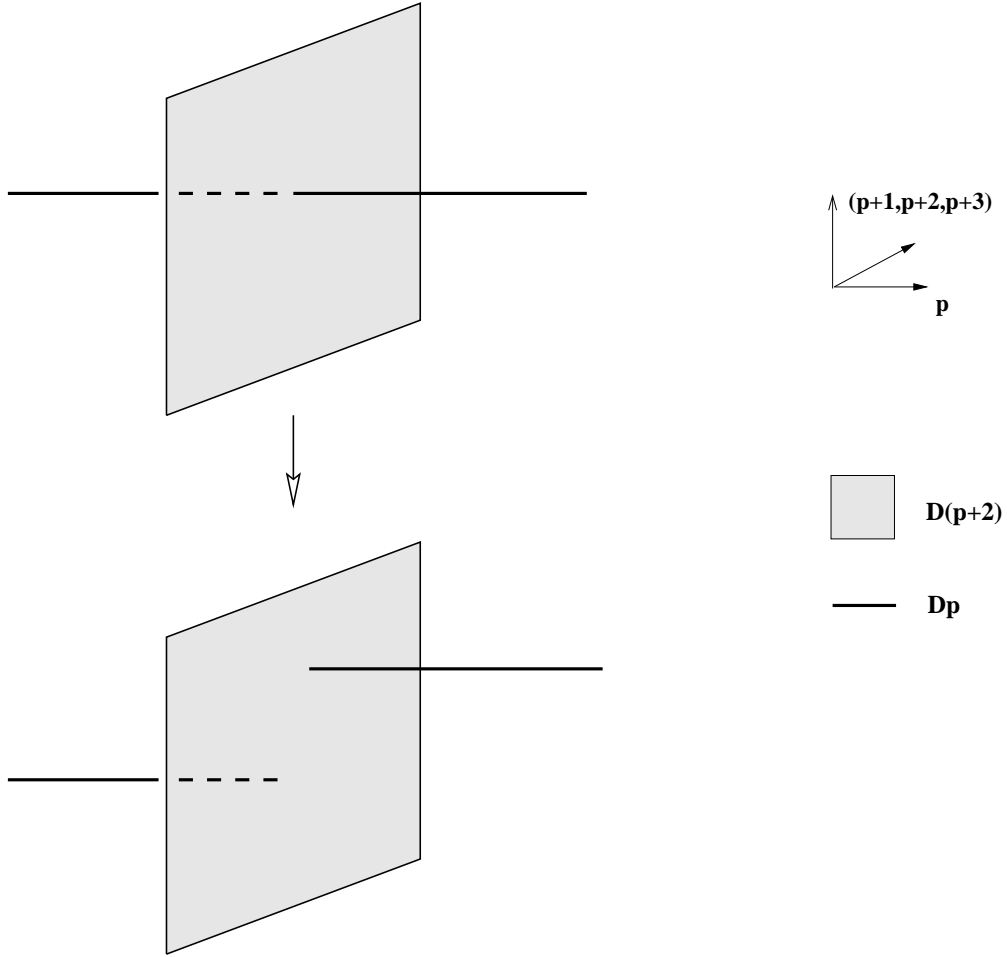}}
\vspace*{1cm}
\caption{A $Dp$-brane intersecting a
$D(p+2)$-brane can split into two
disconnected parts which separate
along the $D(p+2)$-brane.}
\label{five}
\end{figure}
\smallskip

The intersecting brane configurations discussed
earlier in this section are intimately
related to the configurations of branes ending on
branes discussed here. As an example, when the
\ddp\ and $D(p+2)$-branes
of the previous subsection~\footnote{There
we actually considered $p+1$ and $p+3$ branes;
replace $p\to p-1$ there to get the system discussed
here.}
meet in the transverse space $(x^{p+4}, \cdots, x^9)$,
the $p$-brane can split into two parts
$x^p<0$ and $x^p>0$, which can then separate
along the $(p+2)$-brane in the
$(x^{p+1}, x^{p+2}, x^{p+3})$ directions.
Locally, one has then a configuration of
a $p$-brane ending on a $(p+2)$-brane from the
right in $x^p$ and another one ending on it
from the left at a different place 
as shown in Fig.~\ref{five}.

In the gauge theory on the intersection
of the \ddp\ and $NS5$-branes
this realizes geometrically the Higgs branch
of the theory on the D-brane. This will be discussed
in detail in the applications below.

\section{Four Dimensional Theories With $N=4$ SUSY}
\label{D4N4}

At low energies the dynamics on the worldvolume
of $\nc$ parallel $D3$-branes in type IIB string
theory is described by
four dimensional $N=4$ SYM with gauge group $U(\nc)$.
Symplectic and orthogonal
groups can be studied by considering $D3$-branes
near a parallel $O3$-plane. The brane description
provides a natural interpretation of the strong-weak
coupling duality of $N=4$ SYM theories and leads
to a simple geometrical description of BPS saturated
dyons.
In this section we describe this circle of ideas,
starting with the unitary case.

\subsection{Montonen-Olive Duality And Type IIB
S-Duality}
\label{MOS}

Four dimensional $N=4$ supersymmetric gauge
theory with gauge group $G$
can be obtained by dimensionally
reducing $N=1$ SYM from $9+1$ to $3+1$ dimensions.
SUSY (with sixteen supercharges)
places strong constraints on
the structure. The moduli space of vacua
is $6r$ dimensional, where $r$ is the rank of $G$.
It is parametrized by expectation values in the
Cartan subalgebra of the six adjoint scalars
in the $N=4$ multiplet. Generically in moduli
space the gauge symmetry is broken to $U(1)^r$,
but at certain singular subspaces some of the
non-abelian structure is restored. The classical
and quantum moduli spaces are identical in
$N=4$ SYM (in contrast with $N=2$
SYM where the metric on the Coulomb branch
is generally corrected by quantum
effects, and $N=1$ SYM where some or
all of the classical moduli space
can be lifted; these cases will
be discussed later). The leading quantum corrections
modify certain non-renormalizable terms with four
derivatives.

The most singular point
in the moduli space is the origin, where the
full gauge symmetry is unbroken. The theory
at that point is conformal and the gauge
coupling $g_{SYM}$ is an exactly marginal
deformation parametrizing a line of fixed points.
The theory also depends on a parameter
$\theta$ which together
with $g_{SYM}$ forms a complex coupling
\beq
\tau={\theta\over 2\pi}+{i\over g^2_{SYM}}
\label{N41}
\eeq
The theory at the origin of moduli space
is conformal for all $\tau$.

Not all values of $\tau$ correspond to
distinct theories. Since $\theta$ is periodic,
taking $\tau\to\tau+1$ leads to the same theory.
In addition, $N=4$ SYM has a less obvious symmetry,
Montonen and Olive's strong-weak
coupling duality, which takes $\tau\to-1/\tau$,
and exchanges
the gauge algebra ${\cal G}$ with the dual
algebra~\footnote{$\widehat{su}(N_c)=su(N_c)$,
$\widehat{so}(2r)=so(2r)$, $\widehat{so}(2r+1)=sp(r)$.}
${\widehat {\cal G}}$~\cite{MO} 
(see~\cite{Olive,Har,DiVec}
and references therein).
It also acts as electric-magnetic duality
on the gauge field and thus
interchanges electric and
magnetic charges.
Together, the two symmetries generate an $SL(2,Z)$
duality group~\footnote{This was first 
recognized in lattice models~\cite{CR}
and in string theory~\cite{FILQ}.},
which acts on $\tau$ by fractional
linear transformations with integer coefficients:
\beq
\tau\longrightarrow{a\tau+b\over c\tau+d}; \;\;a,b,c,d\in Z,\;
ad-bc=1
\label{N411}
\eeq
We will mostly consider the case of an $SU(2)$
gauge group here, where states carry electric and magnetic
charge under the single Cartan generator, and assemble into
multiplets of $SL(2,Z)$ which contain states with
electric and magnetic charges $(e,m)$ transforming
under $SL(2,Z)$ as:
\beq
\left(
\begin{array}{c}
e\\
m\end{array}\right)
\longrightarrow
\left(\begin{array}{cc}
a&b\\
c&d
\end{array}\right)
\left(\begin{array}{c}
e\\
m\end{array}\right)
\label{N415}
\eeq
For example,
the charged gauge bosons $W^\pm$ with charge $(\pm1,0)$
belong to the same multiplet as the magnetic monopole
with charge $(0, \pm1)$ and various dyons.

To study $N=4$ SYM
with gauge group $U(\nc)$ using branes,
consider $N_c$ parallel $D3$-branes, whose
worldvolumes stretch in $(x^0,x^1,x^2,x^3)$.
The $U(\nc)$ gauge bosons $A_{\mu}^{a\bar b}(x^{\nu})$,
$\mu,\nu=0,1,2,3$,
$a, \bar b=1,\cdots,N_c$, correspond to the ground states of
oriented $3-3$ strings
connecting the $a$'th and $b$'th
threebranes (Fig.~\ref{one}(b)). 
The six scalars  $X^I_{a\bar b}(x^{\mu})$
($I=4,\cdots,9$) in the adjoint representation
of $U(N_c)$ also correspond to $3-3$ strings
describing fluctuations of the threebranes in the
transverse directions $(x^4,x^5,x^6,x^7,x^8,x^9)$.
Together with the ground state fermionic fields
they form an $N=4$ gauge supermultiplet.

The bosonic part of the low energy Lagrangian is given by
(\ref{BST56}, \ref{BST7}), with the $U(\nc)$ gauge coupling
given by $g^2_{SYM}=g_s$ (\ref{BST6}).
The conventional
SYM scalar fields $\Phi^I$ which have dimensions
of energy are related to the scalars $X^I$
which appear naturally in the brane construction
via:
\beq
\Phi^I=X^I/l_s^2
\label{PhiX}
\eeq
The limit in which the theory on the threebrane
decouples from gravity and the four dimensional dynamics
becomes exactly that of $N=4$ SYM at all energy scales
is $l_s\to 0$ with $\gs$,
$\Phi^I$ held fixed. By the latter
it is meant that the energy scale 
studied, $E$, and the scale set
by the expectation values, $\Phi^I$, which typically
are comparable, must be much smaller
than the string scale $1/l_s$, and the Planck
scale $1/l_p$ (which for $\gs\sim 1$ is
comparable to the string scale).
In particular, the transverse separations
of the threebranes parametrizing
the Coulomb branch must satisfy
$\delta x^i\ll l_s, l_p$.

\begin{figure}
\centerline{\epsfxsize=80mm\epsfbox{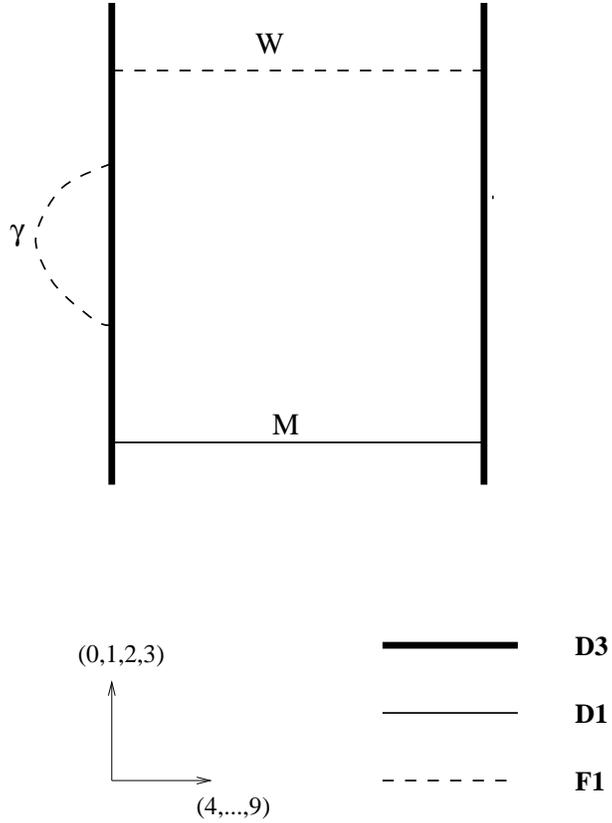}}
\vspace*{1cm}
\caption{$U(2)$ $N=4$ SYM on a pair of
$D3$-branes broken to $U(1)
\times U(1)$ by the separation of
the branes. Dyons in SYM, such as
the photon $\gamma$, the charged gauge
boson $W$, and the magnetic monopole $M$,
are described by $(p,q)$ strings ending
on the threebranes.}
\label{six}
\end{figure}
\smallskip

In the brane picture, the $SL(2,Z)$ Montonen-Olive duality
can be thought of as a remnant of the $SL(2,Z)$
S-duality group of type IIB string theory in the limit
$l_s\to0$. The threebrane is self-dual under S-duality.
The complex worldvolume gauge coupling (\ref{N41})
is the expectation value of the complex type IIB dilaton
$\tau$ (\ref{BST205}) on which S-duality acts by fractional
linear transformations (\ref{N411}), and the type IIB
charges $(p,q)$ which transform under S-duality in an
analogous way to (\ref{N415}) are related to the SYM
charges $(e,m)$. In what follows we will study this
correspondence in more detail in the case $\nc=2$.

An $N=4$ SYM gauge theory with gauge group $G=SU(2)$
is obtained in the brane description by studying the
dynamics on two parallel $D3$-branes~\cite{Tse,GG96}.
Actually, the
gauge group in this case is $U(2)$ but the diagonal
$U(1)\subset U(2)$ will play no role in the discussion
as all fields we will discuss are neutral under it;
therefore it can be ignored.
The six dimensional Coulomb branch 
of the $SU(2)$ SYM theory is
parametrized by the transverse
separation of the two branes
${\vec x}_2-{\vec x}_1$, where $\vec x\equiv (x^4,\cdots, x^9)$.
Displacing the two threebranes from the origin by $\pm \vec x$
(keeping their center of mass corresponding to the decoupled
$U(1)$ fixed at the origin) is equivalent to turning
on a diagonal expectation value for the adjoint scalar
$\vec X$:
\beq
\langle\vec X\rangle=\left(\begin{array}{cc} \vec x&0\\
                          0&-\vec x\end{array}\right)
\label{N416}
\eeq
which breaks $SU(2)\to U(1)$.

The resulting configuration 
is depicted in  Fig.~\ref{six}.
A fundamental string stretched 
between the two $D3$-branes
corresponds to a charged gauge boson in the broken
$SU(2)$ with mass given by (\ref{BST9}).
In the $N=4$ SYM theory it transforms under
electric-magnetic duality (\ref{N411},\ref{N415})
into a dyon. In the brane description
S-duality takes a fundamental string to a $(p,q)$ string;
thus we learn that a dyon with electric-magnetic charge
$(p,q)$ corresponds in the string language to a $(p,q)$
string stretched between the two $D3$-branes.

Note that this is
consistent with our discussion of branes
ending on branes in section \ref{WB} where
we saw that a fundamental string ending on
a $D3$-brane can be thought of as an electric
charge in the worldvolume theory on the
threebrane (\ref{BST39}). Since S-duality
acts on the threebrane as electric-magnetic
duality, this implies that a D-string ending
on a $D3$-brane provides a magnetic source for
the threebrane worldvolume gauge field. The
energy of a D-string stretched between the two
$D3$-branes is $E=2|\vec x|/\gs l_s^2$
or in SYM variables $E=2|\vec\phi|/g^2_{SYM}$
as expected from gauge theory (the mass of the
monopole is $M_{\rm mon}=M_W/g_{SYM}^2$ where
$M_W$ is the mass of the charged $W$-boson).

\subsection{Nahm's Construction Of Monopoles {}From
Branes}
\label{Nahm}

\begin{figure}
\centerline{\epsfxsize=80mm\epsfbox{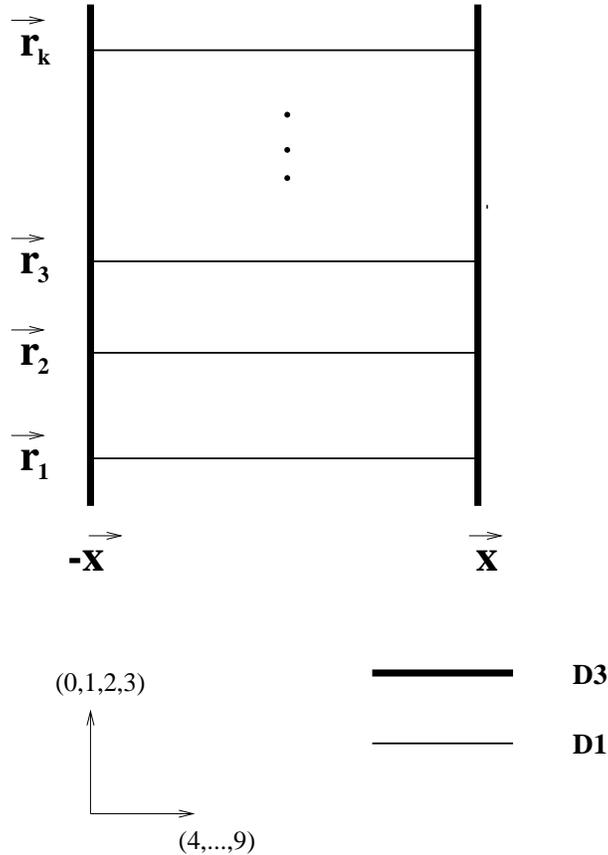}}
\vspace*{1cm}
\caption{A point in the moduli space of $k$ $SU(2)$
monopoles represented by D-strings stretched
between $D3$-branes.}
\label{seven}
\end{figure}
\smallskip

One application of this construction is
to the study of the moduli space of monopoles in gauge
theory. To describe the moduli space of $k$ monopoles
$\MM_k$ one is instructed to study a configuration
of $k$ parallel $D1$-branes
stretched between the two parallel $D3$-branes
(Fig.~\ref{seven}),
say in the $x^6$ direction~\cite{Dia}.
It is easy to check that the configuration
preserves eight of the sixteen supercharges
of the threebrane theory, in agreement with
the fact that the monopoles are $1/2$ BPS objects.
The monopole moduli space $\MM_k$
is the $4k$ real dimensional
space labeled by the locations in $(x^1, x^2, x^3)$
of the $k$ D-strings, and the Wilson lines of the
$k$ $U(1)$ gauge fields along the D-strings, $A_6$.

The brane configuration suggests
an alternative point of view on the space $\MM_k$.
When viewed from the point of view of the
$D3$-branes it describes a moduli space of $k$ monopoles;
from the point of view of the D-strings it
can be thought of as the moduli space of vacua
of the non-abelian gauge theory on the $k$ D-strings
stretched between the $D3$-branes! That theory
lives in the $1+1$ dimensions $(x^0, x^6)$ and,
since the spatial direction $x^6$ is confined to
a finite line segment, it reduces
at low energies to Supersymmertic Quantum Mechanics.
Of course, SQM does not have a moduli space of vacua,
but there is an approximate Born-Oppenheimer notion
of a space of vacua, which arises after integrating
out all the fast modes of the system. The low energy
dynamics is described by a sigma model on the moduli
space $\MM_k$.

The theory on the D-strings has eight supercharges
and the following matter content. The $U(k)$ gauge
field $A_0$ and five adjoint scalars $(\Phi^4,
\Phi^5, \Phi^7,
\Phi^8, \Phi^9)$ have Dirichlet boundary conditions at
$x^6=\pm x$ (the locations of the two threebranes).
The remaining component of the D-string
worldvolume gauge field
$A_6$ and the three adjoint scalars $(\Phi^1,
\Phi^2, \Phi^3)$
have (formally) Neumann boundary
conditions~\footnote{As before, $\Phi^I=X^I/l_s^2$ (\ref{PhiX}).}.

To study the dynamics on the worldvolume of the
D-string we can set to zero all the fields which
satisfy Dirichlet boundary conditions, and the
gauge field $A_6$ (by a gauge choice).
{}From the Lagrangian for $\Phi^1, \Phi^2, \Phi^3$
(\ref{BST56}, \ref{BST7})
\beq
\LL\sim{\rm Tr}\Big( \sum_{I=1}^3
\partial_s \Phi^I\partial_s \Phi^I-\sum_{I,J}
[\Phi^I,\Phi^J]^2\Big)
\label{N417}
\eeq
(where we have denoted $x^6$ by $s$)
it is clear that ground states satisfy
\beq
\partial_s \Phi^I+{1\over2}\epsilon^{IJK}[\Phi^J,
\Phi^K]=0
\label{N418}
\eeq
The boundary conditions of the fields $\Phi^I$ at the edges
of the interval $s=\pm x$ are interesting.
Naively, one would expect that at least as long
as the $k$ D-strings are widely separated in
$\vec r=(x^1, x^2, x^3)$,
we should be able to think of
their locations $\vec r_i$ as the
expectation values of the diagonal components of the
matrix fields $\vec \Phi_{ii}=\vec \phi_i=\vec r_i/l_s^2$
(see (\ref{BST8})).
The off-diagonal components of $\vec \Phi$ are massive
and could be integrated out in the Born-Oppenheimer
approximation. This would lead us to deduce that the
boundary conditions for the matrices $\vec \Phi$ are
\beq
\vec \Phi(s=\pm x)={\rm diag} (\vec \phi_1,\cdots,
\vec \phi_k)
\label{bbb}
\eeq
However, this picture does not make sense for
finite separations of the D-strings. We saw
(after (\ref{BST40})) that
the ``classical'' picture of D-strings attached to the
threebranes at $k$ points
$\vec r=\vec r_1, \cdots, 
\vec r_k$ ($\vec r=(x^1, x^2, x^3)$)
has to be replaced
by a curved threebrane with $s=s(\vec \phi)$ which
approaches the classical location $s=x$
at $|\vec \phi|\to\infty$,
but is actually described asymptotically by:
\beq
s\simeq \sum_{i=1}^k{1\over |\vec \phi-\vec \phi_i|}+x
\label{N419}
\eeq
Each D-string creates a disturbance in
the shape of the threebrane of size
\beq
|\vec \phi-\vec \phi_i|\simeq{1\over s-x}
\label{disturb}
\eeq
which {\em diverges}~\footnote{Note that the
asymptotic expression (\ref{disturb})
becomes more and more reliable in this regime.}
as $s\to x$.

Therefore, for any finite $|\vec\phi_i-\vec\phi_j|$
(as measured in the middle of the $s$ interval)
the different D-strings in fact overlap close to
the edges of the $s$ interval. Hence,
the off-diagonal components of the matrices
$\Phi^I$ ($I=1,2,3$) are light and cannot be
integrated out, and one expects the matrices
$\Phi^I(s\to x)$
not to commute. The only boundary conditions for
$\Phi^I$ that are consistent with (\ref{N418}, \ref{N419})
are (for notational simplicity we have set the center of
mass of the $k$ monopoles $\vec r_0$ to zero)
\beq
\Phi^I\simeq {T^I\over s-x}
\label{N4191}
\eeq
where the $k\times k$ matrices
$T^I$ must satisfy (\ref{N418})
\beq
[T^I,T^J]=\epsilon_{IJK}T^K
\label{N44}
\eeq
and, therefore, define a $k$ dimensional
representation of $SU(2)$. The representation
$T^I$ must furthermore be irreducible;
reducible representations correspond
to splitting the $k$ monopoles into
smaller groups that are infinitely far apart.

As a check, we can compute the size of
the bound state~\footnote{The
$k$ dimensional representation of $SU(2)$
corresponds to $j=(k-1)/2$ and has quadratic
Casimir $T^IT^I=j(j+1)=(k-1)(k+1)/4$.}:
\beq
R^2=\Phi^I\Phi^I\simeq{T^IT^I\over(s-x)^2}={(k-1)(k+1)\over
4(s-x)^2}
\label{N45}
\eeq
\ie\ $R\simeq k/2(s-x)$, roughly the size
of the $k$ D-string system, as given by
(\ref{N419}), $|\vec \phi|\simeq k/(s-x)$.
Clearly a similar analysis
holds at the other boundary of the $s$ interval,
$s=-x$.

Interestingly, we have arrived
(\ref{N418}, \ref{N4191}) at Nahm's description
of the moduli space of $k$ $SU(2)$ monopoles~\cite{Nah}!
The brane realization provides a new perspective
and, in particular, a physical rationale for the
construction. It also makes it easy to describe
generalizations, \eg\ to the case of
the moduli
space of monopoles in higher rank groups.

Monopoles in (broken) $SU(\nc)$ gauge theory can
be discussed by considering a configuration
of $N_c$ $D3$-branes separated in the $x^6$ direction,
and $k_a$ D-strings stretched in $x^6$ between the $a$'th
and the $a+1$'st threebrane, $a=1,\cdots, N_c-1$. Such
configurations preserve eight supercharges and describe
BPS magnetic monopoles of $SU(N_c)$. The magnetic charge
under the natural Cartan subalgebra is
$(k_1,k_2-k_1,\cdots, -k_{N_c-1})$.
The moduli space of such monopoles can be described
by using a generalization of the discussion above.

\subsection{Symplectic And Orthogonal Groups {}From
Orientifolds}
\label{SOG}

To study symplectic and orthogonal groups
we add an orientifold threeplane parallel
to the $N_c$ threebranes. As described in
section \ref{BWC} the low energy worldvolume
dynamics of the $O3$ -- $D3$ system is:
\begin{itemize}
\item
$Sp(N_c/2)$ ($N_c$ even),
$N=4$ SYM in $4d$ if
$Q_{O3}=+{1\over 2}Q_{D3}$.
\item
$SO(N_c)$, $N=4$ SYM in $4d$
if $Q_{O3}=-{1\over 2}Q_{D3}$.
\end{itemize}
In this case we can use the correspondence
between gauge theory and brane theory to
learn about strong coupling properties of
orientifold planes, by using the correspondence
between Montonen-Olive duality in gauge theory
and S-duality in string theory.
{}From gauge theory we expect $SO(2r)$ to be
self-dual under $SL(2,Z)$ while $SO(2r+1)$
and $Sp(r)$ should be dual to each other.
The $SO(2r)$ case works in the obvious way:
the $D3$-branes and $O3$ plane are self-dual
under $SL(2,Z)$. In the non-simply-laced case
there is a new element. Consider a weakly coupled
$SO(2r+1)$ gauge theory. The orientifold
charge is $-Q_{D3}/2$; the $6r$ dimensional
Coulomb branch corresponds to removing $r$
pairs of threebranes from the orientifold
plane. A single threebrane which does not have a
mirror remains stuck at the orientifold.

When the gauge coupling becomes large there are
two ways of thinking about the system. We can
either continue thinking about it as a (strongly
coupled) $SO(2r+1)$ gauge theory, or
relate it to a weakly coupled theory by performing
a strong-weak coupling S-duality transformation.
{}From gauge theory we know that the result should
be a weakly coupled $Sp(r)$ theory, which is described
by an orientifold with charge $+Q_{D3}/2$.

Thus, Montonen-Olive duality of gauge theory teaches
us that a ``bound state'' of an $O3$-plane with negative
Ramond charge and a single $D3$-brane embedded in it
(a configuration with Ramond charge $(-1/2+1)Q_{D3}$)
transforms under S-duality of type IIB string theory
into an $O3$-plane with Ramond charge $+Q_{D3}/2$~\cite{EGKT}.

Monopoles in broken $SO/Sp$ gauge theory are described
as before by $D$-strings stretched between different
$D3$-branes. Consider for example the rank one case
$\nc=2$. For positive  orientifold charge the gauge
group is $Sp(1)\simeq SU(2)$ and the moduli space of
$k$ $SU(2)$ monopoles that we have discussed previously
can be studied by analyzing the worldvolume theory
of $k$ $D$-strings connecting the single ``physical''
$D3$-brane to its mirror image. The gauge
group $U(k)$ is replaced by $SO(k)$, and the matrices
$\Phi^I$ (\ref{PhiX}) and $A_6$ become now symmetric
$k\times k$ matrices. The discussion (\ref{N417}-\ref{N44})
can presumably be repeated, although this has not
been done in the literature.

For negative orientifold charge the gauge group is
$SO(2)\simeq U(1)$ and one does not expect non-singular
monopoles to exist. This means that $D$-strings cannot
connect the single physical $D3$-brane to its mirror
image. This is related by S-duality to the fact,
which was discussed in section \ref{OR},
that for negative orientifold charge the
ground states of fundamental strings
stretched between the $D3$-brane and its image are
projected out.

\subsection{The Metric On The Moduli Space Of Well-Separated
Monopoles}
\label{metr}

The explicit form of the moduli space
metric for $k$ well
separated monopoles in $SU(2)$ gauge theory
is known. Setting
$\gs=1$, $2|\vec x|=1$ (\ref{N416}), and
denoting the locations of the monopoles in
$(x^1, x^2, x^3)$ by $\vec r^i$ and the Wilson lines
$A_6$ by $\theta^i$, so that the $4k$ dimensional
monopole moduli space is labeled by $(\vec r^i,
\theta^i)$, it is~\cite{GM}:
\beq
ds^2=g_{ij}d\vec r^i\cdot d\vec r^j+(g^{-1})_{ij}
d\tilde\theta^i d\tilde\theta^j
\label{metwide}
\eeq
where
\bea
&g_{jj}=1-\sum_{i\not=j}{1\over r_{ij}};\;\;\;
({\rm no\;sum\;over\;}\;j) \nonumber\\
&g_{ij}={1\over r_{ij}};\;\;\;i\not=j\nonumber\\
&d\tilde\theta^i=d\theta^i+\vec W_{ik}\cdot
d\vec r^k\nonumber\\
&\vec W_{jj}=-\sum_{i\not=j} \vec w_{ij};\;\;\;
({\rm no\;sum\;over\;}\;j)\nonumber\\
&\vec W_{ij}=\vec w_{ij};\;\;\; (i\not=j)
\label{defsmon}
\eea
$r_{ij}=|\vec r_i-\vec r_j|$
and $\vec w_{ij}$ is the vector potential of a Dirac
monopole located at the point $\vec r_i$
evaluated at the point $\vec r_j$ (\ref{KK3}).

In the brane language, one can think of the
metric (\ref{metwide}, \ref{defsmon}) as
the perturbative metric on the ``Coulomb
branch'' of the $U(k)$ gauge theory on the
D-strings. Classically, $g_{jj}=1$,
$g_{ij}=0$ (for $i\not=j$).
The corrections proportional
to $1/r_{ij}$ in (\ref{defsmon}) arise at
one loop and can be naturally interpreted
as due to the asymptotic curving of the
threebranes by the D-strings (\ref{BST40}).
For example, the diagonal components
$g_{jj}$ can be thought of
as describing the motion of the $j$'th
D-string in the background of the other
$k-1$ strings which curve the two threebranes
such that the $\vec r$ dependent
distance between them (for large $|\vec r|$) is
\beq
\delta x^6=1-\sum_{i\not=j}{1\over |\vec r-\vec r_i|}
\label{delx}
\eeq
{}From the point of view of the D-string
theory one can interpret (\ref{delx}) as an
$\vec r$ dependent gauge coupling. As we will
see in section \ref{D4N2},
for systems with eight supercharges
the metric $g$ is related to the
gauge coupling by SUSY. This explains the relation
between (\ref{delx}) and the first line of
(\ref{defsmon}).

Due to
$(4,4)$ SUSY, there are no further
perturbative corrections to the metric
beyond one loop. Non-perturbatively,
(\ref{defsmon}) cannot be exact, \eg\ since
the diagonal components of the metric
are not positive definite. In the brane
language, the formula for the curving
of the branes (\ref{delx}) is only
valid asymptotically for large $|\vec r|$
while for $|\vec r-\vec r_i|\to 0$, $x^6$ is clearly
modified; instead of diverging, the two
threebranes effectively ``meet in the middle''
of the $x^6$ interval. Thus,
(\ref{delx}) must be modified.

\begin{figure}
\centerline{\epsfxsize=80mm\epsfbox{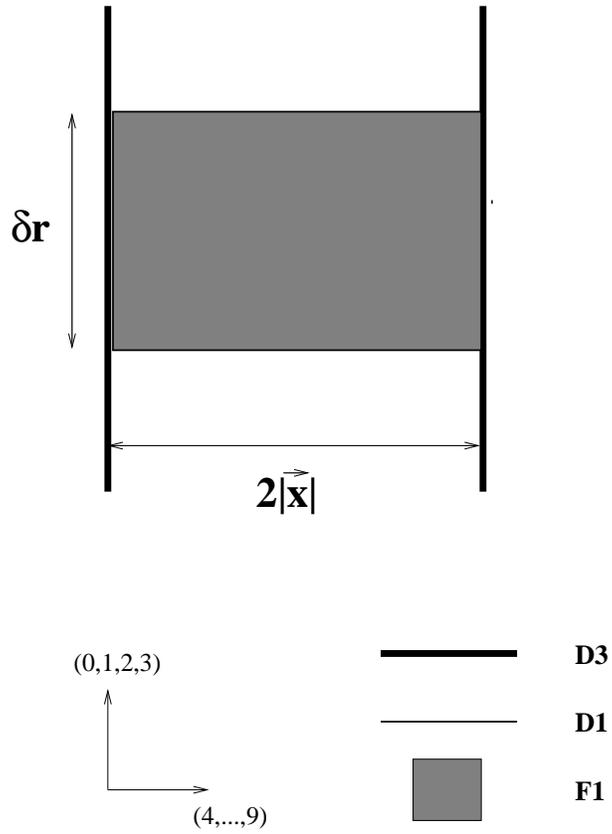}}
\vspace*{1cm}
\caption{Non-perturbative corrections to
the metric on moduli space are due
to Euclidean fundamental
strings stretched between the threebranes
and adjacent D-strings.}
\label{eight}
\end{figure}

One can think of the non-perturbative
corrections to the metric (\ref{defsmon})
as due to Euclidean fundamental strings
stretched between the two $D3$-branes and two
adjacent D-strings (see  Fig.~\ref{eight}).
The action of such an instanton
is proportional to its area,
\beq
S=2|\vec x|\delta r/l_s^2=2|\vec\phi|\delta r=M_w\delta r
\label{nonprtcor}
\eeq
where $2\vec \phi$ is the Higgs expectation value
in the broken
$SU(2)$ gauge theory and $\delta r$ is the separation
between adjacent monopoles.
The corresponding non-perturbative corrections
go like $\exp(-S)\sim\exp(-M_W\delta r)$ where $M_W$
is the mass of the charged $W$-boson.
This is consistent with the fact that the
size of the magnetic monopole in broken $SU(2)$
gauge theory is $M_W^{-1}$, much larger than
its Compton wavelength $M_{\rm mon}^{-1}=
g_{SYM}^2M_W^{-1}$ for weak coupling
$g_{SYM}$.

\smallskip

Note that the instanton effects (\ref{nonprtcor})
are non-perturbative in $l_s^2=\alpha^\prime$,
but they survive in the classical
string limit $\gs\to 0$. Thus they can be thought
of as worldsheet instanton corrections to the
metric (\ref{defsmon}).

\section{Four Dimensional Theories With $N=2$ SUSY}
\label{D4N2}
\subsection{Field Theory Results}
\label{RFTR}
The $N=2$ supersymmetry algebra contains
eight supercharges transforming as two
copies of the ${\bf 2}+{\bf \bar 2}$ of $Spin(1,3)$.
All $N=2$ theories have a global $SU(2)_R$ symmetry
which acts on the two supercharges. Scale invariant
theories have in addition a $U(1)_R$ symmetry under
which the chiral supercharges have charges $\pm1$.

To study $N=2$ supersymmetric gauge theory with
gauge group $G$ one is interested in two kinds of
multiplets. The vectormultiplet
contains a gauge field $A_\mu$,
two Weyl fermions $\lambda_\alpha$, $\psi_\alpha$
and a complex scalar $\phi$, all in the adjoint
representation of $G$. The fermions
$\lambda$, $\psi$ transform in the ${\bf 2}$ of
$SU(2)_R$; $A_\mu$ and $\phi$ are singlets.
Under $N=1$ SUSY the vectormultiplet decomposes
into a vector superfield~\footnote{We use the notations
of~\cite{WB}, except for replacing $v_m$ by $A_\mu$.}
\beq
V=-\theta\sigma^\mu\bar\theta A_\mu-i\bar\theta^2
(\theta\lambda)+i\theta^2(\bar\theta\bar\lambda)
+{1\over2}\theta^2\bar\theta^2D
\label{RFTR1}
\eeq
with the gauge covariant field strength
\beq
W_\alpha={\bar \DD}^2\left(e^{2V}\DD_\alpha e^{-2V}
\right)
\label{RFTR2}
\eeq
and a chiral superfield
\beq
\Phi=\phi+{\sqrt2} \theta\psi+\theta^2 F
\label{RFTR3}
\eeq
In $N=1$ superspace, the low energy
Lagrangian describing the vectormultiplet
is
\beq
\LL_{\rm vec}={\rm Im} \;{\rm Tr}\; \left[\tau\left(
\int d^4\theta\Phi^\dagger e^{-2V}\Phi+
\int d^2\theta W_\alpha W^\alpha\right)\right]
\label{RFTR4}
\eeq
where the trace runs over the group $G$ and
$\tau$ is the complex coupling (\ref{N41}).
The Lagrangian (\ref{RFTR4}) is invariant
under the $U(1)_R$ symmetry $\Phi\to
e^{2i\beta}\Phi(e^{-i\beta}\theta)$,
which is a consequence of its (classical)
conformal invariance. Thus $\Phi$ has $R$-charge
two.

In components, the bosonic part of the
Lagrangian $\LL_{\rm vec}$ includes
kinetic terms for the  fields  (\ref{BST56})
and a potential for the adjoint scalars $\phi$,
$\phi^\dagger$ analogous to (\ref{BST7}),
\beq
V\sim {\rm Tr} [\phi^\dagger,\phi]^2
\label{RFTR6}
\eeq
$N=2$ SUSY
gauge theories in four dimensions
can be obtained from $N=1$
SUSY theories in six dimensions by dimensional
reduction, \ie\ dropping the dependence
of all fields on two of the coordinates,
say $(x^4, x^5)$. The adjoint chiral superfield in the
vectormultiplet $\Phi$ (\ref{RFTR3}) corresponds
to $A_4$, $A_5$;
the potential (\ref{RFTR6}) arises from
the commutator terms in the
action (\ref{BST55}).

The second multiplet of interest is the
hypermultiplet, which in $N=1$ notation
consists of two chiral superfields, $Q$,
$\tilde Q$ in a representation $R$ of the
gauge group (and thus contains $2{\rm dim}R$
complex scalars and Weyl fermions).
The scalar components of $Q$, $\tilde Q$
transform as a doublet under $SU(2)_R$ and
carry no charge under $U(1)_R$; the fermions
are $SU(2)_R$ singlets and carry $U(1)_R$
charge one. The low energy
Lagrangian describing the hypermultiplet is
(in $N=1$ superspace):
\beq
\LL_{\rm hyper}=
\int d^4\theta \left(Q^\dagger e^{-2V}Q+
\tilde Q^\dagger e^{-2V}\tilde Q\right)+
\int d^2\theta\tilde Q \Phi Q +{\rm c.c.}
\label{RFTR7}
\eeq
where $V=V_aT^a$, $a=1,\cdots, {\rm dim}\; G$,
and $T^a$ are generators of $G$ in the representation
$R$.

The theory described by (\ref{RFTR4}, \ref{RFTR7})
has a Coulomb branch corresponding to matrices
$\phi$ satisfying (\ref{RFTR6})
$[\phi, \phi^\dagger]=0$. It is parametrized
by $r={\rm rank}\; G$ complex moduli corresponding
to $\phi$ in the Cartan subalgebra of $G$,
$\phi=\sum_{i=1}^r\phi_iT^i$. The gauge group is
generically broken to $U(1)^r$ and the low energy
dynamics is that of $r$ $U(1)$ vectormultiplets.
$N=2$ SUSY ensures that the moduli space of
vacua is not lifted by quantum effects but
the metric on it is in general modified. The general
form of the action consistent with $N=2$ SUSY 
is~\cite{ST,dWitt,Gates,Sei88}:
\beq
\LL_{\rm vec}={\rm Im} {\rm Tr}\left[\int d^4\theta
{\partial\FF(\Phi)\over\partial \Phi_i}\bar\Phi_i+
{1\over2}\int d^2\theta{\partial^2 \FF(\Phi)\over
\partial \Phi_i\partial \Phi_j}W_\alpha^i W^\alpha_j
\right]
\label{RFTR8}
\eeq
$\FF$ is a holomorphic function of the moduli
known as the prepotential. It determines the
low energy $U(1)^r$ gauge coupling matrix
$\tau_{ij}$:
\beq
\tau_{ij}={\partial^2\FF\over\partial
\phi_i\partial \phi_j}
\label{RFTR9}
\eeq
and the metric on the moduli space
\beq
ds^2={\rm Im} \;
\tau_{ij}\;d\phi_i d\bar\phi_j
\label{RFTR10}
\eeq
Comparing to (\ref{RFTR4}) we see that,
classically, the prepotential is quadratic,
\beq
\FF_0={1\over2}\tau_{cl}\Phi_i\Phi^i
\label{RFTR11}
\eeq
After adding the one loop corrections~\footnote{For
simplicity, we only give the result for gauge
theory without matter.},
it takes the form:
\beq
\FF_1={i\over4\pi}\sum_{\vec \alpha>0}
(\vec\alpha\cdot\vec\Phi)^2
\log{(\vec\alpha\cdot\vec\Phi)^2\over\Lambda^2}
\label{RFTR12}
\eeq
where the sum runs over positive roots of the
Lie algebra of $G$.
The logarithm breaks $U(1)_R$ and is related
through the multiplet of anomalies to the
one loop beta function. Higher order perturbative
corrections are absent due to a
non-renormalization theorem, but non-perturbatively
(\ref{RFTR12}) receives a series of instanton
corrections that fall off algebraically at large
$\Phi$ but are crucial for the structure at small
$\Phi$.

Seiberg and Witten showed that the prepotential $\FF$
can be computed exactly and, in fact, its second
derivative $\tau_{ij}$ (\ref{RFTR9}) is the period
matrix of a Riemann surface~\cite{SW9407}. The moduli space of
vacua of the $N=2$ SYM theory is thus parametrized
by the complex structure of an auxiliary two
dimensional Riemann surface whose physical role seems
mysterious. One of our main goals in this section
will be to elucidate the meaning of this surface
by embedding the problem in string theory.

The prepotential is also important for determining
the mass of BPS states in the theory. {}From the
supersymmetry algebra one can deduce that the mass
of BPS saturated states with electric charge
$(e_1,\cdots, e_r)$ and magnetic charge
$(m^1,\cdots, m^r)$ under the $r$ unbroken
$U(1)$ gauge fields is
\beq
M={\sqrt2} |Z|;\;\; Z=\phi^i e_i + \phi_i^D m^i
\label{RFTR13}
\eeq
where $Z$ is the central charge and
\beq
\phi_i^D={\partial\FF\over\partial\phi^i}
\label{RFTR14}
\eeq

In general, $N=2$ SYM theories have also
Higgs branches in which the rank of the unbroken
gauge group is decreased. The full phase
structure is in general rather rich
(see~\cite{APSei,APS2} for a more detailed
discussion).
In the remainder of this section
we will describe it in a few examples
using branes.

\subsection{Threebranes Near Sevenbranes}
\label{TBSB}

As a first example of four dimensional
$N=2$ SYM on branes we
consider the low energy worldvolume theory on
threebranes in the presence of sevenbranes
and an orientifold sevenplane in type IIB
string theory~\cite{BDS}. It can be
thought of as a special case of the $Dp$ -- $D(p+4)$
system of section (\ref{BST}) with a few special
features due to the fact that $p+4=7$ is
sufficiently large.
In particular, the RR flux of $1+7$ dimensional
objects does not have enough non-compact transverse
directions to escape. Therefore, we should consider
configurations with vanishing total RR charge.
In this section we study a particular configuration
of this sort.

Consider an $O7$-plane with worldvolume in
the $(x^0,x^1,\cdots, x^7)$ directions,
at a point in the $(x^8,x^9)$ plane. Its
RR charge $Q_{O7}$ is (\ref{BST11}) $Q_{O7}=-8Q_{D7}$.
To cancel this charge we can add $N_f=4$ $D7$-branes
and their four mirror images (a total of 8 $D7$-branes
with charge $Q_{D7}$ each) parallel to the orientifold
sevenplane. When the $D7$-branes coincide with the
orientifold plane there is an $SO(8)$ gauge symmetry
on their $1+7$ dimensional worldvolume; when they are
separated from the orientifold this symmetry is generically
broken to $U(1)^4$. In a complex parameterization of the
$(x^8, x^9)$ plane,
\beq
w\equiv x^8+ix^9
\label{D3D71}
\eeq
we can choose the location of the orientifold
plane to be
\beq
w(O7)=0
\label{D3D72}
\eeq
The locations of the four $D7$-branes and their
mirror partners in the $(x^8, x^9)$ plane
will be denoted by $m_i$, $-m_i$, respectively.

In addition, we place a $D3$-brane and its mirror
image at~\footnote{The location of the threebrane in the
directions along the $O7/D7$ are not important for
what follows.}
$w$, $-w$, respectively.
As explained in section \ref{BST},
the low energy $1+3$ dimensional worldvolume dynamics on
the threebranes is an $Sp(1)\simeq SU(2)$ gauge theory
with eight supercharges, namely, an $N=2$ supersymmetric
gauge theory in four dimensions. The neutral gauge boson
$W_\mu^3$ corresponds to the ground state of an open string
both of whose ends terminate on the threebrane. The charged
gauge bosons $W_\mu^\pm$ correspond to the ground states
of strings stretched between the $D3$-brane and its mirror
image. The $D7$-branes are heavy objects; thus from
the point of view of $1+3$ dimensional physics, their
$SO(8)$ gauge symmetry is ``frozen,'' \ie\ the corresponding
gauge coupling vanishes. Moduli in the sevenbrane theory
give rise to parameters in the $1+3$ dimensional
Lagrangian.

The location of the threebrane in the $(x^8, x^9)$
plane corresponds to the expectation value of the
complex chiral field in the adjoint of $SU(2)$
\beq
\Phi_{ab}(x^{\mu})\equiv X^8_{ab}+iX^9_{ab}; \qquad a,b=1,2;
\qquad {\rm Tr}\; \Phi=0
\label{D3D77}
\eeq
which belongs to the $SU(2)$ vectormultiplet
(\ref{RFTR3}). It can be diagonalized to
\beq
\langle\Phi\rangle=\left(\begin{array}{cc} w&0\\
                          0&-w\end{array}\right)
\label{D3D78}
\eeq
When $w=0$, the minimal length of strings
stretched between the $D3$-branes vanishes
and the charged gauge bosons are massless.
When $w\neq 0$,
(\ref{D3D78}) breaks $SU(2)$
to $U(1)$. Correspondingly, the strings
stretched from the threebrane to its mirror
image have minimal length $2|w|$ -- the mass
of $W^{\pm}$ (in string units).

The $D7$-branes give rise to $N_f=4$
fundamental hypermultiplets
$(Q^i,\tilde{Q}_i)$. Their locations
$m_i$ are the bare complex masses of quarks.
Analyzing configurations of strings stretched
between the $D3$ and $D7$-branes we see that
at tree level the  superpotential is just that
expected on the basis of $N=2$ SUSY:
\beq
W=\sum_{i=1}^4 (Q^i\Phi\tilde{Q}_i-m_iQ^i\tilde{Q}_i)
\label{D3D79}
\eeq
The effective masses of the quarks $m_i-w$ and $m_i+w$
are the locations of the four $D7$-branes and
their mirror images, respectively, relative to
the $D3$-brane. The $SO(8)$ gauge symmetry on
the worldvolume of the $D7$-branes turns into
a global symmetry of the four dimensional gauge
theory on the threebranes. It is broken when
$m_i\neq 0$.

Just like in the $N=4$ SYM case discussed
in the previous section,
the complex gauge coupling of the $N=2$ SYM
theory on the threebrane
corresponds to the complex dilaton
(\ref{BST205}) of type IIB string theory.
The $D7$-branes and $O7$-plane carry charge
under the complex dilaton field. Thus it is
non-trivial in their presence and, in particular,
when we go once around a $D7$-brane,
$\tau$ has a monodromy: $\tau\to\tau+1$.
Far from the $D7$-branes that are
located at $w=m_i$ and from the $O7$-plane located
at $w=0$ we expect $\tau$ to behave as:
\beq
\tau(w)=\tau_0+{1\over 2\pi i}\Big[\sum_{i=1}^4
\Big(\log(w-m_i)+\log(w+m_i)\Big)-8\log\, w\Big]
\label{D3D710}
\eeq
since there is charge $+1$ at each $w=\pm m_i$ and
charge $-8$ at $w=0$.

Note that one can use the
above analysis to understand
the identification of the complex dilaton of IIB
string theory (\ref{D3D710}) with the gauge coupling
of the theory on the $D3$-brane.
The metric on the $(x^8, x^9)$ plane implied by (\ref{D3D710})
can be interpreted either as the metric
induced by the $O7$-plane and $D7$-branes or
as the metric on the
Coulomb branch of the $N=2$ SYM theory
on the worldvolume of the threebrane.
In the first interpretation this metric is determined
by the complex coupling of type IIB string theory;
in the second, it is related by (\ref{RFTR9},
\ref{RFTR10}) to the complex gauge coupling
$\tau$. This establishes the relation between the
two $\tau$'s.

The complex coupling $\tau$ is gauge invariant;
this is made manifest by rewriting
(\ref{D3D710}) as
\beq
\tau(u)=\tau_0+{1\over 2\pi i}\Big[\sum_{i=1}^4
\log(u-m_i^2)-4\log\, u\Big]
\label{D3D711}
\eeq
where $u$ is the gauge invariant modulus:
\beq
u={1\over 2}{\rm Tr}\; \Phi^2 =w^2
\label{D3D712}
\eeq
The semiclassical result (\ref{D3D711})
corresponds in gauge theory to
the one loop corrected prepotential (\ref{RFTR12}).
As in the brane picture, semiclassically,
the $SU(2)$ gauge symmetry is restored
at the origin $u=0$ where $W^{\pm}$ become massless, while
quarks $(Q^i,\tilde{Q}_i)$ become massless when $u=m_i^2$.
The appearance of new massless states
is the reason for the singularities at $u=0,m_i^2$
in eq. (\ref{D3D711}).
The coefficient $4$ in front of $\log\, u$ is due to the
fact that $W^{\pm}$ carry twice the electric charge of a quark,
and the relative sign between the two logs in (\ref{D3D711})
is due to the fact that the $W^{\pm}$ belong to a
vectormultiplet whose contribution to the beta function has
an opposite sign to that of a hypermultiplet.

The one loop result (\ref{D3D711}) is
not corrected perturbatively, but it
cannot be exact, \eg\ since
it does not satisfy ${\rm Im}\,\tau\geq 0$
everywhere in the $u$ plane; for small
$u$, ${\rm Im}\,\tau$  becomes large and negative.
Therefore, we expect that strong coupling effects
will modify the solution for finite $u$~\cite{SW9407,SW9408,Sen9605}.
Indeed, as discussed in section \ref{RFTR},
in the $N=2$ SYM analysis one
finds that instanton corrections modify (\ref{D3D711}).
The exact effective coupling is a modular parameter $\tau(u)$
of a torus described by the elliptic curve
\beq
y^2=x^3+f(u,\tau_0)x+g(u,\tau_0)
\label{D3D713}
\eeq
where $x,y,u\in CP^1$, $f(u)$ is a polynomial of degree
two, $g(u)$ is a polynomial of degree three in the
gauge invariant modulus $u$, and
$\exp(i\pi\tau_0)\equiv\Lambda$
is the ``QCD scale'' of the theory.
In the semiclassical limit, namely, for large ${\rm Im}\,\tau_0$
and $|u|\gg |\Lambda|^2$,
the exact $\tau (u)$ can be rewritten as (\ref{D3D711}).
Strong coupling dynamics splits
the singularity at the origin
into two singularities at $u=\pm\Lambda^2$
corresponding to a monopole or dyon becoming massless.

The full non-perturbative description
of the type IIB vacuum discussed above
involves threebranes in F-theory on $K3$~\cite{Vaf}
-- a compactification of the
IIB string on $CP^1$ labeled by the coordinate $u$,
with a non-trivial complex dilaton describing a
two-torus with modular parameter $\tau(u)$ for each
point on $CP^1$. This elliptically fibered surface
is given by the algebraic equation (\ref{D3D713}).

In the F-theory description the
threebrane moves in the background of six sevenbranes
located at the singularities of the curve (\ref{D3D713}).
In the weak string-coupling limit four of these
branes can be regarded as $(1,0)$ sevenbranes, namely,
conventional D-branes,
while the other two are a $(0,1)$ sevenbrane and a
$(2,1)$ sevenbrane related to a $D7$-brane
by $SL(2,Z)$ S-duality transformations.
The $(1,0)$ sevenbranes are free to move
in the $u$ plane, while the $(0,1)$ and $(2,1)$
sevenbranes are stuck at $u=\pm\Lambda^2$. As
$\gs\to0$ these branes approach each other
and are described at weak coupling by an $O7$-plane.
BPS saturated states with
electric and magnetic charges $(e,m)=(p,q)$
in the four dimensional $N=2$ SYM theory
on the threebrane can be described by $(p,q)$
strings stretched between the $(p,q)$ sevenbranes
and the $D3$-brane~\cite{Sen9608}.

\subsection{Branes Suspended Between Branes}
\label{BSB}

The fact that branes can end on other branes
was deduced in subsection \ref{BEB} by starting
from a fundamental string ending on a D-brane
and applying U-duality. Following the same
logic we can deduce that branes can be suspended
between other branes by starting with a configuration
of fundamental strings stretched between two D-branes
and applying a chain of duality transformations.
As in subsection  \ref{BEB}, one can get this way
\ddp-branes stretched between two $D(p+2)$
or $NS5$-branes. The special case of D-strings
stretched between two $D3$-branes was used in section
\ref{D4N4} to describe BPS saturated
`t Hooft-Polyakov monopoles of a broken $SU(2)$ gauge
theory.

In this section we will study similar configurations
with eight supercharges describing $3+1$  dimensional
physics, and use them to learn about $N=2$ SYM.
The starting point of our discussion
will be brane configurations in type IIA string theory
consisting of solitonic (NS) fivebranes, $D4$-branes and
$D6$-branes as well as  orientifold planes $O4$ and $O6$
parallel to the D-branes. Using eqs. (\ref{BST4}, \ref{BST14},
\ref{BST15}) it is not difficult to check that any combination
of two or more of the following objects:

\beq
\begin{array}{ll}
\mbox{$NS5:$}&  \mbox{$(x^0, x^1, x^2, x^3, x^4, x^5)$}\\
\mbox{$D4/O4:$}&  \mbox{$(x^0, x^1, x^2, x^3, x^6)$}\\
\mbox{$D6/O6:$}&  \mbox{$(x^0, x^1, x^2, x^3, x^7, x^8, x^9)$}
\end{array}
\label{BSB1}
\eeq
preserves eight of the thirty two supercharges
of IIA string theory (\ref{BST2}). In eq.
(\ref{BSB1}) we specified the directions
in which each of the branes is stretched.

The Lorentz group $SO(1,9)$ is broken
by the presence of the branes to:
\beq
SO(1,9)\longrightarrow SO(1,3)\times SO(2)\times SO(3)
\label{BSB2}
\eeq
where the $SO(1,3)$ factor acts on
$(x^0, x^1, x^2, x^3)$, the $SO(2)$ on
$(x^4, x^5)$ and the $SO(3)$ on $(x^7, x^8, x^9)$.
We will be interested in physics in the
$1+3$ dimensional spacetime common to all
the branes labeled by $(x^0, x^1, x^2, x^3)$;
thus we interpret the $SO(1,3)$ factor in
(\ref{BSB2}) as Lorentz symmetry and the
$SO(2)$, $SO(3)$ factors as global symmetries.
Due to the ten dimensional origin of these
global symmetries, the supercharges transform
as doublets under $SO(3)$, and are charged
under $SO(2)$. Thus, these are $R$-symmetries.
In fact, the $SO(3)$ can be identified with
the global $SU(2)_R$ of $N=2$ SYM described
in section \ref{RFTR} while the $SO(2)$ can
be identified with the $U(1)_R$ symmetry.

To study situations with interesting $1+3$
dimensional physics some of the branes must
be made finite. We next turn to a discussion
of some specific configurations and their
physics~\cite{HW,EGK}. We start with a description of the
``classical'' type IIA string picture in a
few cases involving unitary, symplectic and
orthogonal groups with matter in the fundamental
representation of the gauge group, as well as a
few more complicated examples, and then
study quantum effects.

\subsubsection{Unitary Gauge Groups}
\label{TBCS}

\begin{figure}
\centerline{\epsfxsize=60mm\epsfbox{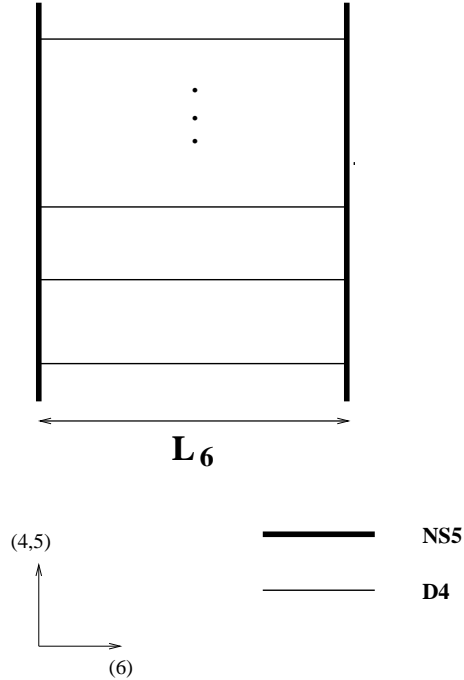}}
\vspace*{1cm}
\caption{$\nc$ $D4$-branes stretched between
$NS5$-branes describe $N=2$ SYM with $G=U(\nc)$.}
\label{nine}
\end{figure}

Consider two infinite $NS5$-branes oriented
as in (\ref{BSB1}), separated by a distance
$L_6$ in the $x^6$ direction and located at the
same point in $(x^7, x^8, x^9)$. We can stretch
between them in the $x^6$ direction $\nc$
$D4$-branes oriented as in (\ref{BSB1}); see 
Fig.~\ref{nine}.

\smallskip

To analyze the low energy physics corresponding to
this configuration it is important to distinguish
between three kinds of excitations of the vacuum:
1) modes that live in the ten dimensional bulk of
spacetime; 2) modes that live on the two $NS5$-branes;
3) modes that live on the fourbranes.
For an observer living in the $1+3$ dimensions
$(x^0, x^1, x^2, x^3)$ the first two sets of excitations
appear at low energies and in the weak string
coupling limit to be frozen at their classical
values by an argument similar to that given around eq.
(\ref{BST37}). Essentially, since they correspond to
higher dimensional excitations any long wavelength
fluctuations away from the classical values of these
fields are suppressed by infinite volume factors.

Excitations attached to the
$\nc$ fourbranes do not have this property. Despite living
in one higher dimension ($x^6$), they are dynamical in
$1+3$ dimensions since the fourbranes are finite in $x^6$.
Thus excitations of the fourbranes can be thought of as
fields living in the $1+4$ dimensional space $R^{1,3}\times I$
where $I$ is a finite interval (of length $L_6$).
Just like in Kaluza-Klein theory, for distance scales
much larger than $L_6$ their physics looks $1+3$ dimensional.
Depending on the boundary conditions at the ends of the
interval, the different fields do or do not give rise to
light fields in $1+3$ dimensions.

The analysis of the boundary conditions can be done
using the results of subsection \ref{BEB}. The light
excitations on a stack of $\nc$ infinite $D4$-branes
(\ref{BSB1}) are a five dimensional $U(\nc)$ gauge
field $A_\mu$, $\mu=0,1,2,3,6$ and five scalars in
the adjoint of $U(\nc)$ corresponding to transverse
fluctuations of the fourbranes,
$(X^4, X^5, X^7, X^8, X^9)$. As we saw in subsection
\ref{BEB}, when the $\nc$ fourbranes end on fivebranes,
$(X^7, X^8, X^9)$ as well as $A_6$ satisfy
Dirichlet boundary conditions on both ends of the
interval $I$. Therefore, they give rise in $1+3$
dimensions to states with masses of order $1/L_6$,
which are invisible in the low energy limit of interest,
$E\ll 1/L_6$. The remaining light degrees of
freedom, the $U(\nc)$ gauge field  $A_\mu$,
$\mu=0,1,2,3$ and the adjoint scalars $(X^4, X^5)$,
satisfy free boundary conditions on $I$ and, therefore,
contribute a $U(\nc)$ vectormultiplet.

Thus, we conclude that the light excitations of 
the brane configuration above describe an $N=2$
SUSY gauge theory with gauge group $U(\nc)$ and no
matter~\footnote{This
theory can be thought of as a reduction
of $N=1$ SYM in $1+5$ dimensions down to $1+3$
dimensions. In the brane description this
process of dimensional reduction is described by
compactification, followed by T-duality and subsequent
decompactification of $(x^4, x^5)$. The six dimensional
version of the theory is obtained by replacing the
fourbranes in (\ref{BSB1}) by sixbranes stretched also
along $(x^4, x^5)$. This enhances the unbroken Lorentz
symmetry (\ref{BSB2}) to $SO(1,5)\times SO(3)$ which is
indeed the global symmetry of $N=1$ SYM in $1+5$
dimensions. The $SO(3)$ corresponds to the $SU(2)_R$
symmetry of $1+5$ dimensional $N=1$ SYM, under which
the two supercharges in the $\bf 4$ of $Spin(1,5)$
transform as a doublet. Upon reduction to $1+3$ dimensions
another global $SO(2)$ symmetry appears (\ref{BSB2}).
As we will see later, the consistency constraints in
six dimensions are more restrictive than in $4d$; thus
only some of the consistent models in $4d$ can be lifted
to $6d$.}.
The gauge coupling of the $4+1$ dimensional gauge theory
on $\nc$ fourbranes is given by standard D-brane techniques
(\ref{BST6}) to be $g_{D4}^2=\gs l_s$. Upon Kaluza-Klein reduction
on a line segment of length $L_6$ we find a $3+1$ dimensional
gauge theory with coupling
\beq
{1\over g^2}={L_6\over \gs l_s}
\label{BSB3}
\eeq
It is interesting to ask in what regime the
brane configuration is well approximated by an $N=2$
SUSY $3+1$ dimensional gauge theory. There are
several issues that need to be addressed in this
regard. First, the physics on the fourbranes
looks $3+1$ dimensional only at distances much
larger than $L_6$. At shorter distances
Kaluza-Klein excitations on the fourbranes,
whose typical energy is $1/L_6$,
begin to play a role and the dynamics becomes $4+1$
dimensional.
Furthermore, in general there are
couplings of the light fields on the fourbranes
to light fields living on the $NS5$-branes, to
massive excited states of the $4-4$ strings living
on the fourbranes, and
to fields living in the bulk of spacetime, such as
gravitons. These coupling are small in the limit
$\gs\to0$ and at distances  much larger than $l_s$.

Thus to study gauge theory dynamics we are led
to consider the brane configuration above in the
limit
\beq
\gs\to 0,\;\;\; L_6/l_s\to0
\label{limvalid}
\eeq
with the ratio
corresponding to $g$ (\ref{BSB3}) held fixed.
If the gauge coupling at some scale
$L$ satisfying $L\gg l_s\gg L_6$ is small but finite,
at larger distances the dynamics of the brane
configuration will be governed by gauge theory
with the other effects mentioned above providing
small corrections.

The classical gauge theory limit is obtained when
in addition to sending $L_6$ and $\gs$ to zero, we also
send the combination (\ref{BSB3}) to zero. The theory
simplifies in this limit, since
when $g=0$ we can ignore the
effects of the ends of the fourbranes on the $NS5$-branes.
In this subsection we will discuss this classical limit;
later we will describe the modifications that take
place when quantum effects are turned on.

Classical $U(N_c)$ $N=2$ SYM in $3+1$ dimensions
has an $N_c$ (complex) dimensional moduli space of vacua
parametrized by diagonal expectation values of
the complex adjoint scalar $\Phi$ that belongs to
the vectormultiplet. At a generic point in the moduli
space, $U(\nc)$ is broken to $U(1)^{\nc}$ and the
charged gauge bosons are massive.

In the brane description, the complex adjoint field
$\Phi$ (\ref{RFTR3})
can be thought of as describing fluctuations
of the fourbranes along the fivebranes,
$X\equiv\Phi l_s^2=X^4+iX^5$.
Turning on an expectation value for $\Phi$ corresponds
to translations of the $N_c$ fourbranes in $x^4$,
$x^5$. For generic positions of the fourbranes along
the fivebranes, the $4-4$ strings connecting different
fourbranes (corresponding to vectormultiplets
charged under a pair of $U(1)$'s) have finite
length and, therefore, describe massive states. Note
also that $\Phi$ has the correct global charges.
Turning on an expectation value for $\Phi$
breaks the $SO(2)$ symmetry (\ref{BSB2}). Thus,
$\Phi$ carries charge (which is two if we normalize
the charge of the supercharges
to one) under $U(1)_R$. Similarly, it is clear that
it transforms as a singlet under $SO(3)\simeq SU(2)_R$;
both facts are in agreement with the field theory
discussion of section \ref{RFTR}.

To have a consistent four dimensional interpretation
of the Coulomb branch we have to require that in
the limit (\ref{limvalid}) the Higgs expectation value 
$\langle\Phi\rangle$ must remain well below
the Kaluza-Klein scale $1/L_6$. This means that
the typical separation between the fourbranes
in the $(x^4, x^5)$ plane $\delta x$ must satisfy
$\delta x\ll l_s^2/L_6$. One should also require
that $\delta x \ll l_s, L_6$ to decouple 
massive string modes on the fourbranes. The resulting
hierarchy of scales in the gauge theory limit
(\ref{limvalid}) is
\beq
\delta x\ll L_6\ll l_s\ll {l_s^2\over L_6}
\label{hierscale}
\eeq

\begin{figure}
\centerline{\epsfxsize=120mm\epsfbox{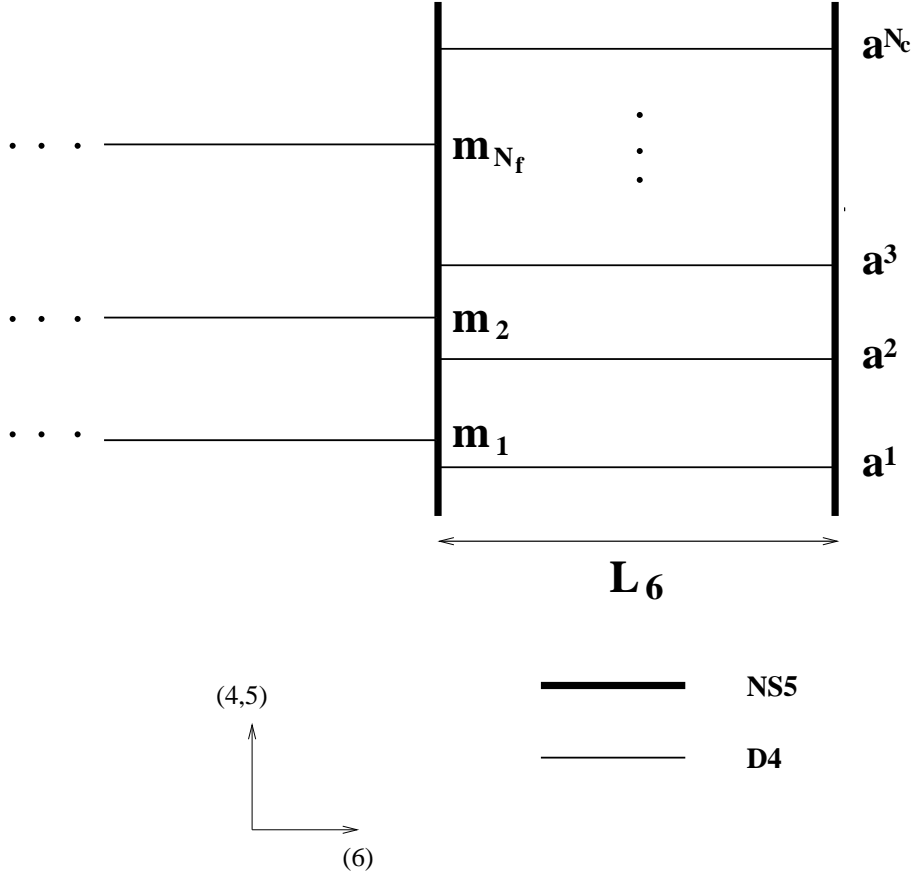}}
\vspace*{1cm}
\caption{$\nc$ ``color fourbranes'' stretched
between $NS5$-branes in the presence of $\nf$
semi-infinite ``flavor fourbranes'' describe
$N=2$ SYM with $G=U(\nc)$ and $\nf$
fundamental hypermultiplets. }
\label{ten}
\end{figure}
\smallskip

To add matter in the fundamental representation of
the gauge group we can proceed in one of a number of
related ways. One is to add semi-infinite fourbranes
attached to one of the $NS5$-branes. For example, one
can add $\nf$ fourbranes ending on the left $NS5$-brane
from the left, extending to $x^6\to-\infty$, as shown
in  Fig.~\ref{ten}. Adding the
semi-infinite fourbranes gives rise to $\nf$ hypermultiplets
in the fundamental representation of $U(\nc)$ corresponding
to strings stretched between the $\nc$ suspended fourbranes
and the $\nf$ semi-infinite ones. The locations at which
these semi-infinite fourbranes attach to the fivebrane in
the $(x^4, x^5)$ plane are $\nf$ complex numbers
$m_1, \cdots, m_{\nf}$ which can be thought of as the
masses of the quarks~\footnote{Up to a factor of $l_s^2$
which is needed to fix the dimensions; we will usually set
$l_s=1$ from now on.}. Note that these locations correspond to
expectation values of scalar fields living on the
worldvolume of the semi-infinite fourbranes. As before,
since these
fourbranes are ``heavy'' objects, they are frozen at their
classical values and give rise to couplings rather
than moduli in the four dimensional 
low energy theory.

A generic point
in the Coulomb branch is 
parametrized by the $\nc$ complex
numbers $a^1, \cdots, a^{\nc}$ 
corresponding to the locations
in  the $(x^4, x^5)$ plane of 
the suspended $D4$-branes. {}From gauge
theory we know that due to 
the superpotential (\ref{D3D79})
(with the sum running over all $N_f$ flavors)
which is required by $N=2$ SUSY, the mass of the quark $Q^i_\alpha$
corresponding to the $i$'th flavor and the $\alpha$'th
color is $m_i^\alpha=|m_i-a^\alpha|$. In the brane
picture, the mass $m_i^\alpha$ is given by the minimal
energy of a fundamental string stretched between the
$\alpha$'th suspended brane and the $i$'th semi-infinite
one. Just like the adjoint field $\Phi$, the mass
parameters $m_i$ (\ref{D3D79}) carry $U(1)_R$ charge two.
Turning on masses breaks $U(1)_R$ (as well as conformal
invariance).

While the above way of introducing fundamental matter
is appropriate for describing the Coulomb branch
of $U(\nc)$ gauge theory with $\nf$ flavors, it
does not provide a geometric description of the
Higgs branches. Recall that the gauge theory in question
has a number of branches of the moduli space of vacua
along which some of the quarks $Q$, $\tilde Q$ get
expectation values and the rank of the unbroken
gauge group decreases. For $\nf \ge 2\nc$
the gauge group can be completely Higgsed and
the complex dimension of the corresponding branch
of moduli space is $2\nc\nf-2\nc^2$.

\begin{figure}
\centerline{\epsfxsize=120mm\epsfbox{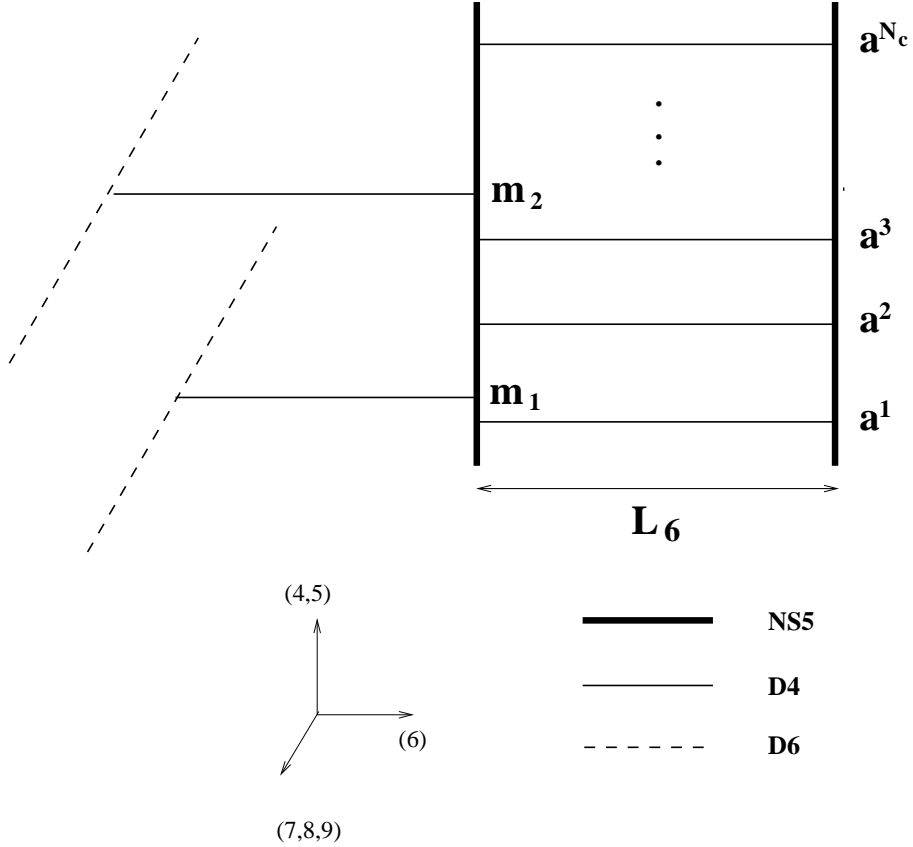}}
\vspace*{1cm}
\caption{Replacing the semi-infinite flavor
$D4$-branes of Fig. 10 by
$D4$-branes ending on $D6$-branes allows
one to explore the full phase structure
of the theory.}
\label{eleven}
\end{figure}

To recover the Higgs branches it is convenient
to generalize the above construction of matter
in a way shown in Fig.~\ref{eleven}. 
Replace the semi-infinite
$D4$-branes to the left of the $NS5$-branes by
finite $D4$-branes each ending on a different
$D6$-brane oriented as in (\ref{BSB1}).
This opens up the possibility of having additional
dynamical degrees of freedom living on these
fourbranes.

\smallskip

For generic masses $\{m_i\}$ (which can now
be thought of as positions of the 
$\nf$ $D6$-branes in the $(x^4, x^5)$ plane) and
points in the Coulomb branch $\{a^\alpha\}$,
there are no new massless states of this kind.
Indeed, all potentially light fields living
on a fourbrane stretched between an $NS5$-brane
and a $D6$-brane have Dirichlet boundary conditions
on one or both boundaries and hence do not
lead to massless degrees of freedom. That is consistent
with the absence of Higgs branches of $N=2$ SYM when
all the masses $m_i$ are distinct.

When two or more masses coincide, say $m_1=m_2$
we expect from gauge theory to be able to enter
a Higgs branch by turning on an expectation value
to quarks $Q$, $\tilde Q$. Furthermore, to enter the
Higgs branch one needs to go to a particular point in
the Coulomb branch for which for some $1\le \alpha\le \nc$,
$m_1^\alpha=m_2^\alpha=0$. To reproduce this in the brane
description one notes that when two masses $m_i$ coincide,
two $D6$-branes are at the same position in the $(x^4, x^5)$
plane. In general they are still separate in the $x^6$
direction and each is connected to the same $NS5$-brane
by a $D4$-brane. Thus the fourbrane connecting the ``far''
$D6$-brane to the $NS5$-brane meets in space and intersects
the ``near'' $D6$-brane.
{}From our discussion of brane intersections in subsection \ref{BEB}
one might conclude at this point that this fourbrane can
break into two pieces, one stretched between the $NS5$-brane
and the near $D6$-brane and the other between the two $D6$-branes.
While the first piece would as before give rise to no massless
degrees of freedom in $3+1$ dimensions, the second one would
in fact give rise to a neutral (under the gauge group)
massless hypermultiplet. The scalars in the hypermultiplet
would correspond to displacements of the $D4$-brane
along the $D6$-branes in the $(x^7, x^8, x^9)$ directions,
and the compact component of the gauge field $A_6$.
This would provide a candidate for a brane realization
of the phase transition from the Coulomb to the Higgs
phase.

However, the picture we got so far is inconsistent with
gauge theory. The process described above appears to be
possible for any values of the Coulomb moduli; in gauge
theory we have to tune to a particular point in the Coulomb
branch in order to be able to enter the Higgs branch.
This and many related puzzles are resolved by 
noting~\cite{HW} that
the following ``s-rule'' holds in brane dynamics:

\medskip

\noindent
{\it A configuration in which an NS fivebrane and a D
sixbrane are connected by more than one $D4$-brane
is not supersymmetric.}

\medskip

\begin{figure}
\centerline{\epsfxsize=130mm\epsfbox{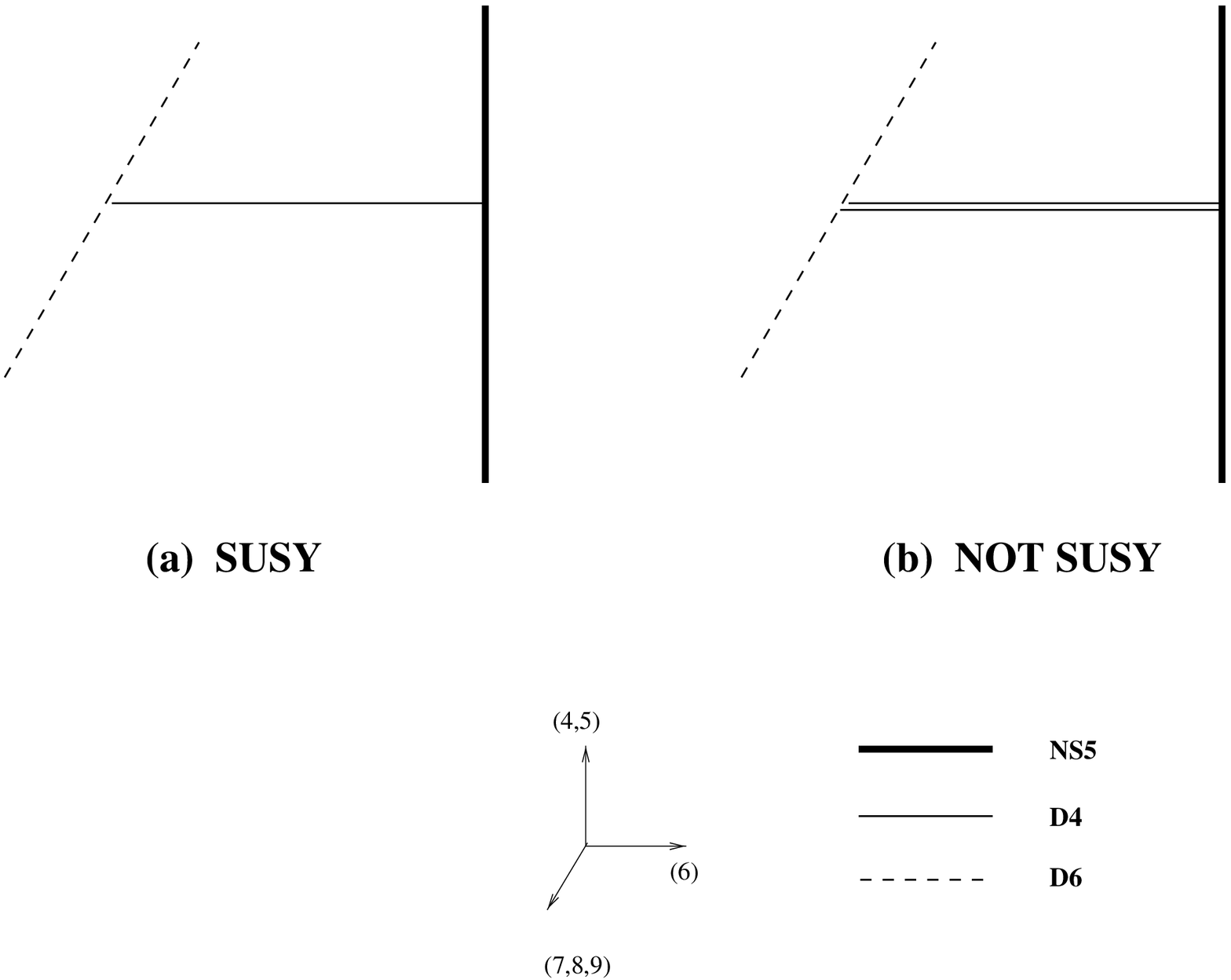}}
\vspace*{1cm}      
\caption{(a) A supersymmetric configuration containing
an $NS5$-brane connected to a $D6$-brane by a
single $D4$-brane. (b) A non-supersymmetric
configuration
in which the two are connected by two $D4$-branes.}
\label{twelve}
\end{figure}

\noindent
The s-rule, which is illustrated in
Fig.~\ref{twelve}, 
is a phenomenological rule of brane
dynamics which has been recently discussed from
various points of view, for example, 
in~\cite{OV97,HOO,BGS,BG}.
It seems to have to do with the fact
that two or more fourbranes connecting
a given NS fivebrane to a given D sixbrane are
necessarily on top of each other -- a rather
singular situation. In particular, in~\cite{BGS,BG}
it has been related by U-duality to Pauli's
exclusion principle.

\smallskip

\begin{figure}
\centerline{\epsfxsize=130mm\epsfbox{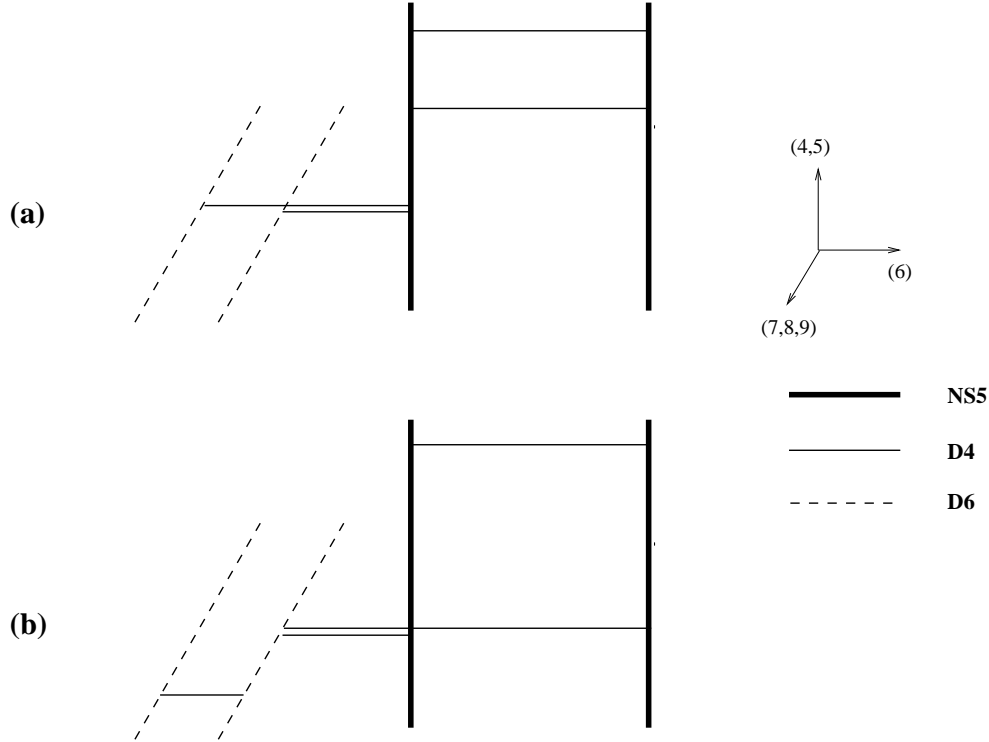}}
\vspace*{1cm}      
\caption{The transition from Coulomb (a) to Higgs (b)
branch.}
\label{thirteen}
\end{figure}

The s-rule explains why the process described
above is forbidden, but the comparison to
the gauge theory picture suggests a way out.
If in addition to having two $D6$-branes coincide
in the $(x^4, x^5)$ plane we also go to a point
in the Coulomb branch where one of the $\nc$
$D4$-branes suspended between the NS-branes
is at the same value of $(x^4, x^5)$ as well,
the above phase transition becomes 
possible (Fig.~\ref{thirteen}).

\smallskip

All we have to do is first reconnect the
$D4$-brane stretched between the left $NS5$-brane
and the far $D6$-brane, combining it with the
coincident $D4$-brane suspended between the NS-branes
such that it now connects the {\it right} $NS5$-brane
to the far $D6$. Now there is no conflict with the
s-rule in breaking the resulting stretched $D4$ into
two pieces as described above. As expected
in the Higgs mechanism, in the process we replace a
massless $U(1)$ vectormultiplet corresponding to a
$D4$-brane stretched between two $NS5$-branes by a
massless neutral hypermultiplet $\tilde Q Q$
corresponding to a $D4$-brane stretched between
adjacent (in $x^6$) $D6$-branes. The expectation
value $\langle \tilde Q Q\rangle$ is parametrized
by the location of this $D4$-brane along the $D6$-branes
in $(x^7, x^8, x^9)$ and the Wilson line of $A_6$.
This is consistent with the expected transformation
of $Q$, $\tilde Q$ under $SU(2)_R\times U(1)_R$:
they are not charged under $U(1)_R$ and transform
as a doublet under $SU(2)_R$. Thus the ``mesons''
$\tilde Q Q$ transform as ${\bf 2}\times {\bf 2}=
{\bf 3}+{\bf 1}$.

\begin{figure}
\centerline{\epsfxsize=160mm\epsfbox{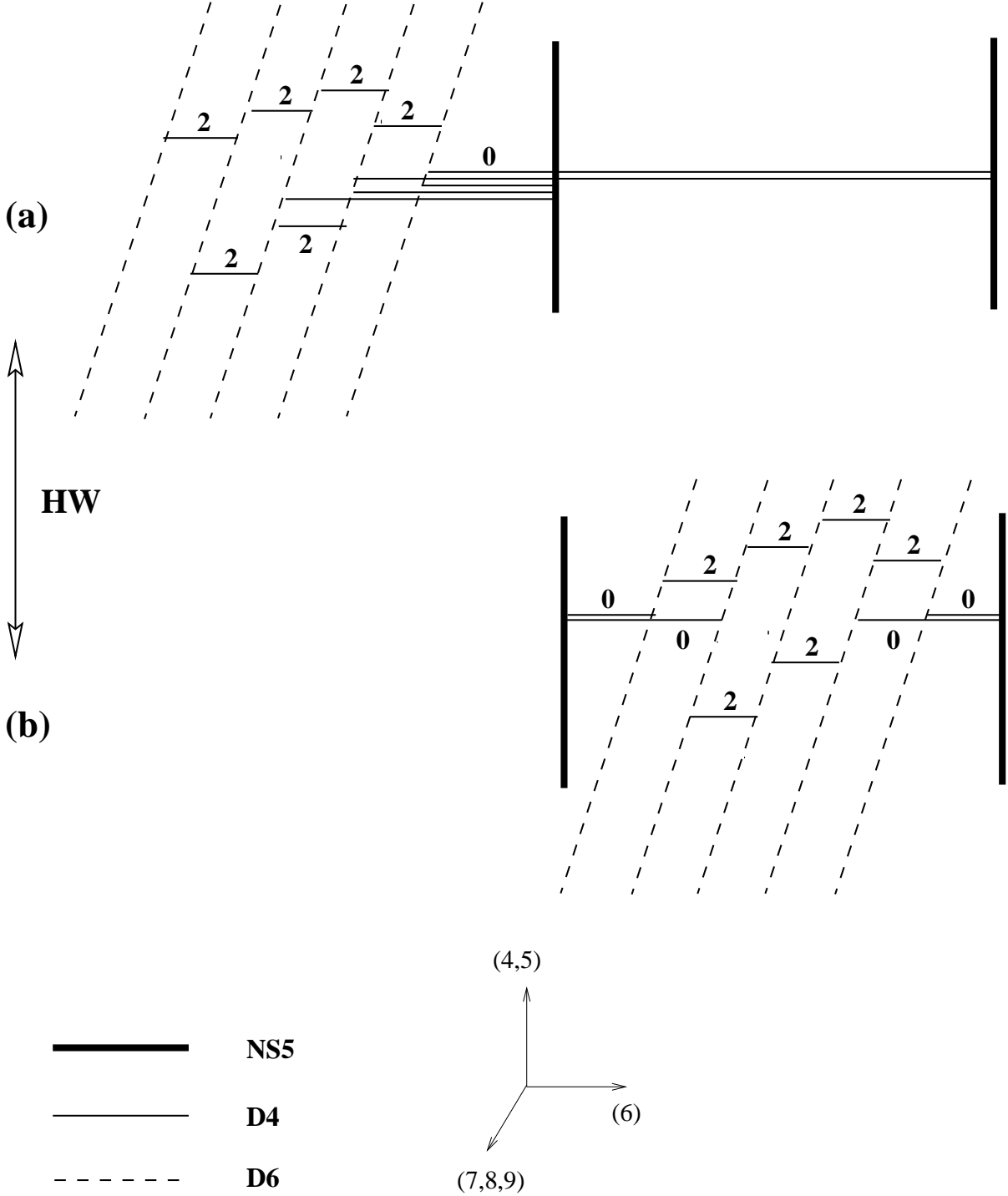}}
\vspace*{1cm}
\caption{The fully
Higgsed branch of moduli space for $\nf=5$,
$\nc=2$; the two equivalent descriptions
are related by a series of HW transitions.}
\label{fourteen}
\end{figure}
\smallskip

Thus we learn that the Higgs mechanism is described in
brane theory as the reconnection (or breaking) of $D4$-branes
stretched between $NS5$-branes which give rise to
vectormultiplets into ones stretched between $D6$-branes
and/or $NS5$-branes consistently with the s-rule.
Performing all such breakings
in the general case of $\nc$ colors and $\nf$
flavors gives rise to the correct (classical) phase
structure of $N=2$ SYM.

As an example, to compute
the dimension of the maximally Higgsed branch
(where the gauge symmetry is completely broken)
for $\nf\ge2\nc$ we proceed as follows (see
Fig.~\ref{fourteen}(a)):
\begin{itemize}
\item Set all the masses equal to each other,
\ie\ bring all $\nf$ $D6$-branes to the same
point in $(x^4, x^5)$, say the origin.
They are still at
different positions in $x^6$.
\item Reconnect the $D4$-brane stretching between
the left $NS5$-brane and the leftmost $D6$-brane
to stretch between the right $NS5$ and the leftmost
$D6$, by going to the point in the Coulomb branch
where one of the $\nc$ ``color'' $D4$'s is at the origin
in the $(x^4, x^5)$ plane as well.
\item Break the resulting $D4$ into $\nf$ pieces stretching
between the right $NS5$ and the rightmost $D6$ and
the different adjacent $D6$'s. This leads to
$\nf-1$ massless
hypermultiplets corresponding to fluctuations of the
$\nf-1$ segments of the $D4$ stretched between the
$D6$'s. The expectation values of these are $2(\nf-1)$
complex moduli.
\item By bringing in another color $D4$, reconnect the ``second
longest'' $D4$ stretching between the left $NS5$ and the
next to leftmost $D6$ to the right $NS5$. Repeat the breaking
procedure. The s-rule applied to the {\it right} $NS5$
implies that there are now $\nf-3$ massless hypermultiplets
and hence $2(\nf-3)$ complex moduli.
\item Continuing this process gives rise to
\beq
\sum_{i=1}^{\nc} (\nf-(2i-1))=\nf\nc-\nc^2
\label{BSB5}
\eeq
massless hypermultiplets and to a $2(\nf\nc-\nc^2)$
complex dimensional Higgs moduli space, in agreement
with gauge theory expectations.
\end{itemize}

\begin{figure}
\centerline{\epsfxsize=120mm\epsfbox{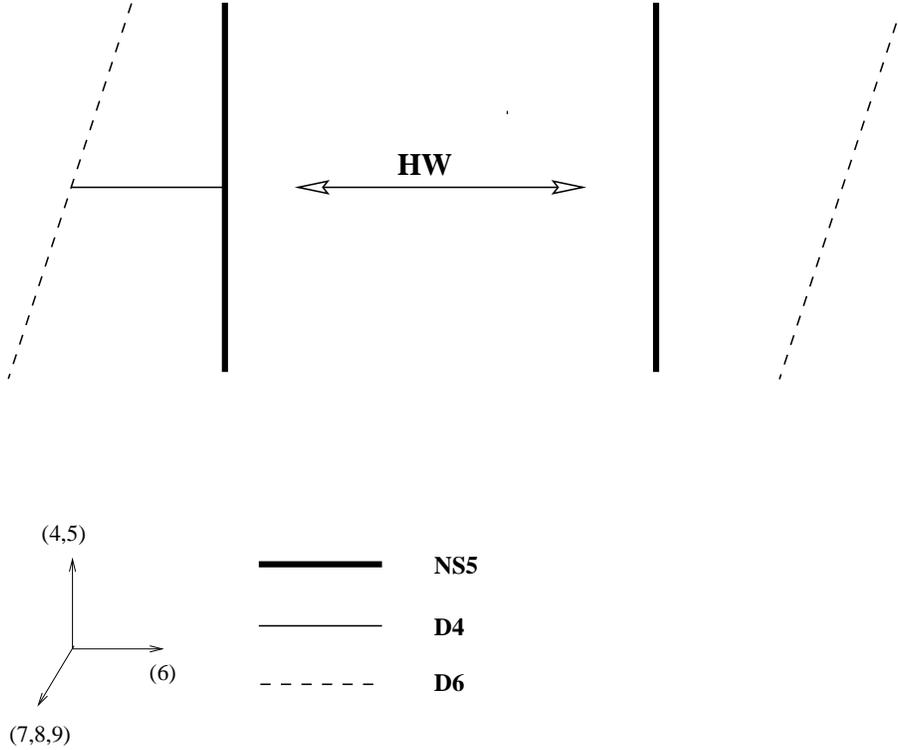}}
\vspace*{1cm}
\caption{The HW transition, in which a fourbrane
is created when an $NS5$-brane and a $D6$-brane
cross in $x^6$ and exchange positions.}
\label{fifteen}
\end{figure}
\smallskip

A peculiar feature of the above analysis is the (lack of)
role of the parameters describing the positions along the
$x^6$ axis of the $\nf$ $D6$-branes providing the
flavors. There are no parameters in the low energy
$N=2$ SQCD corresponding to changing these
parameters and thus they are irrelevant (in the RG
sense). Indeed, the physics of the brane configuration
is independent of the locations of the $D6$-branes
at least when they are all to the left (or equivalently
all to the right) of the two $NS5$-branes.

It is interesting to ask what happens if we try
to vary the $x^6$ positions of the $D6$-branes,
bringing some or all of them inside the interval
between the two NS-branes. Because of the way
the different branes are oriented (\ref{BSB1})
the $D6$ and $NS5$-branes cannot avoid each other,
and when their $x^6$ values coincide they actually
meet in space. What takes place when they switch
positions is known as the ``HW 
transition''~\cite{HW}.
The $D4$-brane connecting the $D6$ and $NS5$ becomes
very short as they approach each other and
{\it disappears} when they cross. Conversely, if
the $D6$ and $NS5$ that approach each other do not
have a $D4$-brane connecting them, one is created
when they exchange positions (Fig.~\ref{fifteen}).

The HW transition is an interesting effect of brane dynamics
which is related by
U-duality to similar transitions for other branes;
it was further investigated from several perspectives,
for example,
in~\cite{OV97,BDG,DFK,BGL9705,Alw,HWu,OSZ,BGL9711,NOYY9711,Yos}.
Heuristically, it is related to conservation of a certain
magnetic charge which can be measured on each brane
known as the ``linking number.''
The total charge measured on each brane is
\beq
L_B={1\over2}(r-l)+L-R
\label{BSB6}
\eeq
For an $NS5$-brane, $r$ and $l$ are the numbers of
$D6$-branes to its right and left, respectively, while
$R$ and $L$ are the  numbers of $D4$-branes ending
on the $NS5$-brane from the right and left, respectively.
Similarly, for a $D6$-brane $r$ and $l$ are the numbers of
$NS5$-branes to its right and left, and $R$ and $L$ are as
above. Right and left here refer to locations along the $x^6$
axis.

As an example, for a $D6$-brane connected to an $NS5$-brane
on its right (in $x^6$) by a $D4$-brane the linking number
is $L_{D6}=-1/2$. For the $NS5$-brane the linking number
is $L_{NS5}=+1/2$. If we try to move the $D6$-brane past
the $NS5$-brane the $D4$ connecting them disappears. The
new configuration includes a $D6$ with a  disconnected
$NS5$ on its left; the linking numbers are seen from
(\ref{BSB6}) to be unchanged.

Taking the HW transition into account we can analyze
what happens when the $D6$-branes are translated in the
$x^6$ direction and placed in the interval between the
two NS-branes. All the $D4$-branes that initially
connected the $D6$-branes to an $NS5$-brane disappear,
and we end up with a configuration of two NS-branes
at different values of $x^6$ connected to each other
by $\nc$ fourbranes,
with $\nf$ sixbranes 
located between them in $x^6$ 
(see Fig.~\ref{fourteen}(b)).

Remarkably, the resulting brane configuration describes
the same low energy physics!
This is a priori surprising since one would in general
expect a phase transition to occur as the two branes
cross; indeed, we will see that such transitions
occur when the crossing branes are parallel.
It is not well understood why there is no phase transition
when non-parallel branes cross.

In any case, in the present setup
the quarks $Q$, $\tilde Q$
that corresponded to $4-4$ strings before are now due
to $4-6$  strings stretched between the $\nc$ suspended
fourbranes and the $\nf$ sixbranes. The locations of the
sixbranes in the $(x^4, x^5)$ plane still correspond to
their masses, and the Higgs branch of the moduli space
is described by breaking $D4$-branes stretched between
the two $NS5$-branes on the $\nf$ $D6$-branes. Taking
into account the s-rule it is easy to see that the
dimension of the Higgs branch is as described above
(\ref{BSB5}).

The $N=2$ SYM theory under consideration has
gauge group $U(\nc)\simeq SU(\nc)\times U(1)$.
Thus one can turn on Fayet-Iliopoulos (FI)
couplings for the $U(1)$.
In $N=1$ superspace they are an $N=1$ FI D-term,
and a linear superpotential for the adjoint chiral
superfield in the vectormultiplet:
\beq
\LL_{FI}={\rm Tr}\;\left(r_3\int d^4\theta V+
r_+\int d^2\theta \Phi + r_-\int d^2\bar\theta
\bar\Phi\right)
\label{BSB7}
\eeq
$r_3$ is real, while $r_+^*=r_-$.
The three FI couplings $r_3, r_\pm$ transform
as the ${\bf 3}$ of $SU(2)_R$ and are neutral
under $U(1)_R$. For $\nf\ge\nc$ the D-terms break
the gauge group completely and force the system
into the Higgs branch. For smaller $\nf$ there
is no supersymmetric vacuum.

\begin{figure}
\centerline{\epsfxsize=100mm\epsfbox{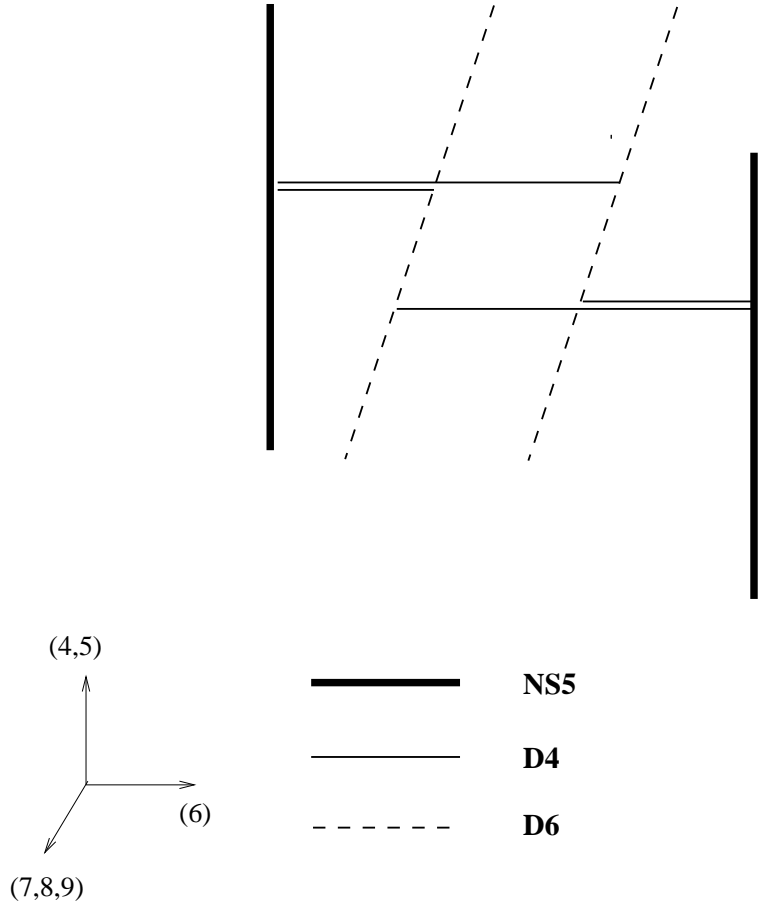}}
\vspace*{1cm}
\caption{FI D-terms in the $N=2$ SYM
theory correspond to relative displacements
of the two $NS5$-branes in $(x^7, x^8, x^9)$.
For non-zero D-terms, all color $D4$-branes
must break and the theory is forced into the Higgs
phase.}
\label{sixteen}
\end{figure}

In the brane language, the FI couplings correspond
to the relative position of the two $NS5$-branes
in the $(x^7, x^8, x^9)$ directions 
(Fig.~\ref{sixteen}); note that
these parameters have the correct transformation
properties under $SO(3)\simeq SU(2)_R$. {}From the
geometry and (\ref{BSB1}) it is clear that only
when the two $NS5$-branes are at the same value of
$(x^7, x^8, x^9)$  can the $D4$-branes stretch
between them without breaking SUSY. For non-zero
D-terms all the $D4$-branes must break on
$D6$-branes. This corresponds to complete Higgsing
of the gauge group; it is consistent with the
s-rule for $\nf\ge \nc$; for smaller $\nf$ there
is no supersymmetric vacuum.

\smallskip

\subsubsection{Orthogonal And Symplectic Groups}
\label{OSG}

In this subsection we will discuss configurations
of branes near orientifold planes. As we saw before,
adding orientifold four and six ($O4$ and $O6$)
planes as in (\ref{BSB1}) does not lead to the breaking
of any further SUSY, or global symmetry (\ref{BSB2}).
In the simplest cases one finds $N=2$ SYM theories with
orthogonal and symplectic gauge groups and matter in the
fundamental representation. We next discuss in turn the
two cases of an $O6$-plane parallel to the $D6$-branes
and of an $O4$-plane parallel to the $D4$-branes.

\medskip
\noindent
{\bf 1. Orientifold Sixplanes}

Consider a configuration in which an $NS5$-brane
is placed at a distance $L_6$ from an orientifold
sixplane (all objects here and below oriented as
in (\ref{BSB1})).
We would like to stretch $D4$-branes
between the $NS5$-brane and the orientifold plane
(a useful way to think about these is as
$D4$-branes stretched
between the $NS5$-brane and its mirror
image with respect to the orientifold).
As discussed in section
\ref{BWC}, there are actually two different kinds of
$O6$-planes, with positive and negative charge.

The first question we have to address is whether we {\it can}
stretch $D4$-branes between the $NS5$-brane and its
image in this situation without breaking
SUSY. One might worry that such fourbranes
are projected out by the orientifold projection
for one of the two possible choices of the
orientifold charge. For example, we will see later
that it is impossible to stretch a BPS saturated
$D4$-brane between a $D6$-brane parallel to
an $O6_-$ plane and its image, and this will
have important consequences for the low energy
gauge theory. We next show that for the case
of $NS5$-branes there is no such obstruction.

U-duality relates the configuration
we have to a more familiar one~\footnote{Below we freely
compactify and decompactify different dimensions. This
should not affect the conclusions as to whether various
brane configurations are allowed.}. Specifically we perform
a T-duality transformation (\ref{TTT1}) $T_{123}$ which
takes the type IIA string theory to a type IIB one, followed
by an S-duality transformation $S$ (\ref{BST30}) on the resulting
IIB string.
$T_{123}$ maps the $NS5$-brane to itself,
the $O6$-plane to an $O3$-plane $(x^0, x^7, x^8, x^9)$
and the $\nc$ $D4$-branes to $\nc$ D-strings stretched in
$x^6$ between the two fivebranes.
The subsequent S-duality transformation acts differently
for positive and negative orientifold charge.

For negative orientifold charge, 
we saw in section \ref{D4N4}
that the $O3$-plane transforms under $S$ to itself.
Thus after performing the transformation, we end up
with a $D5$-brane and its image near an $O3_-$ plane. 
The original $D4$-branes connecting
the $NS5$-brane to its image turn now into fundamental
IIB strings connecting the $D5$-brane to its image
with respect to the $O3$-plane. The low energy
theory on the $D5$-brane is in this case an $SO(2)$
gauge theory; the configuration is T-dual to well
studied systems such as D-strings inside an $O9$-plane
or $D0$-branes near an $O8$-plane. Translations of
the D-brane and its image away from
the orientifold plane (in the $x^6$ direction)
are described by an antisymmetric tensor of
$SO(2)$, \ie\ a singlet.

The original question,
whether one can stretch a $D4$-brane between
the $NS5$-brane and its mirror, is translated in
the U-dual configuration into the question whether
one can stretch a BPS saturated string between the
$D5$-brane and its image. Such a string would correspond
to a state charged under the $SO(2)$ gauge symmetry
on the $D5$-brane, that goes to zero mass as the
$D5$-brane approaches its image. It is well known
that such states exist; they correspond to fields
describing fluctuations of the D-brane along the
orientifold plane, in this case in the
$(x^7, x^8, x^9)$ directions. Such fluctuations
are described by a symmetric tensor of $SO(2)$
which includes a pair of states charged under
$SO(2)$; these states are described by BPS
fundamental strings stretched between the $D5$-branes.
They are U-dual to
$D4$-branes connecting the $NS5$-brane to its image
in the original configuration. In particular,
the latter is clearly possible.

For positive orientifold charge we use the fact
-- explained in section \ref{D4N4} --
that S-duality takes an $O3_+$ plane to an $O3_-$ plane
with a single $D3$-brane (without a mirror partner)
embedded in it.
The resulting system of a $D5$-brane
near an $O3_-$ plane with a $D3$-brane
is similar to that
studied in the previous paragraphs.
The $D3$-brane gives rise
to additional matter which plays
no role in the discussion.
Clearly the rest of the
analysis can be repeated as above,
with the same conclusions --
it is possible to stretch $D4$-branes
between an $NS5$-brane and its image
with respect to an $O6$-plane of either sign.

\begin{figure}
\centerline{\epsfxsize=100mm\epsfbox{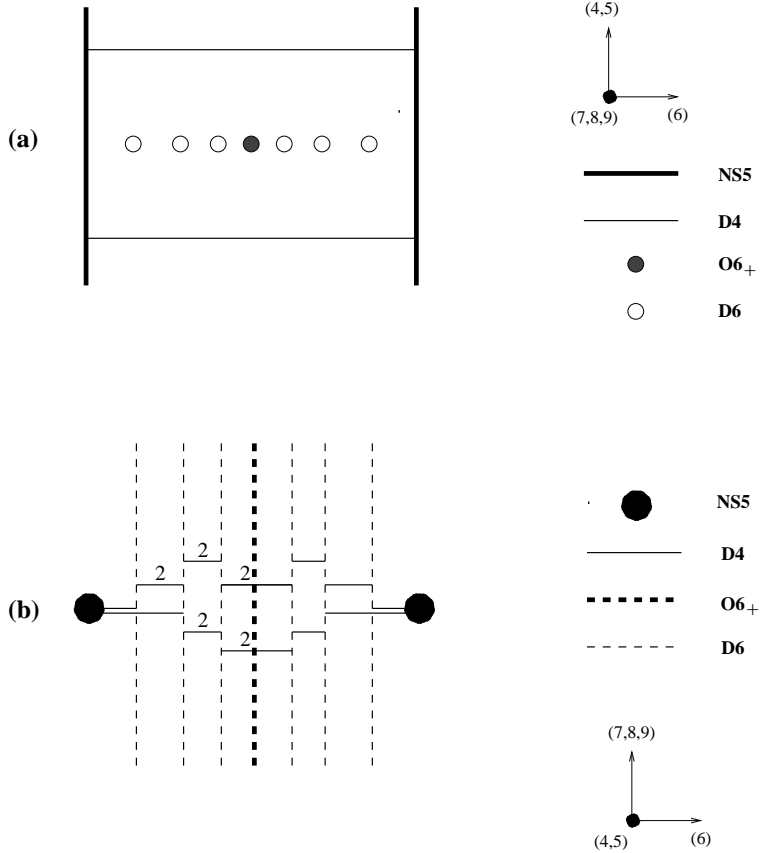}}
\vspace*{1cm}
\caption{The Coulomb, (a), and fully Higgsed, (b),
branches of moduli space of $SO(2)$ gauge theory
with $\nf=3$ charged hypermultiplets.}
\label{seventeen}
\end{figure}

We will next describe the classical gauge
theories corresponding to the two choices
of the sign of the orientifold charge,
starting with the case of positive 
$O6$ charge (Fig.~\ref{seventeen}), which
leads to an orthogonal 
projection on the $D4$-branes.
The case of $O6_-$, which leads to a symplectic
gauge group, will be considered later.

\smallskip

The gauge group on $\nc$ $D4$-branes
connecting an $NS5$-brane to its mirror
image with respect to an $O6_+$ plane
is $SO(\nc)$.
$\nf$ $D6$-branes parallel to the $O6$-plane
located
between the $NS5$-brane and the orientifold
give
$\nf$ hypermultiplets in the fundamental
(${\bf \nc}$) representation of $SO(\nc)$,
arising as usual from $4-6$ strings.
In $N=1$ SUSY notation there are
$2\nf$ chiral multiplets $Q^i$, $i=1,\cdots, 2\nf$,
which are paired to make $\nf$ hypermultiplets.
The global flavor symmetry of this gauge theory
is $Sp(\nf)$, in agreement with the
projection imposed by the positive charge
$O6$-plane on the $D6$-branes.

The Coulomb branch
of the $N=2$ SUSY gauge theory is parametrized
by the locations of the $D4$-branes along the
fivebrane in the $(x^4, x^5)$ plane. Entering the
Coulomb branch involves removing the ends of the
fourbranes from the orientifold plane (which is
located at a particular point in the $(x^4, x^5)$
plane). Since the fourbranes can only leave the
orientifold plane in pairs, the dimension of the
Coulomb branch is $[\nc/2]$, in agreement with
the gauge theory description.

Higgs branches of the gauge theory
are parametrized by all possible breakings of
fourbranes on sixbranes. As for the unitary
case there are many different branches;
as a check that we get the right structure,
consider the fully Higgsed branch
which exists when the number of
flavors is sufficiently large. {}From gauge
theory we expect its dimension to be
$2\nc\nf-\nc(\nc-1)$.

The brane analysis gives
\beq
{\rm dim} \MM_H=\sum_{i=1}^{\nc}
2(\nf+1-i)=2\nf\nc-\nc(\nc-1)=[2\nf\nc-\nc(\nc+1)]+2\nc
\label{OSG1}
\eeq
The number in the square brackets is the number of
moduli corresponding to segments that do
not touch the orientifold, and the additional $2\nc$
is the number of moduli coming
from the segments of the fourbranes connecting
the $D6$-brane closest to the orientifold to its
mirror image. These segments transform to
themselves under the orientifold projection
and thus are dynamical for positive orientifold
charge. An example is given in 
Fig.~\ref{seventeen}(b).

The $2\nc$ moduli coming from fourbranes
connecting a $D6$-brane to its image have
a natural interpretation in the theory on
the $D6$-branes. At low energies this is
an $Sp(1)$ gauge theory with sixteen
supercharges, and the $D4$-branes stretched
between the $D6$ and its mirror can be thought
of as $Sp(1)$ monopoles, as in section
\ref{D4N4}. {}From this point of view the above
$2\nc$ moduli parametrize the space of $\nc$
$Sp(1)$ monopoles.

Thus, the total dimension of moduli space agrees
with the gauge theory result. It is easy
to similarly check the agreement with gauge theory
of the maximally Higgsed branch for small
$\nf$, as well as the structure of
the mixed Higgs-Coulomb branches.

\begin{figure}
\centerline{\epsfxsize=100mm\epsfbox{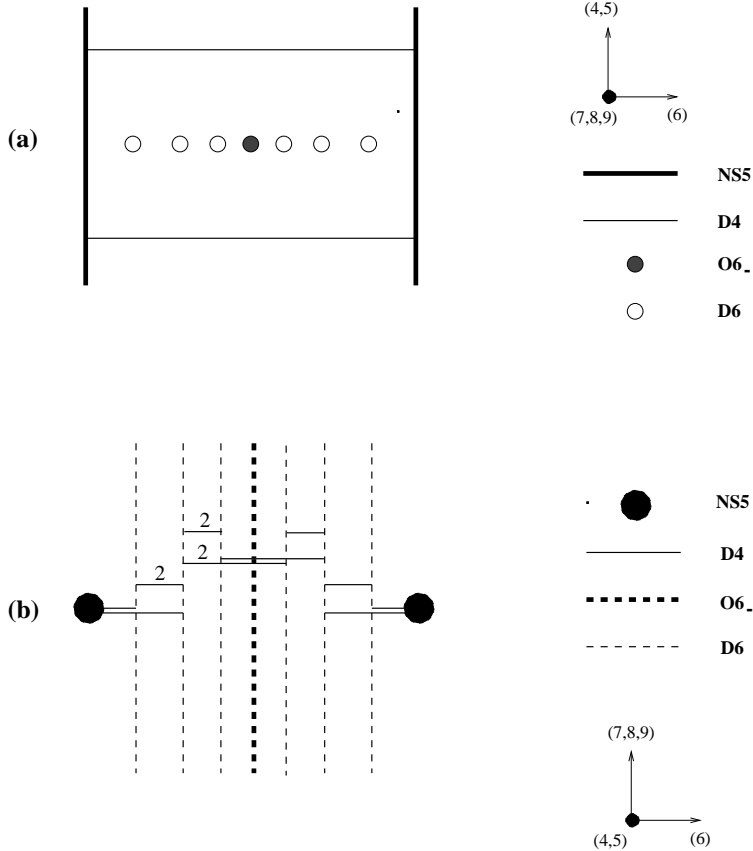}}
\vspace*{1cm}
\caption{The Coulomb, (a), and fully Higgsed, (b),
branches of moduli space of $Sp(1)$ gauge theory
with $\nf=3$ charged hypermultiplets.}
\label{eighteen}
\end{figure}

For negative charge of the $O6$-plane, the configuration
of Fig.~\ref{eighteen} 
describes an $Sp(\nc/2)$ gauge theory
with $\nf$ hypermultiplets in the fundamental
$({\bf \nc})$ representation. Qualitatively,
most of the analysis is the
same as above, but the results are clearly
somewhat different. For example, the dimension of the
fully Higgsed branch is in this case $2\nf\nc-\nc(
\nc+1)$, smaller by $2\nc$ than the $SO$ case discussed
above.

\smallskip

{}From the point of view of the brane construction
the Higgs branch is different because it is no longer
possible to connect a $D6$-brane to its mirror image
by a $D4$-brane. Such fourbranes are
projected out when the $O6$-plane has negative charge.
As in section \ref{D4N4},
this is also clear if we interpret
these fourbranes as magnetic monopoles in the sixbrane
theory. In this case the theory on the $D6$-brane
adjacent to the orientifold and its image
has gauge group $SO(2)$, and there are no non-singular
monopoles.

Therefore, the pattern of breaking of the $D4$-branes
on $D6$-branes near the orientifold is modified. The
result is depicted in Fig.~\ref{eighteen}(b).
We have to stop the usual
pattern (\ref{OSG1}) when we get to the last {\em two}
$D6$-branes before the orientifold, and there we must
perform the breaking as
indicated in Fig.~\ref{eighteen}(b). 
Thus compared to (\ref{OSG1}) we lose $2\nc$ moduli.
Overall, the brane
Higgs branch is $2\nf\nc-\nc(\nc+1)$ dimensional,
in agreement with the gauge theory analysis. One can
again check that the full classical phase structure of the
$Sp(\nc/2)$ gauge theory is similarly reproduced.

Further discussion of
gauge theories on 
brane configurations in the presence
of orientifold sixplanes appears  
in~\cite{EGKRS,LL,EGKT}.

\medskip
\noindent
{\bf 2. Orientifold Fourplanes}

\begin{figure}
\centerline{\epsfxsize=100mm\epsfbox{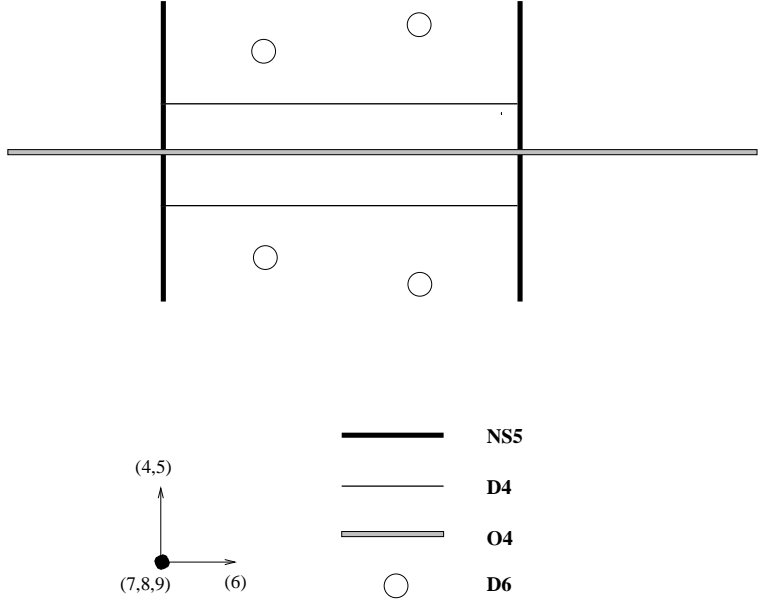}}
\vspace*{1cm}
\caption{Fourbranes stretched between $NS5$-branes
which are stuck on an $O4$-plane in the presence
of $D6$-branes provide an alternative description
of $N=2$ SYM with orthogonal and symplectic
gauge groups.}
\label{nineteen}
\end{figure}

In this case we
suspend $\nc$ fourbranes between a pair of
$NS5$-branes stuck on an orientifold fourplane
at different locations in $x^6$ (Fig.~\ref{nineteen}).
$2\nf$ $D6$-branes placed
between the NS-branes (in $x^6$)
provide $\nf$ fundamental hypermultiplets.
All the branes are as usual
oriented as in (\ref{BSB1}).
Despite much recent
work~\cite{EJS,EGKRS,Tat,Joh,BSTYb,FGP,AOTa,AOTb,AOTc,Ter,LPTb,BHOO98},
such configurations are not well
understood in brane theory,
and the discussion
below should be viewed as conjectural.
The new element in this case, and presumably
the origin of the difficulties,
is the fact that when
an $NS5$ or $D6$-brane intersects an $O4$-plane,
say at $x^6=0$, it splits it into disconnected
components corresponding to $x^6>0$ and
$x^6<0$. This leads to rather exotic behavior
some aspects of which will be described below.

\smallskip

\begin{figure}
\centerline{\epsfxsize=100mm\epsfbox{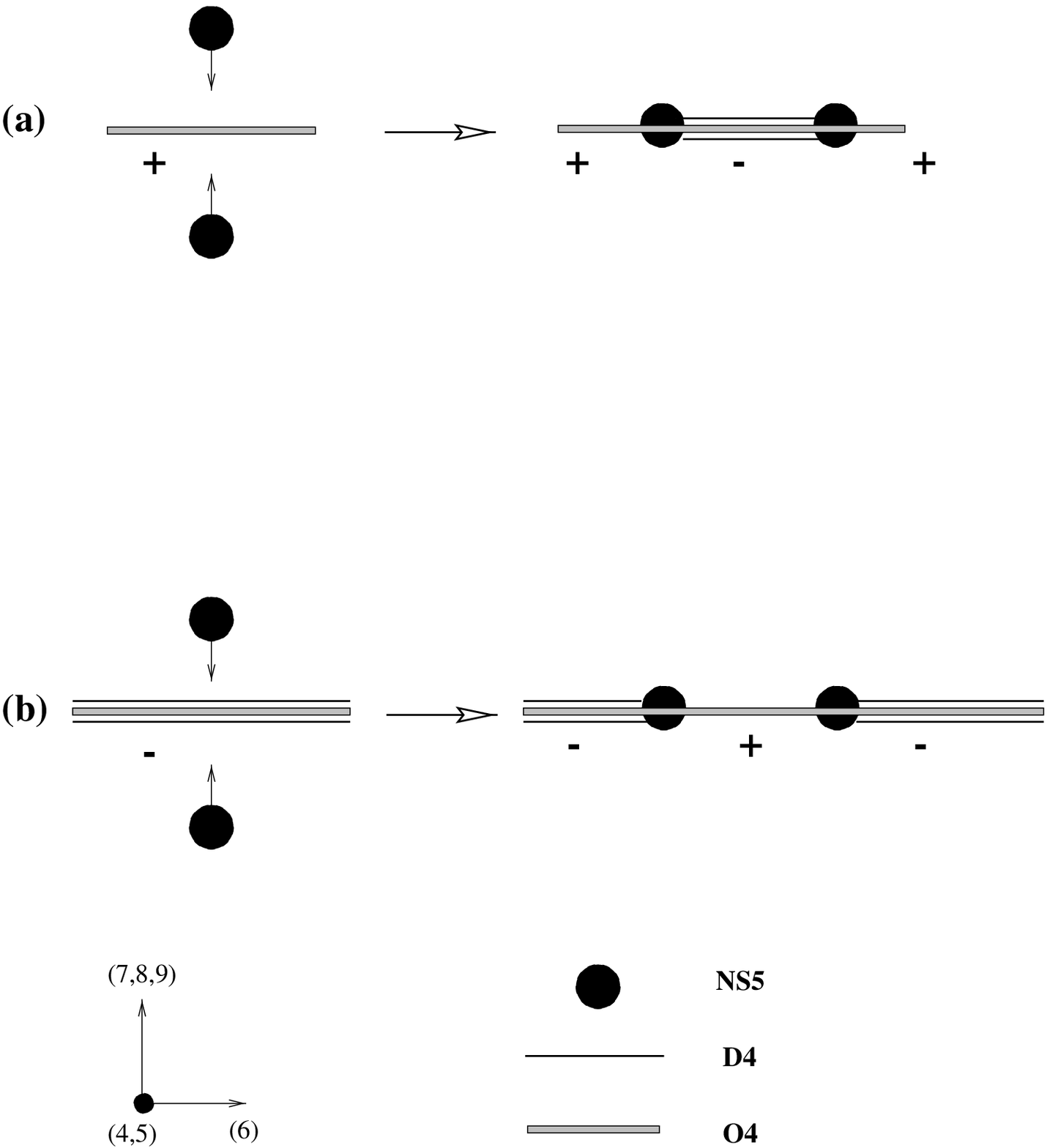}}
\vspace*{1cm}
\caption{(a) An $NS5$-brane and its mirror
image approach an $O4_+$ plane and separate
along it (in $x^6$). The portion of the orientifold
between the fivebranes flips sign in the process,
and a pair of $D4$-branes stretch between the
fivebranes. (b) When an $NS5$-brane and its image
approach an $O4_-$ plane with two adjacent
$D4$-branes, the reverse of (a) happens.}
\label{twentyone}
\end{figure}

One way to study what happens when an $NS5$-brane
intersects an $O4$-plane is to start with a
pair of such fivebranes (\ie\ a fivebrane and its
mirror image) near the orientifold in $(x^7, x^8, x^9)$,
and study the transition in which the pair approaches
each other and the orientifold, and then splits along
the orientifold (in $x^6$). This process is described in
Fig.~\ref{twentyone}.

\smallskip

A closer look reveals that when the charge
of the orientifold is negative, it is in fact
impossible to separate the two $NS5$-branes
along the orientifold.
The low energy worldvolume theory on a pair of
$NS5$-branes near an $O4_-$ plane (more precisely
on the $1+3$ dimensional spacetime
$(x^0, x^1, x^2, x^3)$) has gauge group
$Sp(1)$, and the displacement of the
fivebranes in the $x^6$ direction is described
by an antisymmetric tensor of $Sp(1)$,
\ie\ a singlet. Since it is impossible to Higgs
$Sp(1)$ using a singlet, it is impossible
to separate the two $NS5$-branes on the
$O4_-$ plane.

For positive orientifold charge
the separation of Fig.~\ref{twentyone}(a) 
is attainable. $NS5$-branes near an
$O4_+$ plane are described by an $SO(2)$
gauge theory. The motions in the $x^6$
direction are described by a symmetric tensor
of $SO(2)$, which includes a pair of charged
scalars. Giving an expectation value to the
symmetric tensor completely breaks the $SO(2)$
symmetry and corresponds to displacing
the two fivebranes relative to each other
on the $O4_+$ plane.

What happens when an $NS5$-brane and its
image approach an $O4_+$ plane and, after getting
stuck on it, separate in the $x^6$ direction?
Each of the fivebranes divides the orientifold
into two disconnected parts. One can show that
the parts of the orientifold on different sides
of the fivebrane must carry opposite RR charge.
This has been first shown by~\cite{EJS} by
comparing to gauge theory (see below); a worldsheet
explanation of this effect was given in~\cite{EGKT}.
Since far from the fivebranes the orientifold charge must
(by locality) be positive, between the fivebranes
it is negative. Furthermore, the total RR charge
must be continuous across each fivebrane, since
otherwise the net charge would curve the
fivebrane according to (\ref{BST40}) and, in particular,
change its shape at infinity, again violating
locality. Therefore, one expects to find
two $D4$-branes stretched
between the fivebranes.

Similarly, when a pair of $NS5$-branes approaches
a negatively charged $O4$-plane with two $D4$-branes
embedded in it, it {\em can} split into two fivebranes
at different values of $x^6$, and gives rise to the
configuration depicted in Fig.~\ref{twentyone}(b). Both
possibilities are useful for describing gauge theories
using branes.

Once we understand the behavior of $NS5$-branes near
$O4$-planes, that of $D6$-branes is in principle
determined by U-duality. In particular, it appears
that bringing pairs of $D6$-branes close to
an $O4_+$ plane and separating them in $x^6$ splits
the orientifold into components with alternating
positive and negative charges. 
This might at first sight seem surprising, but
it is related by U-duality to the behavior of
$NS5$-branes intersecting $O4$-planes. 
Compactifying $x^3$ one can use T-duality to
map a $D6$-brane intersecting an $O4$-plane
to a $D5$-brane stretched in $(x^0, x^1, x^2,
x^7, x^8, x^9)$ intersecting an $O3$-plane
stretched in $(x^0, x^1, x^2, x^6)$
and again cutting it into two disconnected
pieces. This system can be analyzed
by using S-duality, and properties of
$NS5$-branes.
 
Indeed, if we replace the $D5$-brane by an
$NS5$-brane, we arrive at a system
similar to that of Fig.~\ref{twentyone},
with the $O4$-plane replaced by an $O3$-plane.
The three dimensional analog of the transition
described in that figure is the following:
a pair of $NS5$-branes approach an $O3_+$
plane, and separate in $x^6$ on it.
The segment of the $O3$-plane between the
fivebranes flips sign and there is 
a single $D3$-brane embedded in it
to make the total RR charge continuous.

S-duality applied to this configuration
gives rise (using the results of section
\ref{D4N4}) to a configuration with
two $D5$-branes intersecting an $O3$-plane
and dividing it into three segments.
The leftmost and rightmost parts of the
orientifold have negative charge and a $D3$-brane
embedded in them, while the segment between
the $D5$-branes has positive RR charge.
Thus we conclude that the RR charge of the
$O3$-plane jumps as we cross a $D5$-brane.
Since the statement is true for any finite
radius of $x^3$, $R$, it is also true
as $R\to\infty$. Therefore, we conclude
that the RR charge of the $O4$-plane jumps
as it crosses a $D6$-brane.

We will also need to understand the generalization
of the s-rule to $D4$-branes stretched between an
$NS5$-brane and a $D6$-brane both of which are
stuck on an $O4$-plane. A natural guess is
the following. The usual s-rule forbids configurations
where two or more $D4$-branes are forced to be right
on top of each other. In the presence of an $O4$-plane,
it is natural to expect that if a part of
the $O4$-plane between an $NS5$-brane and a $D6$-brane
has negative charge and no $D4$-branes, one can connect
the two branes by a {\em pair} of $D4$-branes.
If the part of the orientifold between the two branes
has positive charge, or negative charge with two $D4$-branes
embedded in it (or any combination of these),
one cannot stretch any further fourbranes between them.

\begin{figure}
\centerline{\epsfxsize=140mm\epsfbox{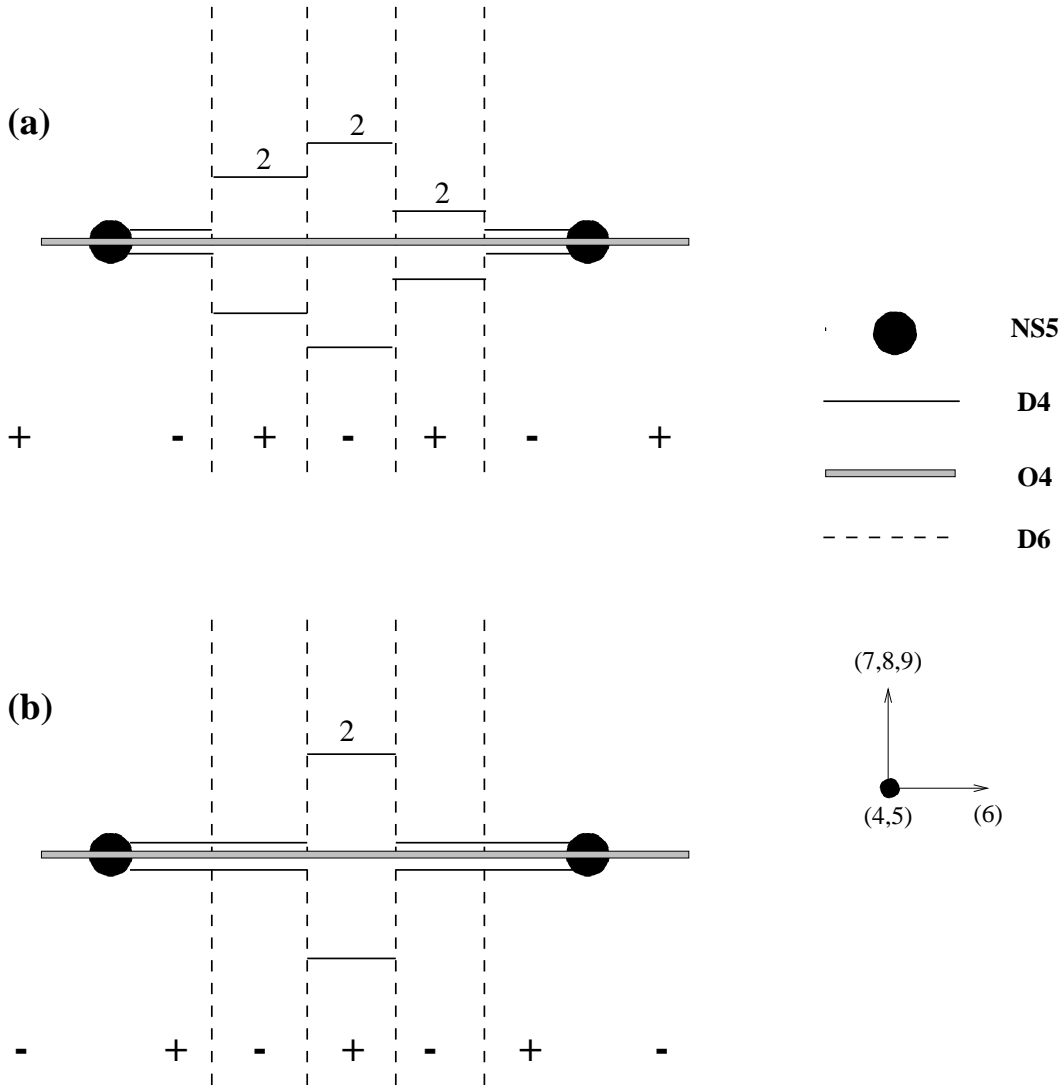}}
\vspace*{1cm}
\caption{The fully Higgsed branches of
$N=2$ SYM with
(a) $G=SO(2)$ and $\nf=2$
charged hypermultiplets;
(b) $G=Sp(1)$
and $\nf=2$ fundamental hypermultiplets.
The orientifold charge flips sign, as
indicated at the bottom, whenever one crosses
a $D6$ or $NS5$-brane.}
\label{twenty}
\end{figure}

We are now finally ready to turn to applications. 
When the charge of the segment of the orientifold
between the fivebranes is negative, the
brane configuration of Fig.~\ref{nineteen} describes an
$SO(\nc)$ gauge theory (we assume that
$\nc$ is even for now).
To describe matter we add $D6$-branes.
Note that when all the $D4$-branes are stretched between
the $NS5$-branes (in the Coulomb branch),
the $D6$-branes sit in pairs that
cannot be separated further, as discussed above.
The number of $D6$-branes must be even; we took it to
be $2\nf$, which corresponds to $\nf$ hypermultiplets
in the $({\bf \nc})$ of $SO(\nc)$. The $\nc/2$
dimensional Coulomb branch is described as usual by
displacing the $D4$-branes along the $NS5$-branes
in pairs away from the orientifold plane. The Higgs branch
is obtained by studying all possible breakings of the
$D4$-branes on $D6$-branes. Taking into account the
s-rule in the presence of an $O4$-plane explained
above leads to the splitting pattern of 
Fig.~\ref{twenty}(a). 

\smallskip

The resulting dimension of the fully Higgsed branch
(for $\nf>\nc$) is
\beq
{\rm dim}\MM_H=2\sum_{i=1}^{\nc\over2}\left[
2\nf-(4i-3)\right]=2\nf\nc-\nc(\nc-1)
\label{OSG2}
\eeq
in agreement with the gauge theory analysis.
It is instructive to verify that one also gets
the correct pattern of breaking and vacuum
structure for low numbers of flavors where the
gauge group cannot be completely Higgsed. We
leave this as an exercise to the reader.
One outcome of this exercise is a description
of the case of odd $\nc$, which can be obtained
from even $\nc$ by Higgsing one hypermultiplet
and breaking $SO(\nc)\to SO(\nc-1)$.

$Sp(\nc/2)$ gauge theory with $\nf$
fundamental hypermultiplets is
described by the configuration
of Fig.~\ref{nineteen} with
positive RR charge between the fivebranes
and negative outside. The charge reversal
changes the counting (\ref{OSG2})
in precisely the right way to reproduce the
appropriate gauge theory results. We
illustrate the structure of the fully Higgsed
branch in Fig.~\ref{twenty}(b).

\subsubsection{Some Generalizations}
\label{SGG}

\begin{figure}
\centerline{\epsfxsize=100mm\epsfbox{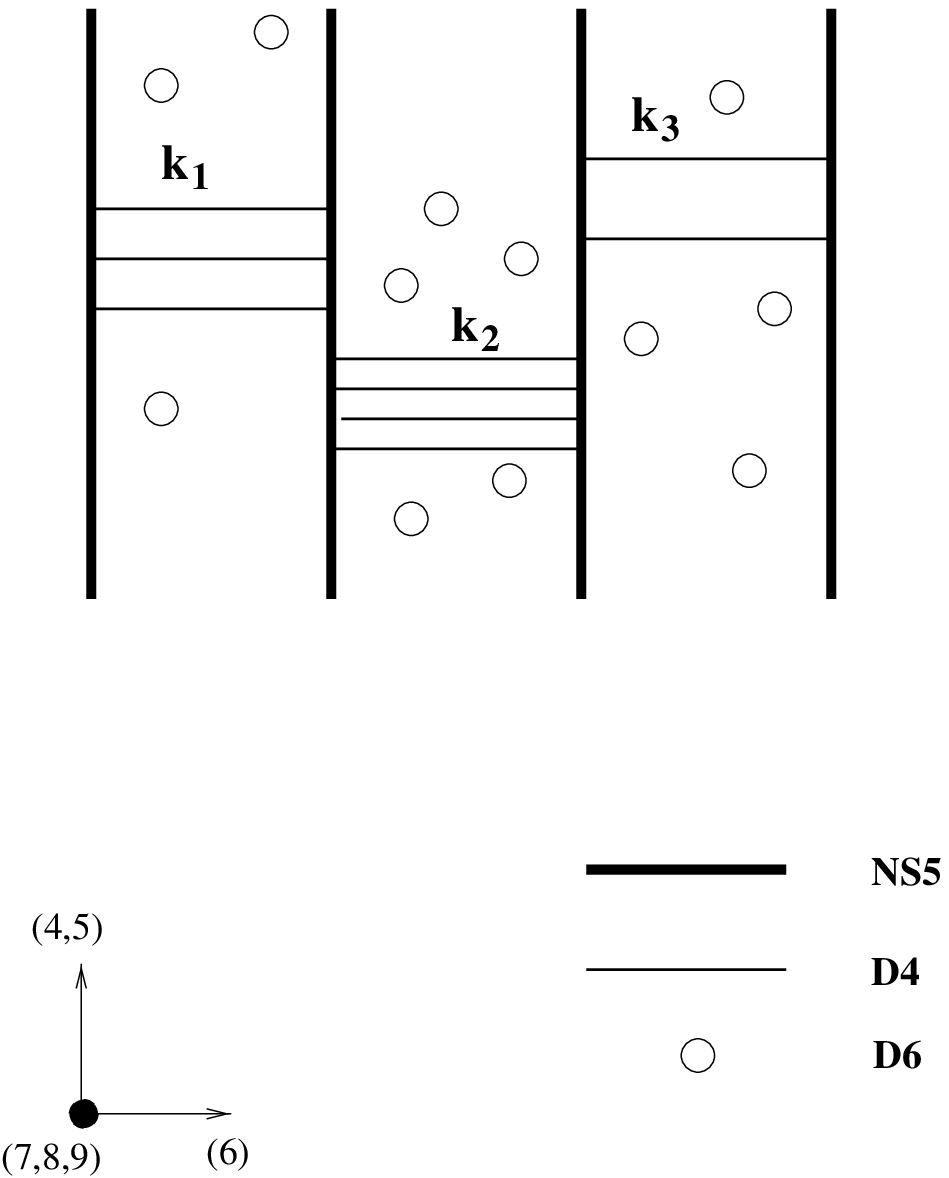}}
\vspace*{1cm}
\caption{A theory
with a product gauge group and matter
in the bifundamental of adjacent factors
of the gauge group (as well as fundamental
matter of individual factors).}
\label{twentytwo}
\end{figure}

Once the physics of the basic brane constructions
has been understood one can generalize them in
many different directions. One obvious example
is to increase the number of $NS5$-branes.
Consider, for example, a chain of $n+1$ fivebranes
labeled from $0$ to $n$, with the $(\alpha-1)$'st
and $\alpha$'th fivebrane connected by $k_\alpha$
$D4$-branes. In addition, let $d_\alpha$
$D6$-branes be localized at points
between the $(\alpha-1)$'st and $\alpha$'th
$NS5$-branes (see Fig.~\ref{twentytwo} for
an example).

The gauge group is in this case
$G=\prod_{\alpha=1}^n U(k_\alpha)$.
The matter hypermultiplets are the
following: $4-4$ strings connecting
the $k_\alpha$ fourbranes in the
$\alpha$'th interval to the $k_{\alpha+1}$
fourbranes in the $(\alpha+1)$'st interval
$(\alpha=1,\cdots, n-1)$
give rise to (bifundamental) hypermultiplets
transforming in the $(k_\alpha, \bar k_{\alpha+1})$
of $U(k_\alpha)\times U(k_{\alpha+1})$. In
addition we have $d_\beta$ hypermultiplets
in the fundamental representation of
$U(k_\beta)$ $(\beta=1,\cdots, n)$.
We leave the analysis of the moduli space of
vacua and the space of deformations to the reader.

If we add to the previous configuration an orientifold
fourplane the gauge group becomes an alternating $SO/Sp$ one.
For example, for even $n$ (an odd number of $NS5$-branes)
and negative RR charge of the segment of the $O4$-plane
between the first and second $NS5$-brane, the
gauge group is
$G=SO(k_1)\times Sp(k_2/2)\times SO(k_3)\times\cdots\times Sp(k_n/2)$
with bifundamental matter charged under adjacent factors of the
gauge group.

\smallskip

Brane configurations corresponding to theories with
such product gauge groups were considered
in~\cite{BSTYa,Tat,LLL97,GP,ENR}.

\begin{figure}
\centerline{\epsfxsize=100mm\epsfbox{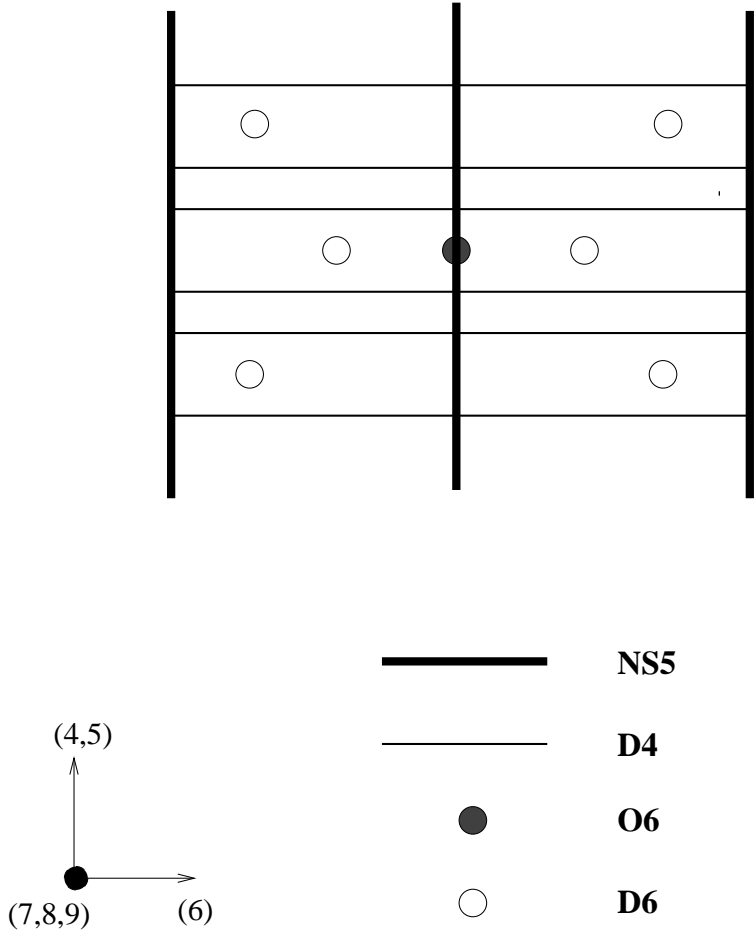}}
\vspace*{1cm}
\caption{A theory
with $G=U(\nc)$ ($\nc=6$ in this case),
one hypermultiplet in the
(anti-) symmetric representation and $\nf=3$
fundamentals.}
\label{twentythree}
\end{figure}
\smallskip

Another example is a generalization of the
configuration involving an orientifold sixplane
that we discussed previously in the context of
orthogonal and symplectic gauge groups. Consider
a configuration in which one $NS5$-brane is placed
at a distance $L_6$ from the $O6$-plane as before,
and another $NS5$-brane is placed so that it
intersects the orientifold plane. $\nc$ $D4$-branes
are stretched between the two fivebranes and $\nf$
$D6$-branes parallel to the $O6$-plane are placed
between the two $NS5$-branes 
(Fig.~\ref{twentythree}). 

The gauge theory describing this configuration
has $G=U(\nc)$ and a matter hypermultiplet $Z$
in the symmetric or antisymmetric
representation of $G$ (depending on the sign
of the orientifold), as
well as the usual $\nf$ fundamental hypermultiplets~\cite{LL}.
The two-index tensor hypermultiplet corresponds to
$4-4$ strings stretched from one side of the orientifold
to the other side, across the stuck $NS5$-brane.

\subsubsection{Quantum Effects: I}
\label{QEFF}
So far in this section we have described the
{\em classical} brane configurations and the
corresponding classical gauge theory dynamics.
For finite $g$ (\ref{BSB3})
there are important qualitative
new effects, which are the subject of this subsection.

We will first discuss these effects in the context
of $U(\nc)$ gauge theory, 
following~\cite{W9703}, which
was shown earlier to be described by a system
of two $NS5$-branes in type IIA string theory
with $\nc$ $D4$-branes stretched between 
them (see Fig.~\ref{nine}).
In the next subsection we will comment on the
generalization to some of the other cases
mentioned above.

For finite $\gs$, the type IIA string theory becomes
eleven dimensional at short distances. The radius
of the eleventh dimension $x^{10}$ is proportional
to $\gs$, $R_{10}=l_s\gs$ (\ref{BST19}). Furthermore,
as we saw in section \ref{MI}, the $D4$-brane can
be thought of as an M-theory fivebrane wrapped around
$x^{10}$. Thus, $D4$-branes stretched between $NS5$-branes
are reinterpreted in M-theory on $R^{10}\times S^1$
as describing a single fivebrane with a curved
worldvolume. Since all IIA branes are extended in the
$1+3$ dimensions $(x^0, x^1, x^2, x^3)$ and are located
at a point in $(x^7, x^8, x^9)$, the worldvolume of the
M-theory fivebrane is $R^{1,3}\times \Sigma$ where $\Sigma$
is a two dimensional surface embedded in the four dimensional
space $Q=R^3\times S^1$ labeled by $(x^4, x^5, x^6, x^{10})$.

It is convenient to parametrize the space $Q$
using the natural complex coordinates
\bea
s=&x^6+ix^{10}\nonumber\\
v=&x^4+ix^5
\label{SGG1}
\eea
In the classical type IIA string limit, the $NS5$-branes
of (\ref{BSB1}) are described by $s=$ constant, while the
$D4$-branes correspond to $v=$ constant. If we place the
two $NS5$-branes at $s=s_1, s_2$ and the $\nc$ fourbranes
stretched between them at $v=v_1, v_2, \cdots, v_{\nc}$,
the ``classical'' surface $\Sigma_{cl}$ is described by
$(s-s_1)(s-s_2)\prod_{a=1}^{\nc}(v-v_a)=0$, with
${\rm Re}(s_1)\le {\rm Re}(s)\le {\rm Re}(s_2)$.
$\Sigma_{cl}$ is a singular
surface with different components which meet at the singular
points $s=s_i$, $v=v_a$ ($i=1,2$; $a=1,\cdots, \nc$).

As we will see shortly, for finite $R_{10}$
and at generic points in moduli space
the singularities in $\Sigma$ are eliminated.
To determine the shape of the smooth surface
$\Sigma$ we next consider its large $v$
asymptotics. Classically we see at
large $v$ the two $NS5$-branes at fixed
$s=s_1, s_2$. However, we know from the discussion
of subsection \ref{BEB} that the ends of the
fourbranes on the fivebrane look like
charges~\footnote{The theory on the IIA fivebrane
is not a gauge theory, but rather a mysterious
non-abelian theory of self-dual $B_{\mu\nu}$
fields. However, the ends of fourbranes on the
fivebranes are codimension two objects and,
therefore, the relevant theory is the fivebrane
theory compactified down to $2+1$ dimensions,
which {\em is} a gauge theory, to which the
discussion of subsection \ref{BEB} can be applied.}.
More precisely, $q_L$ fourbranes ending on the
fivebrane from the left at $v=a_1,\cdots, a_{q_L}$,
and $q_R$ fourbranes  ending on if from the right
at $v=b_1,\cdots, b_{q_R}$ curve it asymptotically
according to
(\ref{BST40}) which in this case is:
\beq
x^6=l_s\gs\sum_{i=1}^{q_L}\log|v-a_i|-
l_s\gs\sum_{i=1}^{q_R}\log|v-b_i|
\label{SGG2}
\eeq
The fact that the coefficient of the $\log$
is proportional to $\gs$ can be understood
at weak coupling as a consequence of
properties of the $NS5$-brane.
Equations (\ref{BST12}, \ref{BST16}) imply that
the SYM coupling of the theory on the fivebrane
(reduced to $2+1$ dimensions) has no $\gs$
dependence, while the kinetic term of $X^6$ is
proportional to $1/\gs^2$. Thus, the BPS saturated
solution describing a fourbrane ending on the
fivebrane has gauge field $A\simeq Q\log|v|$
with the charge quantum $Q$ independent of
$\gs$, and $X^6\simeq Q\gs\log|v|$. The
factor of $l_s$ in (\ref{SGG2}) is required
by dimensional analysis.

At strong coupling (\ref{SGG2}) can be alternatively
derived by identifying $l_s\gs$ with $R_{10}$,
the only scale in the problem. The weak and strong
coupling arguments must agree
because the BPS property of the state in question allows
one to freely interpolate holomorphic properties
between the weak and strong
coupling regimes (and, as we will see,
(\ref{SGG2}) is closely related to a holomorphic
quantity).

Note that the fact that the end of a fourbrane
on a fivebrane looks like a codimension two
charged object implies that unlike the case
$p>2$ in (\ref{BST40}) one cannot, in general,
define quantum mechanically ``the location of the
$NS5$-brane'' by measuring $x^6$ at $|v|\to\infty$,
since the effects of the fourbranes (\ref{SGG2})
are not small for large $v$. $x^6$ approaches
a well defined value, which can then be interpreted
as the location of the fivebrane, only when the total
charge on the fivebrane vanishes, $q_L=q_R$.

The scalar field $x^6$ is related by $N=2$ SUSY to
$x^{10}$; together the two form a complex scalar
field $s$ (\ref{SGG1}) that belongs to the vector
multiplet of $N=2$ SUSY in the $3+1$ dimensional
spacetime. For consistency with SUSY,
(\ref{SGG2}) must be generalized to a holomorphic
equation for $s$ (\ref{SGG1}),
\beq
s=R_{10}\sum_{i=1}^{q_L}\log(v-a_i)-
R_{10}\sum_{i=1}^{q_R}\log(v-b_i)
\label{SGG3}
\eeq
The real part of this equation is (\ref{SGG2});
the imaginary part implies that $x^{10}$ jumps
by $\pm 2\pi R_{10}$ when we circle $a_i$ or
$b_i$ in the complex $v$ plane. Thus, the ends
of fourbranes on the fivebranes look like vortices.
Since $x^{10}$
is compact, it will be convenient for later
purposes to define
\beq
t=\exp(-{s\over R_{10}})
\label{SGG4}
\eeq
in terms of which  (\ref{SGG3}) takes the form
\beq
t={\prod_{i=1}^{q_R}(v-b_i)\over
\prod_{j=1}^{q_L}(v-a_j)}
\label{SGG5}
\eeq

We are now ready to determine the full shape of the
surface $\Sigma$ and thus the embedding of the
fivebrane corresponding to the
classical brane configuration realizing
pure $N=2$ SYM (Fig.~\ref{nine})
in the eleven dimensional spacetime.
SUSY requires $\Sigma$ to be given by a holomorphic
curve in the two complex dimensional space labeled by
$t, v$. It can be described by a holomorphic equation
$F(t,v)=0$ for some function $F$.
Viewed as a function of $t$ for large $v$ we expect
to see two branches corresponding to the two
``$NS5$-branes'' at (\ref{SGG5}):
$t_1\simeq v^{\nc}$ and $t_2\simeq v^{-\nc}$.
Therefore, the curve should be described by
setting to zero a second order polynomial in $t$,
\beq
F(t,v)=A(v)t^2+B(v)t+C(v)=0
\label{SGG6}
\eeq
where $A$, $B$ and $C$ are polynomials of
degree $\nc$ in $v$, so that for fixed $t$
there will be $\nc$ solutions for $v$
corresponding to the ``$D4$-branes''
stretched between the fivebranes.

As we approach a zero of $C(v)$, a solution
of the quadratic equation (\ref{SGG6}) goes
to $t=0$, \ie\
(\ref{SGG4}) $x^6=\infty$. Thus
zeroes of
$C(v)$ correspond to locations
of semi-infinite
fourbranes stretching to the right of the
rightmost $NS5$-brane. Similarly, $A(v)$
describes semi-infinite fourbranes
stretching to $x^6=-\infty$ from the
left $NS5$-brane. In the $N=2$ gauge
theory application semi-infinite fourbranes
give rise to fundamental matter and as
a first step we are not interested in them.
Thus we set~\footnote{We set the QCD scale
$\Lambda=1$ here. Restoring dimensions,
since $v$, $t^{1/\nc}$ scale like energy,
if we set $A(v)=1$ then
$C(v)=\Lambda^{2\nc}$.}
\beq
A(v)=C(v)=1
\label{acone}
\eeq

$B(v)$ is taken to be the most general polynomial
of degree $\nc$ which can, by rescaling and
shifting $v$, be brought to the form
\beq
B(v)=v^{\nc}+u_2v^{\nc-2}+u_3v^{\nc-3}\cdots
+u_{\nc}
\label{SGG7}
\eeq
$u_2, \cdots, u_{\nc}$ are complex constants
parametrizing the polynomial $B$.
The curve (\ref{SGG6}) with the choice
(\ref{acone}, \ref{SGG7}) of $A$, $B$
and $C$ has the right structure:
for fixed $t$ it has $\nc$ roots $v_i$
corresponding to the $\nc$ ``fourbranes.''
Note that while classically there should
only be such solutions for $t$ between
the $NS5$-branes, because of the bending
(\ref{SGG5}) there are in fact $\nc$
solutions for $v$ for any $t\not=0$.
Similarly, for all $v$ there are two solutions
for $t$, which for large $v$ behave like
\beq
t_\pm\simeq v^{\pm \nc}
\label{SGG8}
\eeq
in agreement with the general structure
expected from (\ref{SGG5}).

To recapitulate, just like classically there is
a one to one correspondence between configurations
of $D4$-branes stretched between $NS5$-branes
and vacua of classical $N=2$ SYM, quantum
mechanically there is a one to one correspondence
between vacua of quantum $N=2$ SYM and supersymmetric
configurations of an $M5$-brane
with worldvolume $R^{3,1}\times
\Sigma$, with $\Sigma$ described by
(\ref{SGG6}-\ref{SGG7}). Roots of the polynomial
$B$ (\ref{SGG7}) correspond to
``the locations of the
$D4$-branes''  and label different
points in the quantum Coulomb branch of
the $N=2$ SYM theory.

It is interesting that there are
only $\nc-1$ independent roots, labeled by
the moduli $\{u_i\}$.
This appears to be in contradiction
with the fact that there are $\nc$ massless
vectormultiplets living on the fourbranes
for generic values of the moduli $\{u_i\}$
corresponding to the unbroken $U(1)^{\nc}
\subset U(\nc)$.
In fact, the number of vectormultiplets
is $\nc-1$, in agreement with (\ref{SGG7}).
The $U(1)\subset U(\nc)$ has vanishing
coupling and is ``frozen.''

This can be understood semiclassically 
by evaluating
the kinetic term of the $U(1)$. The $\nc$
``fourbranes'' ending on an $NS5$-brane
from the left (say), bend it according
to (\ref{SGG3}). $a_i$ are the moduli,
and to probe their dynamics one should
allow them to slowly vary as a function
of $(x^0, x^1, x^2, x^3)$.
The kinetic energy of the fivebrane behaves
for such slowly varying configurations as
\beq
S\simeq \int d^4x\int d^2v|\partial_\mu s|^2
\simeq R_{10}^2\int d^4x\int d^2v
|\sum_i\partial_\mu a_i{1\over v-a_i}|^2
\label{sva}
\eeq
At large $v$, where (\ref{sva}) is
expected to be accurate,
we find
\beq
S\simeq R_{10}^2\int {d^2v\over|v|^2}
\int d^4x |
\sum_i\partial_\mu a_i|^2
\label{svadiv}
\eeq
The logarithmically divergent $v$
integral (\ref{svadiv})
leads to a vanishing coupling
for the $U(1)\subset U(\nc)$
\beq
{1\over g_1^2}\simeq R_{10}^2\int{d^2v\over
|v|^2}\to\infty
\label{onegsym}
\eeq

Interestingly, equations (\ref{SGG6}-\ref{SGG7})
describe the Seiberg-Witten curve for $SU(\nc)$
gauge theory with no matter! In gauge theory,
the low energy coupling matrix $\tau_{ij}$
(\ref{RFTR9}) is the period matrix of the
corresponding Riemann surface. This is also
the case in the fivebrane construction.
The worldvolume theory of a flat fivebrane
includes a self-dual $B_{\mu\nu}$ field
$(H=dB=*dB)$. Upon compactification on $\Sigma$,
$B$ gives rise to $g$ abelian 
vectormultiplets in $3+1$
dimensions, with $g$ the genus of the
Riemann surface $\Sigma$. In our case,
the surface $\Sigma$ can be thought of 
as describing two sheets (the ``fivebranes'')
connected by $\nc$ tubes (the ``fourbranes''),
and hence it has genus $g=\nc-1$. The coupling
matrix of the resulting $U(1)^{\nc-1}$ gauge theory
is the period matrix of $\Sigma$~\cite{EV}.

Another derivation of the relation between the
period matrix of the Riemann surface around which the
fivebrane is wrapped and the coupling matrix of the
abelian gauge theory on the brane that emphasizes the
role of the scalar fields living on the brane appears 
in~\cite{HLW}.

To summarize, the brane analysis agrees with
SW theory. It offers a rationale as to {\em why}
the low energy gauge coupling matrix and metric
on moduli space of $N=2$ SYM are described
by a period matrix of a Riemann surface. The
natural context for studying SW theory appears
to be as a compactification of the $(2,0)$
field theory of an $M5$-brane (the low energy
limit of the theory of $M5$ or 
type IIA $NS5$-branes)
on the Riemann surface $\Sigma$.

At this point we would like to pause for a few
comments on the foregoing discussion:

\medskip
\noindent
{\bf 1. Global Symmetry, Conformal Invariance And The
Shape Of The Fivebrane}

As we discussed in section \ref{RFTR}, classical
$N=2$ SYM has at the origin of moduli space a
global symmetry $SU(2)_R\times U(1)_R$. The
$SU(2)_R$ symmetry is part of $N=2$ SUSY; the
$U(1)_R$ reflects the classical conformal
invariance of the theory and is broken at one
loop by the chiral anomaly to $Z_{2\nc}$.

In the brane description, the $U(1)_R$ symmetry
is realized as the $SO(2)$ rotation group
of the $v$ plane, which acts as $v\to v\exp(i\alpha)$.
The classical configuration of $\nc$ $D4$-branes
-- all at $v=0$ -- stretched between the two $NS5$-branes
(say, at $s=0$) is invariant under this $SO(2)$ symmetry.
The brane analog of one loop effects is
the leading quantum correction, which
is the asymptotic bending (\ref{SGG3}).
It breaks
the $U(1)_R$ symmetry by curving the left and right
fivebranes~\footnote{For large $v$ it makes sense to
talk about the left and right fivebranes although
they are connected at small $v$.} to
\bea
s_L=&-\nc R_{10}\log v\nonumber\\
s_R=&+\nc R_{10}\log v
\label{slr}
\eea
This configuration is no longer invariant
under
\beq
v\to v\exp(i\alpha)
\label{via}
\eeq
Under (\ref{via})
\bea
s_L\to& s_L-\nc R_{10}i\alpha\nonumber\\
s_R\to& s_R+\nc R_{10}i\alpha
\label{slsr}
\eea
For generic $\alpha$ (\ref{slsr}) is clearly
not a symmetry, however, there are residual
discrete symmetry transformations
corresponding to $\alpha=2\pi n/2\nc$
due to the fact that
${\rm Im} s=x^{10}$ lives on a circle of radius
$R_{10}$. Thus a $Z_{2\nc}$ subgroup of $U(1)_R$
remains unbroken, in agreement with the
gauge theory analysis.

\medskip
\noindent
{\bf 2. Adding Flavors}

It is easy to add fundamental hypermultiplets
to the discussion. As we have noted above,
to describe the Coulomb branch of a model with
$\nf$ fundamentals of $SU(\nc)$ we can add
$\nf$ semi-infinite fourbranes, say to the right
of the $NS5$-branes. These are described by
turning on $C(v)$ in (\ref{SGG6}):
\beq
C(v)=\prod_{i=1}^{\nf}(v-m_i)
\label{cvmi}
\eeq
$m_i$ are the locations of the
semi-infinite fourbranes in the
$v$ plane which, as we have seen,
correspond to the masses of the
fundamental ``quarks.'' Thus
$N=2$ SQCD with $G=SU(\nc)$
and $\nf$ fundamentals is described
by the Riemann surface
\beq
t^2+B(v)t+C(v)=0
\label{sqcdnf}
\eeq
with $B(v)$ given by (\ref{SGG7}) and $C(v)$
by (\ref{cvmi}). This agrees
with the gauge theory results of~\cite{HOz,APS}.

\medskip
\noindent
{\bf 3.} ${\bf SU(\nc)}$ {\bf Versus} ${\bf U(\nc)}$

We have argued that the brane configuration
of Fig.~\ref{nine}, which classically describes
a $U(\nc)$ gauge theory, in fact corresponds
quantum mechanically to an $SU(\nc)$ one;
the coupling of the extra $U(1)$ factor vanishes.
This observation appears to be in contradiction
with the fact that the moduli and
deformations of the brane configuration
discussed above seem to be those of a $U(\nc)$
theory. This issue remains unresolved as of this
writing; below we explain the specific puzzles.

We saw previously that the moduli space of brane
configurations seems to match the
classical Higgs branch of $U(\nc)$ gauge theory
with $\nf$ hypermultiplets.
If the gauge group is
$SU(\nc)$, the complex dimension of the Higgs
branch is $2\nf\nc-2(\nc^2-1)$ and the brane
counting misses two complex moduli. By itself,
this need not be a serious difficulty; we saw
previously examples where some or all of the
field theory moduli were not seen in the brane
analysis. Unfortunately, the mismatch in the
structure of the Higgs branch is related to
a more serious difficulty having to do with
the field theory interpretation of
certain deformations of the brane configuration.

In the classical discussion
we have interpreted the relative location of the
two $NS5$-branes in $(x^7, x^8, x^9)$ as a
FI D-term for the $U(1)\subset U(\nc)$
(\ref{BSB7}). If the
gauge group is $SU(\nc)$, we have to modify
that interpretation, as the theory no longer
has a FI coupling. The question is whether
in the quantum theory the parameters corresponding
to a relative displacement of the two asymptotic
parts of the $M5$-brane in $(x^7, x^8, x^9)$
are moduli in the $3+1$ dimensional field theory
on the brane, or whether -- like the $U(1)$ vectormultiplet --
they are decoupled. There seem to be two logical
possibilities, each of which has its 
own difficulties~\cite{GP}.

An argument similar to that outlined in equations
(\ref{sva}-\ref{onegsym}) would suggest that
these parameters are frozen and correspond to
couplings in the $3+1$ dimensional gauge theory.
The kinetic energy of the scalar fields
$X^I$, $I=7,8,9$, seems to diverge (assuming
an asymptotically flat metric on the fivebrane,
as we have for $X^6(x^\mu)$, (\ref{sva}))
as
\beq
\LL_{\rm kin}\simeq \int d^2v (\partial_\mu X^I)^2
\label{kinenerg}
\eeq
Thus, the kinetic energy of the fields
$X^I$ is infinite and
we must find a coupling in
the Lagrangian of $SU(\nc)$ gauge theory that
has the same effect on the vacuum
structure as a FI D-term (\ref{BSB7}).
It is not known (to us) how to write such a coupling.
In order for such a coupling to exist, the
$U(1)$ factor would have to be unfrozen, and
the estimate of the kinetic energy
(\ref{sva}-\ref{onegsym}) would have to be
invalid.

Alternatively, one might imagine that
the parameters corresponding to
$(x^7, x^8, x^9)$ are in fact moduli
in the gauge theory. This would
apparently be consistent with gauge
theory; these moduli would
provide three of the four missing
moduli parametrizing the baryonic
branch of the theory. However, for
this interpretation to be valid
we have to come up with a mechanism
for rendering the naively divergent
kinetic energy (\ref{kinenerg}) finite
(without spoiling (\ref{sva})).
This sounds even more implausible
than the first scenario as one has
to cancel a more divergent kinetic
energy. It is not clear to us what is
the resolution of this problem.

\medskip
\noindent
{\bf 4.} ${\bf \nf\ge 2\nc}$

For $\nf=2\nc$, at the
origin of the Coulomb branch and for
vanishing quark masses, the curve
(\ref{sqcdnf}) is:
\beq
t^2+a v^{\nc}t+bv^{2\nc}=0
\label{specurv}
\eeq
or equivalently
\beq
t_\pm=\lambda_\pm v^{\nc}
\label{tpmv}
\eeq
with $\lambda_\pm$ the two solutions
of
\beq
\lambda^2+a\lambda+b=0
\label{lab}
\eeq
The $U(1)_R$ symmetry
\bea
v\to&e^{i\alpha}v\nonumber\\
t\to&e^{i\nc\alpha}t
\label{vtconf}
\eea
is unbroken.
Thus the theory at the origin is an interacting
non-trivial $N=2$ SCFT. This is consistent with
the fact that
the curve (\ref{tpmv}) is singular
at $t=v=0$ -- a hallmark of a non-trivial
SCFT. The ratio $w=\lambda_+\lambda_-/
(\lambda_+-\lambda_-)^2$ is invariant
under rescaling of $t$, $v$ and can be thought
of as parametrizing the coupling constant
of the SCFT. For weak coupling, $w\simeq0$,
one has $w=\exp(2\pi i\tau)$ ($\tau$ is the
complex gauge coupling), but more generally,
due to duality, $\tau$ is a many valued function
of $w$.

For $\nf>2\nc$ the description
(\ref{sqcdnf}) seems to
break down. Both solutions for $t$ behave at
large $v$ as $t\sim v^{\nf/2}$, while (\ref{SGG3})
(in the presence of $\nf$ semi-infinite ``fourbranes''
stretching to $x^6\to\infty$) predicts $t_1\sim
v^{\nc}$, $t_2\sim v^{\nf-\nc}$. Not coincidentally,
in this case the gauge theory is not asymptotically
free and must be embedded in a bigger theory to make
it well defined in the UV. And, in any case,
it is free in the IR. It is in fact possible
to modify (\ref{sqcdnf}) to accommodate these cases
(see~\cite{W9703} for details).

\medskip
\noindent
{\bf 5. BPS States}

The fivebrane description of $N=2$ SYM
can also be used to study massive BPS saturated
states. Examples of such states in SYM include
charged gauge boson vectormultiplets and magnetic
monopole hypermultiplets. In the classical IIA
limit, massive gauge bosons are described by
fundamental strings stretched between different
fourbranes. For finite $R_{10}$ these fundamental
strings are reinterpreted as membranes wrapped
around $x^{10}$. Thus, charged $W$ bosons are
described in M-theory by minimal area membranes
ending on the fivebrane. Clearly, the topology
of the resulting membrane is cylindrical.

Monopoles are described in the IIA limit by
$D2$-branes stretched between the two $NS5$-branes
and two adjacent $D4$-branes, as in Fig.~\ref{eight}. 
In M-theory they correspond to membranes
with the topology of a disk ending on the
fivebrane.

There are other BPS states such as quarks and
various dyons, all of which are described in M-theory
by membranes ending on the fivebrane. Membranes
with the topology of a cylinder always seem to
give rise to vectormultiplets, while those
with the topology of a disk give hypermultiplets.
We will not describe the corresponding membranes
in detail here, referring instead to~\cite{HY,Mik}.

\medskip
\noindent
{\bf 6. Compact Coulomb Branches And Finite} 
${\bf N=2}$ {\bf Models}

The fact that the gauge coupling of the $U(1)\subset
U(\nc)$ vanishes is related to the infinite
area of the $v$-plane (\ref{onegsym}).
One might wonder what would happen if we
compactified $(x^4, x^5)$ on a two-torus.

Already classically we see that in this situation
the Coulomb branch of the theory, labeled by
locations of $D4$-branes stretched between
fivebranes, becomes compact.
Quantum mechanically we see that
since (\ref{SGG3}) is a solution
of a two dimensional Laplace equation
\beq
\partial_v\partial_{\bar v} s=R_{10}
\sum_{i=1}^{q_L}\delta^2(v-a_i)-
R_{10}\sum_{i=1}^{q_R}\delta^2(v-b_i)
\label{tdl}
\eeq
on a compact surface, the total
charge on each fivebrane must vanish: $q_L=q_R$.
This means that there must be
$\nc$ semi-infinite fourbranes
attached to each fivebrane and the
total number of flavors must thus
be $\nf=2\nc$. The solution of
(\ref{tdl}) that generalizes
(\ref{SGG3}) to the case of a two-torus
is ($q=q_L=q_R$):
\beq
s=R_{10}\sum_{i=1}^q\left[
\log\chi(v-a_i|\rho)-\log\chi(v-b_i|\rho)
\right]
\label{newsrten}
\eeq
where $\rho$ is the modular parameter
(complex structure) of the $v$-plane
torus ($v\sim v+1$, $v\sim v+\rho$), and
\beq
\chi(z|\rho)={\theta_1(z|\rho)\over
\theta^\prime_1(0|\rho)}
\label{chizrho}
\eeq
$\log\chi$ is related to the propagator of
a two dimensional scalar field
on a torus with modulus $\rho$
(see~\cite{GSW} for notation and references). 
Note that $\chi$ itself is not well
defined on the torus; its periodicity
properties are:
\bea
&\chi(z+1|\rho)=-\chi(z|\rho)\nonumber\\
&\chi(z+\rho|\rho)=-e^{-i\pi\rho-2i\pi z}\chi(z|
\rho)
\label{perpro}
\eea
However, we only require that the curve
built using $\chi$ should exhibit the
periodicity. To construct this curve,
start with the infinite volume curve
(\ref{SGG6}) describing this situation~\footnote{Recall
that $v_i$ are moduli parametrizing the Coulomb
branch of the theory, while $m_i^{(1)}$,
$m_i^{(2)}$ are masses of flavors corresponding to
semi-infinite fourbranes extending to the left and right,
respectively.}
\beq
t^2\prod_{i=1}^{\nc}(v-m_i^{(1)})+t\prod_{i=1}^{\nc}
(v-v_i)+\prod_{i=1}^{\nc}(v-m_i^{(2)})=0
\label{infvtm}
\eeq
and replace each $(v-a_i)$ by $\chi(v-a_i|\rho)$.
This gives
\beq
t^2\prod_{i=1}^{\nc}\chi(v-m_i^{(1)}|\rho)+t\prod_{i=1}^{\nc}
\chi(v-v_i|\rho)+\prod_{i=1}^{\nc}\chi(v-m_i^{(2)}|\rho)=0
\label{finchivmi}
\eeq
Using (\ref{perpro}) and the fact that the moduli
$v_i$ and masses $m_i$ satisfy the relations
\bea
&\sum_{i=1}^{\nc} (v_i-m_i^{(1)})={\rm const}\nonumber\\
&\sum_{i=1}^{\nc} (v_i-m_i^{(2)})={\rm const}
\label{eqchrg}
\eea
we find that the curve (\ref{finchivmi})
indeed has the right periodicity properties.

At first sight the generalization
of $N=2$ SYM with compact Coulomb branch
seems mysterious, but in fact it can be
thought of as the moduli space of vacua
of a six dimensional ``gauge theory''
compactified on a two-torus. Indeed, if
$v$ parametrizes a two-torus $T^2$, we
can T-dualize our classical configuration
of $D4$-branes ending on $NS5$-branes, and
using the results of section \ref{BST}
reach a configuration of $D6$-branes
wrapped around the dual torus $\tilde T^2$
and ending on the $NS5$-branes in the
$x^6$ direction.

{}From the six dimensional point of view
it is clear that we must require
$\nf=2\nc$, since the only configuration
consistent with RR charge conservation
involves in this case $\nc$ infinite
$D6$-branes extending to infinity in
$x^6$ and intersecting the
two $NS5$-branes. Wilson lines around
the $\tilde T^2$ give rise to the
parameters $m$, $v$ (\ref{finchivmi}).
{}From the gauge theory
point of view, $\nf=2\nc$ is necessary
due to the requirement of cancellation
of six dimensional chiral anomalies.

The curve (\ref{finchivmi}) exhibits
an $SL(2,Z)$ duality symmetry
corresponding to modular transformations
$\rho\to(a\rho+b)/(c\rho+d)$ under which
\beq
\chi\left({z\over c\rho+d}|{a\rho+b\over c\rho+d}
\right)=
{\eta \exp({i\pi cz^2\over c\rho+d})
\over c\rho+d}\chi(z|\rho)
\label{modtran}
\eeq
$\eta$ is an eight-root of unity.
This $SL(2,Z)$ symmetry
provides a geometric realization of the
duality symmetry of finite $N=2$ SYM models
(which are anomaly free in $6d$ and thus
can be lifted to $6d$). Note that the area
of the $v$-plane torus does not appear in
(\ref{finchivmi}). This is essentially because
$v$ has been rescaled to absorb a factor
of the area. The four dimensional limit
of (\ref{finchivmi}) is obtained by taking
$v\ll 1,\rho$ where $\chi(v|\rho)$ reduces to $v$.

\medskip
\noindent
{\bf 7. SQCD Versus MQCD}

As we discussed before, the limit
that one needs to take to study
decoupled gauge dynamics on the
fivebrane is $R_{10},L_6\to0$
holding $R_{10}/L_6=g_{SYM}^2$
fixed. In this limit the fivebrane
becomes singular although its
complex structure (\ref{sqcdnf})
is regular. To fully understand
gauge dynamics in this limit one
needs to study the fivebrane
theory in the IIA limit.

Witten has suggested to study the theory
in the opposite limit $R_{10},L_6\to\infty$,
$R_{10}/L_6$ fixed, observing that in that
limit (\ref{sqcdnf}) describes a large
smooth fivebrane and thus can be accurately
studied by using low energy M-theory, \ie\
eleven dimensional supergravity (the fivebrane
dynamics in this limit is sometimes referred
to  in the literature as ``MQCD'').

For holomorphic properties of the vacuum,
such as the low energy gauge couplings and metric
on moduli space (\ref{RFTR8}-\ref{RFTR10}),
the two limits must agree due to SUSY. However,
non-holomorphic low energy features are quite
different in the two limits. In particular,
in the MQCD
limit the fivebrane dynamics
is no longer effectively four dimensional,
and there is large mixing between gauge
degrees of freedom
and other excitations. Thus, it is
misleading to refer to the M-theory limit
as QCD (M or otherwise).

The situation is similar to the well
known worldsheet duality in open plus
closed string theory. The physics
can be viewed either in the open string
channel (where light states
are typically gauge fields) or as due to
closed string exchange (gravitons, dilatons, etc.).
Worldsheet duality implies that the two
representations must agree, but one may
be simpler than the other. In some situations
the open string representation
is dominated by the massless
sector, but then in the closed string channel
one needs to sum over exchanges of arbitrarily
heavy string states.
In such cases, the relevant physics is that
of gauge theory.

In other cases, the closed string channel
is dominated by exchange of massless modes
such as gravitons, but then the open string
calculation receives contributions from
arbitrarily heavy states and there is no
simple gauge theory interpretation of the physics.

The only known cases (with the possible exception
of DLCQ matrix theory -- reviewed in~\cite{Ban,BS} --
whose status is unclear as of
this writing) where there is a simple interpretation
in both the open and closed string channels involve
quantities protected by supersymmetry.

In our case, the analog of the closed string
channel is the eleven dimensional ``MQCD''
limit $R_{10}, L_6\to\infty$ where physics
is dominated by gravity, while the analog
of the open string channel is the IIA limit
$R_{10}, L_6\to 0$. Low energy features
that are not protected by SUSY need not
agree in the two limits (except perhaps
in certain large $N$ limits). SQCD corresponds
to the latter.

\medskip
\noindent
{\bf 8. Non-Trivial Infrared Fixed Points}

At generic points in the Coulomb branch,
the infrared dynamics of $N=2$ SYM 
is described by $r$ massless photons
whose coupling matrix is the period
matrix of the Riemann surface $\Sigma$.
At points where additional matter
goes to zero mass, the infrared dynamics
changes, and in many cases describes
a non-trivial SCFT~\cite{AD,APSW}. 
These situations correspond
to degenerate Riemann surfaces $\Sigma$.

Whenever that happens, the supergravity 
description breaks down, even if
$R_{10}$ and $L_6$ are large. Thus,
eleven dimensional supergravity 
provides a useful description of the
fivebrane wrapped on $\Sigma$ only
sufficiently far from any points
in moduli space where the infrared
behavior changes; in particular,
it cannot be used to study (beyond
the BPS sector) the non-trivial SCFTs
discussed by~\cite{AD,APSW}.

\subsubsection{Quantum Effects: II}
\label{QEFFII}

The analysis of the previous subsection
can be easily generalized to the chain
of fivebranes connected by fourbranes
mentioned above~\cite{W9703}. Specifically, consider
the IIA configuration of $n+1$ $NS5$-branes
labeled by $\alpha=0,1,\cdots, n$, with
$k_\alpha$ $D4$-branes connecting the
$(\alpha-1)$'st and $\alpha$'th 
fivebranes (Fig.~\ref{twentytwo}).
For simplicity, we assume that there are
no semi-infinite fourbranes at the edges.

Classically we saw that the gauge group
was $\prod_{\alpha=1}^nU(k_\alpha)$, but
the $n$ $U(1)$ factors are frozen as before.
Thus the gauge group is
\beq
G=\prod_{\alpha=1}^n SU(k_\alpha)
\label{gska}
\eeq
with matter in the bifundamental representation
$(k_\alpha, \bar k_{\alpha+1})$ of
adjacent factors of the gauge group.
We will further assume that all factors
in (\ref{gska}) are asymptotically free:
\beq
2k_\alpha-(k_{\alpha_+1}+k_{\alpha-1})\ge 0,\;
\;\;\;\;\forall\alpha
\label{kaka}
\eeq
Following the logic of our
previous discussion we expect
the Riemann surface $\Sigma$
to be described in this case
by the holomorphic equation
\beq
F(t,v)=t^{n+1}+P_{k_1}(v)t^n
+P_{k_2}(v)t^{n-1}+\cdots+P_{k_n}(v)t+1=0
\label{nfkf}
\eeq
The fact that (\ref{nfkf}) is a polynomial
of degree $n+1$ in $t$ ensures that there
are $n+1$ solutions for $t$ corresponding
to the $n+1$ $NS5$-branes. The $v$ independence
of the coefficients of $t^{n+1}$ and $1$
implies the absence of semi-infinite fourbranes.
The degrees of the polynomials in $v$
$P_{k_\alpha}=c_\alpha v^{k_\alpha}+\cdots$
are determined by the fact that rewriting
\beq
F(t,v)=\prod_{\alpha=0}^n(t-t_\alpha(v))
\label{ftvrew}
\eeq
the locations of the $n+1$ fivebranes $t_\alpha(v)$
must behave for large $v$ as (\ref{SGG3}-\ref{SGG5}):
\beq
t_\alpha(v)\sim v^{k_{\alpha+1}-k_\alpha}
\label{tavk}
\eeq
(where $k_0=k_{n+1}=0$); to check that this
leads to (\ref{nfkf}) one has to use (\ref{kaka}).
The roots of $P_{k_\alpha}(v)$ correspond to
the positions of the $k_\alpha$ fourbranes
connecting the $(\alpha-1)$'st and $\alpha$'th
fivebranes.

As we have seen in the classical brane construction,
semi-infinite fourbranes provide a convenient
tool to describe the Coulomb branch of SQCD with
fundamental matter, but to study the full moduli
space of vacua (in particular, to see the Higgs
branches) it is necessary to introduce $D6$-branes.
Our next task is to understand models with sixbranes
at finite $R_{10}/L_6$.

Recall that the $D6$-brane corresponds
in M-theory to a KK monopole magnetically
charged under $G_{\mu 10}$. The (hyper-K\"ahler)
metric around a collection of KK monopoles is the
multi Taub-NUT metric (\ref{KK1}-\ref{KK3}).
We do not actually need the metric around
a KK monopole, but only its complex structure.
The hyper-K\"ahler manifold (\ref{KK1}-\ref{KK3})
in fact admits three independent complex structures,
any of which is suitable for our purposes.

The typical situation we will be interested in
is one where there are $\nf$ KK monopoles at
$v=m_1,\cdots, m_{\nf}$. One of the complex
structures of the corresponding multi Taub-NUT
space can be described by embedding it in a three
complex dimensional space with coordinates
$y,z,v$. It is given by:
\beq
yz=\prod_{i=1}^{\nf}(v-m_i)
\label{comstr}
\eeq
When all the KK monopoles coincide,
(\ref{comstr}) approaches an $A_{\nf-1}$
singularity $yz=v^{\nf}$. The symmetry
$y\to \lambda y$, $z\to \lambda^{-1}z$ of
(\ref{comstr}) corresponds to $t\to \lambda t$.
Thus one can think of $y$ as corresponding to
$t$ (with $z$ fixed) or of $z$ as corresponding
to $t^{-1}$ (with $y$ fixed).

Note that the complex structure (\ref{comstr})
is insensitive to the $x^6$ location of the
$\nf$ KK monopoles. That information resides
in the K\"ahler class of the metric (\ref{KK1}-\ref{KK3})
which does depend on $x^6$.
Thus even when different $m_i$ in (\ref{comstr})
coincide, the corresponding $A_{\nf-1}$ singularity
may still be resolved by separating the centers
of the monopoles in $x^6$.

Consider as an example $N=2$ SQCD with gauge group
$G=SU(\nc)$ and $\nf$ flavors, realized classically
as $\nf$ $D6$-branes situated between the two
$NS5$-branes in the $x^6$ direction (Fig.~\ref{fourteen}).
At finite $L_6/R_{10}$ we need again to find the shape
of an $M5$-brane, except now it lives not in
$Q=R^3\times S^1$, but rather in $\tilde Q=$ resolved
$A_{\nf-1}$ multi Taub-NUT space (\ref{comstr}).

We can again describe the fivebrane by a curve of the
form
\beq
A(v)y^2+B(v)y+C(v)=0
\label{dsixc}
\eeq
with some polynomials $A$, $B$, $C$.
As before, $A(v)=1$, since otherwise
$y$ (and, therefore, $t$) diverges at
roots of $A(v)$. Rewriting (\ref{dsixc})
in terms of $z=\prod(v-m_i)/y$ and requiring
that there should be no singularities of $z$
for finite $v$ (these too would correspond
to semi-infinite fourbranes) one finds that
$C=a\prod(v-m_i)$ (see~\cite{W9703} for a more 
detailed analysis). 
Finally, $B(v)$ (\ref{dsixc}) is a polynomial
of degree $\nc$ as before (\ref{SGG7}).

Thus we recover the solution found before
using semi-infinite fourbranes. The fact
that the result (\ref{dsixc}) is independent
of the $x^6$ position of the $D6$-branes is
consistent with our discussion in subsection
\ref{TBCS} where this was deduced as a consequence
of the HW transition.

The description of $N=2$ SQCD
with sixbranes (KK monopoles) can be used
to describe the Higgs branch of the theory
as well. We refer to~\cite{W9703} for a detailed
discussion of this.

Finally, $N=2$ gauge theories on $N_c$ fourbranes in the
presence of sixbranes and orientifold planes can be
lifted to M-theory, and used to derive the curves
and describe the Higgs branches of $SO(N_c)$ and $Sp(N_c/2)$
theories as well as many product
groups~\cite{LLL97,BSTYb,FS9706,NOYY9707,LL,NOY,ENR}.

\section{Four Dimensional Theories With $N=1$ SUSY}
\label{D4N1}

In this section we turn our attention
to four dimensional $N=1$ supersymmetric
gauge theories which typically
have the richest dynamics among the
different SYM theories and are the
closest to phenomenology.

As we saw in the previous sections,
$N=4$ supersymmetric gauge theory
has the simplest dynamics and
phase structure. The theory is specified
by the choice of a gauge group; all
matter is in the adjoint representation.
The vacuum structure consists of
a Coulomb branch with singularities
corresponding to points of enhanced
unbroken gauge symmetry. The most singular
point is the origin of moduli space, which
corresponds to a non-trivial CFT
parametrized by the exactly marginal
gauge coupling $g$. The form of the effective
action up to two derivatives is completely
determined by the symmetry structure; in particular,
the metric on the Coulomb branch is flat.
The leading terms that receive quantum corrections
are certain non-renormalizable (=irrelevant)
four derivative terms, and these corrections
can be controlled since they receive contributions
only from BPS states.
The most interesting
feature of the dynamics of $N=4$ SYM is
the discrete identification of theories
on the line of fixed points labeled by
$g$ provided by Montonen-Olive duality
(which acts as $g\leftrightarrow 1/g$).
Another interesting phenomenon is the appearance
of non-trivial infrared fixed points of the RG
at which electrically and magnetically charged
particles become massless at the same time.

In the $N=2$ SYM case
there are some new features.
Theories are now labeled by the choice
of a gauge group and a set of
matter representations.
Non-trivial quantum corrections
to the two derivative terms in the vectormultiplet
action lead to a modification of the
metric on the Coulomb branch,
described by Seiberg and Witten.
In addition,
Higgs branches in which the rank
of the gauge group is reduced
appear; as we saw before, $N=2$ theories
typically have a rather rich phase structure.

$N=1$ dynamics generally leads to yet
another host of new phenomena 
(see~\cite{AKMRV,Sei95,IS95,Giv,Peskin,Shifman}
and references therein).
It is now possible to write a classical
(tree level) superpotential. Furthermore,
the superpotential can in general receive quantum
corrections which modify the potential of the light
fields. At the same time these corrections are
often under control since they are holomorphic,
taking the form of a {\em superpotential}
on the classical moduli space. The effect
of the quantum superpotential may be to
lift a part of the classical moduli space,
change its topology,
or in some cases break SUSY completely,
a possibility with obvious phenomenological
appeal. $N=1$ SYM theories may also have a
chiral matter content and exhibit confinement,
possibilities that
do not exist in $N\ge2$ theories, and are
clearly desirable in a realistic theory.
Another interesting phenomenon is the infrared
equivalence between different $N=1$ SUSY
gauge theories discovered by Seiberg.
It provides a generalization
of Montonen-Olive duality to theories with
a non-trivial beta function. As we
discuss below, despite the running of the
coupling there is a sense in which
Seiberg's duality can be sometimes thought
of as a strong-weak coupling duality, and
in many cases it allows one to analyze
the strongly coupled dynamics of $N=1$
SYM theories.

In this section
we will describe $N=1$ SYM theories
using branes~\cite{EGK}. We will see that just like
in theories with more SUSY, embedding
$N=1$ SYM in brane theory provides a useful
qualitative and quantitative guide
for studying the classical and quantum
vacuum structure of these theories. In particular,
brane dynamics can be used to understand
Seiberg's infrared duality and a host of
other interesting strong coupling
effects. We start with a brief summary
of some field theory results (for more details see
the reviews cited above and references therein), and then
move on to the brane description.

\subsection{Field Theory Results}
\label{SFTR}

Pure $N=1$ SYM theory with a simple
gauge group $G$ describes
a vectormultiplet
$V$ (\ref{RFTR1}) transforming in the adjoint
representation of $G$.
The classical theory has a
single vacuum and a $U(1)_R$
symmetry, discussed in section \ref{RFTR}.
Just like the $N=2$ case, the existence
of the classical $R$-symmetry is related
to the classical superconformal invariance
of pure $N=1$ SYM.

Quantum mechanically, the theory develops
a $\beta$-function which breaks
conformal invariance. Accordingly,
the $U(1)_R$ symmetry is broken
by the gaugino condensate:
\beq
\langle \left({\rm Tr}
\lambda\lambda\right)^{C_2}\rangle
\sim \left(\nc\Lambda^3\right)^{C_2}
\label{gacon}
\eeq
to a discrete
subgroup $Z_{2C_2}$.
$\Lambda$ is the dynamically generated QCD scale and
$C_2$ is the second Casimir in the adjoint representation;
\eg\ $C_2=N_c$ for $G=SU(N_c)$, $C_2=\nc-2$ for $G=SO(\nc)$,
$C_2=\nc+2$ for $G=Sp(\nc/2)$.
The theory
has $C_2$ vacua corresponding to
different values of the condensate
consistent with (\ref{gacon}):
\beq
\langle {\rm Tr}\lambda\lambda\rangle =
{\rm const}\times\nc\Lambda^3
e^{2\pi i k\over C_2};\;\;\; k=0,1,2,\cdots, C_2-1
\label{FT1}
\eeq
which spontaneously breaks the discrete chiral
symmetry $Z_{2C_2}\rightarrow Z_2$. Each
of the $C_2$ vacua contributes $1$ to the Witten
index, ${\rm Tr} (-)^F=C_2$.
It is useful to note for future use that
(\ref{gacon}, \ref{FT1}) are equivalent
to a constant non-perturbative superpotential
\beq
W_{\rm eff}={\rm const}\times \nc^2\Lambda^3
\label{nonprtsup}
\eeq

Matter is described by chiral multiplets
$Q_f$ in representations $R_f$ of $G$.
The classical potential for the scalars
in the multiplets (which will be denoted
by $Q_f$ as well) includes
a ``D-term'' contribution
(the analogue of (\ref{RFTR6})):
\beq
V_D(Q)=\sum_a\Big(\sum_fQ_f^{\dagger}T_f^aQ_f\Big)^2
\label{FT2}
\eeq
$a=1,\cdots, {\rm dim}\;G$ runs over the
generators of the gauge group,
$f$ labels different ``flavors''
or representations;
$T_f^a$ are the generators of $G$ in the
representation $R_f$.
The only other contribution to the scalar
potential comes from the superpotential
\beq
\int d^2\theta W(Q)+\int d^2\overline{\theta}
W^*(Q^\dagger)
\label{supp}
\eeq
which leads after performing the $\theta$ integrals
to a potential
\beq
V_W(Q)\sim \sum_{f}|{\partial W\over\partial Q_f}|^2
\label{potsup}
\eeq
Classically there are often flat directions in field
space along which the potential vanishes. They correspond
(\ref{FT2}, \ref{potsup}) to solutions of $V_D=V_W=0$,
\ie:
\beq
\sum_fQ_f^{\dagger}T_f^aQ_f=
{\partial W\over\partial Q_f}=0
\label{flatdir}
\eeq
When the superpotential vanishes,
one can show that the space of solutions
of $V_D=0$ (\ref{flatdir}) is parametrized
by holomorphic gauge invariant combinations
of the matter fields $Q_f$. When $W\not=0$
one has to mod out that
space by the second constraint in (\ref{flatdir}).

Quantum corrections in general modify the
superpotential (\ref{supp}) and consequently
lift some or all of the classical moduli
space. Because
$W$ is a holomorphic function of
$Q$, in many cases the form of the exact
quantum superpotential can be determined
exactly. The quantum corrections to the K\"ahler
potential are in general more complicated
and are not under control. Fortunately, to study
the vacuum structure it is not important what
the K\"ahler potential is precisely,
as long as it is non-singular (and SUSY
is not broken).
Thus, below we will usually ignore the K\"ahler
potential, assuming it is non-singular in
the variables we will be using. Usually,
there is some circumstantial evidence
that this is the case (which we will not
review).

In the following we shall discuss
a few examples,
starting with $N=1$ SQCD -- an $SU(N_c)$
SYM theory with $N_f$ flavors
$Q^i,\tilde{Q}_i$, $i=1,...,N_f$, in the fundamental
and antifundamental representation, respectively.
In the absence of a superpotential,
the classical global symmetry of the theory is
\beq
SU(N_f)_L\times SU(N_f)_R\times U(1)_B \times U(1)_a \times U(1)_x
\label{FT3}
\eeq
The two $SU(N_f)$ factors rotate the quarks $Q^i$,
$\tilde Q_i$, respectively; $U(1)_B$ is a vectorlike
symmetry, which assigns charge $+1$ ($-1$) to $Q$ ($\tilde Q$).
$U(1)_a$ and $U(1)_x$ are $R$-symmetries
under which the gaugino is assigned
charge one, and the quarks $Q$, $\tilde Q$ have charge 0 or 1.
Only one combination of the two $R$-symmetries
is anomaly free -- we will refer to
it as $U(1)_R$.
The anomaly free global symmetry
of $N=1$ SQCD (with vanishing superpotential)
is
\beq
SU(N_f)_L\times SU(N_f)_R\times U(1)_B \times U(1)_R
\label{FT35}
\eeq
The $U(1)_R$ charge of
the quarks is
\beq
R(Q)=R(\tilde Q)=1-N_c/N_f
\label{FT4}
\eeq
The $U(1)_R$ symmetry (\ref{FT4}) plays
an important role in analyzing the strongly
coupled quantum dynamics of SQCD. At long
distances the theory flows to a fixed point
in which $N=1$ supersymmetry is enhanced to
$N=1$ superconformal symmetry. The
$U(1)_R$ symmetry (\ref{FT4}) becomes
part of the superconformal algebra
in the infrared. This is important
because the superconformal algebra
implies that for chiral operators
the scaling dimension at the infrared
fixed point $D$ is related to their $R$-charge
$Q$ via the relation $D=3Q/2$. The fact that
the symmetry (\ref{FT4}) is a good symmetry
throughout the RG trajectory allows
one to compute ``critical exponents''
at a non-trivial IR fixed point by
calculating charges of operators at the free
UV fixed point.

\subsubsection{Classical $N=1$ SQCD}

For $N_f<N_c$ massless flavors of
quarks the moduli space of vacua
is $\nf^2$ dimensional. It is labeled by
the gauge invariant meson fields
\beq
M^i_{j} \equiv Q^i\tilde Q_j \, , \qquad
i,j=1,...,N_f
\label{FT6}
\eeq
The gauge group can be maximally broken to
$SU(\nc-\nf)$. As a check, the quarks have
$2\nc\nf$ complex components, out of which
$\nc^2-(\nc-\nf)^2$ are eaten by the Higgs
mechanism, leaving $\nf^2$ massless degrees
of freedom $M^i_j$.
In various subspaces of the classical
moduli space, part or all of the broken gauge
symmetry is restored, and the classical
moduli space is singular -- one has to
add additional degrees of freedom
corresponding to massless quarks and
gluons to describe the low energy dynamics.

For $\nf\ge \nc$ new gauge invariant
fields appear in addition to (\ref{FT6}),
the baryon fields
\beq
B^{i_1i_2\cdots i_{\nc}}=
\epsilon^{\alpha_1\alpha_2\cdots\alpha_{\nc}}
Q_{\alpha_1}^{i_1} Q_{\alpha_2}^{i_2}\cdots
Q_{\alpha_{\nc}}^{i_{\nc}}
\label{baryons}
\eeq
There are $\nf\choose\nc$ baryon fields.
In particular, for $\nf=\nc$ there is a unique
baryon field $B$,
\beq
B=\epsilon_{i_1\cdots i_{N_c}}Q^{i_1}\cdots Q^{i_{N_c}}, \qquad
\label{FT9}
\eeq
This structure is doubled since
there are also fields $\tilde B$ constructed
out of the antifundamentals $\tilde Q$ in
an analogous way to (\ref{baryons}, \ref{FT9}).

Since for $\nf\ge\nc$ the gauge group can be
completely broken by the Higgs mechanism, the
complex dimension of the classical moduli space
is $2\nc\nf-(\nc^2-1)$. That means that there are
many constraints relating the baryon and meson
fields. For example, for $\nf=\nc$ the constraint
is
\beq
\det M - B\tilde B =0
\label{FT8}
\eeq
which gives the correct dimension of
moduli space $\nc^2+2-1=\nc^2+1$.
As before, the manifold (\ref{FT8}) describing
the classical moduli space is
singular, with additional massless
fields (gluons and quarks) coming
down to zero mass when $B$, $\tilde B$
and/or $\det M$ go to zero.

For general $\nf>\nc$ the classical
moduli space of vacua is rather
complicated.
The full set of classical constraints
among the mesons and baryons for general
$\nf$, $\nc$ has not been written down.

\subsubsection{Quantum $N=1$ SQCD}

For $\nf<\nc$ the classical picture
of an $\nf^2$ dimensional moduli space
labeled by the mesons $M^i_j$, with
singularities corresponding to enhanced
unbroken gauge symmetry, is drastically
modified due to the fact that the theory
generates a non-perturbative superpotential
for $M$. The unique superpotential
(up to an overall scheme dependent
constant) which is compatible with
the symmetries is
\beq
W_{\rm eff}=(N_c-N_f)
\left({\Lambda^{3N_c-N_f}\over \det M}\right)^{1\over N_c-N_f}
\label{FT5}
\eeq
where $\Lambda$ is the dynamically generated QCD scale. It
has been shown that the
superpotential (\ref{FT5}) is indeed generated
by gaugino condensation in the unbroken
gauge group $SU(N_c-N_f)$ for $N_f\leq N_c-2$,
and by instantons for $N_f=N_c-1$.

Using (\ref{potsup})
we see that $W_{\rm eff}$ gives rise to a
potential with no minimum at a finite
value of the fields. Thus the quantum
theory exhibits runaway behavior to
$M\to\infty$. Adding masses to all the
quarks,
\beq
W=W_{\rm eff}-m_i^j Q^i\tilde Q_j
\label{FT7}
\eeq
where the rank of the mass matrix is
$\nf$, stabilizes the runaway
behavior and gives rise to the
$N_c$ vacua
of pure $N=1$, $SU(N_c)$ SYM
mentioned above. To see that,
one integrates out the massive
fields $M^i_j$, which leads to
a superpotential
\beq
W_{\rm eff}={\rm const}\times
\left(\Lambda^{3\nc-\nf}\det m\right)^{1\over\nc}
\label{detmm}
\eeq
Using the scale matching relation
between the high energy theory
with $\nf$ flavors, $\Lambda_{\nc,\nf}$, and
the low energy theory with no
flavors, $\Lambda_{\nc,0}$,
\beq
\Lambda_{\nc,0}^{3\nc}=\Lambda_{\nc,\nf}^{3\nc-\nf}
\det m
\label{scalematch}
\eeq
leads to the pure SYM superpotential
(\ref{nonprtsup}).

For $N_f=N_c$ the superpotential (\ref{FT5}) is
singular. One finds that
$W_{\rm eff}=0$, but there are still important
quantum effects. In particular, the classical
constraint (\ref{FT8}) is modified to
\beq
\det M - B\tilde B = \Lambda^{2N_c}
\label{FT10}
\eeq
Thus, in this case quantum effects smoothen
the singularities in the classical moduli
space; in particular, there is no point
in moduli space where quarks and gluons
go to zero mass and the physics is well
described by the mesons and baryons subject
to the constraint (\ref{FT10}). This means
that color is confined. Note also that
since the point $M=B=\tilde B=0$ is not
part of the quantum
moduli space, there is no point
where the full chiral anomaly free global
symmetry (\ref{FT35})
is unbroken. Thus in this case
SQCD is confining and breaks chiral symmetry.
The moduli space (\ref{FT10}) can be
thought of as the moduli space of vacua
of a sigma model for a set of fields
$M^i_j$, $B$, $\tilde B$ and $\lambda$
with the superpotential
\beq
W_{\rm eff}=\lambda\left(\det M - B\tilde B -
\Lambda^{2N_c}\right)
\label{nfnc}
\eeq
$\lambda$ is a Lagrange multiplier
field, which is massive and hence does not
appear in the low energy dynamics. Integrating
it out leads to (\ref{FT10}).

For $N_f=N_c+1$ the baryons (\ref{baryons})
can be dualized to fields with one flavor
index, $B_i=\epsilon_{i i_1\cdots i_{\nc}}
B^{i_1\cdots i_{\nc}}$. Classically, the low
energy degrees of freedom $M^i_j$, $B_i$,
$\tilde B_i$ satisfy the constraints:
\beq
\det M (M^{-1})_i^j-B_i\tilde B^j=M_j^i B_i=
M^i_j\tilde B^j=0
\label {FT12}
\eeq
Quantum mechanically, the classical constraint
is lifted and the mesons and baryons can be thought
of as independent fields, governed by
the superpotential
\beq
W_{\rm eff}=-{\det M - M^i_j B_i \tilde B^j\over \Lambda^{2N_c-1}}
\label{FT11}
\eeq
The vacuum equations $\partial_M W_{\rm eff}=
\partial_B W_{\rm eff}
=\partial_{\tilde B} W_{\rm eff}=0$
give the classical constraints (\ref{FT12}).

It is at first sight surprising that the
quantum meson and baryon fields satisfy
the classical constraints (\ref{FT12})
only as equations
of motion, when in the classical limit
they are ``Bianchi identities.''
Two comments are useful to clarify the
situation. First, the classical limit
corresponds here to
$\Lambda\to 0$; in that case
the path integral
is dominated by configurations satisfying
the constraints (\ref{FT12}).
Second, the situation is analogous
to what happens under electric-magnetic
duality. In the electric variables,
$\partial_\mu F^{\mu\nu}=0$ is an
equation of motion while
$\partial_\mu \tilde F^{\mu\nu}=0$
is a Bianchi identity, while in the magnetic
variables the roles are reversed.
In fact, as we will discuss later,
the situation here is not only analogous
but identical to this example.
The relation between $N=1$ SQCD with
$\nf=\nc+1$ and the $\sigma$-model (\ref{FT11})
is a special case of a non-abelian generalization
of electric-magnetic duality, which indeed
exchanges Bianchi identities and equations
of motion.

The resulting quantum moduli space
for $N_f=N_c+1$
is identical to the classical one.
In particular, it has the same singularity
structure, but the interpretation of the
singularities is different. While in the
classical theory the singularities are
due to massless quarks and gluons, in
the quantum one they are due to massless
mesons and baryons. The theory again
confines, but this time the point
$M=B=\tilde B=0$ {\em is} in the
moduli space and chiral symmetry is
unbroken there. It is not difficult to
see that adding a mass to one or more of
the flavors (\ref{FT7}) gives rise to
the results (\ref{FT10}, \ref{FT5}), respectively.

For $N_f>N_c+1$ there is no known description
of the quantum moduli space in terms of a
$\sigma$-model for the gauge invariant
degrees of freedom $M$, $B$. Attempts to
write superpotentials consistent with the
symmetries typically lead to singularities,
suggesting the presence of additional
light degrees of freedom. For $\nf\ge 3\nc$
it is clear what the relevant degrees of
freedom are. In that case the theory is
not asymptotically
free and at low energies the quarks and gluons
are free (up to logarithmic corrections), much
like in QED. We refer to the theory as being in
a free electric phase, since electrically
charged sources have a QED-like potential
$V(R)\sim 1/R\log R$ in this case.

For $\nf<3\nc$ the theory is
asymptotically free. If $\nf$ is very close to
$3\nc$ (a possibility that exists \eg\ if $\nc, \nf$
are large) there is a weakly coupled infrared
fixed point
that can be studied perturbatively and
describes interacting quarks and gluons.
Electrically charged sources have a potential
$V(R)\sim \alpha^*/R$, and we say that the
theory is in a non-abelian Coulomb phase.
As $\nf$ is decreased, the infrared coupling
$\alpha^*$ increases, and perturbation theory
breaks down. For most values of $\nf$ in the
region $\nc+1<\nf<3\nc$ the theory is strongly
coupled and it is not clear what is the infrared
dynamics.

The degrees of freedom needed to describe
low energy $N=1$ SQCD in this range were uncovered by
Seiberg, who has shown that there is another
gauge theory -- with a different high energy
behavior -- that flows to the same infrared
fixed point as SQCD. Specifically, he discovered:

\noindent
{\bf Seiberg's Duality}

The following two theories
flow at long distances to the same fixed point:
\begin{enumerate}
\item ``Electric'' SQCD, with gauge group
$G_e=SU(\nc)$, and $\nf$ flavors of quarks
$Q^i$, $\tilde Q_i$.
\item ``Magnetic'' SQCD, with gauge group
$G_m=SU(\nf-\nc)$, $\nf$ magnetic quarks
$q_i$, $\tilde q^i$ and a gauge singlet
``magnetic meson'' chiral superfield $M^i_j$
which couples to the magnetic quarks via the
superpotential
\beq
W_{\rm mag}=M^i_jq_i\tilde q^j
\label{FT13}
\eeq
\end{enumerate}
The singlet mesons $M^i_j$ are the magnetic
analogs of the composite mesons $Q^i\tilde Q_j$
of the electric theory. Other operators can be
mapped as well, but it is not understood in the
context of gauge theory how to perform directly
a transformation from the
electric to the magnetic path integral.
In particular, the magnetic quarks and gluons
must be rather non-local functions of their
electric counterparts. For example, the mapping
of the baryons (\ref{baryons}) implies that (suppressing
flavor indices) $q^{\nf-\nc}\sim Q^{\nc}$.

Seiberg's duality allows one to study the low energy
dynamics of the electric theory in the regime
$\nc+1<\nf<3\nc$, by passing to the magnetic variables.
The magnetic, $SU(\nf-\nc)$ gauge theory is not
asymptotically free when $\nf<3\nc/2$; in this regime,
Seiberg's duality predicts that the strongly interacting
electric $SU(\nc)$ gauge theory is in fact free
in the appropriate variables! Since the weakly
coupled variables in this case are the dual,
magnetic variables, we refer to
the electric theory as being in
a free magnetic phase.

For $\nf>3\nc/2$ the magnetic theory is asymptotically
free, but just like in the electric case, when
$\nf$ is sufficiently close to $3\nc/2$ it describes
weakly interacting magnetic quarks and gluons
(as well as the fields $M$) in the IR. As we increase
$\nf$, the coupling in the IR increases. We see that
the electric and magnetic descriptions provide
complimentary pictures of the non-abelian
Coulomb phase. As $\nf$
increases, the electric description becomes more
weakly coupled (and thus more useful)
while the magnetic one becomes more strongly coupled
and vice-versa.

The original SQCD examples constructed by Seiberg
were generalized in a few different directions,
and many additional examples of the basic phenomenon
have been found. There is in general
no proof of Seiberg's duality in the context
of gauge theory but there is a lot of evidence
supporting it. There are three kinds of independent
tests:

\begin{itemize}
\item Members of a dual pair
have the same global symmetries
and the `t Hooft anomaly matching
conditions for these symmetries
are satisfied.
\item The two theories have the same
moduli spaces of vacua,
obtained by giving expectation values
to the first components of chiral
superfields.
\item The equivalence is preserved
under deformations of the theories
by the F-components of chiral operators.
In particular, the moduli spaces and
chiral rings agree as a function of these
deformations.
\end{itemize}
It is important to stress that in every one
of these tests the classical theories are
different and only the quantum theories
become equivalent. For example, in SQCD
the electric theory does not develop a quantum
superpotential (for $\nf>\nc+1$), while in the
magnetic theory
the classical superpotential (\ref{FT13})
is corrected quantum mechanically to
\beq
W_{\rm quantum}\sim {1\over \mu}
Mq\tilde q+\Lambda^{3\nc-\nf\over N_c-N_f}
(\det M)^{1\over N_f-N_c}
\label{FT14}
\eeq
where $\mu$ is some fixed scale.

There is also a crucial
difference in the interpretation of the
deformations of the two theories. Often,
when one theory is Higgsed and becomes
weaker, its dual is confining and becomes
stronger. This is one reason for interpreting
the relation between these theories as
electric-magnetic duality.

\subsubsection{SQCD With An Adjoint Superfield}
\label{AdSup}

An interesting generalization of $N=1$ SQCD
is obtained by adding to the theory a chiral
superfield $\Phi$ in the adjoint representation
The theory without a classical superpotential
is very interesting~\cite{KSS}. Unfortunately, not much
is known about its long distance behavior.
It is known that the quantum moduli space is
identical to the classical one. The only
singularities are at points where classically
the unbroken gauge symmetry is enhanced. The
most singular point in moduli space is the origin.
It is expected that the theory at the origin
is in a non-abelian
Coulomb phase for all $\nf\ge 1$
(for $\nf=0$ it actually has $N=2$ SUSY and
is equivalent to pure $N=2$ SYM, described
in section \ref{D4N2}).
As we saw before, the physical
interpretation of the singularities in the quantum
theory may be different from the classical one.

While the infrared physics at the origin of moduli
space is mysterious, some perturbations of the
theory by tree level superpotentials lead to theories
whose low energy behavior is understood. If we
add the superpotential
\beq
W=\lambda\sum_{i=1}^{N_f}\tilde{Q}_i\Phi Q^i
\label{FT16}
\eeq
we get a theory that can be analyzed easily.
When the Yukawa coupling $\lambda$ is one we recover
the $N=2$ SUSY theory discussed in section \ref{D4N2}.
The moduli space of the theory has a Coulomb branch
which has only massless photons at generic points.
At special singular points on the moduli space there
are more massless particles: massless monopoles, dyons,
massless gluons and quarks, and even points with interacting
$N=2$ superconformal field theories. More quantitatively,
this branch of the moduli space is described by the
hyperelliptic curves discussed in section \ref{D4N2}.

It is easy to extend the curve away from the $N=2$
limit~\cite{EFGR,HOz}.
Using the symmetries of the theory this is achieved by
replacing factors of $\Lambda^{2\nc-\nf}$ in the curve
by $\lambda^{\nf}\Lambda^{2\nc-\nf}$. Therefore, as
$\lambda\to 0$, all the features of the Coulomb branch
approach the origin; this is clearly a singular
limit which is not easy to describe from this point of
view.

Another deformation that simplifies the dynamics
involves turning on a polynomial superpotential
for $\Phi$.
When $\Phi$ is massive,
\beq
W=\mu{\rm Tr}\Phi^2+\lambda\sum_{i=1}^{N_f}\tilde{Q}_i\Phi Q^i
\label{FT17}
\eeq
we can integrate it out and obtain a superpotential
for the quarks of the form
\beq
W\sim {\lambda^2\over \mu} \tilde Q_i Q^j \tilde Q_j Q^i
\label{FT18}
\eeq
In the limit $\mu\to\infty$ the quartic superpotential
(\ref{FT18}) disappears and we recover the $SU(N_c)$
theory with $N_f$ flavors considered above.

An interesting deformation corresponds to
the pure polynomial superpotential
\beq
W_{\rm el}= \sum_{i=0}^k {s_i\over k+1-i}{\rm Tr}\Phi^{k+1-i}
\label{FT19}
\eeq
At first sight the fact that the high order polynomial
appearing in (\ref{FT19}) can have any effect on the
physics is surprising. Indeed, the presence of these
non-renormalizable interactions seems irrelevant
for the long distance behavior of the theory, which
is our main interest. Nevertheless, these operators
have in general strong effects on the infrared
dynamics. They are examples of operators that in
the general theory of the renormalization group
are known as {\em dangerously irrelevant}.

It was shown in~\cite{K,KS,KSS} that in the presence
of the superpotential (\ref{FT19}) there is
a simple dual description.
The magnetic theory has gauge group $SU(kN_f-N_c)$ with
$N_f$ magnetic quarks $q,\tilde q$, an adjoint
field $\varphi$ and
$k$ gauge singlet magnetic
meson fields $M_j$, $j=1,...,k$,
which correspond to the composite operators
\beq
(M_j)^i_l=\tilde Q_l\Phi^{j-1}Q^i
\label{mesons}
\eeq
The magnetic theory has a superpotential
\beq
W_{\rm mag}=-\sum_l{t_l\over k+1-l}{\rm Tr}\varphi^{k+1-l}+
\sum_{l=0}^{k-1}t_l\sum_{j=1}^{k-l} M_j\tilde q \varphi^{k-j-l}q
\label{FT20}
\eeq
where $\{t_i\}$ are coordinates on the space of magnetic
theories, related to the $\{s_i\}$ by a known coordinate
transformation on theory space.

When all the $\{s_i\}$ except for $s_0$ vanish, the
same is true for the magnetic couplings $\{t_i\}$,
and the duality relates in general non-trivial
strongly coupled gauge theories with
$W_{\rm el}\sim {\rm Tr} \Phi^{k+1}$, and
$W_{\rm mag}\sim {\rm Tr} \varphi^{k+1}$. When the
$\{s_i\}$ are generic, the
$k$ solutions of $W^\prime(x)=0$
for both the electric and magnetic theories are
distinct and both theories have a rather rich
vacuum structure. If we place $r_i$ eigenvalues
of $\Phi$ in the $i$'th minimum of the bosonic
potential corresponding to (\ref{FT19}), the
theory describes at low energies
$k$ decoupled SQCD systems with gauge group
$SU(r_i)$, $\nf$ flavors of quarks and gauged
baryon number. The total gauge group is broken as:
\beq
SU(\nc)\rightarrow SU(r_1)\times SU(r_2)\cdots
\times SU(r_k)\times U(1)^{k-1}
\label{breaking}
\eeq
A similar story holds for the magnetic theory;
the electric-magnetic duality between
(\ref{FT19}, \ref{FT20}) reduces in such
vacua to $k$ decoupled versions of
the original SQCD duality due to Seiberg.
More generally, a matrix version of singularity
theory is useful in the analysis of the theory~\cite{KSS}.

\subsection{Branes Suspended Between Non-Parallel Branes}
\label{BSBNPB}
In section \ref{BSB} we discussed configurations
of $NS5$, $D4$ and $D6$-branes
(\ref{BSB1}) which preserve eight
supercharges and are useful for describing
four dimensional $N=2$ SUSY gauge theories.
To describe $N=1$ SYM, we would like to break
four supercharges by changing the orientation
of some of the branes in the configuration.
This problem was encountered and discussed
in section \ref{WBS}. We saw there that performing
complex rotations such as that given by (\ref{BST38})
leads to configurations depending on continuous
parameters, which preserve the same four supercharges
for all values of the parameters. In this section
we will use this basic idea to study $N=1$ SYM
using branes~\cite{EGK,EGKRS}.

Starting with the brane configuration describing
$N=2$ SQCD with $G=SU(\nc)$ and $\nf$ fundamental
hypermultiplets (Fig.~\ref{eleven}), we can apply
complex rotations of the general form
(\ref{BST38}) to one or both of the $NS5$-branes,
or one or more of the $D6$-branes, such that
$N=1$ SUSY is preserved. Of course, only
the relative orientation in the $(v,w)$ plane of
all these objects is meaningful.
Recall that $NS5$-branes are located at
some particular value of $w$ and
are stretched in the $(x^{\mu},v)$ directions,
where $\mu=0,1,2,3$ and
\bea
v=&x^4+ix^5\nonumber\\
w=&x^8+ix^9
\label{D4N12}
\eea
while $D6$-branes are located at a particular value
of $v$ and are stretched in $(x^{\mu},w)$ (as well
as $x^7$).

If we rotate (say) the rightmost $NS5$-brane
in Fig.~\ref{eleven} by the angle $\theta$~\cite{Bar}
$(v,w)\to (v_{\theta},w_{\theta})$ (\ref{BST38}),
where
\bea
v_{\theta}=&v\cos\theta+w\sin\theta \nonumber\\
w_{\theta}=&-v\sin\theta+w\cos\theta
\label{wthe}
\eea
then the resulting fivebrane, which we may refer to
as the ``$NS5_{\theta}$-brane,'' is
located at $w_{\theta}=0$, or:
\beq
w_{\theta}=0\Rightarrow \;\;\;
w=v\tan\theta\equiv\mu(\theta) v
\label{muthe}
\eeq
Obviously, one can also apply
rotations of the $(x^8, x^9)$
plane, $w\to e^{i\varphi}w$ (or
rotations of the $(x^4, x^5)$
plane, $v\to e^{-i\varphi}v$).
Therefore, generically, $\mu$ is complex
\beq
\mu(\theta,\varphi)=e^{i\varphi}\tan\theta
\label{muthephi}
\eeq
(we will usually ignore this possible $\varphi$ dependence).

$\theta=0$ corresponds to the original NS-brane:
$NS5_0\equiv NS5$. For
$\theta=\pi/2$, the rotated brane is
stretched in $w$ and it is located at $v=0$.
Since this object will be particularly useful
below we give it a name,
the ``\nsp-brane:'' $NS5_{\pi/2}\equiv \nspp$.
Its worldvolume is
\beq
\nspp: \qquad (x^0, x^1, x^2, x^3, x^8, x^9)
\label{D4N11}
\eeq
Note that to be able to rotate one of the $NS5$-branes
relative to the other we need to locate all the $D4$-branes
stretched between them in Fig.~\ref{eleven}
at $v=w=0$, \ie\ approach the origin of the
Coulomb branch. The field describing fluctuations
of the fourbranes along the fivebranes, 
the chiral multiplet in the adjoint
representation of $SU(\nc)$ that belongs
to the vectormultiplet of $N=2$ SUSY, gets
a $\theta$ dependent mass due to the
rotation. Thus, the effect
of the rotation on the low energy field theory
on the $D4$-branes can be parametrized by the
superpotential
\beq
W\sim \mu(\theta)\Phi^2+\sum_{i=1}^{N_f}\tilde{Q}_i\Phi Q^i
\label{mutheta}
\eeq
which is a special case of the
theory discussed in (\ref{FT17}).

The mass of the adjoint chiral superfield
$\mu(\theta)$ in (\ref{mutheta}) clearly
breaks $N=2$ SUSY to $N=1$. The resulting
low energy theory is $N=1$ SQCD with a
superpotential for the quarks (\ref{FT18})
obtained by integrating out the massive
adjoint field $\Phi$. At least on
a qualitative level, the mass $\mu$ in
(\ref{mutheta})  is related to the geometrical
complex rotation parameter given in (\ref{muthe}).
Indeed, both vanish for $\theta=0$
(the $N=2$ SUSY configuration),
while when $\theta\to\pi/2$ we will see later
that the mass $\mu$ must go to infinity
and we recover
SQCD with vanishing superpotential (\ref{FT18}).

What happens when we rotate both $NS5$-branes of the
$N=2$ configuration of Fig.~\ref{eleven} by the same
angle $\theta$?
In the absence of $D6$-branes (\ie\ for $\nf=0$)
the answer is nothing, since there is a symmetry
between $v$ and $w$, so the  
low energy theory is pure $N=2$ SYM
for all $\theta$. In the presence
of $D6$-branes, the relative orientation between
the $NS5$ and $D6$-branes changes, and it is natural to
expect that the Yukawa coupling necessary for
$N=2$ SUSY will change with $\theta$,
\beq
W=\lambda(\theta)\sum_{i=1}^{N_f}\tilde{Q}_i\Phi Q^i
\label{D4N18}
\eeq
This is the model discussed after eq.
(\ref{FT16}).
The massless adjoint chiral superfield
$\Phi$ is now associated with fluctuations along
the $v_{\theta}$ (\ref{wthe}) directions.
The locations of $D4$-branes
along the $NS5_{\theta}$-branes correspond to
the expectation values
$\langle\Phi\rangle$ and parametrize the
Coulomb branch. The quarks are massive on the
Coulomb branch; their mass
$\lambda(\theta) \langle\Phi\rangle$ is due
in the brane description
to open $4 - 6$ strings whose minimal length is
$\langle\Phi\rangle \cos\theta$. We thus learn that
the Yukawa coupling $\lambda$ depends on the
angle $\theta$ via
\beq
\lambda(\theta)=\cos\theta
\label{lathe}
\eeq
$\theta=0$ corresponds to $\lambda=1$, the
$N=2$ configuration, while for $\theta=\pi/2$
(\ie\ after rotating the $NS5$-branes to
\nsp-branes)
the superpotential vanishes.

To recapitulate, the dictionary between
the deformations of the $N=2$ SUSY brane
configuration that preserve $N=1$ SUSY
and their manifestations
in the low energy theory on the fourbranes
is as follows.
Keeping one of the $NS5$-branes
and all the $D6$-branes fixed and rotating
the remaining $NS5$-brane corresponds to
changing the mass of the adjoint chiral
superfield $\Phi$ (\ref{mutheta}).
Rotating both $NS5$-branes relative to
the $D6$-branes, keeping the two fivebranes
and all the sixbranes parallel among themselves,
corresponds to changing the value of the Yukawa
coupling between $\Phi$ and the quarks
(\ref{D4N18}).

The most general configuration of this sort
corresponds to rotating all $\nf+2$
objects (the $\nf$ $D6$-branes and the two
$NS5$-branes) by arbitrary angles $\theta_i$
all of which are different. Since this configuration
breaks the $SU(\nf)$ symmetry between the
$D6$-branes, to describe it
one needs to vary individually the
different Yukawa interaction terms
of the different flavors.
We will next study a few examples that will hopefully
make the general case clear.

Our first example is $N=1$ SQCD. The main goals are to
describe the classical and quantum moduli space
of vacua and explain Seiberg's $N=1$ duality using
branes~\cite{EGK,EGKRS}. To this end, we explain in the next
two subsections the brane realization of the
classical electric and magnetic SQCD theories.
The study of quantum corrections is postponed
to the next section.

\subsubsection{Classical SQCD: The Electric Theory}
\label{D4N1ET}

\begin{figure}
\centerline{\epsfxsize=140mm\epsfbox{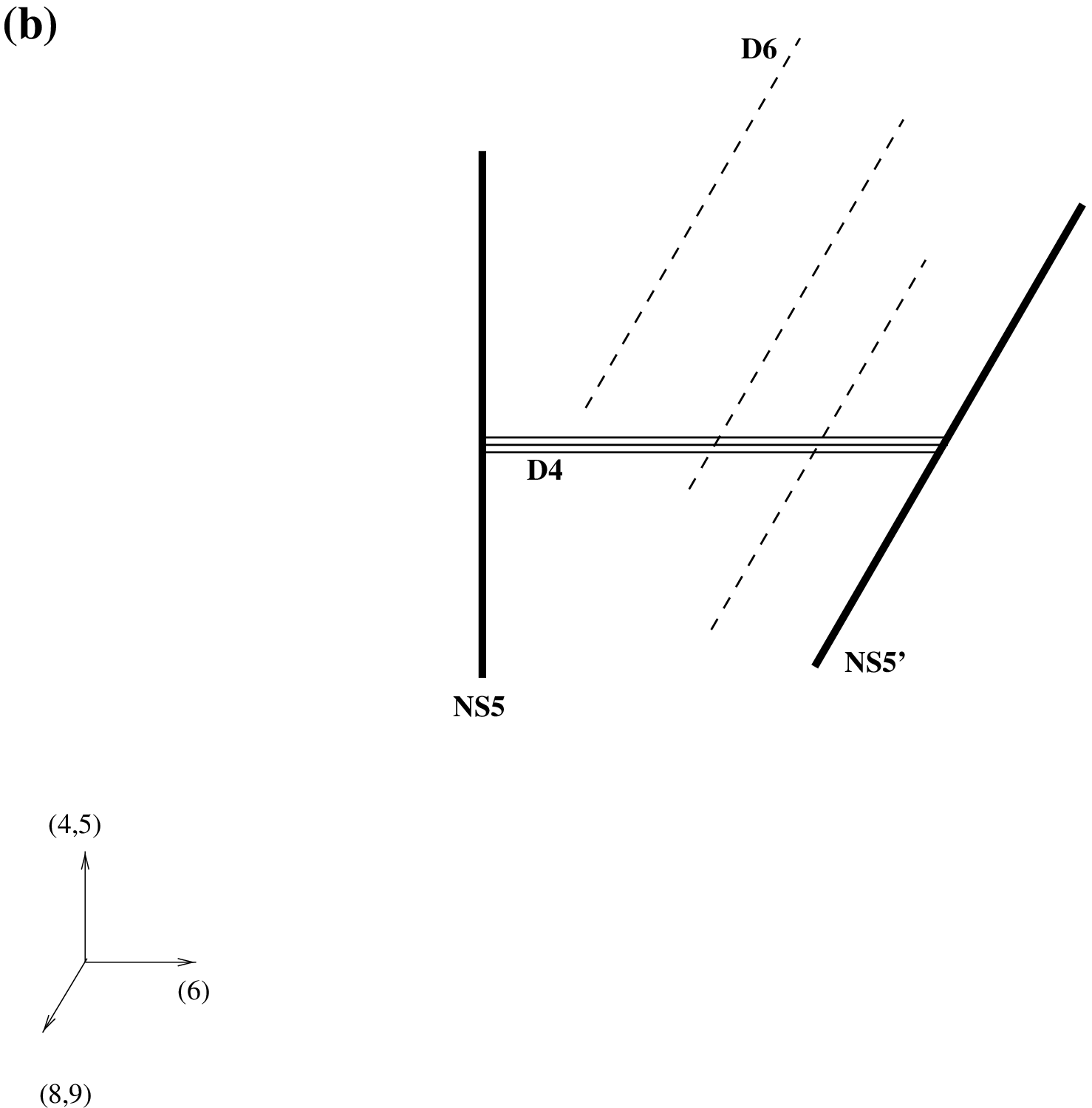}}
\vspace*{1cm}
\caption{Two descriptions of
$N=1$ SQCD with $G=U(\nc)$ and
$\nf$ fundamentals ($\nf=\nc=3$),
related by a series of HW transitions.}
\label{twentyfourb}
\end{figure}

Consider a configuration of $N_c$ $D4$-branes
stretched between an $NS5$-brane and an \nsp-brane
along the $x^6$ direction. The $NS5$ and
\nsp-branes are separated by a distance
$L_6$ in the $x^6$ direction, with $x^6(NS5)
<x^6(\nspp)$. In addition, there are
$N_f$ $D6$-branes to the left of the $NS5$-brane, each
of which is connected to the $NS5$-brane by a
$D4$-brane (see Fig.~\ref{twentyfourb}(a)).
The branes involved are extended in the directions
given in (\ref{BSB1}, \ref{D4N11}).
We call this brane configuration the ``electric theory.''

\begin{figure}
\centerline{\epsfxsize=140mm\epsfbox{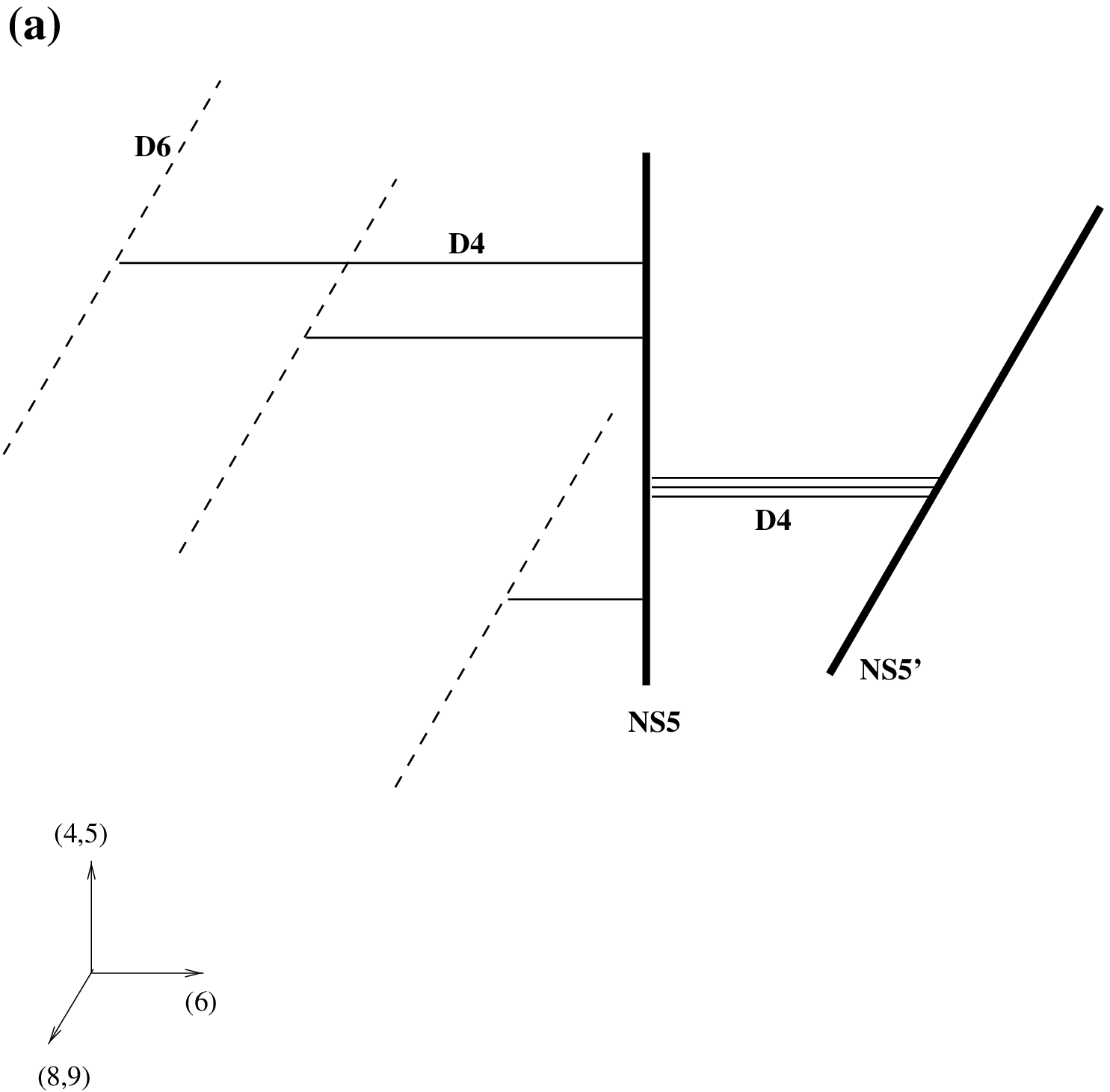}}
\vspace*{1cm}
\label{twentyfoura}
\end{figure}
\smallskip

An equivalent configuration, which
is related to the previous one by
a series of HW transitions (see section \ref{TBCS}),
consists of $N_c$ $D4$-branes
stretched between an $NS5$-brane  and an \nsp-brane
along the $x^6$ direction, with $N_f$ $D6$-branes at
values of $x^6$ that are between those corresponding
to the positions of the $NS5$ and \nsp-branes
(Fig.~\ref{twentyfourb}(b)).

This brane configuration describes classically
$N=1$ SQCD with gauge group $G=U(\nc)$,
$\nf$ flavors of chiral superfields in the
fundamental and antifundamental representations
and vanishing superpotential. Quantum mechanically,
the $U(1)$ factor in $U(\nc)$ will have vanishing
gauge coupling and decouple; we will discuss the
quantum case in the next section.
The gauge theory limit corresponds again to
$L_6, l_s, \gs\to 0$ with fixed gauge coupling
(\ref{BSB3}-\ref{hierscale}).

It is instructive to relate the
supersymmetric deformations of the gauge theory to
parameters defining the brane configuration, using
the dictionary established in the previous sections:

\begin{itemize}
\item
{\em Moduli Space of Vacua:}
The structure
of the moduli space of the gauge theory
was discussed in subsection \ref{SFTR}.
For $N_f<N_c$, the
$U(N_c)$ gauge symmetry can be broken to
$U(N_c-N_f)$. The complex dimension of the
moduli space of vacua is
\beq
N_f<N_c: \qquad
{\rm dim}\MM_H=2N_cN_f-\left( N_c^2-(N_c-N_f)^2\right)=N_f^2
\label{D4N19}
\eeq
For $N_f\geq N_c$ the gauge symmetry can be completely
broken, and the complex dimension of the moduli space is
\beq
N_f\geq N_c: \qquad  {\rm dim}\MM_H=2N_cN_f-N_c^2
\label{D4N110}
\eeq
In the brane description, Higgsing corresponds
to splitting fourbranes on sixbranes.
Consider, \eg, the case $N_f\geq N_c$ (the case
$N_f<N_c$ is similar). A generic point in moduli
space is described as follows (Fig.~\ref{twentyfive}).
The first $D4$-brane
is broken into $N_f+1$ segments connecting the
$NS5$-brane  to the first (\ie\ leftmost)
$D6$-brane, the first $D6$-brane to the second,
etc., with the last segment connecting the rightmost
$D6$-brane to the \nsp-brane. The second $D4$-brane
can now only be broken into $N_f$ segments, because of the
s-rule (see section \ref{TBCS}):
the first segment must stretch between the
$NS5$-brane and the {\it second} $D6$-brane, with the
rest of the breaking pattern as before.

\begin{figure}
\centerline{\epsfxsize=100mm\epsfbox{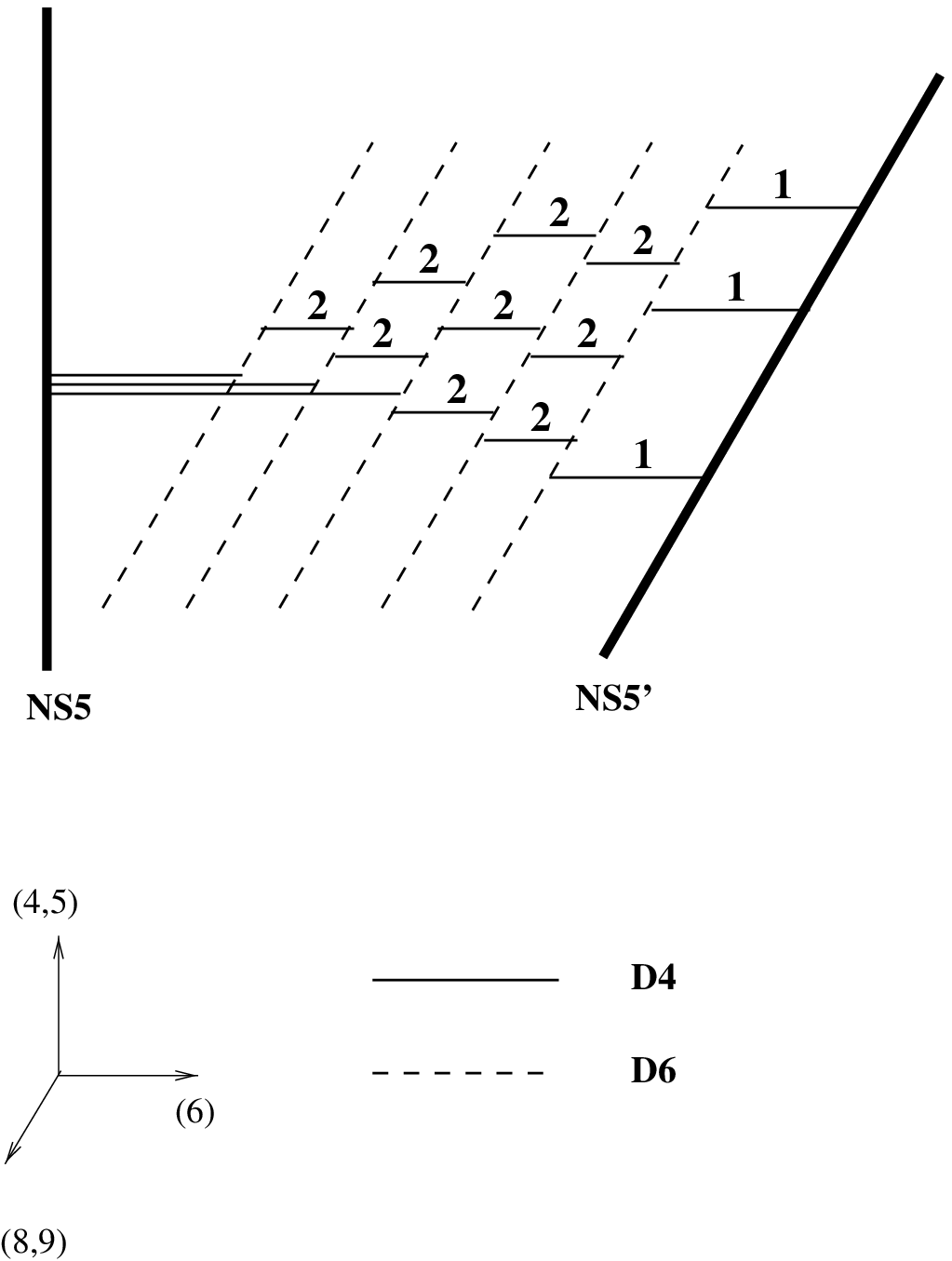}}
\vspace*{1cm}
\caption{The fully Higgsed branch of $N=1$ SQCD
with $G=U(3)$ and $\nf=5$ fundamentals.}
\label{twentyfive}
\end{figure}
\smallskip

We saw in section \ref{TBCS} that
a $D4$-brane stretched between two
$D6$-branes has two complex massless
degrees of freedom. Similarly, it is
geometrically obvious that
a $D4$-brane stretched between a $D6$-brane
and an \nsp-brane has one complex
massless degree of freedom, corresponding
to motions in $w$.
Moreover, one does not expect an analog of
the s-rule (see section \ref{TBCS})
for $D4$-branes stretched between an
\nsp-brane and a $D6$-brane, \eg\ because
two such fourbranes can be separated in the
$(x^8, x^9)$ directions, which are common to
both kinds of branes.

Therefore, the
dimension of moduli space is
\beq
N_f\geq N_c: \qquad  {\rm dim}\MM_H=
\sum_{l=1}^{N_c}\left[2(N_f-l)+1\right] =2N_fN_c-N_c^2
\label{D4N111}
\eeq
in agreement with the gauge theory result (\ref{D4N110}).

\item
{\em Mass Deformations:}
In gauge theory we can turn on a mass matrix
for the (s)quarks, by adding a superpotential
\beq
W=-m_i^jQ^i\tilde Q_j
\label{wmq}
\eeq
with $m$ an arbitrary $N_f\times N_f$ matrix of complex
numbers. In the brane description, masses correspond
to relative displacement of the $D6$ and
$D4$-branes (or equivalently the $D6$ and \nsp-branes)
in the $(x^4, x^5)$ directions. The configuration
can be thought of as realizing
a superpotential of the form (\ref{wmq}), with the mass matrix
$m$ satisfying the constraint
\beq
[m, m^\dagger]=0
\label{mmdag}
\eeq
Thus, we can diagonalize $m, m^\dagger$ simultaneously;
the locations of the $D6$-branes in the $v$-plane
are the eigenvalues of $m$.

Hence, the brane configuration describes only a
subset of the possible deformations of the gauge
theory. We have already encountered such situations
before; they are rather standard in string theory.
In this context the constraint (\ref{mmdag}) can
be ``explained'' by noting that it appears
as a consistency condition in $N=2$
supersymmetric gauge theories. Our theory is clearly
not $N=2$ supersymmetric; nevertheless, it is not surprising
that the condition (\ref{mmdag}) arises, since one can think
of $m$ as the expectation value of a superfield in the
adjoint of the $U(N_f)$ gauge group on the $D6$-branes.
The theory on the infinite sixbranes is invariant under
sixteen supercharges in the bulk of the worldvolume, and
while it is broken by the presence of the other
branes, it inherits (\ref{mmdag}) from the theory with more
supersymmetry.

\item
{\em $x^6(D6)$ -- A Phase Transition:} One important
difference between the $N=2$ configurations considered
in section \ref{D4N2} and the present discussion is that
it is no longer true that the low energy physics is
completely independent of the positions of the $D6$-branes
in $x^6$. If we move one or more of the $D6$-branes of
Fig.~\ref{twentyfourb}(a) 
towards the $NS5$-brane, as the two branes
cross there is no change in the low energy physics;
this is guaranteed by the HW transition. If all the $D6$-branes
move to the other side of the $NS5$-brane we arrive at the
configuration of Fig.~\ref{twentyfourb}(b), which describes the
same low energy physics as Fig.~\ref{twentyfourb}(a) (as for the
$N=2$ case).

When the $D6$-branes are displaced towards the \nsp-brane
and eventually pass it, the physics changes. No branes
can be created in the transition, \eg\ because the
$D6$ and \nsp-branes can avoid each other in space by
going around each other in the $(x^4,x^5,x^6)$ directions.
Therefore, every time a $D6$-brane moves out of the interval
between the two NS-branes by passing the \nsp-brane,
the theory loses one light flavor of $U(\nc)$.

There is an interesting lesson here.
Brane dynamics apparently
has the property that when D and NS-branes that
are {\em not parallel} cross each other, there is no
change in the low energy physics, while crossing of
{\em parallel} branes leads in general to phase
transitions.

\begin{figure}
\centerline{\epsfxsize=90mm\epsfbox{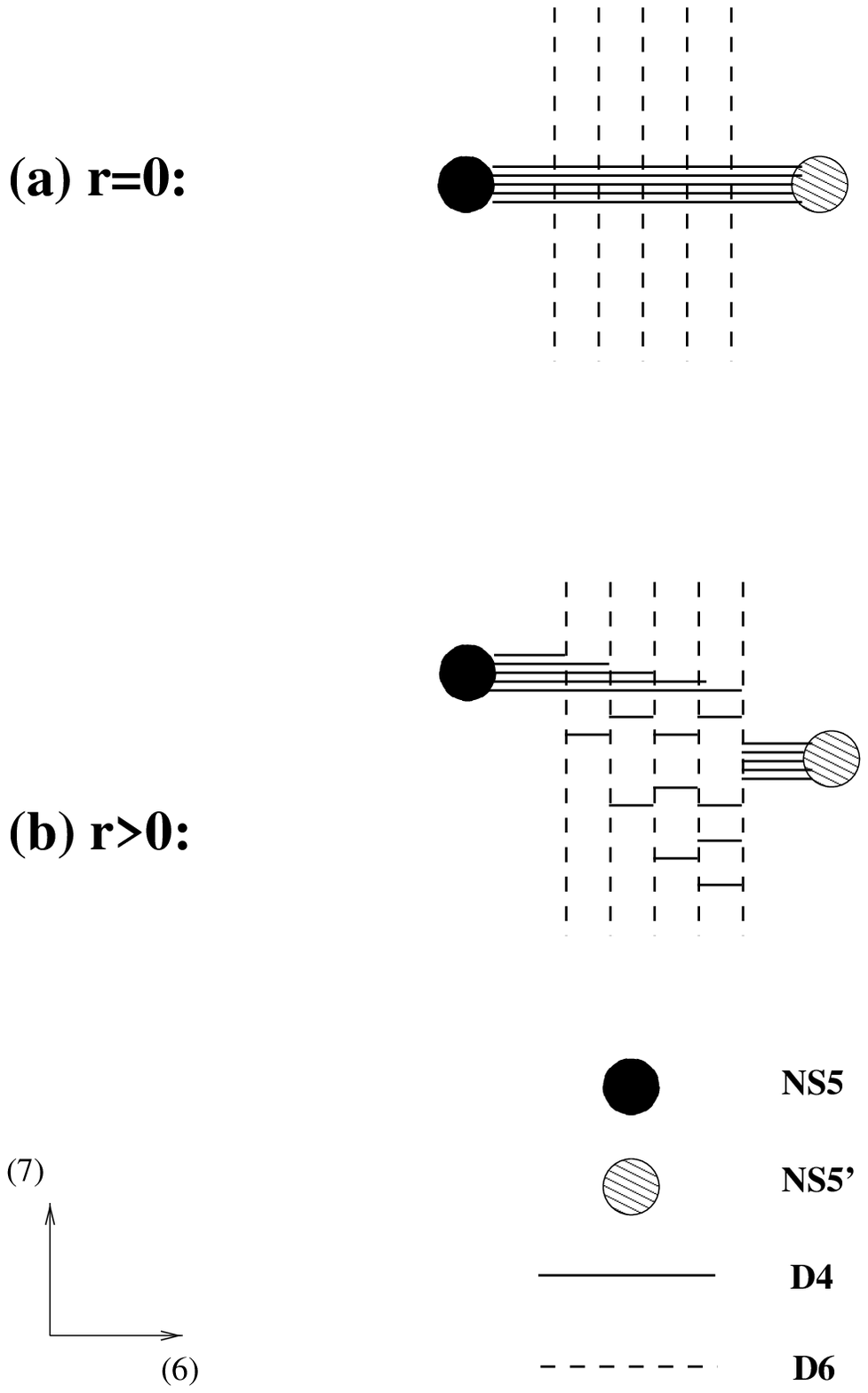}}
\vspace*{1cm}
\caption{Displacing the fivebranes in $x^7$
corresponds to a FI D-term in the worldvolume
gauge theory.}
\label{twentysix}
\end{figure}
\smallskip

\item
{\em FI D-Term:}
In the gauge theory it is possible to turn on a D-term for
$U(1)\subset U(N_c)$
\beq
\LL_{FI}=r\int d^4\theta\, {\rm Tr}\, V
\label{D4N15}
\eeq
Note that -- unlike the $N=2$ SUSY case (\ref{BSB7}) --
here the D-term is a single real number $r$.
For $0<N_f<N_c$ adding (\ref{D4N15}) breaks SUSY. For
$N_f\geq N_c$ there are supersymmetric vacua in which the gauge
symmetry is completely broken
and the system is forced into a Higgs phase.
In the brane description, the role of the FI D-term is played
by the relative displacement of the $NS5$ and \nsp-branes
in the $x^7$ direction (Fig.~\ref{twentysix}). 
Clearly, when the two are at different
values of $x^7$, a fourbrane stretched between them breaks SUSY.
To preserve SUSY, all such fourbranes must break on $D6$-branes,
which as we saw above is only possible for $N_f\geq N_c$ because
of the s-rule. Once all fourbranes have been split, there is no
obstruction to moving the $NS5$ and \nsp-branes to different
locations in $x^7$. At generic points in
the Higgs phase, nothing special happens
when the D-term is turned off.
In the brane construction the reason is that once
all $N_c$ $D4$-branes have been broken on $D6$-branes in a
generic way, nothing special happens when the relative
displacement of the two fivebranes in $x^7$ vanishes.

\item
{\em Global Symmetries:}
Classical supersymmetric QCD
with gauge group $SU(N_c)$ and $N_f$
quarks has the global symmetry (\ref{FT3});
quantum effects break it to (\ref{FT35}).
The anomaly is a quantum effect which is not expected
to be visible in the classical brane construction (we
will exhibit it in the brane description
in section \ref{QEN1}) .

\begin{figure}
\centerline{\epsfxsize=90mm\epsfbox{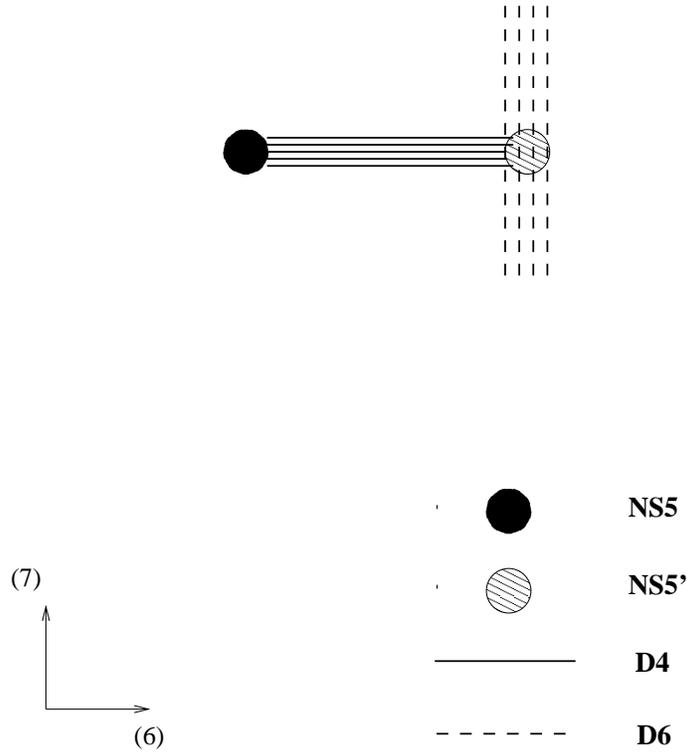}}
\vspace*{1cm}
\caption{Placing the $D6$ branes on the \nsp-brane
enhances the global symmetry at finite energies
from $SU(\nf)$ to $SU(\nf)\times SU(\nf)$.}
\label{twentyseven}
\end{figure}
\smallskip

In our case, the (classical) gauge symmetry is
$U(\nc)\simeq SU(\nc)\times U(1)$, with the
extra $U(1)$ factor in the gauge group
corresponding to gauging baryon number $U(1)_B$.
The brane configuration has a manifest (vector)
$SU(N_f)$ symmetry, which is a gauge symmetry on the
$D6$-branes and a global symmetry on the $D4$-branes.
The other (axial) $SU(N_f)$ symmetry is generically
not an exact symmetry of the brane configuration
of Fig.~\ref{twentyfourb}(b), 
and arises as an effective symmetry
when we take the infrared limit. In the general spirit
of brane theory -- trying to realize as much
as possible of the symmetry structure of the
low energy theory throughout the RG flow --
one might wonder whether it is possible to realize it
too as an exact symmetry of the brane vacuum.

This is indeed the case as shown by~\cite{BH,AH,HZ}.
The main idea is the following. We saw before that
the positions of the $D6$-branes in $x^6$ are
not visible in the low energy theory, but of
course their values influence the high energy
structure. One may thus hope that the full
chiral symmetry may be restored for some
particular value of these parameters. 
When the $D6$-branes
are placed at the same value of $x^6$ as the
\nsp-brane (Fig.~\ref{twentyseven})
the full chiral symmetry is restored~\cite{BH}.

To see that this is geometrically plausible,
note that when that happens, the \nsp-brane
located at (say) $x^7=0$
cuts each $D6$-brane into two disconnected
halves, the $x^7>0$ and $x^7<0$
parts~\footnote{Note that this does not
happen when the $D6$-branes intersect an
$NS5$-brane. This is consistent with the fact
that in the $N=2$ SUSY configurations we do not
expect a chiral enhancement of the global symmetry.}.
The situation
is very similar to that encountered in section
\ref{QEFF} when we discussed compact Coulomb branches.
Using our analysis there, it is clear that there
are now two separate $SU(\nf)$ symmetries acting
on the two disconnected groups of
$\nf$ sixbranes. Just like in section \ref{QEFF},
despite the fact that the two groups of
sixbranes are independent, we cannot remove
one of them from the configuration. {}From the
brane theory point of view this is due to the
fact that, as discussed in section \ref{QEFF},
this would lead to non-conservation of charge.
{}From the point of view of the gauge theory
on the fourbranes the reason is that the resulting
four dimensional gauge theory, with only
fundamentals and no antifundamentals, would
be anomalous.

The symmetries $U(1)_x$, $U(1)_a$ (\ref{FT3}) are
also realized
in the brane picture. They correspond
to rotations in the $(x^4, x^5)$ and $(x^8, x^9)$
planes, $U(1)_{45}$, $U(1)_{89}$.
These rotations are $R$-symmetries
because the four preserved
supercharges of the brane configuration of
Fig.~\ref{twentyfourb}
are spinors of the $Spin(9,1)$ Lorentz
group in ten dimensions and, therefore, are charged
under both $U(1)_{45}$ and $U(1)_{89}$.
{}From the discussion
of the mass deformations and Higgs moduli space above,
it is clear that the mass parameters (\ref{wmq}) are
charged under $U(1)_{45}$, while the quarks
$Q$, $\tilde Q$ are charged under $U(1)_{89}$.
If we assign $U(1)_{45}\times U(1)_{89}$ charges
$(1,1)$ to the superspace coordinates $\theta_\alpha$,
the quarks
$Q$ and $\tilde Q$ have charges $(0,1)$, while the
mass parameters $m$ in (\ref{wmq}) have charges $(2,0)$.
With these assignments, the mass term (\ref{wmq})
is invariant under both global 
symmetries~\footnote{The discussion of global
charges is somewhat oversimplified. A more
precise description of the transformation
properties of gauge invariant observables
requires a detailed mapping of the
brane and gauge theory degrees of freedom.}.

\end{itemize}

\subsubsection{Classical SQCD: The Magnetic Theory}
\label{D4N1MT}
\begin{figure}
\centerline{\epsfxsize=120mm\epsfbox{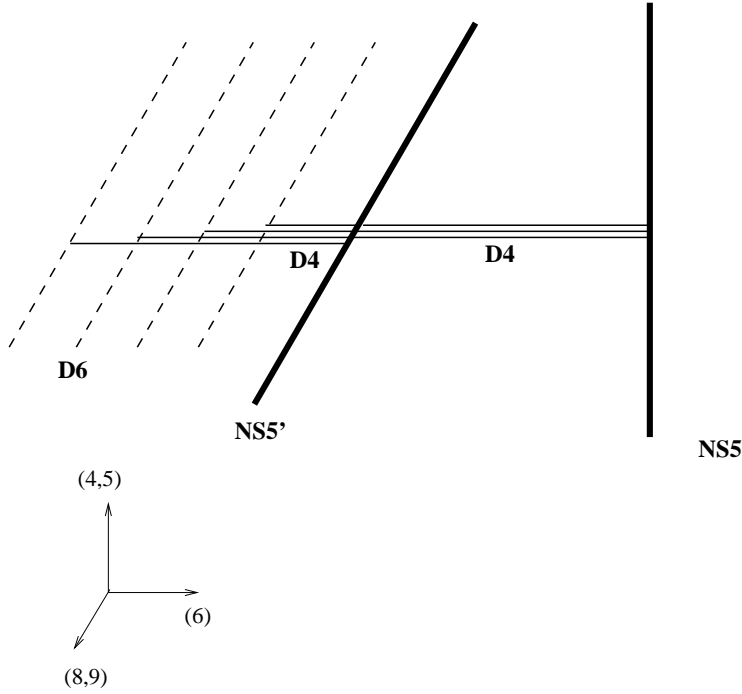}}
\vspace*{1cm}
\caption{The magnetic brane configuration.}
\label{twentyeight}
\end{figure}

The ``magnetic'' brane configuration --
the reason for the
name will become clear soon -- contains
$N_c$ $D4$-branes connecting the \nsp-brane
to an $NS5$-brane on its right (we will refer
to these as ``color fourbranes''), and
$N_f$ $D4$-branes connecting
the \nsp-brane to $N_f$ $D6$-branes on its left,
which we will refer to as ``flavor fourbranes.''
The configuration is depicted in 
Fig.~\ref{twentyeight}. 
As usual, all the branes involved
are stretched in the directions
given in (\ref{BSB1}, \ref{D4N11}).
We will consider the case $N_f\geq N_c$ in what follows.

\smallskip

This configuration describes SQCD with
``magnetic gauge
group'' $G_m=U(N_c)$ (with the gauge bosons
coming as before from $4-4$ strings connecting
different color fourbranes), $N_f$ flavors of
``magnetic quarks''
$q_i$, $\tilde q^i$ ($4-4$ strings connecting the
$N_c$ color fourbranes with the $N_f$
flavor fourbranes). In addition to the
$N=1$ SQCD matter content there are now
$N_f^2$ chiral superfields that are singlets
under the gauge group $G_m$,
arising from $4-4$ strings connecting
different flavor fourbranes. Denoting
these ``magnetic meson'' fields by
$M^i_j$ $(i,j=1,\cdots, N_f)$, it is clear
that the standard coupling of three open strings
gives rise to a superpotential connecting the
magnetic mesons and the magnetic quarks,
\beq
W_{\rm mag}=M^i_j q_i\tilde q^j
\label{wmag}
\eeq
This is precisely the ``magnetic theory'' discussed
in section \ref{SFTR}.

\begin{figure}
\centerline{\epsfxsize=120mm\epsfbox{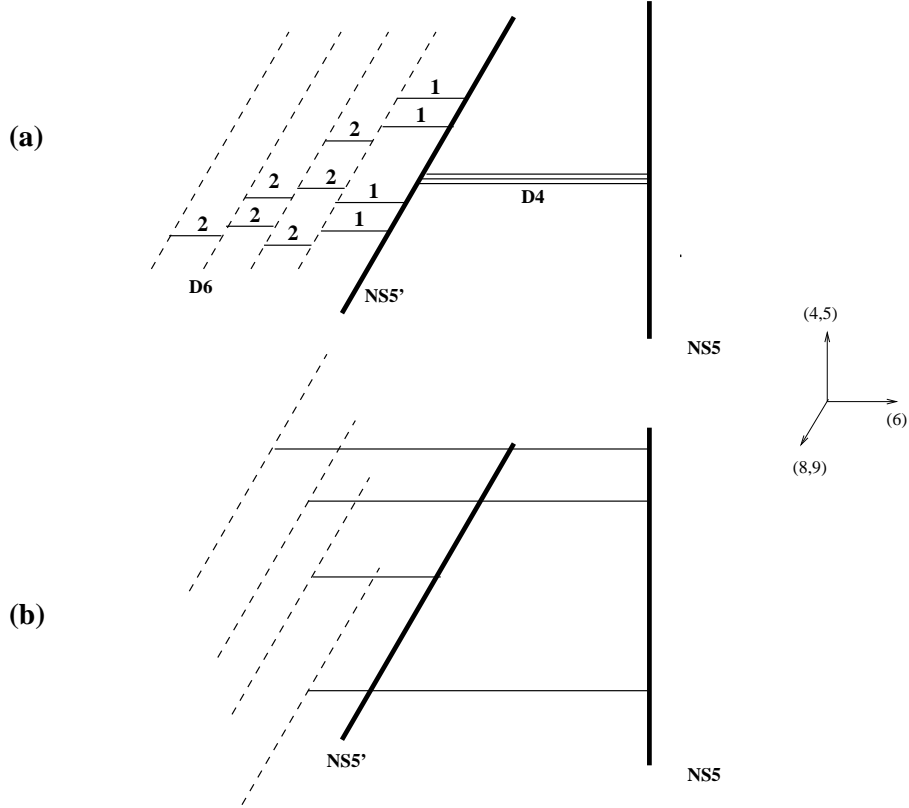}}
\vspace*{1cm}
\caption{(a) The $\nf^2$ dimensional
classical magnetic moduli
space corresponding to unbroken gauge symmetry
and arbitrary expectation values for the singlet
meson $M$. (b) The brane description of adding
a linear superpotential $W=-mM$. The eigenvalues
of $m$ correspond to locations of $D6$-branes
in the $v$ plane.}
\label{twentynine}
\end{figure}

The analysis of moduli space and deformations
of this model is similar to the electric theory,
with a few differences due to the existence of
the superpotential (\ref{wmag}). Consider first mass
deformations. In gauge theory we can add a
mass term to the magnetic quarks, by modifying
the superpotential to
\beq
W_{\rm mag}=
M^i_j q_i\tilde q^j+\delta M^i_j q_i\tilde q^j
\label{wmagmm}
\eeq
The mass parameters $\delta M$ can be absorbed in
the expectation value of the magnetic meson
$M^i_j$, and can be thought of as
parametrizing a moduli space of vacua.
The $N_f^2$ resulting parameters are described
in the brane language by splitting the $N_f$
flavor fourbranes on the $D6$-branes
in the most general way consistent
with the geometry (Fig.~\ref{twentynine}(a)).
This results
in a total of $N_f^2$ massless modes
corresponding to the
$N_f^2$ components of $M$: $N_f$ of them describe
fluctuations in the $(x^8, x^9)$
plane of fourbranes stretched between the \nsp-brane
and the rightmost $D6$-brane, and the
remaining $\sum_{l=1}^{N_f-1} 2l = N_f(N_f-1)$
parametrize fluctuations in
$(x^6,x^7, x^8, x^9)$ of the fourbranes connecting
different sixbranes.

\smallskip

Another interesting deformation of the
magnetic gauge theory
corresponds to adding a linear term
in $M$ to the magnetic superpotential
\beq
W_{\rm mag}=M^i_j (q_i\tilde q^j-m_i^j)
\label{mwmag}
\eeq
Integrating out the massive field $M$ we find that
in the presence of the ``mass parameters'' $m_i^j$
the gauge group is broken; thus the
parameters $m$ play the role of
Higgs expectation values. In the brane description,
these deformations correspond to a process where
color fourbranes are aligned with flavor fourbranes
and reconnected to stretch between
the $NS5$-brane  and a $D6$-brane 
(see Fig.~\ref{twentynine}(b)).
If $m$ has rank
$n(\leq N_c)$, $n$ such fourbranes are reconnected. The
$D6$-branes on which the reconnected fourbranes end
can then be moved in the $(x^4, x^5)$ directions,
taking the fourbranes with them and breaking the
$U(N_c)$ gauge group to $U(N_c-n)$.
The brane description realizes only a subset of the
possible ``mass matrices'' $m$, namely, those which
satisfy (\ref{mmdag}) (the reason is similar to the one
described there). We will soon see that this analogy
is not coincidental.

\begin{figure}
\centerline{\epsfxsize=100mm\epsfbox{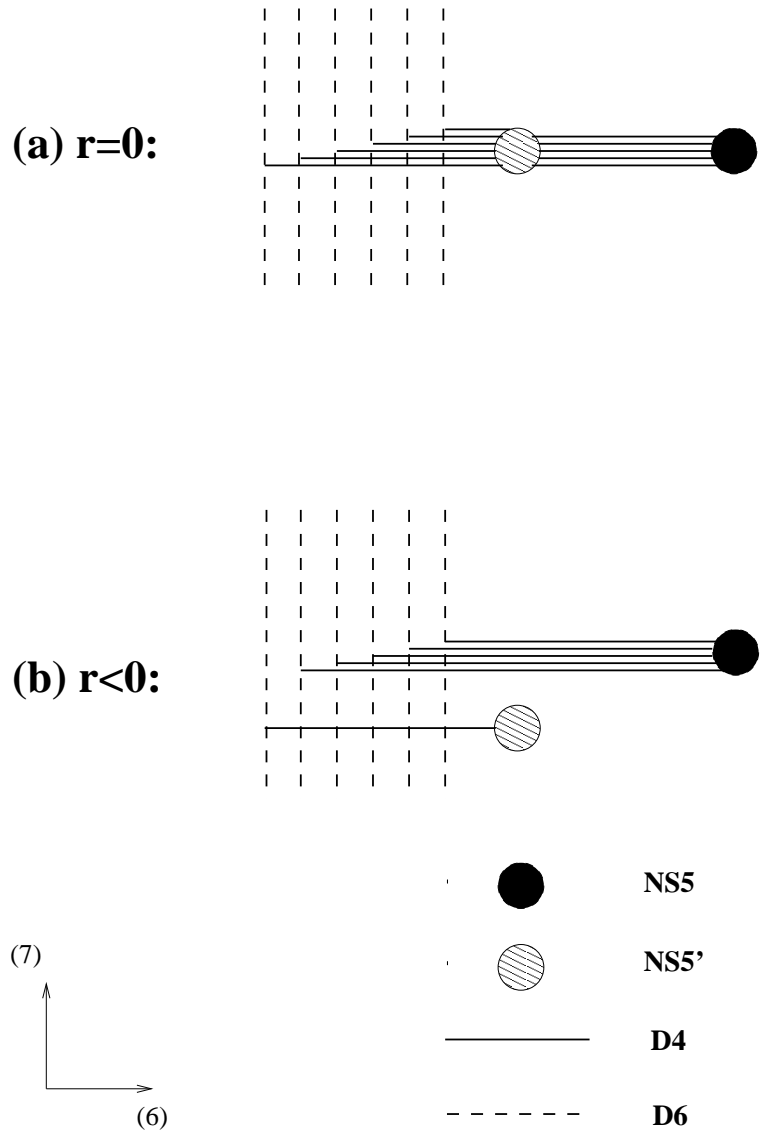}}
\vspace*{1cm}
\caption{The brane description of the FI D-term
in the magnetic theory.}
\label{thirty}
\end{figure}

Another deformation of the magnetic gauge theory
and of the corresponding brane configuration, which
will play a role in the sequel, is switching on
a FI D-term for the $U(1)$ subgroup of $U(N_c)$.
Again, in the brane construction this corresponds to a
relative displacement of the $NS5$ and \nsp-branes
in the $x^7$ direction (Fig.~\ref{thirty}). 
To preserve SUSY, all
$N_c$ color fourbranes have to be
reconnected to $N_c$ of the $N_f$
flavor fourbranes,
leading to a situation where
$N_c$ fourbranes stretch between the $NS5$-brane
and $N_c$ different sixbranes and $N_f-N_c$
fourbranes stretch between the \nsp-brane
and the remaining sixbranes (Fig.~\ref{twentynine}(b)).
Once this occurs,
the two fivebranes can be separated in $x^7$.

\smallskip

Unlike the electric theory, here there
is a jump in the dimension of the classical
moduli space of the theory as we vary the D-term.
For non-vanishing D-term
there are only $N_f-N_c$ fourbranes that give rise
to moduli (the other $N_c$ are frozen because
of the s-rule), and the moduli space is easily checked
to be $N_f^2-N_c^2$ dimensional. When the D-term
vanishes, the previously frozen fourbranes can be
reconnected to yield the original configuration, with
unbroken $U(N_c)$, and we gain access to the full
$N_f^2$ dimensional moduli space of Fig.~\ref{twentynine}(a).
We will see in section \ref{QEN1} that quantum
mechanically this classical jump in the
structure of the moduli space disappears.

The magnetic brane configuration is invariant
under the same global symmetries as the electric
theory described above (\ref{FT3}).
The charge assignments under the
$U(1)_{45}\times U(1)_{89}$ symmetry are as follows:
the magnetic quarks $q$, $\tilde q$ have charges
$(1,0)$, the mass parameters $m$  have charges
$(2,0)$, the magnetic meson $M$ has charges
$(0,2)$, and the superspace coordinates $\theta_\alpha$
have charges $(1,1)$.

\subsubsection{Seiberg's Duality In The Classical Brane Picture}
\label{D4N1SD}
We have now constructed using branes
two $N=1$ supersymmetric gauge theories, the electric
and magnetic theories discussed in the previous
two subsections. Seiberg has shown that
the electric gauge theory with gauge group $U(N_c)$ and
the magnetic theory with gauge group $U(N_f-N_c)$
are equivalent in the extreme infrared~\footnote{Seiberg
actually considered the $SU(N_c)$ and $SU(N_f-N_c)$
theories (see section \ref{SFTR}), but the statement for $U(N_c)$
and $U(N_f-N_c)$ follows from his results by
gauging baryon number.} (\ie\ they flow to
the same infrared fixed point)~\cite{Sei94b}. Seiberg's duality
is a quantum symmetry, but it has classical consequences
in situations where the gauge symmetry is completely
broken and there is no strong infrared dynamics. In
such situations Seiberg's duality reduces to a
classical equivalence of moduli spaces and their
deformations.

In this subsection we show using brane theory
that the moduli spaces of vacua
of the electric and magnetic
theories with gauge groups $U(N_c)$ and $U(N_f-N_c)$
coincide. They provide different parametrizations
of the moduli space of vacua of the
appropriate brane configuration.
This explains the
classical part of Seiberg's duality. As one approaches
the root of the Higgs branch, non-trivial quantum
dynamics appears, and we have to face the resulting
strong coupling problem. This will be addressed in
section \ref{QEN1}.

Start, for example, with the electric theory with gauge
group $U(N_c)$ (the configuration of Fig.~\ref{twentyfourb}(a)).
Now enter the Higgs phase by connecting the $N_c$
original fourbranes stretched between the $NS5$ and \nsp-branes
to $N_c$ of the $N_f$ fourbranes stretched between the
$NS5$-brane  and the sixbranes; we then further
reconnect the resulting fourbranes in the most general
way consistent with the rules described in sections
\ref{TBCS}, \ref{D4N1ET}.
The resulting moduli space is $2N_fN_c-N_c^2$ dimensional,
as described in subsection \ref{D4N1ET}. Note that, generically,
there are now $N_f-N_c$ $D4$-branes attached to the
$NS5$-branes, and $N_c$ $D4$-branes connected to the
\nsp-brane (the other ends of all these fourbranes lie on
different $D6$-branes).

\begin{figure}
\centerline{\epsfxsize=100mm\epsfbox{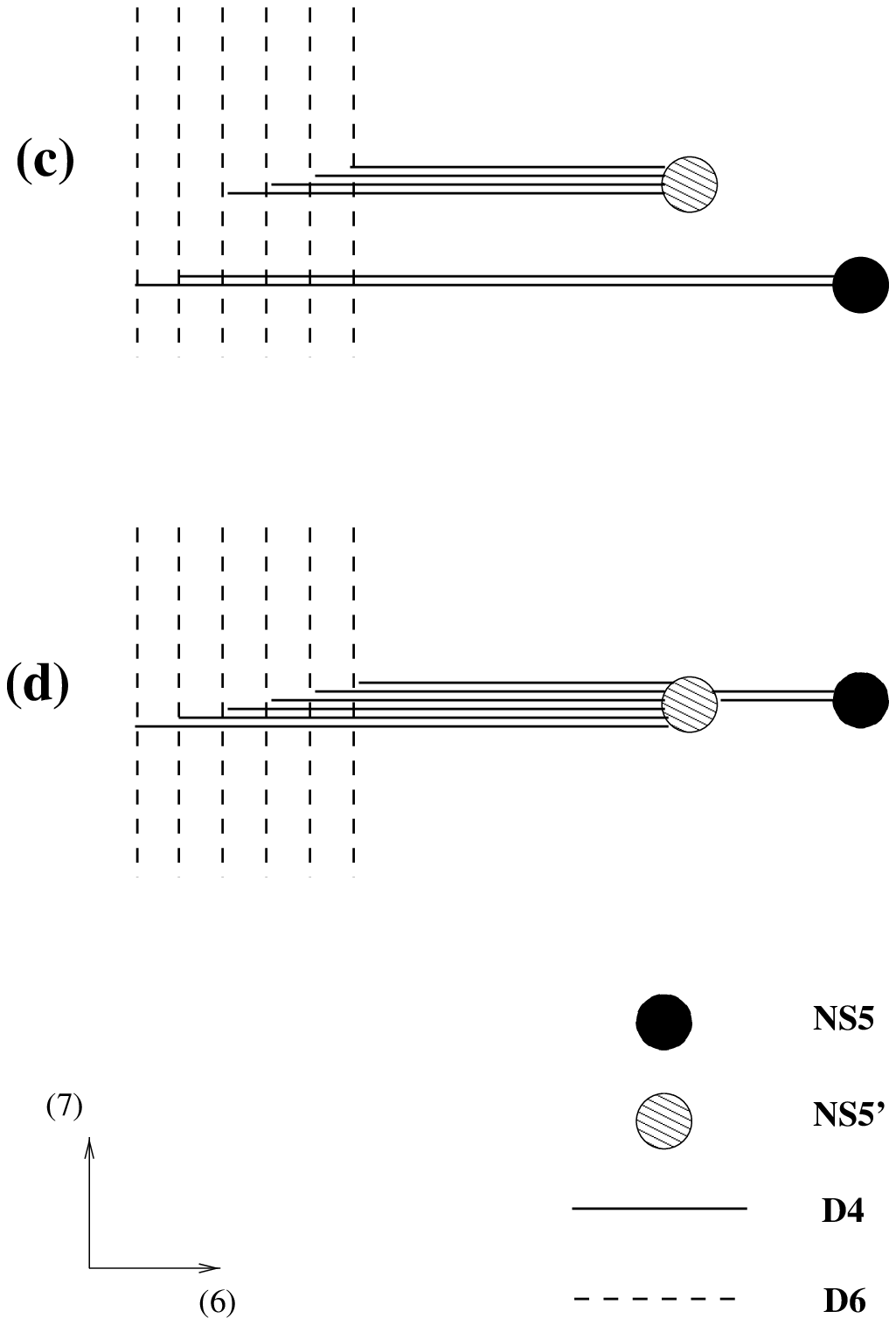}}
\vspace*{1cm}
\caption{The electric and magnetic brane
configurations are continuously connected
in the way indicated here.
Starting with the
electric configuration (a), one can turn on
a FI D-term (b), exchange the fivebranes
in $x^6$ (c), and switch off the D-term,
arriving at the magnetic configuration (d).}
\label{thirtyonecd}
\end{figure}

Once we are in the Higgs phase, we can freely move the
$NS5$-brane  relative to the \nsp-brane and, in
particular, the two branes can pass each other in the
$x^6$ direction without ever meeting in space. This
can be achieved by taking the $NS5$-brane  around the
\nsp-brane in the $x^7$ direction, \ie\ turning
on a FI D-term in the worldvolume gauge theory 
(the process is described in Fig.~\ref{thirtyonecd}). At
a generic point in the Higgs branch of the electric
theory, turning on such a D-term is a completely smooth
procedure; this is particularly clear from the brane
description, where in the absence of $D4$-branes
connecting the $NS5$-brane  to the \nsp-brane,
the relative displacement of the two in the $x^7$
direction can be varied freely.

\begin{figure}
\centerline{\epsfxsize=100mm\epsfbox{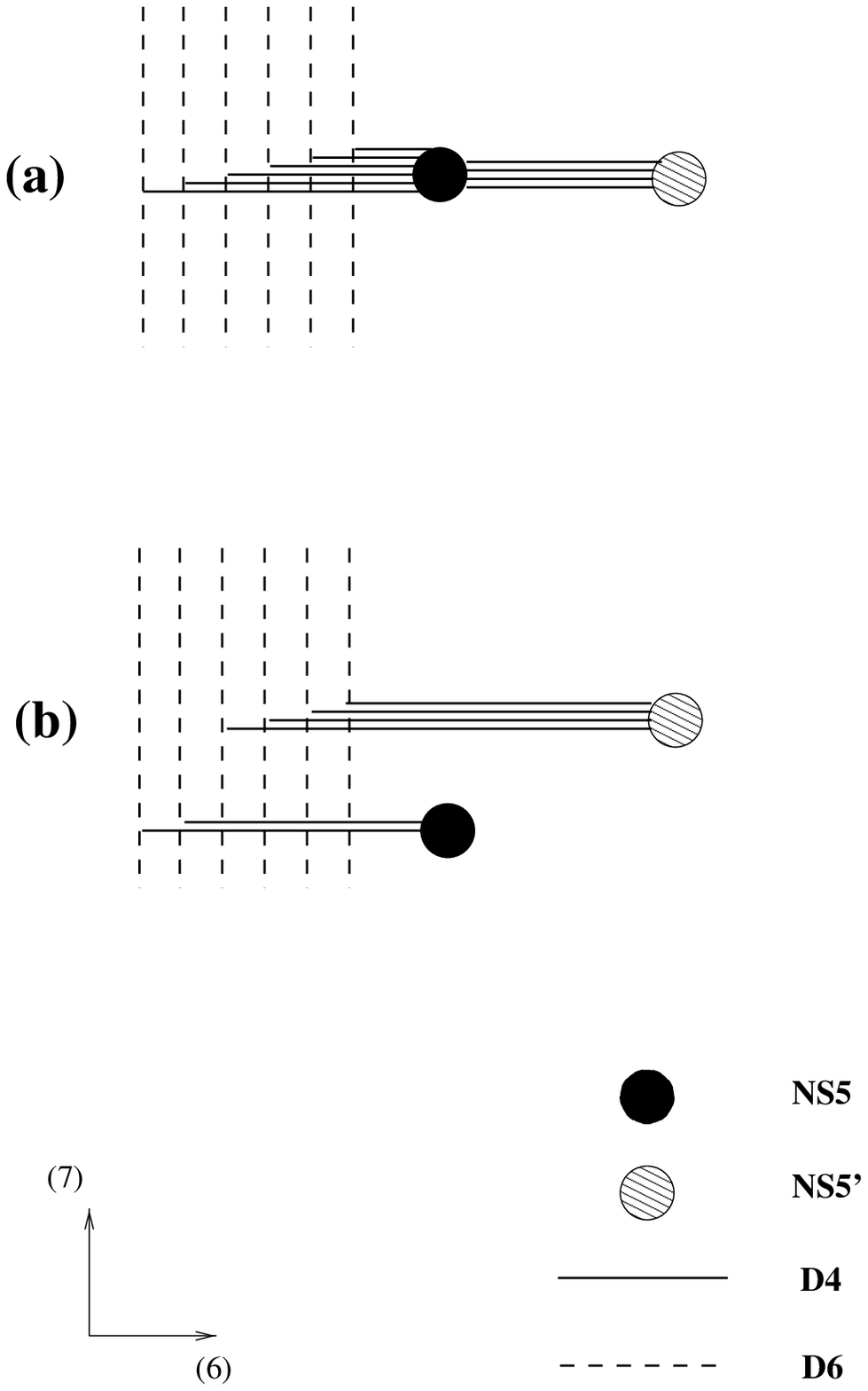}}
\vspace*{1cm}
\label{thirtyoneab}
\end{figure}
\smallskip

After exchanging the $NS5$ and
\nsp-branes, the brane configuration
we find can be interpreted as
describing the Higgs phase of {\it
another} gauge theory. To find out
what that theory is, we approach
the root of the Higgs branch by
aligning the $N_f-N_c$ $D4$-branes
emanating from the $NS5$-brane
with the \nsp-brane, and the
$N_c$ $D4$-branes emanating from
the \nsp-brane with
$D4$-branes stretched between $D6$-branes.

We then reconnect the $D4$-branes to obtain a
configuration consisting of $N_f-N_c$ $D4$-branes
connecting the \nsp-brane to an $NS5$-brane
which is to the right of it; the \nsp-brane
is further connected by $N_f$ $D4$-branes to the
$N_f$ $D6$-branes which are to the left of it
(see Fig.~\ref{thirtyonecd}(d)). This
is the magnetic SQCD of subsection \ref{D4N1MT}, with
gauge group $U(N_f-N_c)$.

To summarize, we have shown that the moduli
space of vacua of the electric SQCD theory
with (completely broken) gauge group $U(N_c)$
and $N_f$ flavors of quarks,
and the moduli space of vacua of the magnetic
SQCD model with (broken) gauge group $U(N_f-N_c)$,
can be thought of as providing different
descriptions of a single
moduli space of supersymmetric brane
configurations. One can smoothly interpolate
between them by varying the scale $\Lambda$
(related to the displacement of the $NS5$ and
\nsp-branes in $x^6$),
keeping the FI D-term fixed but non-zero.
Since the only role of $\Lambda$ in the low
energy theory  is to normalize the
operators~\cite{KSS}, theories with different values of
$\Lambda$ are equivalent. The electric and
magnetic theories will thus share all features,
such as the structure of the chiral ring (which
can be thought of as the ring of functions
on moduli space), that are independent of the
interpolation parameter $\Lambda$.

The above smooth interpolation relies on the
fact that the gauge symmetry is completely
broken due to the presence of the FI D-term.
As mentioned above, it is not surprising that
duality appears classically in this situation
since there is no strong infrared gauge dynamics.

The next step is to analyze what happens as the
gauge symmetry is restored when the D-term goes
to zero and we approach the origin of moduli space.
Classically, we find a disagreement. In the electric
theory, we saw in subsection \ref{D4N1ET} that nothing
special happens when the gauge symmetry is restored.
New massless degrees of freedom appear, but there
are no new branches of the moduli space that we gain
access to.

In the magnetic theory the situation is different.
When we set the FI D-term to zero, we saw in subsection
\ref{D4N1MT} that a large moduli space of previously
inaccessible vacua becomes available. While the
electric theory has a $2N_fN_c-N_c^2$ dimensional
smooth moduli space, the classical magnetic theory
experiences a jump in the dimension of its moduli
space from $2N_fN_c-N_c^2$ for non-vanishing
FI D-term to $N_f^2$ when the D-term is zero.
However, in the magnetic theory when the D-term
vanishes the $U(N_f-N_c)$ gauge symmetry is
restored, and to understand what really happens we
must study the quantum dynamics. We will discuss
this in section \ref{QEN1}, where we shall see that
quantum mechanically the jump in the
magnetic moduli space disappears, and the quantum
moduli spaces of the electric and magnetic theories
agree.

It is instructive to map the deformations of the
classical electric theory to those of the classical
magnetic one. Turning on masses (\ref{wmq}) in the electric
theory corresponds to moving the $D6$-branes away
from the $D4$-branes (or equivalently from the
\nsp-brane) in the $(x^4, x^5)$ directions.
As discussed in subsection \ref{D4N1MT}, in the magnetic
description, the electric mass parameters correspond
to Higgs expectation values  (\ref{mwmag}).

Turning on expectation values to the electric
quarks, which was described in the brane language
in subsection \ref{D4N1ET}, corresponds on the
magnetic side to varying the expectation value of
the magnetic meson $M$ (\ref{wmagmm}).
This gives masses to the magnetic quarks.

The transmutation of masses into Higgs expectation values
and vice-versa observed in the brane construction is one
of the hallmarks of Seiberg's duality.

\subsubsection{Other Rotated $N=2$ Configurations}
\label{ROT}

The brane configurations corresponding to
electric and magnetic SQCD were obtained
above by rotating branes in the $N=2$
SUSY configuration studied in section
\ref{D4N2}. Before moving
on to the study of quantum dynamics of
these theories we would like to discuss
a few additional theories that can
be realized using such rotations.

\medskip
\noindent
{\bf 1. ${\bf U(N_c)}$ With Adjoint, 
${\bf N_f}$ Flavors And ${\bf W=0}$}

\begin{figure}
\centerline{\epsfxsize=100mm\epsfbox{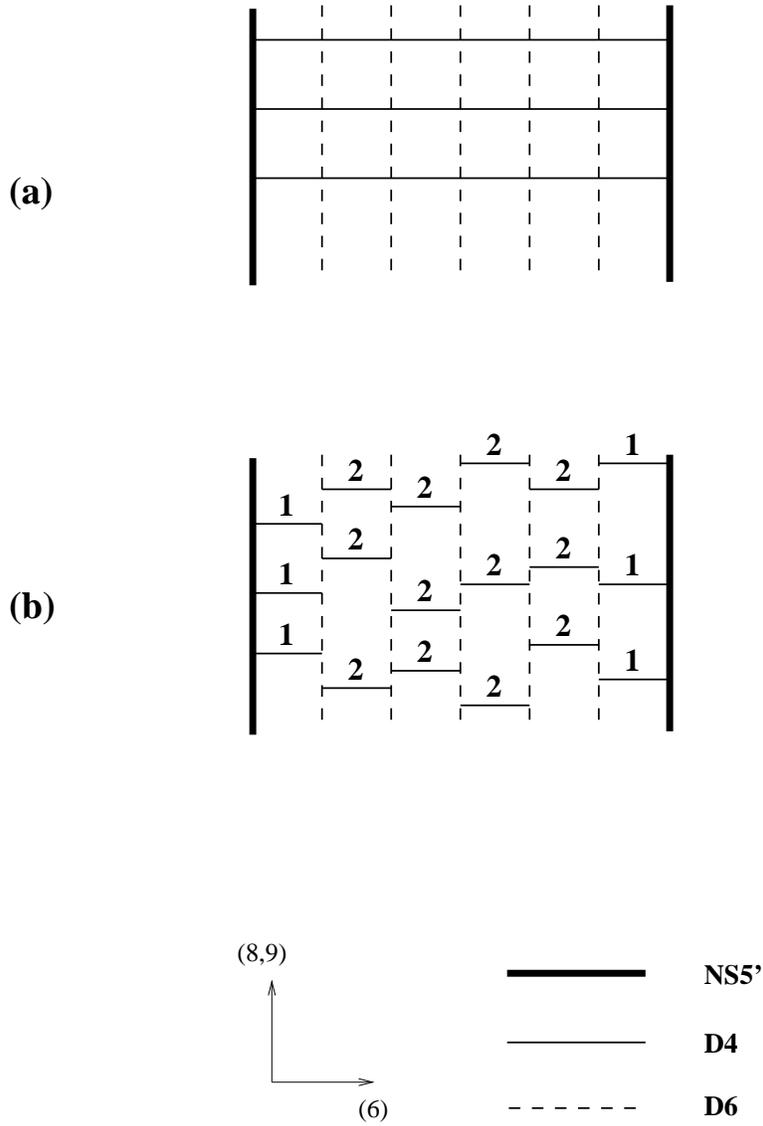}}
\vspace*{1cm}
\caption{(a) $N=1$ SYM with $G=U(\nc)$, $\nf$
fundamentals, and an adjoint superfield
with vanishing superpotential. (b) The fully
Higgsed branch of moduli space.}
\label{thirtytwo}
\end{figure}

Starting with the $N=2$ SUSY brane configuration
of Fig.~\ref{fourteen},
rotate the two $NS5$-branes as in
(\ref{wthe}-\ref{muthephi}) keeping them parallel
to each other. As discussed above
(\ref{D4N18}) the angle of rotation determines
the Yukawa coupling; in particular, when the
two $NS5$-branes are rotated into \nsp-branes
(Fig.~\ref{thirtytwo}(a))
the Yukawa coupling disappears. The resulting
theory has gauge group $U(\nc)$, the matter
content necessary for $N=2$ SUSY, \ie\ an
adjoint chiral multiplet $\Phi$ and $\nf$
fundamentals $Q^i$, $\tilde Q_i$,
but the superpotential
$W=\tilde Q \Phi Q$ required by $N=2$ SUSY in $4d$
is absent here; instead, $W=0$.
This is a model discussed in subsection \ref{AdSup}.

\smallskip

Fluctuations of the $N_c$ fourbranes along the
\nsp-branes
parametrize the Coulomb branch of the model.
Displacements of the $N_f$ $D6$-branes relative
to the \nsp-branes in the $(x^4, x^5)$
directions give masses $m$ to the fundamental multiplets
$Q$, $\tilde Q$
\beq
W=-\sum_{i=1}^{N_f} m_i\tilde{Q}_iQ^i
\label{D4N14}
\eeq
The relative position of the two
\nsp-branes in the $(x^4, x^5)$ directions is
not an independent parameter; it can be compensated
by a change in the positions of the $D6$-branes in the
$(x^4, x^5)$ plane (\ie\ the masses of the fundamentals),
and an overall rotation of the configuration.
The relative displacement of the two \nsp-branes
in the $x^7$ direction plays the
role of a FI D-term (\ref{D4N15}).
Complete Higgsing is possible for all $N_f\geq 1$;
the (complex) dimension of the Higgs branch is
\beq
{\rm dim}\MM_H=2N_fN_c+N_c^2-N_c^2=2N_fN_c
\label{D4N16}
\eeq
The first two terms
on the left hand side are the numbers of components
in the fundamental and adjoint chiral multiplets,
and the negative term accounts for degrees of
freedom eaten up by the Higgs mechanism.

The brane configuration provides
a simple picture of the moduli
space of vacua. As usual,
complete Higgsing
corresponds to breaking all $N_c$
fourbranes on various $D6$-branes
as indicated in Fig.~\ref{thirtytwo}(b).
We find that the dimension of
moduli space of brane configurations with
completely broken $U(N_c)$ gauge symmetry is
\beq
{\rm dim}\MM_H=N_c\left[2(N_f-1)+1+1\right]=2N_cN_f
\label{D4N17}
\eeq
in agreement with the gauge theory analysis (\ref{D4N16}).

\medskip
\noindent
{\bf 2. Mixed Electric-Magnetic Theories}

\begin{figure}
\centerline{\epsfxsize=140mm\epsfbox{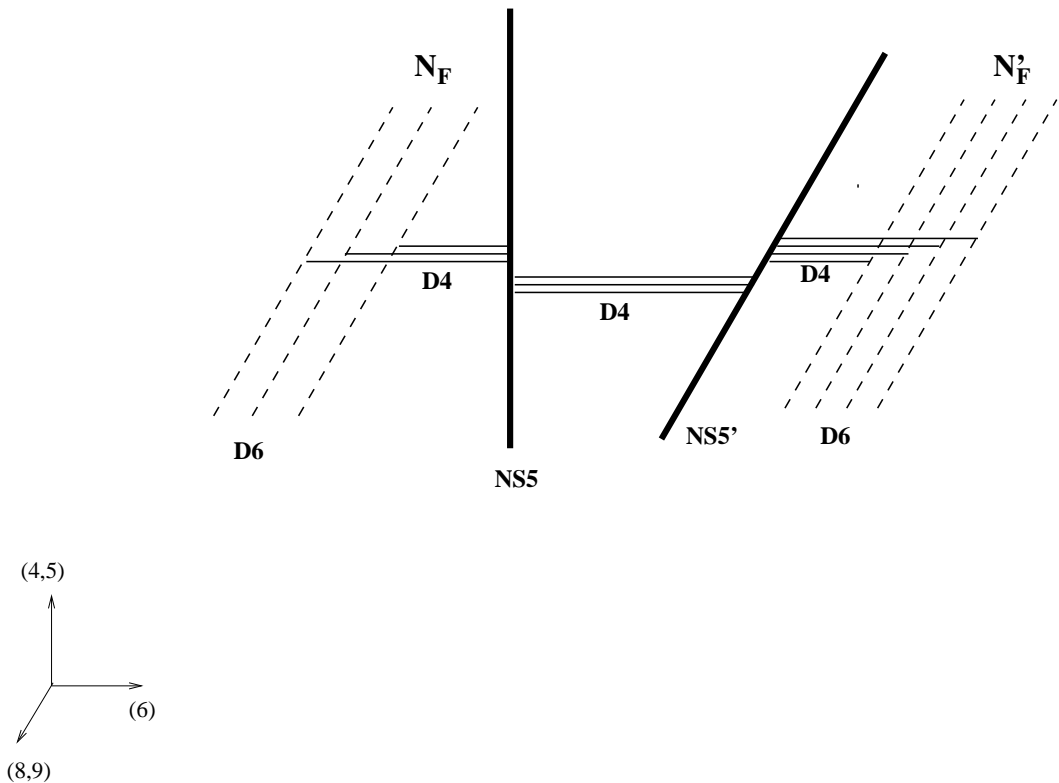}}
\vspace*{1cm}
\caption{A ``mixed electric-magnetic'' theory,
in which some of the fundamentals couple to
singlet mesons and some do not.}
\label{thirtythree}
\end{figure}

A straightforward generalization of the electric
and magnetic SQCD brane configurations
is a configuration which includes both
``electric'' and ``magnetic'' quarks.
Consider the configuration of Fig.~\ref{thirtythree};
an $NS5$-brane connected by $\nc$ $D4$-branes
to an \nsp-brane which is to its right (in $x^6$).
To the left of the $NS5$-brane we put $N_f$ $D6$-branes
each of which is connected by a fourbrane to the $NS5$-brane.
As before, these represent $N_f$ quarks $Q,\tilde Q$.
To the right of the \nsp-brane we put
$N^{\prime}_f$ $D6$-branes
each of which is connected to the \nsp-brane
by a fourbrane. These represent $N^{\prime}_f$
quarks $Q',\tilde Q'$ and $N^{\prime 2}_f$ complex scalars $M'$
with a tree level superpotential $W=M'Q'\tilde Q'$.
The SYM theory thus obtained is a ``mixed electric-magnetic''
$SU(N_c)$ gauge theory with $N_f$ ``electric'' quarks
and $N^{\prime}_f$ ``magnetic'' quarks coupled to
``magnetic mesons.''

The discussion of Seiberg's duality
can be repeated for such theories.
Interchanging the two NS-branes in $x^6$
gives rise to an
$SU(N_f+N^{\prime}_f-N_c)$ theory with $N_f$ magnetic quarks $q,\tilde q$
coupled to $N_f^2$ complex scalars $M$ via $M q\tilde q$,
and $N^{\prime}_f$ electric quarks $q',\tilde q'$.
Of course, the dual theory is also a mixed electric-magnetic
theory where the role of electric and magnetic quarks is
interchanged. In the particular case $N^{\prime}_f=0$ the original
theory is the electric theory studied in section \ref{D4N1ET}
while its dual is the magnetic theory as considered in
section \ref{D4N1SD}.

\smallskip

\medskip
\noindent
{\bf 3. ${\bf D6'}$-Branes}

Another interesting deformation of the
$N=2$ SUSY configuration involves rotating
some of the $D6$-branes as well. Restricting our
attention to ninety degree rotations, for simplicity,
we would like to consider, in addition to the
objects studied above, rotated sixbranes that are
located at $w=0$ and stretched in $v$. We will
refer to them as $D6^\prime$-branes:
\beq
D6^{\prime}: \qquad (x^0,x^1,x^2,x^3,x^4,x^5,x^7)
\label{D6prime}
\eeq
To study brane configurations including both
$D6$ and $D6^\prime$-branes one has to keep
in mind the following interesting feature
of brane dynamics.

Consider a configuration in which a pair of
$D4$-branes connect a $D6$-brane to a
$D6^\prime$-brane.
Naively the configuration preserves
four supercharges and there are two complex
moduli describing the locations of the
two $D4$-branes along the sixbranes
(in $x^7$) together with the compact
component of the gauge field $A_6$.

\begin{figure}
\centerline{\epsfxsize=100mm\epsfbox{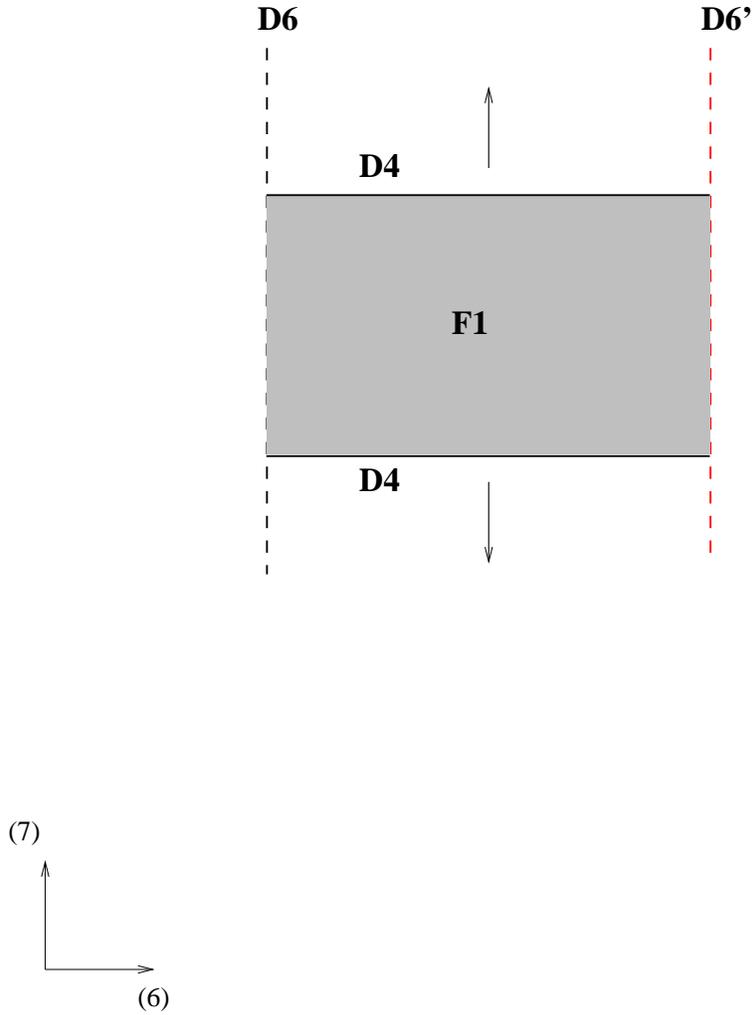}}
\vspace*{1cm}
\caption{The repulsive interaction between
$D4$-branes stretched between $D6$ and
$D6^\prime$-branes is due to Euclidean fundamental
strings stretched between the various branes.}
\label{thirtyfour}
\end{figure}
\smallskip

However, there is a superpotential
due to Euclidean fundamental strings
stretched between the $D4$ and $D6$-branes,
as indicated in Fig.~\ref{thirtyfour}.
If the distance between the sixbranes
is $\delta l_6$ and the separation
between the fourbranes is $\delta l_7$,
the superpotential due to these
Euclidean strings is of order
$\exp(-\delta l_6\delta l_7/l_s^2)$.
This effect is non-perturbative in
$l_s$ but does not go to
zero in the limit $\gs\to0$ -- it is
a worldsheet instanton effect.
In particular, it does not vanish
in the classical gauge theory limit discussed above
and leads to long range repulsive
interactions between the fourbranes.
It is closely related to the non-perturbative
effects discussed in section \ref{D4N4}
in systems with twice as much SUSY,
where they contribute to the metric
on moduli space (see eq. (\ref{nonprtcor})).

We thus arrive at the following classical
rule of brane dynamics:
\medskip

\noindent
{\it There is a long range repulsive interaction
between $D4$-branes stretched between a $D6$-brane and a
$D6^\prime$-brane. This repulsion does not go to
zero in the classical limit $\gs\to0$.}

\medskip
Taking this rule into account allows one to
understand configurations including both
$D6$ and $D6^\prime$-branes. The resulting
physics depends on the ordering of the
sixbranes along the $x^6$ axis. When a
$D6$-brane passes a $D6^\prime$-brane
there is a phase transition; this can be
seen by U-duality which can be used to
map this system to an \nsp-brane and
a $D6$-brane; as we saw before, the
physics certainly changes when we exchange those.
We will next consider the physics
for a particular ordering of the branes;
the generalization to other cases is
straightforward.

\begin{figure}
\centerline{\epsfxsize=100mm\epsfbox{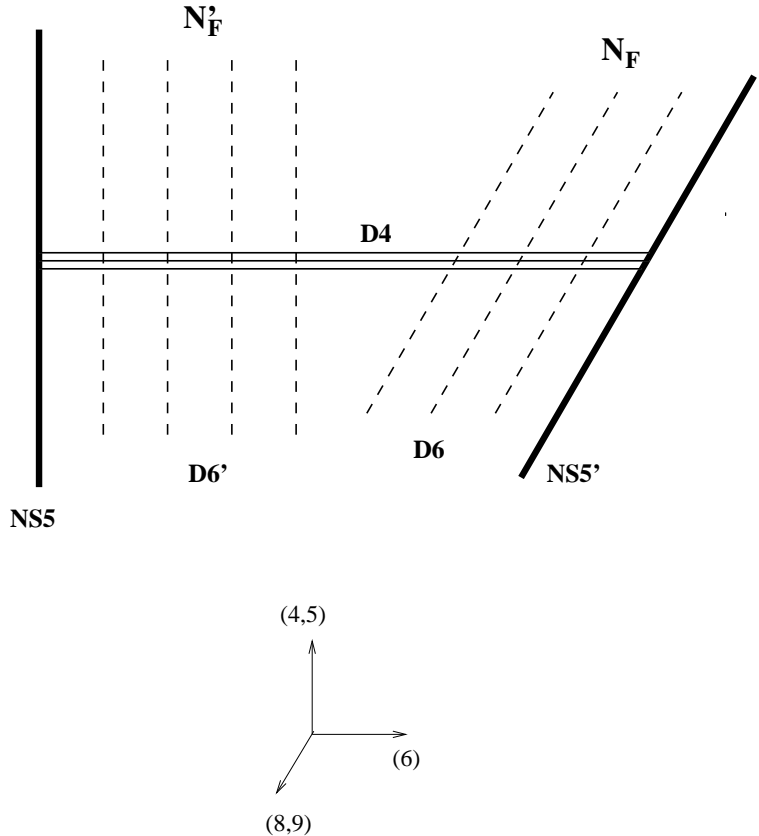}}
\vspace*{1cm}
\caption{$N=1$ SQCD with vanishing superpotential
described by a configuration with both $D6$ and
$D6^\prime$-branes.}
\label{thirtyfive}
\end{figure}

Consider the configuration
of Fig.~\ref{thirtyfive}. In addition to the
usual $\nc$ $D4$-branes stretched between
$NS5$ and \nsp-branes, which give rise to
a $U(\nc)$ gauge group, we have $\nf$ $D6$-branes
located next to the \nsp-brane and
$\nf^\prime$ $D6^\prime$-branes located
next to the $NS5$-brane, which give rise
to $\nf+\nf^\prime$ flavors. Clearly the theory
does not have a massless adjoint field
as there is no Coulomb branch, and by placing
the $D6$-branes on top of the \nsp-brane and
the $D6^\prime$-branes on top of the $NS5$-brane
we deduce that the symmetry of the theory
is at least $SU(\nf)\times SU(\nf)\times
SU(\nf^\prime)\times SU(\nf^\prime)$,
which does not allow a superpotential.

The theory is, therefore, $N=1$ SQCD with gauge
group $U(\nc)$, $\nf+\nf^\prime$ flavors of quarks
and $W=0$, which we have analyzed before.
The analysis of the moduli space gives
the right structure; we leave the details
to the reader. To get the correct structure
it is important
to use the rule stated above, which
implies here that configurations in which 
multiple $D4$-branes connect a given $D6$ and
$D6^\prime$-brane are unstable.

\smallskip

\subsection{Quantum Effects}
\label{QEN1}
In this section we study quantum effects in
$N=1$ SYM using brane theory. We
describe the quantum moduli space of
vacua and complete the
demonstration of Seiberg's duality.
We start by following a similar
route to that taken in section
\ref{D4N2}, and study the form of
the M-theory fivebrane describing the
brane configuration at finite
$R_{10}/L_6$, first semiclassically
and then exactly. Then we present
a qualitative picture of the moduli space
as resulting from certain
quantum interactions between
branes analogous to the classical
interactions encountered in the last
subsection.

\subsubsection{Semiclassical Description}
\label{QEN11}

At finite gauge coupling $g$ (\ref{BSB3}) we should
interpret our brane configurations as describing
fivebranes and sixbranes in M-theory with finite
$R_{10}/L_6=l_s g_s/L_6$. Recall the definitions
\bea
s=& x^6+i x^{10}\nonumber\\
v=& x^4+i x^5\nonumber\\
w=& x^8+i x^9
\label{defsvw}
\eea
Classically, \ie\ ignoring the
size of the $x^{10}$ circle,
the $D4$-brane
is located at $v=w=0$ and is
extended in $s$, the $NS5$-brane
is at $s=w=0$ and is extended
in $v$, while the \nsp-brane
is at $v=0$, $s=L_6$ and is extended in $w$.

Quantum mechanically, the fourbranes and
fivebranes merge into a single fivebrane in
M-theory, as described in the $N=2$ case
in section \ref{D4N2}. The vacuum configuration
of the fivebrane is described by a curve
$\Sigma$ embedded in the space $R^5\times S^1$
(\ref{defsvw}). As before, for large $v$, $w$
we can think of the shape of the
resulting $M5$-brane in terms of the original
$NS5$ and \nsp-branes, appropriately deformed
by the fourbranes ending on them (\ref{SGG3}).
The structure for large $v$, $w$ is the brane
analog of one loop corrections to classical
physics in gauge theory. In this subsection
we will describe these effects~\cite{EGKRS}.

Consider the classical electric configuration 
of Fig.~\ref{twentyfourb}(a). 
According to (\ref{SGG3}), far from the origin this
configuration is deformed
as follows. For large $v$ (and small $w$), the
shape of the $M5$-brane is that of the deformed $NS5$-brane
\beq
s_5=(N_f-N_c)R_{10}\log v
\label{ssff}
\eeq
while for large $w$ (and small $v$)
it looks like the deformed \nsp-brane
\beq
s_{5^\prime}=N_cR_{10}\log w
\label{ssnsp}
\eeq
The two asymptotic regions join in a way that
will be discussed later at small $v$, $w$.

As explained in section
\ref{QEFF}, among other things,
this bending causes
the freezing of the $U(1)\subset U(N_c)$.
Therefore, quantum mechanically we are
dealing with an
$SU(N_c)$ gauge theory.

In subsection \ref{SFTR} we saw that the
classical electric SQCD is invariant
under two $U(1)$ $R$-symmetries
(\ref{FT3}). In gauge theory
only one combination of the two
is preserved quantum mechanically;
the other is broken by the chiral anomaly
or, equivalently, instantons (see section \ref{SFTR}).
We will next exhibit this effect in brane theory.

The two classical $R$-symmetries correspond
in the brane construction to rotations of
the $(x^4, x^5)$ plane, $v\to e^{i\alpha} v$,
and the $(x^8, x^9)$ plane, $w\to e^{i\beta} w$.
The semiclassical configuration
(\ref{ssff}, \ref{ssnsp}) breaks both
symmetries. ${\rm Re}s_5$ and ${\rm Re}s_{5^\prime}$
are invariant under $U(1)_{45}$ and $U(1)_{89}$, but
${\rm Im}s_5$ and ${\rm Im}s_{5^\prime}$ are not invariant.
Overall shifts of ${\rm Im}s$ can be compensated by
a translation in $x^{10}$, hence any combination of
$U(1)_{45}$ and $U(1)_{89}$ which preserves
the relative location of the $NS5$ and
\nsp-branes in $x^{10}$ is a symmetry.
The (semiclassically)
unbroken $R$-symmetry is, therefore,
the one which preserves
\beq
s_5-s_{5^\prime}=R_{10}\log \left(
w^{-N_c}v^{N_f-N_c}\right)
\label{snsnsp}
\eeq
It is not difficult to check that
if (by definition) the $R$-charge of $\theta$
under this symmetry is one, that of $Q$,
$\tilde Q$ is $R(Q)=R(\tilde Q)=1-N_c/N_f$, in agreement
with the gauge theory answer (\ref{FT4}).
Of course, so far all we have checked is
that this symmetry is conserved semiclassically.
In field theory there is no contribution to the
anomaly beyond one loop; brane dynamics
reflects that, and one can check that the
exact form of the fivebrane preserves the symmetry
as well.

One can tell the same story for the brane
construction describing magnetic SQCD. The classical
$R$-symmetry corresponds again to
$U(1)_{45}\times U(1)_{89}$.
The charge assignments
of the various fields ($q$, $\tilde q$, $M$)
agree with those found in gauge theory (see section
\ref{SFTR}) and with the electric configuration.
Quantum mechanically
the fivebranes are deformed due to the presence
of the fourbranes; equations (\ref{ssff}), (\ref{ssnsp})
which were found for the electric configuration,
are valid for the magnetic one as well. In brane
theory this is a consequence of the fact that the
two configurations are related by the smooth
transition discussed earlier. In gauge theory
it is one of the checks of Seiberg's duality.

The foregoing discussion may be used to provide
a heuristic explanation of
a certain scale matching relation between the
electric and magnetic theories used in gauge
theory studies of Seiberg's duality. We can think
of the electric
coupling $(s_5-s_{5^\prime})/R_{10}$
(\ref{snsnsp}) as describing the
electric QCD scale $\Lambda_e$:
\beq
\Lambda_e^{3N_c-N_f}=\mu^{3N_c-N_f}
e^{-(s_5-s_{5^\prime})/R_{10}} =\mu^{3N_c-N_f}
w^{N_c}v^{N_c-N_f}
\label{scdep}
\eeq
where $\mu$ is some fixed scale.
If we start with a large and negative
${\rm Re} (s_5-s_5^\prime )$ the QCD scale $\Lambda_e$
is large. Exchanging the branes as discussed
above leads to a theory with
${\rm Re} (s_5-s_5^\prime )>0$ and, therefore,
small $\Lambda_e$. In this situation
we can continue thinking about
the theory as the electric theory with
a small $\Lambda_e$; alternatively,
we can switch to the magnetic point of
view and define the magnetic QCD scale $\Lambda_m$:
\beq
\Lambda_m^{3\bar N_c-N_f}=\mu^{3\bar N_c-N_f}
e^{+(s_5-s_{5^\prime})/R_{10}}
\label{scmag}
\eeq
where $\bar N_c\equiv N_f-N_c$.
Equations (\ref{scdep}), (\ref{scmag}) lead to the scale
matching relation
\beq
\Lambda_e^{3N_c-N_f}\Lambda_m^{3\bar N_c-N_f}=\mu^{N_f}
\label{scmatch}
\eeq
which has been argued to hold in gauge theory,
with $\mu$ a constant related to the coefficient
of the magnetic superpotential (\ref{FT14})~\cite{KSS}.
The relation (\ref{scmatch}) emphasizes the
strong-weak coupling aspect of Seiberg's duality,
since if $\Lambda_e$ becomes small (thus
making the electric theory weakly coupled)
$\Lambda_m$ is large, and vice-versa.

\subsubsection{Exact Results}
\label{QEN12}

So far we have focused on the
large $v$, $w$ form of the $M5$-brane
into which the IIA fivebranes and fourbranes merge
for finite $R_{10}/L_6$. Following the
logic of section \ref{D4N2} we next
derive its exact form.

We start with the case of pure SYM with
$G=SU(N_c)$ and no matter, described
by the brane configuration of 
Fig.~\ref{twentyfourb} (without sixbranes). 
We can proceed as in the $N=2$ case
studied in section \ref{D4N2}.
The worldvolume of the $M5$-brane is
$R^{3,1}\times \Sigma$, where the
complex curve $\Sigma$ is now embedded
in the three complex dimensional space
$Q\simeq R^5\times S^1$ parametrized by
$(v,w,s)$. The shape of the curve $\Sigma$
can be determined by using the symmetries
and singularity structure~\cite{HOO,W9706}.

Defining the variable $t$ as in (\ref{SGG4}),
we know that as $v\to\infty$ on $\Sigma$
(the region corresponding to the $NS5$-brane),
$t$ diverges (\ref{ssff}) as $t\simeq v^{\nc}$,
while $w$ goes to zero. Similarly, as $w\to\infty$
(the \nsp-brane (\ref{ssnsp})),
$t\simeq w^{-\nc}$ while
$v\to 0$. More generally,
$t$ should be a function of
$v$ that does not have poles or zeroes except
at $v=0$ (which is $w=\infty$)
and $v=\infty$. The unique solution
to all the constraints, up to
an undetermined constant $\zeta$, is:
\bea
&v^{\nc}=t \nonumber\\
&w^{\nc}=\zeta^{\nc}t^{-1} \nonumber\\
&vw=\zeta
\label{FFF7}
\eea
One way of arriving at the curve (\ref{FFF7}),
which also helps to understand the role of the
parameter $\zeta$,
is to start with the $N=2$ SUSY configuration
described in section \ref{D4N2}, and rotate
one of the $NS5$-branes as described in
subsection \ref{BSBNPB}~\cite{HOO,W9706}.

The $N=2$, $SU(N_c)$
brane configuration is given by the curve
(\ref{SGG6}-\ref{SGG7})
\beq
t^2+B(v,u_k)t+\Lambda_{N=2}^{2N_c}=0
\label{FFF1}
\eeq
where we have restored the dependence
on the QCD scale $\Lambda_{N=2}$.
We would like to find the curve corresponding
to a configuration where the right
$NS5$-brane has been rotated as in
(\ref{wthe}-\ref{muthephi}), which corresponds
to turning on a (complex) mass $\mu$
to the adjoint field (\ref{mutheta}), breaking $N=2$
SUSY to $N=1$.
In order to
``rotate the $NS5$-brane'' we must consider
configurations where the genus
$\nc-1$ curve (\ref{FFF1})
degenerates to a genus zero one.
In gauge theory this
is the statement that the adjoint mass lifts
the Coulomb branch, except for isolated
points. In the classical IIA limit
there is one such point, where all the
$D4$-branes are placed together,
corresponding to the
origin of the Coulomb branch. For finite
$R_{10}/L_6$
there are $N_c$ points where the curve (\ref{FFF1})
is completely degenerate. These points are related
by the discrete unbroken $Z_{2N_c}$
subgroup of $U(1)_{45}$ whose action on $v$, $t$ was
described in section \ref{QEFF} (after eq.
(\ref{slsr})). It acts on the QCD scale as:
\beq
\Lambda_{N=2}^2\to e^{2\pi i\over N_c} \Lambda_{N=2}^2
\label{FFF1a}
\eeq
At one of these degenerate points
the curve takes the form
\beq
v=t^{1/N_c}+
\Lambda_{N=2}^2 t^{-1/N_c}
\label{FFF2}
\eeq
Rotating the right $NS5$-brane from
$w=0$ to $w=\mu v$ implies that at
large $t$ we would like the curve to
approach $v^{\nc}=t$ and $w$ to be small,
while for $t\to 0$ we want it to approach
$w=\mu v$ with large $v$, $w$
(corresponding to the
$NS_\theta$ brane).
This is achieved by supplementing
(\ref{FFF2}) by:
\beq
w=\mu\Lambda_{N=2}^2 t^{-1/\nc}
\label{FFF3}
\eeq
To make contact with (\ref{FFF7})
we would like to take the adjoint
mass $\mu\to\infty$. Scale matching
between the high energy theory with
the adjoint field and the low energy
theory obtained by integrating it out,
\beq
\Lambda_{N=1}^3={\mu\over\nc}\Lambda_{N=2}^2
\label{FFF6}
\eeq
implies that at the same time we have
to take $\Lambda_{N=2}\to0$ holding the
$N=1$ SYM scale (\ref{FFF6}) fixed.
Rewriting
equations (\ref{FFF2}, \ref{FFF3})
in terms of $\Lambda_{N=1}$,
\bea
&v=t^{1/N_c}+{\nc\over\mu}\Lambda_{N=1}^3t^{1/N_c}
\nonumber\\
&w=\nc\Lambda_{N=1}^3t^{-1/N_c}
\label{FFF5}
\eea
and dropping the term proportional to
$\mu^{-1}$ in the first equation of (\ref{FFF5}),
leads to the curve (\ref{FFF7})
with~\footnote{We also rename
$\Lambda_{N=1}\to\Lambda$.}
\beq
\zeta=\nc\Lambda^3
\label{FFF8}
\eeq

We saw earlier that pure $N=1$ SYM
with $G=SU(\nc)$
has a $U(1)_R$ symmetry
that is broken at one loop to
$Z_{2\nc}$ by the chiral anomaly,
and is further spontaneously broken
non-perturbatively to $Z_2$, giving
rise to $\nc$ vacua with different
values of the gaugino condensate
(\ref{gacon}, \ref{FT1}).
This pattern of breaking of the chiral
$U(1)_R$ symmetry has a direct analog
in the brane language. In the previous
subsection we saw that the brane analog
of the one loop effect of the anomaly
is the asymptotic
curving of the branes for large $v$,
$w$. Thus studying the fivebrane
(\ref{FFF7}) semiclassically is tantamount
to having access to its large $v$, $w$
asymptotics, described by the first two
equations in (\ref{FFF7}), but not to
the shape of the
fivebrane for small $v$, $w$ which is
described by the last equation in (\ref{FFF7}).

It is therefore interesting that $\zeta$
appears in the first two equations
only in the combination $\zeta^{\nc}$,
while the third equation depends on
$\zeta$ itself. This means that
fivebranes (\ref{FFF7}) related by
the $Z_{\nc}$ transformation
\beq
\zeta\to e^{2\pi i\over N_c} \zeta
\label{FFF9}
\eeq
look the same asymptotically (or semiclassically)
but differ in their detailed shape.
Each of the $\nc$ possible values of
$\zeta$ (\ref{FFF9}) corresponds to a different
fivebrane and, therefore, to a different
vacuum of the quantum theory. The $Z_{\nc}$
symmetry relating them is spontaneously
broken. One can think of $\zeta$ as the
gaugino condensate (\ref{FT1})~\cite{HOO,BIKSY}.

In addition to the $Z_{\nc}$ symmetry
mentioned above, which acts on $v$, $w$
and $t$ as
\bea
&v\to v \nonumber\\
&w\to e^{2\pi i\over\nc}w\nonumber\\
&t\to t
\label{zznc}
\eea
and which -- as explained above -- is a symmetry
of the first two equations in (\ref{FFF7})
but does not leave invariant the third
one (or in other words has to be combined
with (\ref{FFF9}) to become a symmetry),
there are two more global symmetries.
One is a $U(1)$ $R$-symmetry discussed
near eq. (\ref{snsnsp}),
\bea
&v\to e^{i\delta} v\nonumber\\
&w\to e^{-i\delta}w\nonumber\\
&t\to e^{i \nc\delta} t
\label{rchrge}
\eea
As anticipated there, this symmetry
that is preserved semiclassically,
is an exact symmetry of the brane
configuration. For $\nf>0$ it corresponds
to a symmetry of the
low energy SYM theory, becoming
part of the $N=1$ superconformal
algebra in the IR. In the case
considered here, in the absence of
matter ($\nf=0$), the SYM fields
do not carry charge under this
symmetry. It is possible that
this $U(1)$ symmetry is still part
of the $N=1$ superconformal algebra
in the infrared, but pure SYM
theory has a mass gap and does not contribute
to the extreme infrared CFT. If the
brane configuration is to describe
SYM physics at low but non-zero
energies, any states charged
under (\ref{rchrge}) must decouple
from SYM physics.

There is also a $Z_2$ symmetry corresponding
to exchanging $v$ and $w$,
\bea
&v\to w\nonumber\\
&w\to v\nonumber \\
&t\to \zeta^{\nc} t^{-1}
\label{chcon}
\eea
This symmetry reverses the orientation of
$4-4$ strings stretched between different
fourbranes and, therefore, acts as charge
conjugation. The fact that it is an exact
symmetry of the vacuum is in agreement
with gauge theory.

Having understood chiral symmetry breaking in
the brane language we next turn to confinement~\cite{W9706}.
Pure $N=1$ SYM is expected to have the
property that if one introduces a heavy quark
and antiquark into the system, the energy
of the pair will grow with their separation
as if the two were connected by a string
with tension $\Lambda^2$. This ``QCD string''
can thus end on external quarks, but in
the absence of quarks it is stable. Since $\nc$
fundamentals of $SU(\nc)$ can combine into
a singlet, QCD strings can annihilate in
groups of $\nc$. It is expected that large
$\nc$ QCD can be reformulated in terms of
weakly coupled QCD strings. Establishing
the existence and studying the properties of
QCD strings is one of the major challenges
in QCD.

In brane theory it is natural to identify the
QCD string with an
$M2$-brane ending on the $M5$-brane (\ref{FFF7}).
We are searching for a membrane which
looks like a string to a four dimensional
observer and is also a string
in the space $Q$ labeled by $(v,w,t)$.
We can describe the string in $Q$ by an open curve
$C$ parametrized by $0\leq \sigma \leq1$,
such that both of its endpoints
(the points with $\sigma=0,1$)
are in $\Sigma$. It turns out that the
right curve for describing a QCD string is:
\bea
&t=t_0={\rm const}\nonumber\\
&v=t_0^{1/N_c}e^{2\pi i\sigma/N_c} \nonumber\\
&vw=\zeta
\label{FFFb}
\eea

The string in spacetime obtained by wrapping a membrane
around the curve $C$ has the following properties:
\begin{enumerate}
\item Groups of $\nc$ (but not less)
strings can annihilate.
\item The QCD string can end on an external quark.
\item For a particular choice of $t_0$, $C$ has
minimal length.
\end{enumerate}
The fact that QCD strings annihilate in groups
of $\nc$ can be seen by aligning
strings described by curves $C_j$ of
the form (\ref{FFFb}) with $2(j-1)\pi\le 2\pi\sigma\le 2j\pi$
($j=1,\cdots, \nc$). The $\nc$ strings form a long
closed string in $Q$ which can detach from the
fivebrane and shrink to a point. At the same time,
the strings corresponding to different $C_j$ are
all equivalent as they can be mapped into each other
by continuously varying the phase of $t_0$,
$t_0\to t_0\exp(2\pi i\alpha)$ with $0\le\alpha
\le 1$.

To minimize the length of $C$ one notes that
$t$ is constant along it, while $v$ and $w$
change by amounts of order $t_0^{1/\nc}/\nc$ and
$\zeta t_0^{-1/\nc}/\nc$, respectively (for large
$\nc$). The
length is minimized for $t_0\sim \zeta^{\nc/2}$;
it is of order $l_C\sim\zeta^{1/2}/\nc$.
The tension of the QCD string is
obtained by multiplying $l_C$
by the tension of the $M2$-brane
$1/l_p^3$. Restoring dimensions
in (\ref{FFF8}), $\zeta=\nc l_p^6\Lambda^3/R_{10}$,
we find that the tension of the
QCD string is
\beq
T\sim \left({\Lambda^3\over R_{10}\nc}\right)^{1\over2}
\label{tqcd}
\eeq
In SYM one expects the tension of the QCD
string to be of order $T\sim\Lambda^2$.
Comparing to (\ref{tqcd}) we see that
for agreement with SYM we must choose
\beq
R_{10}\sim {1\over\nc\Lambda}
\label{rqcd}
\eeq
For such values of $R_{10}$ there is no
decoupling of the four dimensional SYM
physics from Kaluza-Klein excitations
carrying momentum in the $x^{10}$
direction.
One might think that, due to (\ref{rqcd}),
at least for large $\nc$ the Kaluza-Klein
scale would be much higher than the QCD
scale $\Lambda$. Unfortunately, since
the Riemann surface $\Sigma$
winds $\nc$ times around
the $x^{10}$ direction, the Kaluza-Klein
modes see an effective radius $\nc R_{10}$
and have energies of order $\Lambda$.
Thus decoupling fails even in the
large $\nc$ limit.

{}From the discussion in previous sections
it is clear what went wrong. The QCD string
is not a BPS saturated object and, therefore,
its tension is not protected by the usual
non-renormalization theorems. The estimate
(\ref{tqcd}) of its tension is semiclassical
in nature and is valid when the supergravity
approximation for describing membranes and
fivebranes is applicable. We are discovering
that in this regime the system does not
describe decoupled SYM physics.
The regime corresponding to SYM is
(\ref{BSB3}-\ref{hierscale}); in that
regime it is not clear at present how to study
properties of the QCD string, such as the tension,
but there is no reason for the formula
(\ref{tqcd}) to be valid. It is known~\cite{BHOO97}
that other non-holomorphic
SYM features, such as the K\"ahler potential
for mesons and baryons, depend sensitively
on $R_{10}$, $L_6$, and there is no 
reason to expect that the tension of QCD
strings is any different.

In addition to QCD strings, one can
construct using branes domain walls
separating regions in space corresponding
to different vacua (different values of
$\zeta$). A domain wall occurs when
as $x^3\to-\infty$ the configuration
approaches one value of $\zeta$ while
as $x^3\to\infty$ it approaches another.
The resulting $M5$-brane interpolates
between the two solutions (\ref{FFF7}).
It is known in gauge theory that
such domain walls are BPS saturated
and their tension is the difference
between the values of the superpotential
(\ref{nonprtsup}) between the different
vacua. At large $\nc$ it thus goes
like $T_D\simeq \nc\Lambda^3$.

Unlike the QCD string, the tension
of the BPS saturated domain wall
(or membrane) can be exactly calculated
using branes. Witten has shown that
the tension of the domain wall goes
at large $\nc$ like $T_D\simeq R_{10}
|\zeta|/l_p^6$ which, using the form of
$\zeta$ and $R_{10}$ discussed above, agrees
with the gauge theory analysis. Witten
furthermore pointed out that the domain wall
behaves in large $\nc$ gauge theory like
a Dirichlet two brane in string theory;
its tension goes like
$\nc$, which is the inverse QCD string coupling,
and the QCD string can end on it,
just like the fundamental string can end
on a D-brane.

The above discussion can be generalized by
adding $N_f$ fundamental chiral multiplets
of $SU(\nc)$ with masses $m_i$, $i=1,\cdots,\nf$.
We saw that these can be described by
adding $N_f$ semi-infinite fourbranes
to the left of the $NS5$-brane at
$v=m_i$. The corresponding Riemann surface
$\Sigma$ takes the form~\cite{HOO,W9706,BIKSY}
\bea
&v^{\nc}=
t\prod_{i=1}^{N_f}\left(1-{v\over m_i}\right)
\nonumber\\
&vw=\zeta
\label{FFF7f}
\eea
where
\beq
\zeta^{\nc}=\Lambda^{3N_c-N_f}
\prod_{i=1}^{N_f}m_i
\label{FFF8f}
\eeq
For large $m_i$ the configuration
(\ref{FFF7f}) is essentially
the same as (\ref{FFF7}) and one can
think of the quarks with masses $m_i$
as static sources.

Quarks are confined in this system,
and one expects the energy of a state with a quark and
antiquark separated by a large distance
$\delta x\gg \Lambda^{-1}$ to grow like $T\delta x $
where $T$ is the tension of the QCD string.
Classically, the quark and antiquark
are described by fundamental strings
connecting a flavor fourbrane to
the stack of color fourbranes. Quantum
mechanically these fundamental strings
turn into membranes and the only stable
configuration has them connected by a
long QCD string; thus its energy is indeed
proportional to the separation of the
two quarks as expected from gauge theory.

To study
theories with massless quarks we have to
take the limit $m_i\to 0$ (\ref{FFF8f}).
This was discussed in~\cite{HOO}.
For $0<N_f<N_c$ massless flavors,
the curve one finds in the limit is singular --
it is infinitely elongated in the
$x^6$ direction and, therefore,
the corresponding brane configuration
does not describe a four dimensional field theory.
This is consistent with the field theory analysis:
the gauge theory has no vacuum due to the
non-perturbative superpotential (\ref{FT5}).

For $\nf\ge \nc$ the SYM theories
under consideration have quantum moduli
spaces of vacua that were described
in section \ref{SFTR}. To study them one
needs to replace the semi-infinite
fourbranes by fourbranes ending
on sixbranes, described as in section
\ref{D4N2} by an $M5$-brane in the
background of a resolved $A_{\nf-1}$
multi Taub-NUT space. It is then possible,
by rotating the $N=2$ SYM curves with
matter studied in section
\ref{D4N2}, to describe the roots
of different
branches of the moduli space. As an example,
the root of the baryonic branch, which exists
for all $\nf\ge \nc$, is (formally) described by the
factorized curve
\bea
\Sigma_L:& \qquad t=v^{N_c-\nf}, \qquad w=0
\nonumber\\
\Sigma_R:& \qquad 
t=\Lambda^{3N_c-N_f}w^{-N_c}, \qquad v=0
\label{FFF7fa}
\eea
It can be shown that 
deformations of the curve (\ref{FFF7fa})
lead to a $2N_cN_f-(N_c^2-1)$ complex dimensional
space parametrizing the Higgs branch of the theory,
in agreement with field theory results (with the
caveat discussed in section \ref{QEFF} that one
complex modulus appears to be a parameter in
the brane description).

It should be emphasized that just like in
section \ref{D4N2}, when one approaches
a singular point in moduli (or parameter)
space where the infrared behavior changes,
such as (\ref{FFF7fa}),
the fivebrane degenerates and the supergravity
approximation breaks down, even if overall
the fivebrane (\ie\ $L_6$, $R_{10}$) is large.
Thus, one cannot use supergravity to study
most aspects of the non-trivial SCFT at the
origin of moduli space for $\nf\ge\nc$. 

What is done in practice 
is to resolve the singularity by turning
on a superpotential for the quarks that
lifts all the flat directions, or study
the theory in its fully Higgsed branch.
As is standard in gauge 
theory~\cite{Sei94a}, by computing the
expectation values of chiral fields
as a function of the deformation parameters
one can recover the superpotential
at the origin of moduli space. 

Further study of confinement and extended 
objects in MQCD appear in
several recent
works~\cite{HSZ,NOH,Vol,FS9711,AOTd,Ahn9712}.
The description of the duality 
trajectory of section \ref{D4N1SD}
in M-theory appears in~\cite{Fur,SS,CS}.

\subsubsection{Brane Interactions}
\label{QEN13}

So far we have discussed
the vacuum structure of $N=1$ SQCD by using
properties of the M-theory fivebrane. We saw
that many features of the quantum vacuum
structure can be understood using
fivebranes. In particular, M-theory
techniques provide a very natural
description of the Coulomb branch
of various $N=1,2$
SUSY gauge theories.  They are also
very useful for describing isolated
vacua witha mass gap, such as those of SQCD
with massive quarks, and for studying
properties of BPS states in such vacua.

There are also some drawbacks.
One is that the description in terms
of large and smooth fivebranes
is inapplicable in the SYM limit
(\ref{BSB3}-\ref{hierscale}) where the fivebrane
in fact degenerates, and at the same time most
quantities that one might be interested in
calculating in SYM
depend strongly on $R_{10}$ and $L_6$.
Also, the long distance behavior at the origin
of moduli space is described by singular 
fivebranes for which the supergravity
description is not valid.
Even restricting to the vacuum structure,
the description of the global structure of moduli space
is rather involved in the M-theory
language already for SQCD, which makes it difficult to
extract physical consequences and study more
complicated situations.

Also, one may want
a more uniform description of the physics in
different dimensions. We will discuss later
three dimensional analogs of the theories
studied in this section, which correspond to
brane configurations in type IIB string theory,
where the M-theory construction
is inapplicable.
It is one of the remarkable features of
brane dynamics that rather different
dynamical systems such as three and four dimensional
gauge theories are described by closely
related brane configurations. It is difficult
to believe that when the dynamics of branes
is eventually understood, the story will
be drastically different in different
dimensions.

To really solve QCD using webs of branes
one needs a much better understanding of
the theory on fourbranes stretched between
fivebranes in the appropriate scaling
limit. Already for a stack of flat
parallel NS fivebranes, the worldvolume
dynamics is not understood 
(see~\cite{ABKSS,ABS,GS} and references therein
for recent work on this problem). It is
even less clear what happens when one suspends
fourbranes between the fivebranes and studies
the system in the limit (\ref{BSB3}-\ref{hierscale}).

In the absence of understanding of the theory
on the fivebrane one may proceed as follows~\cite{EGKRS}.
The quantum vacuum structure
of different brane configurations can be thought
of as a consequence of interactions between
different branes. For
theories with eight supercharges 
such interactions
modify the metric on moduli space, while
for systems with four supercharges they
give rise to forces between
different branes that sometimes
lift some or all
of the classical moduli space.

When the interacting branes are nearby,
one expects the resulting forces to be rather
complicated and a more detailed understanding
of fivebrane dynamics is necessary. 
For widely separated branes, \ie\ far
from the origin of moduli space,
the interactions should simplify.
The purpose of this subsection is to describe the
quantum moduli space of vacua of SQCD with
$G=SU(\nc)$ and $\nf$ fundamentals by postulating
certain long range
interactions between different branes.
In the next section we will show that these
interactions also explain the vacuum structure
of $N=2$ SUSY gauge theories in three dimensions.

Of course, these interactions are not derived
from ``first principles'' but rather guessed
by comparing to the gauge theory results, so
on the level of the present discussion they
do not necessarily have much predictive power.
However, as usual in brane theory, the interactions
are local in the sense that they do not
depend on the global structure of the
configuration in which the
branes are embedded. Therefore, once the local
rules are formulated one can use them in
more complicated situations, and even
different dimensions, to learn more
about gauge dynamics. And, of course, once
one is convinced that these rules
are valid, they teach us about
brane dynamics as well, and need to be
eventually reproduced by the theory of
the fivebranes.

\begin{figure}
\centerline{\epsfxsize=60mm\epsfbox{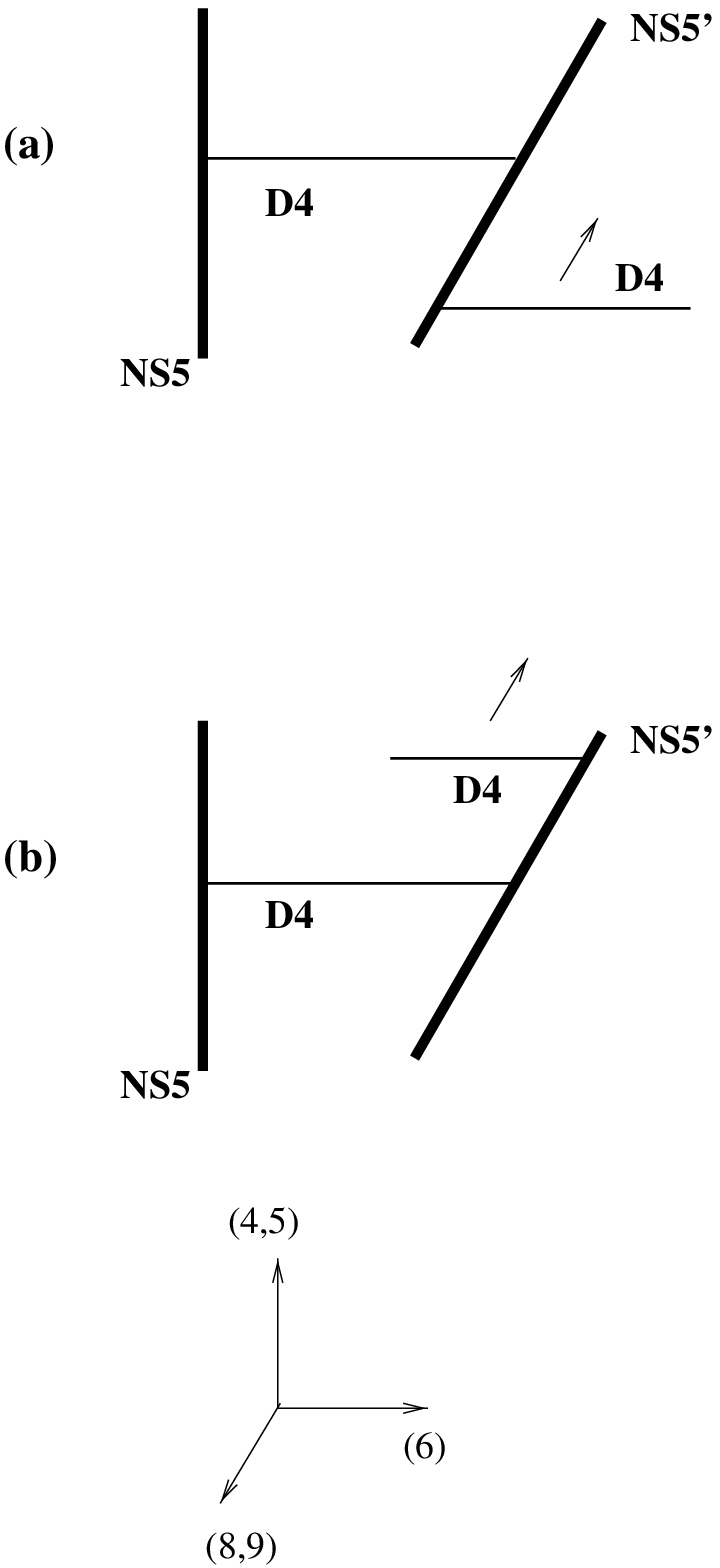}}
\vspace*{1cm}
\caption{Widely separated fourbranes in configurations
with $N=1$ SUSY act as charged particles: branes
that end on opposite sides of an NS fivebrane attract
each other (a), while those that end on the same side
have a repulsive long range interaction (b).}
\label{thirtysix}
\end{figure}
\smallskip

The quantum rule of brane dynamics
that we will postulate is
(see Fig.~\ref{thirtysix}):
\medskip

\noindent
{\it There is a long range interaction between
a $D4$-brane stretched between an $NS5$ and
an $NS5'$-brane, and any other $D4$-brane
ending on one of the fivebranes. It is repulsive if
the $D4$-branes are on the same side of the
fivebrane, and attractive if they are on different
sides.}

\medskip
\noindent
Comments:
\begin{enumerate}
\item
U-duality relates the above rule to many other
cases. For example, the classical interaction
between $D4$-branes stretched between $D6$ and
$D6^\prime$-branes -- discussed in subsection
\ref{ROT} -- is related to it
by compactifying (say) $x^3$ and applying the
U-duality transformation $U=T_3ST_3$.
$U$ relates quantum interactions
to classical ones in this case
because it involves
a strong-weak coupling duality transformation
($S$). As another example,
in the next section we will discuss
the consequences of the above quantum
interactions for
systems related to the current setup
by applying $T_3$ (\ie\
$D3$-branes ending on $NS5$ and
\nsp-branes). In the rest of this section
we use the quantum interactions to describe
the moduli space of vacua of SQCD.
\item The quantum rules are useful in
describing situations where the different
branes that interact are widely separated.
They provide a qualitative picture
of the quantum moduli space and can be
used to understand the semiclassical 
corrections to the superpotential. 
One can thus 
see using the quantum rules when runoff
to infinity in moduli space will occur,
and in situations with unlifted quantum
moduli spaces, the quantum rules allow one
to study the structure of the moduli space
far from the origin. The origin of moduli
space and, in general, situations where
the branes are close to each other need to
be studied by different techniques.
\end{enumerate}

We start with electric SQCD described
by the brane configuration of 
Fig.~\ref{twentyfourb}.
For $\nf=0$ the system
contains $\nc$ $D4$-branes stretched
between an $NS5$ and an \nsp-brane.
The quantum rule formulated above cannot
be applied to this case.
The $D4$-branes repel each other
but are restricted by the
geometry to lie on top of each other,
and the vacuum structure is determined
by short distance properties of the brane
system. In the previous subsection we
saw that the M-theory analysis
gave a good description of the vacuum
structure for this case.

For $1\le\nf\le \nc-1$ massless flavors
the system develops an instability that
can be understood using the quantum rule.
Describing the flavors by $D6$-branes intersecting
the $D4$-branes (Fig.~\ref{twentyfourb}(b)), there is
now the possibility for $D4$-branes to break
on the $D6$-branes, and the segments of the broken
$D4$-branes connecting the \nsp-brane to the nearest
$D6$-brane are repelled from the remaining
color $D4$-branes, which are still stretched
between the $NS5$ and \nsp-branes. Since the
repulsion is presumed to be long range, these
segments run off to $w\to\infty$, and there is
no stable vacuum at a finite value of the
moduli.

For $\nf\ge\nc$ the situation changes.
Now there do exist stable configurations
of the branes with no repulsive interactions.
They correspond to breaking all $\nc$ color
fourbranes on $D6$-branes, which effectively
screens the repulsive interactions and gives
rise to a quantum moduli space that looks
qualitatively the same as the classical one.
Interesting effects occurring
near the origin of the quantum moduli space,
such as the quantum modification (\ref{FT10})
for $\nf=\nc$, again correspond to a regime
where the brane interactions are not
well understood and have to be studied by
different techniques. One consequence of
this discussion is that the dimension
of the quantum
moduli space of electric SQCD is seen
in brane theory to be $2\nf\nc-\nc^2$
(presumably $+1$ to account for the difference
between $SU(\nc)$ and $U(\nc)$ as discussed
above), just like that of the classical theory.

A similar analysis can be performed
for the magnetic configuration of Fig.~\ref{twentyeight} 
with gauge group $G=SU(\bar\nc)$ and $\nf$ flavors.
As before, we will restrict to the case $\nf\ge\bar\nc$.
We saw before that the classical moduli space is
$\nf^2$ dimensional, corresponding to
giving expectation values to the components
of the magnetic meson field $M$ (\ref{wmag}),
without breaking the gauge group. We also
saw that turning on a FI D-term (\ref{D4N15})
changes the form of the moduli space discontinuously.
In particular, for $r\not=0$ the moduli space
is $\nf^2-\bar\nc^2$ dimensional.

\begin{figure}
\centerline{\epsfxsize=100mm\epsfbox{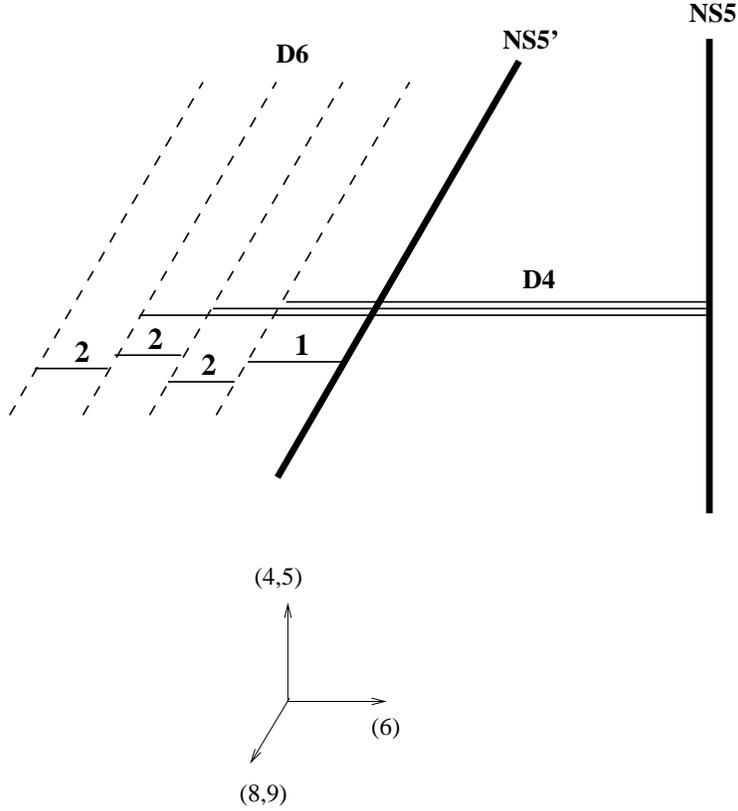}}
\vspace*{1cm}
\caption{Quantum brane interactions lift
a part of the magnetic moduli space
Fig. 29, leaving an unlifted
$\nf^2-\bar\nc^2$ dimensional quantum
moduli space.}
\label{thirtyseven}
\end{figure}
\smallskip

Quantum mechanically the discontinuity
in the structure of the moduli space
is eliminated. The $\bar\nc$ color
fourbranes are attracted to the $\nf$
flavor fourbranes. Hence,
$\bar N_c$ of the $N_f$ flavor fourbranes
align with the color fourbranes
and reconnect, giving rise to $\bar N_c$
fourbranes stretched between the $NS5$-brane
and $\bar N_c$ different sixbranes (in agreement
with the s-rule). The remaining $\nf-\bar\nc$
flavor fourbranes are easily seen to give
rise to an $\nf^2-\bar\nc^2$ dimensional
moduli space (see Fig.~\ref{thirtyseven}). 
Furthermore,
as is obvious from Fig.~\ref{thirtyseven}, 
the part of the classical moduli space that
remains unlifted in the quantum theory
is precisely the part that is smoothly
connected to the structure at non-zero
FI D-term $r$ (or to the baryonic
branch of moduli space, if the gauge group
is really $SU(\bar\nc)$ and $r$ describes
the baryonic branch).

In gauge theory, the lifting of a part of
the classical moduli space in the quantum
magnetic theory follows from the fact
that the classical magnetic
superpotential (\ref{wmag}) is corrected quantum
mechanically~\footnote{Equation
(\ref{FT14}) corresponds to $\bar\nc=\nf-\nc$,
the value relevant for $N=1$ duality.}
to (\ref{FT14}).
The second term in
(\ref{FT14}) is due to
the fact that when $M$ gets
an expectation value, the magnetic quarks
become massive due to the classical coupling
(\ref{wmag}), and a superpotential of the form
(\ref{detmm}) with $\nc\to\bar\nc$ is generated.

It is not difficult to show that the
moduli space corresponding to (\ref{FT14})
is the same as the quantum moduli space of
brane configurations (Fig.~\ref{thirtyseven}).
Thus we see that the quantum brane interactions
described above know about non-perturbative
superpotentials in SYM theory.

\subsubsection{Quantum $N=1$ Duality
And Phase Transitions}
\label{QEN14}

After understanding the form of the
quantum moduli spaces of vacua of the electric
and magnetic theories we can complete the
demonstration of Seiberg's duality using branes.
We saw before that classically the moduli spaces
of the electric and magnetic theories agree for
non-zero $r$ (\ref{D4N15}), but there is a discrepancy
between the structures for $r=0$. We have now
seen that quantum mechanically the discrepancy
disappears. The electric moduli space is not
modified quantum mechanically, while in the magnetic
theory quantum effects lift part of the classical
moduli space, leaving behind precisely the subspace
that connects smoothly to the electric theory
via the construction of subsection \ref{D4N1SD}!
This completes the proof of the equivalence
of the quantum moduli spaces of the electric and
magnetic theories and, therefore, also of the
corresponding chiral rings.

In gauge theory one distinguishes between
two notions of $N=1$ duality. The weaker
version is the statement that members of
a dual pair share the same quantum chiral
ring and moduli space of vacua, as a function
of all possible deformations. In Seiberg's
original work this statement has been
proven for supersymmetric QCD, and we have
now rederived it using branes.
The stronger version
of Seiberg's duality asserts that the full
infrared limits of the electric and magnetic
theories coincide. In field theory, no proof
of this assertion has been given, but it is
believed to be correct. One may ask whether
the embedding of the problem in brane theory
helps to settle the issue.

To show the equivalence of
the (in general) non-trivial
infrared theories at the origin of
the electric and magnetic
moduli spaces, one would like to continuously
interpolate between them while staying
at the origin of moduli space
and only varying $\Lambda$, or $x^6$. In the
process we pass through
a region where the $NS5$ and
\nsp-branes cross. We will next
discuss this region.

In fact, one can ask more generally, what happens
to the low energy physics on webs of
branes as some of the branes (which are
in general connected by other branes
to each other) meet in space and
exchange places. We discussed a few examples of
such transitions at various points in the review.
Let us summarize the results.

The low energy physics is smooth
when non-parallel NS branes
connected by fourbranes cross (in which
case the smoothness of the transition is
equivalent to the strong version of Seiberg's
duality), and when non-parallel NS and D-branes
cross (the HW transition of Fig.~\ref{fifteen}).
When parallel NS
fivebranes connected by fourbranes cross,
the transition relates $N=2$ SYM theories with
different rank gauge groups, \eg\ $U(\nc)$
and $U(\nf-\nc)$. By construction, these theories
have the same fully Higgsed branch but in general
different mixed and
Coulomb branches, and even different numbers
of massless fields. Thus in that case there is a phase
transition. Similarly, when parallel D and NS branes
cross, there is a phase transition. For example, we saw
that as a $D6$-brane passes an \nsp-brane, we lose
or gain a light matter multiplet.

In both of the above cases, phase transitions occur
in situations where a configuration containing
parallel coincident branes is deformed in
different directions. An interesting
example that superficially
shows a different
behavior is configurations
with rotated sixbranes $D6_\th$,
discussed in subsection \ref{ROT},
where the low energy
physics
depends on the order in which
different non-parallel
sixbranes appear along the $x^6$ axis
(different orders corresponding
to different superpotentials).
A closer look reveals that, in fact,
this example follows the same pattern as the
others. When {\em all} branes (the two NS branes and
$\nf$ sixbranes) are non-parallel, there
is in fact {\em no} phase transition as different
sixbranes cross. It is only when some of the sixbranes
are parallel to other sixbranes or to one
or more of the NS fivebranes, as in
the configuration of Fig.~\ref{thirtyfive},
that changing the
order of the sixbranes influences the low energy dynamics.

For the case when some of the sixbranes
become parallel, it is easy to
understand the mechanism for the phase
transition. Imagine first placing all
$\nf$ sixbranes at the same value of
$x^6$. In this case, fundamental
strings connecting
different sixbranes give rise
to {\em massless} fields which we will
collectively denote by $A$.
Quarks $Q$ are as usual described
by $4-6$ strings.
The standard three open string
coupling gives rise to cubic
superpotentials of the form
$W=\tilde Q A Q$. As we displace
the sixbranes relative to each other
in $x^6$ the fields $A$ become massive,
and integrating them out gives rise
to quartic superpotentials for the
quarks, of the general form
$W\sim (\tilde QQ)^2$. It is rather
easy to see that in the generic
case, when no branes are parallel,
the superpotential generated this way
is the most general one, and the
low energy theory is insensitive to
the precise coefficients. When
some of the sixbranes are parallel,
different deformations give
superpotentials with inequivalent
long distance behaviors.

The lesson from this example
is the following. When branes meet
in space, additional degrees of freedom
in the theory in general become massless.
If these degrees of freedom couple to
the gauge theory on the fourbranes,
it is possible that different deformations
of the singular point in which
branes touch produce different
low energy behaviors. Otherwise
the transition is smooth.

What happens in the other cases
described above? When two parallel
NS branes approach each other,
degrees of freedom corresponding
to membranes stretched between them
go to zero mass and eventually
become tensionless BPS saturated
strings trapped in the fivebrane(s).
The usual three membrane vertex
in eleven dimensions implies that
these tensionless strings interact
with the degrees of freedom describing
the gauge theory on the fourbranes
and, therefore, it is not surprising
as in the previous case to find that
different deformations of the system
correspond to different phases.

When two non-parallel fivebranes,
$NS_{\theta_1}$ and $NS_{\theta_2}$
with $\theta_1\not=\theta_2$,
approach each other in $x^6$,
membranes stretched between
the two fivebranes do not lead to BPS
saturated strings inside the fivebrane.
Hence, there is no mechanism for a phase
transition to occur as the two fivebranes
are exchanged.

It is important to emphasize that the above
argument does not prove full infrared equivalence
of members of a Seiberg dual pair. The fact
that membranes stretched between non-parallel
NS fivebranes are not BPS saturated provides
another proof of the fact that the {\em vacuum
structure} is smooth. To rule out a change
in the full infrared CFT, one needs to
understand the interactions of all the
light non-BPS modes of a membrane stretched
between the $NS_{\theta_1}$ and $NS_{\theta_2}$
fivebranes with the gauge theory degrees of
freedom. This is beyond the reach of available
methods.

\subsection{Generalizations}
\label{GEN}

Branes can be used to study
the dynamics of a wide variety of $N=1$
supersymmetric gauge theories with different
matter contents and superpotentials. In this
subsection we briefly describe a few constructions
that appeared in the
recent literature. In situations where a
good brane description exists, it leads
to new insights both on gauge theory and
on brane dynamics. Therefore, it is
important to enlarge the class of models that
can be described this way. This may also
provide clues towards the formulation of
the fivebrane theory.

\subsubsection{Product Groups}
\label{SFPG}

In section \ref{SGG} we discussed $N=2$ SUSY
theories with product gauge groups
$G=\prod_{\alpha=1}^n SU(k_\alpha)$, by
considering $n+1$ parallel $NS5$-branes
connected by fourbranes. $N=1$ configurations
of this sort are obtained by performing relative
rotations (\ref{wthe}) of the fivebranes.

\begin{figure}
\centerline{\epsfxsize=100mm\epsfbox{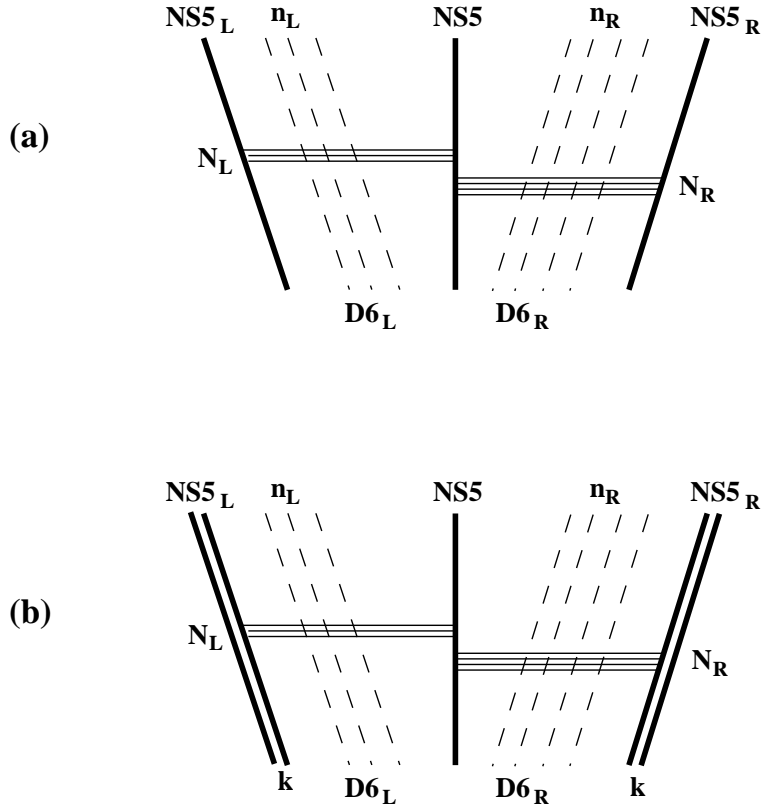}}
\vspace*{1cm}
\caption{An $N=1$ SUSY theory
with $G=U(N_L)\times U(N_R)$ and matter
in the bifundamental and fundamental representations.
The bifundamental has a quartic (a) or higher
order polynomial (b) superpotential.}
\label{thirtyeight}
\end{figure}
\smallskip

As an example, consider the configuration of
Fig.~\ref{thirtyeight}(a), 
which was studied in~\cite{BH,GP}.
Three NS fivebranes denoted by $NS5_L$, $NS5$
and $NS5_R$ are
ordered in the $x^6$ direction
such that the $NS5_L$-brane
is the leftmost while
the $NS5_R$-brane
is the rightmost. We can choose to orient
the (middle) $NS5$-brane as in (\ref{BSB1}),
and rotate the other two relative to it
by $(\theta_L,\varphi_L)$ and
$(\theta_R,\varphi_R)$ (see (\ref{muthephi})).
$N_L$ ($N_R$) $D4$-branes are stretched in the $x^6$
direction between the $NS5_L$ and $NS5$-branes
(the $NS5$ and $NS5_R$-branes).

The theory on the fourbranes is an $SU(N_L)\times SU(N_R)$
gauge theory with two chiral multiplets in the adjoint
of the respective gauge groups $\Phi_L,\Phi_R$,
and bifundamentals $F,\tilde F$ in the
$(N_L,\bar N_R),(\bar N_L,N_R)$. The classical
superpotential is:
\beq
W=\mu_L {\rm Tr}\Phi_L^2+\mu_R {\rm Tr}\Phi_R^2 +
  {\rm Tr}\tilde F\Phi_L F+{\rm Tr}F\Phi_R\tilde F
\label{gen1}
\eeq
where (see (\ref{muthephi}))
\beq
\mu_L=e^{i\varphi_L}\tan\theta_L, \qquad
\mu_R=e^{i\varphi_R}\tan\theta_R
\label{gen2}
\eeq
Integrating out the massive adjoints we obtain
(for generic rotation angles)
\beq
W\sim {\rm Tr}(F\tilde F)^2
\label{gen3}
\eeq
We can add fundamental quarks to the theory by adding
to the configuration sixbranes and/or
semi-infinite fourbranes.

A qualitative identification between the parameters and
moduli of the field theory on the fourbranes
and the parameters determining the brane configuration
can be made along the lines of this section.
The quantum vacuum structure can be studied by
starting with the $N=2$ curve (\ref{nfkf}) and
rotating it, following the logic of the discussion
of section \ref{QEN1} for a simple group.
This was done in~\cite{GP}.

It is straightforward to find Seiberg dual
configurations by interchanging the order
of the NS-branes, a procedure that is expected
to preserve the long distance physics as long
as the fivebranes being exchanged are not parallel
(which is the case for generic $\theta_L$,
$\theta_R$ (\ref{gen2})). In particular, if we start
in the ``electric'' configuration of 
Fig.~\ref{thirtyeight}(a) 
with $n_L$ and $n_R$ flavors of $SU(N_L)$
and $SU(N_R)$, respectively, exchanging
the $NS5_L$ and $NS5_R$-branes leads~\cite{BH} to
a magnetic theory with
$G=SU(n_L+2n_R-N_R)\times SU(n_R+2n_L-N_L)$ and the
same number of flavors, in agreement with
the field theory results~\cite{ILS}.

\subsubsection{Landau-Ginzburg Superpotentials}
\label{SFLGS}

Brane configurations containing $D4$-branes
ending on a stack of parallel NS fivebranes
are interesting since the theory on the
fivebranes is in this case non-trivial in the IR
(it is the (2,0) theory discussed before),
and it is interesting to see how this is reflected
in the structure of the theory on the fourbranes.

\begin{figure}
\centerline{\epsfxsize=140mm\epsfbox{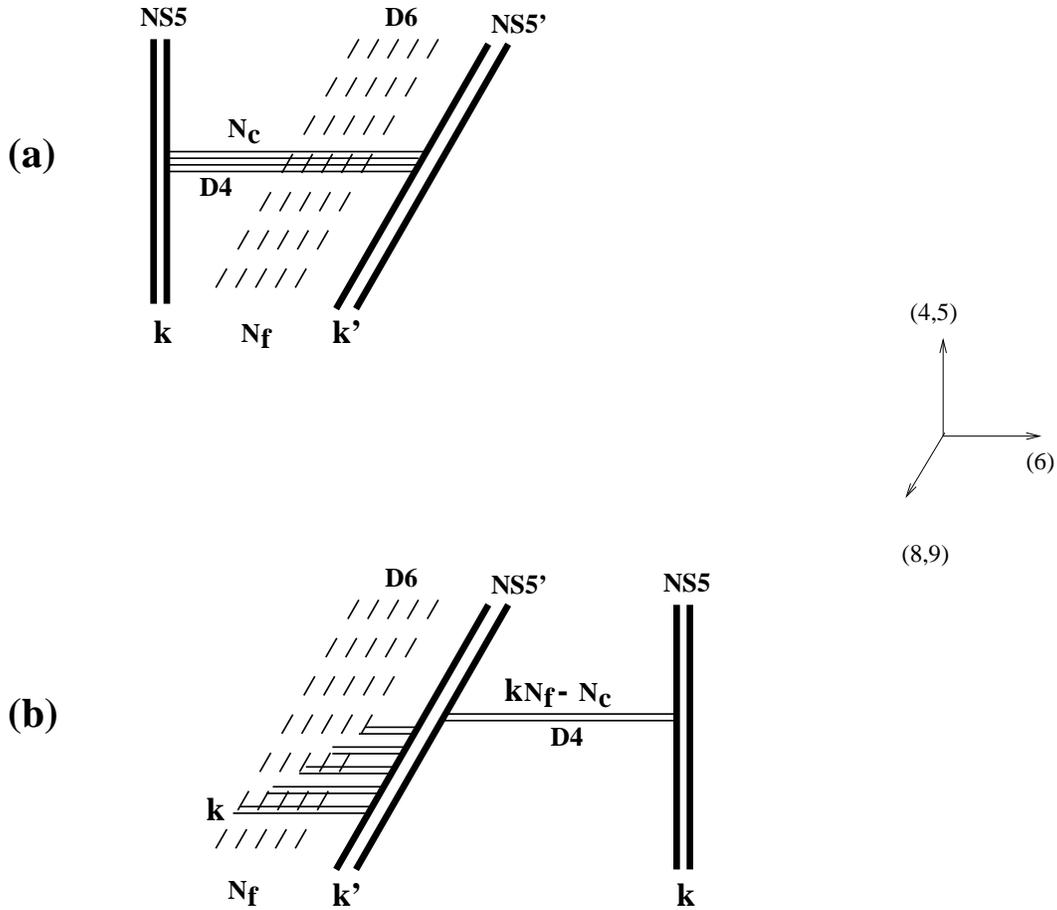}}
\vspace*{1cm}
\caption{(a) A theory with
$G=U(\nc)$, two adjoint superfields with
polynomial superpotentials and fundamentals. (b) The
Seiberg dual model with $\bar G=k\nf-\nc$.}
\label{thirtynine}
\end{figure}
\smallskip

Consider~\cite{EGK,EGKRS}, as an example, a configuration
of $k$ coincident $NS5$-branes connected by $N_c$
$D4$-branes to $k^\prime$ coincident \nsp-branes,
with $N_f$ $D6$-branes located between
the $NS5$ and \nsp-branes (see Fig.~\ref{thirtynine}(a)).
$N=1$ SQCD corresponds
to the case $k=k^\prime=1$.
The classical low energy theory on the fourbranes is in this
case $N=1$ SYM with gauge group $U(N_c)$, $N_f$
fundamental flavors $Q^i$, $\tilde Q_i$, and two
adjoint superfields $\Phi$, $\Phi^\prime$. The classical
superpotential is
\beq
W={s_0\over k+1}{\rm Tr} \Phi^{k+1}+{s_0^\prime\over k^\prime +1}
{\rm Tr}{\Phi^\prime}^{k^\prime+1}+
{\rm Tr}\left[\Phi,\Phi^\prime\right]^2
+\tilde Q_i\Phi^\prime Q^i
\label{wxx}
\eeq
$\Phi$ and $\Phi^\prime$ can be thought of as describing
fluctuations of the fourbranes in the $w$
and $v$ directions, respectively.
They are massless, but the superpotential (\ref{wxx}) implies
that there is a polynomial potential for the corresponding
fluctuations, allowing only infinitesimal deviations
from the vacuum at $\Phi=\Phi^\prime=0$. The couplings
$s_0$, $s_0^\prime$ should be thought of as very
large: $s_0, s_0^\prime\to\infty$. This can be deduced
\eg\ on the basis of the transformation properties
of (\ref{wxx}) under the $R$-symmetries $U(1)_{45}$
and $U(1)_{89}$.

To see that the configuration of
branes constructed above indeed
describes a gauge theory with the stated matter
content and, in particular, to see the
origin of the adjoint fields $\Phi$,
$\Phi^\prime$ one matches the
deformations of the brane
configuration with those of the gauge theory (\ref{wxx}).
Consider first the case $k^\prime=1$
for which $\Phi^\prime$ is massive and
can be integrated out. For large $s_0^\prime$
this amounts to putting $\Phi^\prime=0$ in
(\ref{wxx}).

\begin{figure}
\centerline{\epsfxsize=60mm\epsfbox{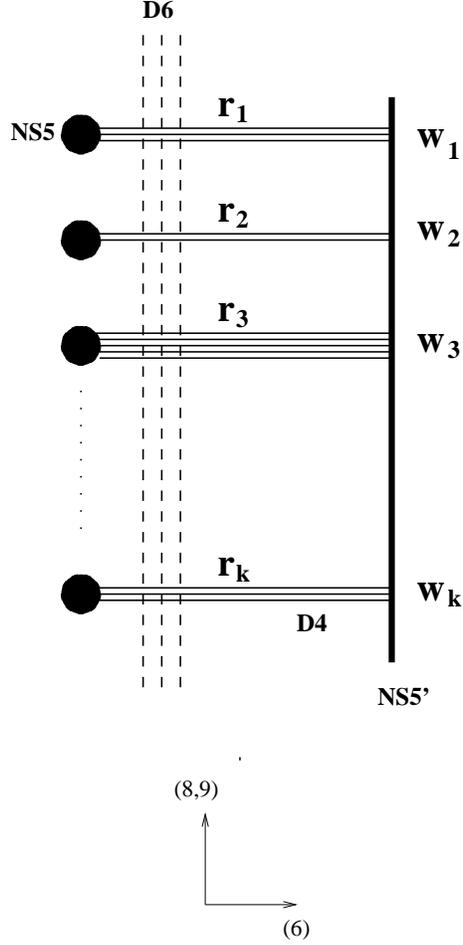}}
\vspace*{1cm}
\caption{Displacing the $k$ $NS5$-branes
in $w$ gives rise to a rich vacuum structure
labeled by the numbers of $D4$-branes attached
to the different $NS5$-branes, $r_j$.}
\label{fourty}
\end{figure}

An interesting deformation of the brane
configuration of Fig.~\ref{thirtynine}(a)
corresponds
to displacing the $k$ $NS5$-branes
in the $(x^8, x^9)$ plane
to $k$ different points $w_j$, $j=1, \cdots, k$.
Since the \nsp-brane is extended in $w$, this
gives rise to many possible
supersymmetric configurations,
labeled by sets of non-negative
integers $(r_1, \cdots, r_k)$ with $\sum_j r_j=N_c$,
which specify the number of fourbranes
stretched between the $j$'th $NS5$-brane
and the \nsp-brane (Fig.~\ref{fourty}).

When all the $\{w_j\}$ are distinct,
the low energy physics described by
the configuration of Fig.~\ref{fourty} 
corresponds to $k$ decoupled SQCD
theories with gauge groups $U(r_i)$
and $\nf$ flavors of quarks.
As we approach the origin of parameter
space, $w_j=0$, the full $U(\nc)$ gauge
group is restored.

To translate the above discussion to
the language of gauge theory on the fourbrane
one notes that displacing the $NS5$-branes
in $w$, the fourbranes attached to them are
displaced as well. The locations of
the fourbranes in $w$ correspond to
the expectation value of an adjoint
of $U(\nc)$, $\Phi$,
describing fluctuations of the
fourbranes in $(x^8, x^9)$. In a vacuum
labeled by $(r_1, \cdots, r_k)$
the expectation value of $\Phi$ is
$\langle\Phi\rangle={\rm diag} (w_1^{r_1},
\cdots, w_k^{r_k})$. Furthermore, in
the brane construction the $\{w_j\}$
correspond to locations of heavy objects
(the fivebranes) and thus they are expected
to appear as parameters rather than moduli
in the gauge theory description.

\smallskip

The gauge theory that achieves all of the
above is the one described by (\ref{wxx}).
Generic $\{w_j\}$ correspond to a
polynomial superpotential for $\Phi$,
\beq
W=\sum_{j=0}^k{s_j\over k+1-j} {\rm Tr} \Phi^{k+1-j}
\label{wadj}
\eeq
For generic $\{s_j\}$ the superpotential has $k$
distinct minima $\{w_j\}$ related to the parameters in
the superpotential via the relation
\beq
W^\prime(x)=\sum_{j=0}^k s_j x^{k-j}\equiv
s_0\prod_{j=1}^k(x-w_j)
\label{wprime}
\eeq
The integers
$(r_1, \cdots, r_k)$ introduced above are the
numbers of eigenvalues of the matrix $\Phi$
residing in the different minima of the
potential $V=|W^\prime(x)|^2$. Thus, the
set of $\{r_j\}$ and $\{w_j\}$ determines
the expectation value of the adjoint field $\Phi$,
in agreement with the brane picture. When all
$\{w_j\}$ are distinct the adjoint field is
massive, the gauge group is broken
\beq
U(N_c)\to U(r_1)\times U(r_2) \times\cdots\times U(r_k)
\label{brokk}
\eeq
and the theory splits in the infrared into $k$ decoupled
copies of SQCD with gauge groups $\{U(r_i)\}$ and
$N_f$ flavors of quarks. The brane description makes
this structure manifest.

For $k^\prime >1$,
the above discussion can be repeated for the
parameters corresponding to the locations of
the $k^\prime$ \nsp-branes in the $v$
plane.
These $k^\prime$ complex numbers can
be thought of as parametrizing the extrema of a
polynomial superpotential in $\Phi^\prime$ of order
$k^\prime+1$, in complete analogy to (\ref{wadj}, \ref{wprime}).
The only new element is that when we displace the
$k^\prime$ \nsp-branes in the $v$
directions, leaving the $N_f$ $D6$-branes fixed,
we make the quarks $Q$, $\tilde Q$ massive with
masses of order $\langle \Phi^\prime\rangle$. This
is the origin of the Yukawa coupling  in the
superpotential (the last term on the r.h.s. of
(\ref{wxx})).
One can also consider situations where both $NS5$ and
\nsp-branes are displaced in the $w$ and $v$
directions, respectively, and study the moduli space
of vacua of the theory (\ref{wxx}) for general $k$ and
$k^\prime$.

A Seiberg dual of the system
(\ref{wxx}) can be obtained
by interchanging the $NS5$ and
\nsp-branes in $x^6$ (Fig.~\ref{thirtynine}(b)).
For $k\ge1$, $k^\prime=1$ one derives this way~\cite{EGK}
the dual description (\ref{FT20})
obtained in field theory by~\cite{K,KS}.
For $k=1$, $k^\prime>1$ one finds
a perturbation of this duality
which was discussed in field theory by~\cite{ASY}.
For general $k$, $k^\prime$ the brane
construction predicts a new duality that was not
previously known in field theory~\cite{EGKRS}.

Quantum mechanically,
the type IIA configuration of 
Fig.~\ref{thirtynine}(a) 
is again replaced by a smooth $M5$-brane.
For $k=1$ and general $k'$ this
fivebrane was obtained by~\cite{BO}
by rotating an $N=2$ SUSY configuration.
It was shown that monopole and meson
expectation values computed from M-theory match the
results obtained in field theory via confining phase
superpotentials~\cite{EFGIR}.

More generally, one may consider chains of
stacks of coincident fivebranes, separated
in the $x^6$ direction as before, and rotated
with respect to each other. An example
that was discussed in~\cite{BH} and is depicted
in Fig.~\ref{thirtyeight}(b) involves an $NS5$-brane
connected to $k$ $NS5_L$-branes
on its left by $N_L$ fourbranes, and to
$k$ $NS5_R$-branes on its right
by $N_R$ fourbranes.
$n_L$ and $n_R$ sixbranes are located to the left and the
right of the $NS5$-brane, respectively.

For generic orientations $\theta_L$, $\theta_R$,
this brane configuration corresponds to an
$SU(N_L)\times SU(N_R)$ gauge theory with $n_L$ ($n_R$)
fundamental quarks of $SU(N_L)$ $(SU(N_R))$, and bifundamentals
$F,\tilde F$, with the classical superpotential
\beq
W\sim (F\tilde F)^{k+1}
\label{gen4}
\eeq
The dual configuration is obtained by interchanging the
$NS5_{\theta_L}$ and $NS5_{\theta_R}$-branes.
The magnetic gauge group is
$SU((k+1)(n_L+n_R)-n_L-N_R)\times
SU((k+1)(n_L+n_R)-n_R-N_L)$, in
agreement with field theory~\cite{ILS}.
The case $k=1$ was discussed after (\ref{gen3}).

Finally, note that configurations
containing coincident NS fivebranes
provide an example of a phenomenon
mentioned above: different deformations
of the configuration describe different
low energy theories. For example,
the configuration of
$k$ coincident $NS5$-branes connected
by fourbranes to an \nsp-brane
(Fig.~\ref{thirtynine}) can be deformed
in two different directions. Separating
the fivebranes in $w$ we find a theory
that is well described by the gauge
theory with an adjoint superfield $\Phi$
and a polynomial superpotential (\ref{wxx})
described in this subsection.
On the other hand, separating the
$NS5$-branes in $x^6$ leads to a
low energy description
in terms of a product group of the general
sort described in the previous subsection.
The two configurations are clearly
inequivalent and are continuously
connected through a transition which
involves crossing parallel $NS5$-branes.
We conclude that as in the other examples
mentioned above, a phase transition occurs
when the $NS5$-branes coincide. This
transition is apparently related to the
non-trivial CFT on $k>1$ fivebranes;
to understand the nature of the transition
a better understanding of the $(2,0)$ theory on
$k$ parallel fivebranes will probably be required.

\subsubsection{Orthogonal And Symplectic Gauge Groups
{}From Orientifolds}
\label{OSSS}

Just like for $N=2$ SUSY
configurations, many new
theories are obtained by adding
an orientifold plane.
In this subsection we list a few
examples of such theories and mention
some of their properties. We start
with an $O6$-plane and then move on
to an $O4$-plane.

\medskip
\noindent
{\bf 1. Orientifold Sixplane}

\begin{figure}
\centerline{\epsfxsize=100mm\epsfbox{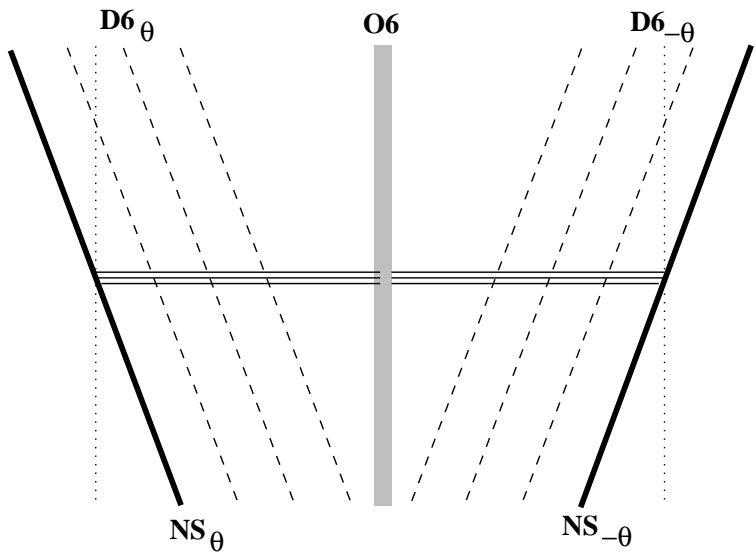}}
\vspace*{1cm}
\caption{$N=1$ SQCD with orthogonal and symplectic
groups can be realized using rotated NS fivebranes
near an orientifold sixplane.}
\label{fourtyone}
\end{figure}

The simplest configurations to consider
are again rotated $N=2$ ones. Starting
with the configuration of Fig.~\ref{seventeen} 
and rotating the $NS5$-brane by
a generic angle $\theta$, to an
$NS_\theta$-brane~\footnote{ The mirror
image of the NS-brane is necessarily
rotated by the angle $-\theta$
and becomes an $NS_{-\theta}$-brane.},
gives a mass to the adjoint chiral
multiplet. The resulting
configuration has $N=1$ SUSY
and light matter in the fundamental
representation of the gauge group,
which we recall is $SO(\nc)$ for
positive orientifold charge
and $Sp(\nc/2)$ for negative charge.
If we leave the $D6$-branes parallel
to the orientifold, we find a theory
with a quartic superpotential for the
quarks. To switch off the superpotential
we rotate the $D6$-branes as well until
they are parallel to the $NS_\theta$ brane
(and their mirrors are parallel to the mirror
$NS_{-\theta}$-brane; see Fig.~\ref{fourtyone}).

\smallskip

The moduli space of vacua can be studied
by combining the discussion of section
\ref{OSG} of orientifolds in theories
with $N=2$ SUSY, and the results of this
section on the reduction to $N=1$. We
leave the details to the reader.

One can also analyze Seiberg's duality
for these systems by exchanging fivebranes
and orientifolds in $x^6$.
In the absence of orientifolds a quick way
to find the dual is to exchange the branes
while requiring the conservation of the
linking number (\ref{BSB6}).
The linking number for fivebranes
near an orientifold sixplane
is also given by (\ref{BSB6}) with
the understanding that an $O6_\pm$
plane contributes to $L_{NS}$
like $\pm 2$ $D6$-branes.
Using this, it is not difficult to
show that the electric configuration
of Fig.~\ref{fourtyone} 
is connected by duality
to a magnetic one with gauge group
$SO(\bar\nc)$ with $\bar\nc=\nf-\nc+4$
for $O6_+$,
and $Sp(\bar\nc/2)$ with $\bar\nc=\nf-\nc-4$
for $O6_-$.

If the rotation angle $\theta$ above is tuned
to $\theta=\pi/2$, the fivebrane and its
mirror image turn into \nsp-branes and become
parallel to the orientifold.
The resulting SYM theory
on the fourbranes is an $SO(N_c)$ gauge theory
with $2N_f$ chiral superfields in the vector representation,
a chiral superfield $S$ in the symmetric representation
and $W=0$. Motions of $D4$-branes along
the \nsp-brane (in $w$) correspond to
expectation values 
of $S$ which parametrize an $N_c$ dimensional
moduli space along which
$SO(N_c)$ is generically completely broken.
Reversing the charge of the orientifold
replaces $SO(\nc)\rightarrow Sp(\nc/2)$
and $S\rightarrow A$, with $A$ a chiral
multiplet in the antisymmetric
tensor representation of $Sp(\nc/2)$.

The last two models are direct analogs
of the $SU(\nc)$ theory with an adjoint,
fundamentals and $W=0$, discussed in subsection
\ref{ROT}, with the symmetric of an orthogonal
group ($S$) or antisymmetric of a symplectic
group ($A$) playing the role of the adjoint
field $\Phi$. Just like there, one can
turn on a polynomial superpotential for
the (anti-) symmetric tensor. For example,
for the case of an $SO(\nc)$ gauge group
this is obtained by studying the following
configuration: $k$ coincident $NS5_{\theta}$-branes
to the left (in $x^6$) of an $O6_{+}$ plane,
connected to their mirror images (which are
$k$ coincident $NS5_{-\theta}$-branes)
by $N_c$ fourbranes, with $N_f$
sixbranes parallel to the
$NS5_{\theta}$-branes placed between the
fivebranes and the orientifold.
The SYM on the fourbranes is an
$SO(N_c)$ gauge theory with $2N_f$ vectors,
a symmetric flavor $S$ and
\beq
W\sim {\rm Tr} S^{k+1}
\label{gen5}
\eeq
The magnetic theory in the brane picture is obtained
by interchanging the $k$ fivebranes with their mirror
images while preserving the linking
number (\ref{BSB6})~\cite{EGKRS}.
The magnetic theory has $G_m=SO(k(2N_f+4)-N_c)$,
$2N_f$ magnetic quarks, magnetic mesons and an
appropriate superpotential, in agreement with
field theory~\cite{ILS}.

\medskip
\noindent
{\bf 2. Orientifold Fourplane}

As for the $N=2$ SUSY case discussed in section
\ref{OSG}, the situation is less well understood
than that for $O6$-planes, so we will be brief.

\begin{figure}
\centerline{\epsfxsize=100mm\epsfbox{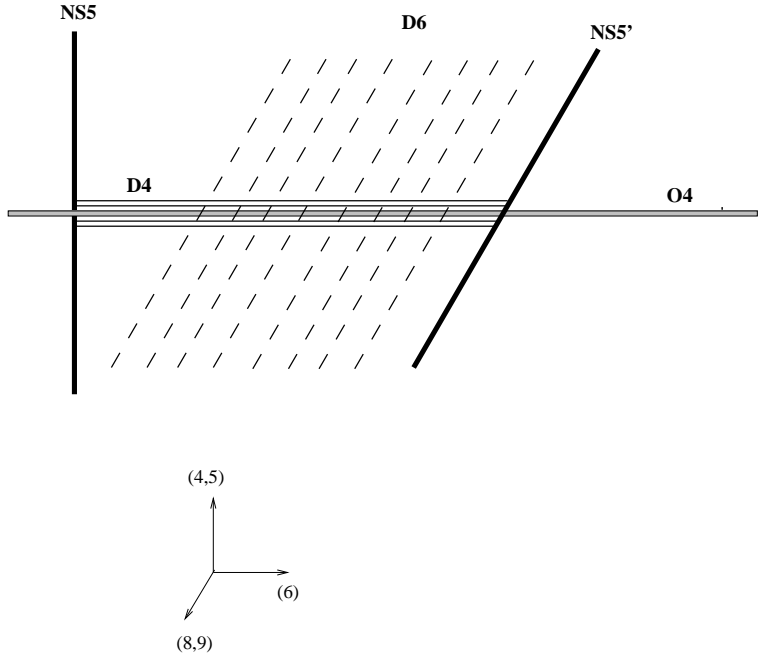}}
\vspace*{1cm}
\caption{Branes near an $O4$-plane provide
an alternative description of $N=1$ SYM
theories with orthogonal and symplectic
gauge groups.}
\label{fourtytwo}
\end{figure}

\begin{figure}
\centerline{\epsfxsize=120mm\epsfbox{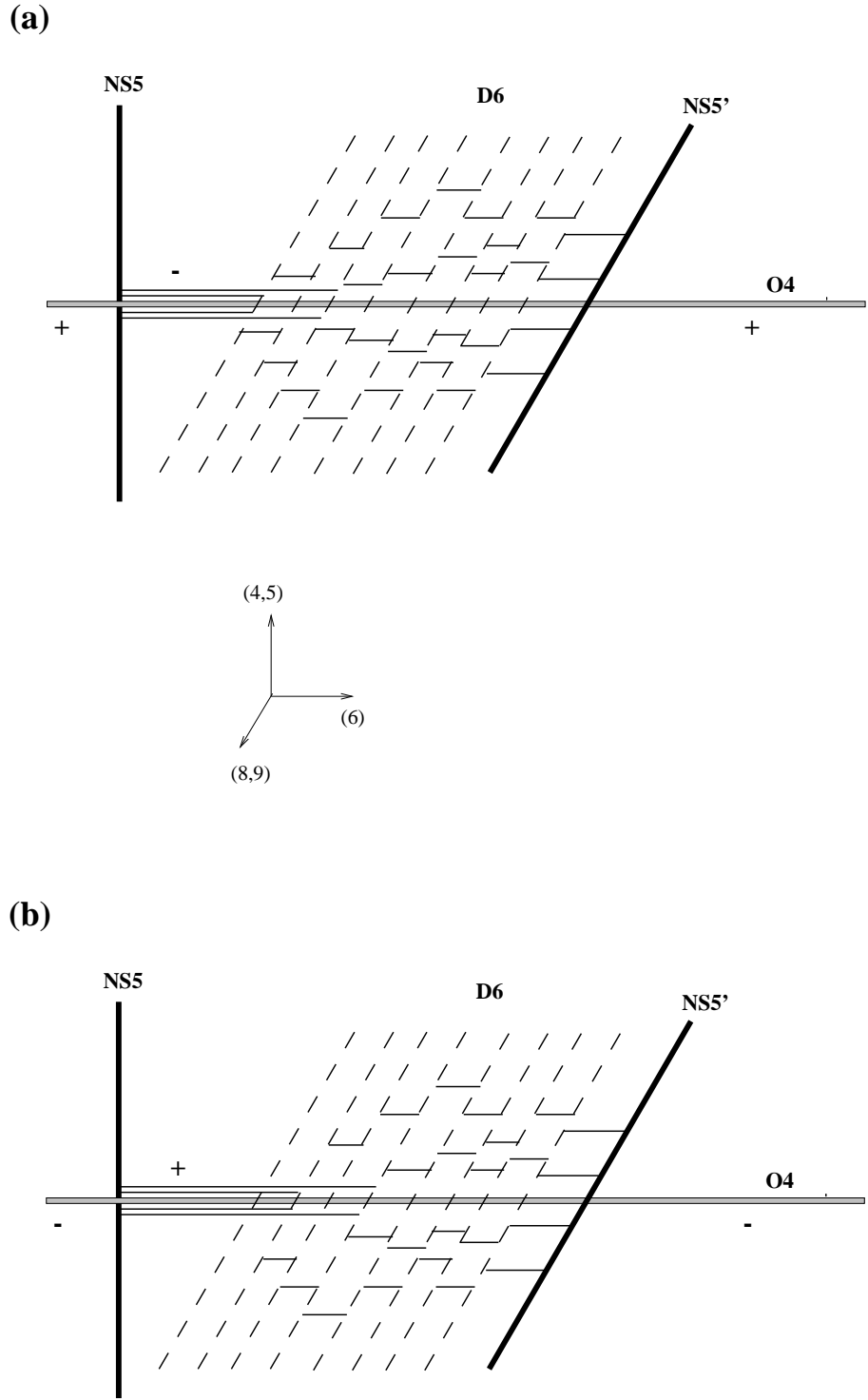}}
\vspace*{1cm}
\caption{The fully Higgsed branch of moduli space
corresponding to (a) $G=SO(4)$, and $\nf=8$ fundamentals;
(b) $G=Sp(2)$, and $\nf=8$ fundamentals.}
\label{fourtythreeb}
\end{figure}

The basic configuration that describes
$N=1$ SQCD with an orthogonal or symplectic gauge
group and matter in the fundamental representation
includes $\nc$ fourbranes stretched between an $NS5$
and an \nsp-brane, with $D6$-branes
between them; all objects are stuck
on an $O4$-plane, although the $D6$
and $D4$-branes could leave it in pairs
(see Fig.~\ref{fourtytwo}).
As discussed in section \ref{OSG},
the charge of the orientifold flips
sign each time it passes through an
NS fivebrane. If the charge between the
$NS5$ and \nsp-branes is positive the
gauge group is $Sp(\nc/2)$; negative
charge corresponds to $SO(\nc)$.
The moduli space of vacua can be analyzed
as in section \ref{OSG}; we will not
describe the details here. The fully
Higgsed branch for both signs of the
orientifold plane is illustrated in
Fig.~\ref{fourtythreeb}. 

\smallskip

To analyze the
smooth transition that corresponds in brane
dynamics to Seiberg's duality we need to
understand how to compute linking numbers
in the presence of the $O4$-plane.
Again, eq. (\ref{BSB6}) is essentially
correct as long as we take into account
the contributions of the $O4$-plane.
An $O4_\pm$ plane contributes like $\pm1$
$D4$-branes. Using this result, one can verify
that Seiberg's duality is reproduced in this
system~\cite{EGK,EJS,EGKRS}.

\begin{figure}
\centerline{\epsfxsize=120mm\epsfbox{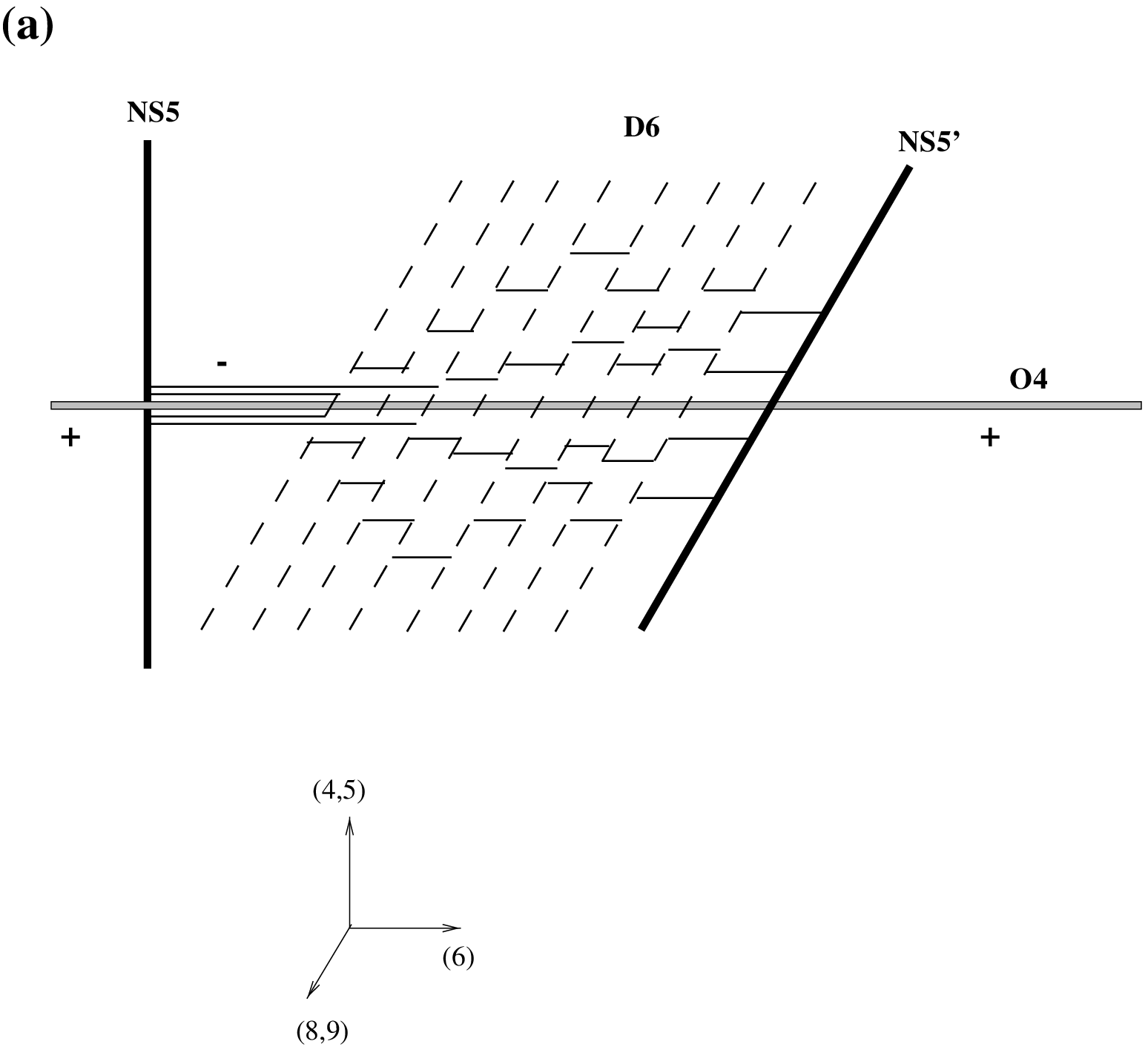}}
\label{fourtythreea}
\end{figure}
\smallskip

One can also study generalizations,
\eg\ replacing the single
$NS5$-brane by $2k+1$ fivebranes
leads to orthogonal and symplectic
gauge theories with a massless adjoint
field, with the polynomial superpotential
$W\sim{\rm Tr}\Phi^{2(k+1)}$.
Placing a sequence of $NS_\theta$-branes
with different $\theta=\theta_i$ along
the orientifold leads to theories with
product gauge groups of the form
$SO(k_1)\times Sp(k_2/2)\times SO(k_3)\times\cdots$~\cite{Tat}.

\subsubsection{Unitary Gauge Groups With Two-Index Tensors}
\label{UMTIT}

$N=1$ SYM theories with an $SU(N_c)$ gauge group
and chiral superfields in the (anti-) symmetric
tensor representation can be constructed
by starting with an $N=2$ configuration
of branes near an $O6$-plane --
mentioned at the end of subsection \ref{SGG} --
and applying to it all the operations described
in other cases. It is again sufficient
to describe the theory for one sign of the
orientifold charge (we will choose
the case of positive sign). To get the theory
corresponding to the other sign, one simply
replaces symmetric tensors by antisymmetric ones,
or vice-versa.

Consider an $NS5$-brane which
is stuck on an $O6_{+}$ plane.
An \nsp-brane located to the
left of the orientifold (in $x^6$)
is connected to the $NS5$-brane
by $N_c$ fourbranes. As usual,
we place $N_f$ sixbranes
between the fivebranes.
The theory on the fourbranes
is classically a $U(N_c)$ gauge theory
with $N_f$ fundamental flavors,
two symmetric flavors $S$, $\tilde S$
and $W=0$. The analysis of the brane moduli
space is easily seen to reproduce that of
the proposed gauge theory. In particular,
motions of the $D4$-branes in $w$, away
from the $NS5$-brane, parametrize the
$\nc$ dimensional moduli space of the theory
along which $S$, $\tilde S$ get expectation values
and the gauge group is typically completely
broken. When all the fourbranes meet at a point
in the $w$ plane
that is not the position of the $NS5$-brane,
an $SO(\nc)$ gauge symmetry is restored and one
recovers the theory with $G=SO(\nc)$, a symmetric
tensor, fundamentals and $W=0$, described in the
previous subsection. Turning on the FI D-term
in the $U(\nc)$ theory (or entering the
baryonic branch of the moduli space of the
$SU(\nc)$ one) has a similar effect.

Rotating the external \nsp-brane to
an $NS5_{\theta}$-brane, and at the
same time rotating the $D6$-branes
so that they are parallel to the
$NS5_{\theta}$-branes, leads to a
theory with the same matter content
as before, but now with a classical
superpotential for the symmetric tensor,
\beq
W\sim {1\over \mu}{\rm Tr}(S\tilde S)^2
\label{gen8}
\eeq
The previous case corresponds to $\mu=\infty$.
Rearranging the branes leads to a dual
configuration with gauge group
$G_m=SU(3N_f+4-N_c)$ and matter that
can be easily analyzed as above.
The resulting theory agrees with
the field theory analysis~\cite{ILS}.

If there are $k$ coincident $NS5_{\theta}$-branes
outside the orientifold, one finds
a similar theory but with (\ref{gen8})
replaced by
\beq
W\sim {\rm Tr}(S\tilde S)^{k+1}
\label{gen9}
\eeq
Brane rearrangement leads to
the Seiberg dual gauge group
$SU((2k+1)N_f+4k-N_c)$, again
in agreement with field theory.

\subsubsection{Chiral Models}
\label{CHIRAL}

Generic $N=1$ SYM theories
are chiral. Such theories are
interesting both because of their relevance
to phenomenology and because of their rich dynamics.
Their exploration using branes is in its infancy.
Here we discuss a few families of brane configurations
in the presence of orientifolds and orbifolds
leading to chiral models that appeared in the recent
literature.

\begin{figure}
\centerline{\epsfxsize=80mm\epsfbox{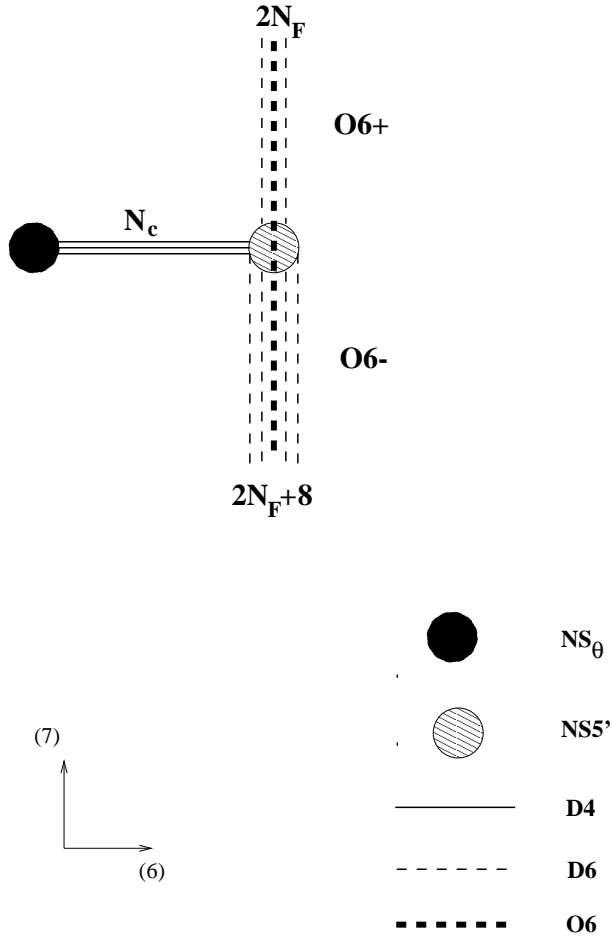}}
\vspace*{1cm}
\caption{A chiral brane configuration in which
an \nsp-brane is stuck at an $O6$-plane and is
connected to an $NS_\th$-brane outside of the orientifold.}
\label{fourtyfour}
\end{figure}

The first family was studied in~\cite{LLL98,BHKL,EGKT}.
The brane configuration shown in 
Fig.~\ref{fourtyfour}
involves an \nsp-brane which is
embedded in an $O6$-plane, say at
$x^7=0$. The \nsp-brane divides
the $O6$-plane into two disconnected
regions, corresponding to positive
and negative $x^7$.
As we saw before,
in this situation
the RR charge of the orientifold jumps,
from $+4$ to $-4$, as we cross the
\nsp-brane. The part of the orientifold
with negative charge (which we will
take to correspond to $x^7<0$) has furthermore
eight semi-infinite $D6$-branes
embedded in it.
The presence of these $D6$-branes
is required for charge conservation
or, equivalently, vanishing of the
six dimensional anomaly.

In addition to the eight semi-infinite
$D6$-branes, we can place on the
orientifold any number of parallel
infinite $D6$-branes extending all the
way from $x^7=-\infty$ to $x^7=\infty$.
We will denote the number of such
$D6$-branes by $2\nf$.

Then, an $NS_\th$ fivebrane~\footnote{An
$NS_\th$ fivebrane is an $NS5$-brane rotated
(\ref{BST38}) by the angle $\th$.} located
at a distance $L_6$ in the $x^6$ direction
from the \nsp-brane, but at the same value
of $x^7$, is connected to the
\nsp-brane by $\nc$ $D4$-branes stretched
in $x^6$.
$\nc$ must be even for consistency.
The mirror image
of the $NS_{\th}$-fivebrane, which is an
$NS_{-\th}$-fivebrane, is necessarily also
connected to the \nsp-brane.

\smallskip

We can also place any number of $D6$-branes
oriented at arbitrary angles $\theta_i$
(\ref{BST38}) between the $NS_\th$ fivebrane
and the orientifold (in $x^6$). We will
mainly discuss the case where such branes are
absent, but it is easy to incorporate them.

We will next describe the gauge theory
described by the above brane configuration.
Before studying
the general case we describe the
structure for $\theta=0$ (when
the external $NS_{\pm \th}$-fivebranes
are $NS5$-branes), and $\theta=\pi/2$
(when they are \nsp-branes). We will
only state the result, referring the
reader to~\cite{EGKT}
for further discussion and derivations.

\medskip
\noindent
{\bf 1. The Case $\theta=0$}

The theory on the $D4$-branes
has classical gauge group
$U(N_c)$ with a symmetric tensor $\tilde S$, an antisymmetric
tensor $A$, $2N_f+8$ quarks $Q$ in the fundamental
representation and $2N_f$ quarks $\tilde Q$
in the antifundamental representation. The
superpotential is:
\beq
W=Q\tilde S Q +\tilde Q A \tilde Q
\label{gen10}
\eeq
Fundamental chiral multiplets
of the gauge group come from
$4-6$ strings connecting the $D4$-branes
to $D6$-branes ending on the \nsp-brane
from below (in $x^7$), while antifundamentals
arise from $D6$-branes ending on the
\nsp-brane from above.
The global symmetry
of the system is determined by the gauge symmetry
on the $D6$-branes, $Sp(\nf)\times SO(2\nf+8)$.
The superpotential (\ref{gen10}) is the unique
one consistent with this symmetry.

The theory is chiral and potentially
anomalous as there are eight more fundamental
than antifundamental chiral multiplets.
The superpotential
(\ref{gen10}) implies that
the symmetric tensor $\tilde S$ is in fact
a symmetric bar (\ie\ a symmetric
tensor with two antifundamental indices).
Thus the total anomaly
$(2\nf+8)-2\nf+(\nc-4)-(\nc+4)$ vanishes,
as one would expect for a consistent
vacuum of string theory.

As a further check on the identification of the
brane configuration and the chiral gauge theory
one can analyze the moduli space of vacua
as a function of various parameters one can add to
the Lagrangian. An example is the FI D-term, which
from the brane point of view corresponds to
displacements in $x^7$ of the $NS5$-brane
relative to the \nsp-brane.
In the gauge theory, adding to the Lagrangian
a FI D-term
for the
$U(1)$ vectormultiplet ${\rm Tr} V$,
$r\int d^4\theta\, {\rm Tr} V$,
modifies the
D flatness vacuum conditions:
\beq
A A^\dagger -\tilde S \tilde S^\dagger
+Q Q^\dagger -\tilde Q \tilde Q^\dagger
=-r
\label{dtermflat}
\eeq
Setting the quarks $Q$, $\tilde Q$
to zero we see that
when $r$ is positive, $S$ gets an expectation
value which breaks $U(\nc)\to SO(\nc)$.
Due to the superpotential
(\ref{gen10}) the
$2\nf+8$ chiral multiplets $Q^i$
as well as $\tilde{S}$ become
massive and one if left with the
$N=2$ spectrum and interactions
for gauge group $SO(\nc)$,
with the antisymmetric
tensor $A$ playing the role of the adjoint
of $SO(\nc)$. All of this is
easily read off the brane configuration.
In particular, the fact that the $2\nf+8$
quarks $Q^i$ are massive is due to the fact
that the corresponding $4-6$ strings
have finite length (proportional to $r$).

Similarly, for negative $r$ (\ref{dtermflat})
implies that $A$ gets an expectation value,
breaking $U(\nc)$ to $Sp(\nc/2)$. The quarks
$\tilde Q$ get a mass and we end up with
an $N=2$ gauge theory with $G=Sp(\nc/2)$
and $2N_f+8$ light quarks.

\medskip
\noindent
{\bf 2. The Case $\theta=\pi/2$}

In this case the external fivebrane and its mirror
image are \nsp-branes.
In addition to the matter discussed for the previous
case there is now an adjoint field $\Phi$ parametrizing
fluctuations of the fourbranes in the $w$ plane.
The classical superpotential is:
\beq
W={\rm Tr} \tilde S\Phi A+Q\tilde S Q +\tilde Q A \tilde Q
\label{gen11}
\eeq
As a check on the gauge theory we can again
study the D-term perturbation corresponding
to relative displacement in $x^7$
of the \nsp-branes. For
positive $r$ we now find an $SO(\nc)$ gauge theory
with $2\nf$ fundamental chiral multiplets, a symmetric
tensor and vanishing superpotential. This can be understood
by analyzing the D-flatness conditions (\ref{dtermflat})
in the presence of the superpotential (\ref{gen11}).
As before, the symmetric tensor $\tilde S$ gets
an expectation value, which for unbroken $SO(\nc)$
must be proportional to the identity matrix.
The first term in the superpotential (\ref{gen11})
then gives rise to the mass term $W\sim \Phi A$.
Since $A$ is antisymmetric, this term gives a mass
to the antisymmetric part of $\Phi$ (as well as to
$A$). The symmetric part of $\Phi$ becomes the
symmetric tensor mentioned above. Clearly, it
does not couple to the $2\nf$ fundamental chiral multiplets.
In the brane description the fact that fluctuations
of the fourbranes in $w$ are described by a symmetric
tensor is a direct consequence of the action
of the orientifold projection~\cite{GPo}.

\medskip
\noindent
{\bf 3. The General Case}

For generic rotation
angle $\theta$ (\ref{BST38}) the adjoint field
$\Phi$ discussed in the previous point is
massive. Its mass $\mu(\th)$ varies smoothly
between zero at $\th=\pi/2$ and $\infty$ for
$\theta=0$. The superpotential describing this
system is
\beq
\label{eq:sup4}
W=Q\tilde{S}Q+\tilde QA\tilde{Q}+\Phi A\tilde{S}+
\mu(\theta)\Phi^2
\eeq
For non-zero $\mu$ we can integrate $\Phi$ out
and find the superpotential
\beq
\label{eq:sup5}
W=Q\tilde{S}Q+\tilde QA\tilde{Q}+\frac{1}{\mu(\theta)}
(A\tilde{S})^2
\eeq
for the remaining degrees of freedom. When $\theta\to 0$,
$\mu\to\infty$, and (\ref{eq:sup5}) approaches
(\ref{gen10}). When $\theta=\frac{\pi}{2}$, the mass
$\mu$ vanishes and it is inconsistent to integrate
$\Phi$ out.

For generic $\theta$ none of the NS fivebranes
in the configuration are parallel, and one can
interchange them to find a dual magnetic theory.
The magnetic gauge group one finds is
$U(2N_f+4-N_c)$. A careful field theory analysis
leads to the same conclusion~\cite{EGKT}.

\bigskip

A second family of chiral models
was studied by~\cite{LPTa}.
It has a gauge group which is a product of unitary
groups with matter in the bifundamental of
different pairs. It is obtained
from brane configurations in $Z_n$ orbifold backgrounds in the
following way.
Start with $nN_c$ fourbranes stretched
between two $NS5$-branes. The low energy theory on the
fourbranes is $N=2$ SYM with gauge group $G=SU(nN_c)$.
We now  mod out by the $Z_n$ symmetry
acting on $v$ and $w$ as
\beq
(v,w)\to (v\exp(2\pi i/n),w\exp(-2\pi i/n))
\label{gen13}
\eeq
Orbifolding breaks half of the supercharges and leads
to an $N=1$ SUSY gauge theory with gauge group
$SU(N_c)_1\times SU(N_c)_2\times SU(N_c)_3 \times
\cdots \times SU(N_c)_n$ with matter fields $F_i$, $i=1,...,n$,
in the bifundamental $({\bf N}_c, {\bf \bar N}_c)$ of
$SU(N_c)_i\times SU(N_c)_{i+1}$
(where $SU(N_c)_{n+1}\equiv SU(N_c)_1$).
This theory is chiral for $n> 2$.
The curve describing its moduli
space was obtained by~\cite{LPTa}.

An interesting variant of this theory is obtained by
stretching $nN_C$ $D4$-branes between an $NS5$-brane
and $n$ rotated fivebranes located at
\beq
v=\mu w, \quad v=\mu e^{4\pi i\over n}w,
\quad v=\mu e^{8\pi i\over n}w,
\quad \cdots  \quad v=\mu e^{4(n-1)\pi i\over n}w
\label{gen14}
\eeq
(of course these fivebranes are identified
after orbifolding by (\ref{gen13}), and so
really describe a single fivebrane
on $R^4/Z_n$).
After modding out by the $Z_n$ group (\ref{gen13})
one finds a gauge theory that is similar to that
described above, but with a tree level superpotential
\beq
W=\mu{\rm Tr}F_1\cdots F_n
\label{gen15}
\eeq
This superpotential lifts the moduli space,
in agreement with the brane picture where for $\mu\neq 0$
the fourbranes are stuck at $v=w=0$. Adding $nN_f$ sixbranes and
interchanging the $NS5$-brane with the $n$ rotated fivebranes
leads to a magnetic $SU(N_f-N_c)^n$ dual gauge theory.

A third class of models was also studied by~\cite{LPTb}.
It corresponds to webs of branes in the presence
of orientifold planes and orbifold fixed points.
As an example, 
one can start with the
configuration of Fig.~\ref{nineteen},
that was shown
in subsection \ref{OSG} to describe $SO$ or
$Sp$ theories with $N=2$ SUSY (depending
on the sign of the orientifold charge), and then
mods out by the $Z_3$ symmetry (\ref{gen13})
with $n=3$. The resulting gauge group is either
$SO(N+4)\times SU(N)$ or $Sp(M)\times
SU(2M+4)$, with matter in the
following representations. For the first
case (an $SO\times SU$ gauge group)
there is an antisymmetric tensor
field $A$ in the ${\bf
{1\over2}N(N-1)}$
of $SU(N)$ (it is a singlet under
$SO(N+4)$), a field $\bar Q$ in the
bifundamental $({\bf N+4, \overline{N}})$
of $SO(N+4)\times SU(N)$,
and fundamentals of both groups, whose number
is partly constrained by anomaly cancellation.
The second case (an $Sp(M)\times SU(2M+4)$
gauge group) is related to the first one
by replacing the antisymmetric tensor
$A$ by a symmetric one $S$ but is otherwise
similar.

The theories obtained this way have
a vanishing superpotential.
Rotating one of the $NS5$-branes in a way
compatible both with the $Z_2$ orientifold
projection and the $Z_3$ orbifold one, as in
(\ref{gen14}) (with $n=3$), leads to the appearance
of a superpotential of the form
$W\sim (\bar Q A \bar Q)^2$ or
$W\sim (\bar Q S \bar Q)^2$ for the two cases.
In the presence of a superpotential
one can study $N=1$ duality,
recovering results first obtained in field
theory by~\cite{ILS}.

\section{Three Dimensional Theories}
\label{D3}
So far in this review we have focused on
brane configurations realizing four dimensional
physics, however, it is clear that the framework
naturally describes field theory dynamics in
different dimensions. In the remainder of the
review we will study some brane configurations
describing field theories in two, three, five and
six dimensions.

We will see that these theories exhibit many
interesting phenomena which can be studied using
branes. Apart from the intrinsic interest in
strongly coupled dynamics of various field theories
in different dimensions and its realization in
string theory, the main reason for including
this discussion here is that it adds
to the ``big picture'' and, in particular,
emphasizes the generality and importance of:

\begin{enumerate}
\item {\em ``Universality:''} one of the interesting
features of the four dimensional analysis was the
fact that understanding a few local properties
of branes allowed the study of a wide variety
of models with various matter contents and numbers
of supersymmetries. These were obtained by combining
branes in different ways in a sort of flat space
``geometric engineering.''
We will in fact see that this universality
may allow one to understand~\footnote{This program
is not complete as of this writing; we will
mention some open problems in the discussion
section.} in a uniform way theories in different
dimensions. This should be contrasted with the
situation in field theory where the physics
is described in terms of perturbations of
weakly coupled fixed points, whose nature
depends strongly on the dimensionality.

\item{\em Hidden relations between different
theories:} in section \ref{D4N4}
we saw how viewing a brane configuration from
different points of view provides a relation
between gauge theories in different dimensions
with different amounts of SUSY. In that case,
a relation between four dimensional $N=4$ SYM
and two dimensional $N=(4,4)$ SYM provided an
explanation of Nahm's construction of multi-monopole
moduli space. In this and the next two sections
we will see that this is an example of a much
more general phenomenon.
\end{enumerate}

In this section we discuss three dimensional
field theories, starting with the case of
eight supercharges ($N=4$ SUSY), followed by four
supercharges ($N=2$ SUSY). In the next two sections
we discuss five, six and two dimensional theories.
The presentation is more condensed
than in the four dimensional case above. We only
explain the basic phenomena in the simplest examples,
referring the reader to the original papers for
more extensive discussion.

\subsection{$N=4$ SUSY}
\label{D3N4}
The main purpose of this subsection is to describe
the explanation using branes of two interesting
field theory phenomena:

\begin{enumerate}
\item The Coulomb branch of a three dimensional
$N=4$ SUSY gauge theory is often identical to
the moduli space of monopoles in a {\em different}
gauge theory.

\item Three dimensional $N=4$ SUSY gauge theories
often have ``mirror partners'' such that the Higgs branch
of one theory is the Coulomb branch of its mirror
partner and vice-versa.
\end{enumerate}

To study three dimensional gauge dynamics we
consider, following~\cite{HW},
configurations of $D3$-branes suspended
between $NS5$-branes in the presence of $D5$-branes.
Using eqs. (\ref{BST4}, \ref{BST14},
\ref{BST15}) it is not difficult to check that any combination
of two or more of the following objects

\beq
\begin{array}{ll}
\mbox{$NS5:$}&  \mbox{$(x^0, x^1, x^2, x^3, x^4, x^5)$}\\
\mbox{$D3:$}&  \mbox{$(x^0, x^1, x^2, x^6)$}\\
\mbox{$D5:$}&  \mbox{$(x^0, x^1, x^2, x^7, x^8, x^9)$}
\end{array}
\label{ndd3}
\eeq
in type IIB string theory
preserves eight of the thirty two supercharges,
and gives rise to an $N=4$ SUSY theory in the $1+2$
dimensional spacetime common to all branes
$(x^0, x^1, x^2)$. One can think of the branes (\ref{ndd3})
as obtained from (\ref{BSB1}) by performing T-duality
in $x^3$.

As a first example, consider a configuration containing
$k$ $D3$-branes stretched between two $NS5$-branes
separated by a distance $L_6$ in $x^6$. As discussed
at length above, the low energy theory on the threebranes
is a three dimensional $N=4$ SUSY  gauge theory
with gauge group $G=U(k)$ and no additional light
matter. The three dimensional gauge coupling is
\beq
{1\over g^2}={L_6\over \gs}
\label{gc3d}
\eeq
Motions of the $k$ threebranes along the
$NS5$-branes in $(x^3, x^4, x^5)$ together
with the duals of the $k$ photons, corresponding to
the Cartan subalgebra of $G$,
parametrize the $4k$ dimensional
Coulomb branch of the $N=4$ SUSY gauge theory
$\MM_k$.
Relative displacements of the two $NS5$-branes
in $(x^7, x^8, x^9)$ are interpreted as in
(\ref{BSB7}) as FI D-terms. Note that the theory
under consideration here can be
thought of as a dimensional reduction
of four dimensional $N=2$ SYM, or six dimensional
$N=1$ SYM, in both cases without hypermultiplets, and
thus much of the discussion of section
\ref{TBCS} applies to it. The $R$-symmetry, which is
$SU(2)_R$ in six dimensions and $SU(2)_R\times U(1)$
in four dimensions, is enhanced by the reduction to three
dimensions to $SU(2)_R\times SU(2)_{R^\prime}$, where
$SU(2)_{R^\prime}$ acts as an $SO(3)$ rotation
symmetry on $(x^3, x^4, x^5)$.

{}From the point of view of the
theory on the fivebranes, the $4k$
dimensional moduli space of BPS saturated deformations
of the brane configuration $\MM_k$ has a rather
different interpretation. The situation
is very similar to that discussed in section \ref{D4N4}.
The worldvolume theory on the fivebranes is a gauge theory
with $N=(1,1)$ SUSY (sixteen supercharges) and gauge group
$G=U(2)$, broken down to $U(1)\times U(1)$ by an expectation
value of one of the worldvolume scalars on the IIB fivebrane
discussed in section \ref{BST}. This expectation value
is proportional to the separation of the fivebranes $L_6$.

The massive $SU(2)$ gauge bosons correspond to D-strings
connecting the two $NS5$-branes. $D3$-branes stretched
between the $NS5$-branes are magnetic $SU(2)$ monopoles
charged with respect to the unbroken $U(1)\subset SU(2)$
(the other $U(1)$ corresponding to joint motion
of the fivebranes does not play a role and will be ignored
below). In compact space they are U-dual to D-strings
stretched between $D3$-branes, which were shown in
section \ref{D4N4} to describe monopoles in a broken
$SU(2)$ gauge theory. The $4k$ dimensional moduli
space of brane configurations $\MM_k$ is, from the point
of view of the fivebrane theory, the moduli space
of $k$ monopoles.

Thus we learn that the two spaces in question -- the Coulomb
branch of $N=4$ SUSY $U(k)$ gauge theory in $2+1$
dimensions and the moduli space of $k$ monopoles
in $SU(2)$ gauge theory broken to $U(1)$ -- are
closely related;
both are equivalent to the moduli space of SUSY
brane configurations of Fig.~\ref{nine}. 
The $U(1)\subset U(k)$ corresponding to the
center of mass of the $k$ monopole system
gives rise to a  trivial $R^3\times S^1$ part of
the moduli space. The space of vacua of the
remaining $SU(k)$ gauge theory corresponds to the
moduli space of centered monopoles.

A closer inspection reveals that the two spaces related
above are actually not identical; rather they
provide descriptions of the moduli space of brane
configurations in two different limits, which we
describe next.

As we saw before, to
study gauge physics using branes one needs
to consider a limit in which gravity and massive
string modes decouple. The relevant limit in this
case is
\beq
L_6,l_s, \gs\rightarrow 0
\label{limlsg}
\eeq
with $L_6/\gs$ (\ref{gc3d}) held fixed.

{}From the point of view of the theory
on the threebrane, the typical energy scale is
set by the Higgs expectation values
parametrizing the Coulomb branch.
These are related using eq. (\ref{BST5}) to the relative
displacements of the threebranes along the fivebranes
$\delta x$ by $\langle\phi\rangle\sim \delta x/l_s^2$.
Thus, the typical distances between the threebranes
in the gauge theory limit are
\beq
\delta x\sim \left({l_s^2 \gs\over L_6^2}\right) L_6
\label{septhree}
\eeq
To have a reliable $2+1$ dimensional picture one would
like to require $\delta x \ll L_6$, \ie:
\beq
Y\equiv\left({l_s\over L_6}\right)^2\gs\ll 1
\label{dxla}
\eeq
The parameter $Y$ is clearly arbitrary in the limit
(\ref{limlsg}) and when it satisfies (\ref{limlsg},
\ref{dxla})
the brane configuration is well described by
$2+1$ dimensional field theory.

The scale (\ref{septhree}) is natural from the point
of view of the fivebrane theory as well. The (massive)
charged $W$ bosons correspond to D-strings stretched
between the two $NS5$-branes. Their mass is:
\beq
M_W={L_6\over \gs l_s^2}
\label{mww}
\eeq
The magnetic monopoles are much heavier. The gauge
coupling of the $5+1$ dimensional fivebrane theory
is (\ref{BST16}) $g_{SYM}^2=l_s^2$; thus the effective
coupling in the $1+3$ dimensional spacetime $(x^0, x^3,
x^4, x^5)$ is
\beq
{1\over g^2}={V_{12}\over l_s^2}
\label{effcoup}
\eeq
where $V_{12}$ is the volume of the $(x^1, x^2)$
plane which is eventually taken to infinity.
Magnetic monopoles have mass
\beq
M_{\rm mon}\simeq {M_W\over g^2}={L_6V_{12}\over
\gs l_s^4}
\label{monmass}
\eeq
in agreement with their interpretation as $D3$-branes
stretched between the $NS5$-branes. Recall that the
{\em size} of a magnetic monopole is $\simeq M_W^{-1}$,
much larger than its Compton wavelength $M_{\rm mon}^{-1}$
for weak coupling.

Thus we see that the scale $\delta x$ (\ref{septhree})
is nothing but the Compton wavelength of a charged
$W$ boson (\ref{mww}) or, equivalently, the size
of a magnetic monopole. The five dimensional description
as the moduli space of monopoles is appropriate when
the scale of $SU(2)$ breaking $M_W$ (\ref{mww}) is much
smaller than the scale of Kaluza-Klein excitations
of the strings and threebranes stretched between fivebranes
$1/L_6$. Requiring $M_W\ll 1/L_6$ leads to the constraint
\beq
Y\gg 1
\label{dxlb}
\eeq
on the parameter $Y$ defined in eq. (\ref{dxla}). This is the
opposite limit from that in which the $2+1$ dimensional
picture is valid (\ref{dxla}).

We see that rather than being identical, the three and
five dimensional descriptions of the brane configuration
are appropriate in different limits. As $Y\to 0$
the description of the space of vacua as the Coulomb branch
of a three dimensional $SU(k)$ gauge theory becomes
better and better, while as $Y\to\infty$ the five dimensional
description becomes the appropriate one.

The dependence of the metric on $Y$ has not been
analyzed. Presumably,
as in other cases considered in previous sections, SUSY
ensures that the metric on $\MM_k$
does not depend on $Y$ and, therefore, its form for large
$Y$ (where it is interpreted as the metric on the moduli
space of $k$ monopoles) and for small $Y$ (where it is
thought of as the metric on the Coulomb branch of
a $d=2+1$, $N=4$ SUSY $SU(k)$ gauge theory)  must coincide.
It would be interesting to make this more precise.

The relation between monopoles and vacua of $2+1$
dimensional field theories can be generalized in many
directions. To study monopoles in higher
rank gauge theories we can consider, as in section
\ref{SGG}, chains of $NS5$-branes connected by $D3$-branes.
For example, the configuration of Fig.~\ref{twentytwo}
with the
$D4$-branes (\ref{BSB1}) replaced by $D3$-branes (\ref{ndd3})
and no $D5$-branes ($d_\beta=0$)
describes monopoles in broken $U(n+1)$ gauge theory.
The monopoles carry charges under the $n$ unbroken
$U(1)$'s in $SU(n+1)$. In a natural basis the magnetic
charge of the configuration is $(k_1, k_2-k_1, k_3-k_2,
\cdots, k_n-k_{n-1}, -k_n)$.

{}From the point of view of the threebranes the
configuration describes a $2+1$ gauge theory
with gauge group $G=U(k_1)\times U(k_2)\times
\cdots\times U(k_n)$ with
hypermultiplets transforming in the
bifundamental representation
of adjacent factors in the gauge
group,
$(k_\alpha,\bar k_{\alpha+1})$ of $U(k_\alpha)\times
U(k_{\alpha+1})$ ($\alpha=1,\cdots, n-1$).
The moduli space of vacua of this gauge theory
is identical to the space of monopoles in broken
$SU(n+1)$ gauge theory as discussed above.

The second field theory phenomenon that we would like
to understand using branes is mirror
symmetry~\cite{IS96}, which was
studied in string theory and M-theory
in~\cite{BHOO96,PZ,HW,Gom96,BHOOY}.
As pointed out by Hanany and Witten, this symmetry
is a manifestation of the S-duality of the underlying
$9+1$ dimensional IIB string theory.
We will next illustrate the general idea in an example.

$N=4$ supersymmetric gauge theory with $G_e=U(\nc)$
and $\nf$ hypermultiplets in the fundamental
representation of the gauge group can be studied
as in section \ref{D4N2}. We consider $\nc$
$D3$-branes stretched between two $NS5$-branes,
in the presence of $\nf$ $D5$-branes  placed between
the $NS5$-branes. All branes are oriented
as in (\ref{ndd3}).

This theory has, like its
four dimensional $N=2$ SUSY analog, a rich
phase structure of mixed Higgs-Coulomb phases
which can be studied classically as in section
\ref{D4N2}.

\begin{figure}
\centerline{\epsfxsize=120mm\epsfbox{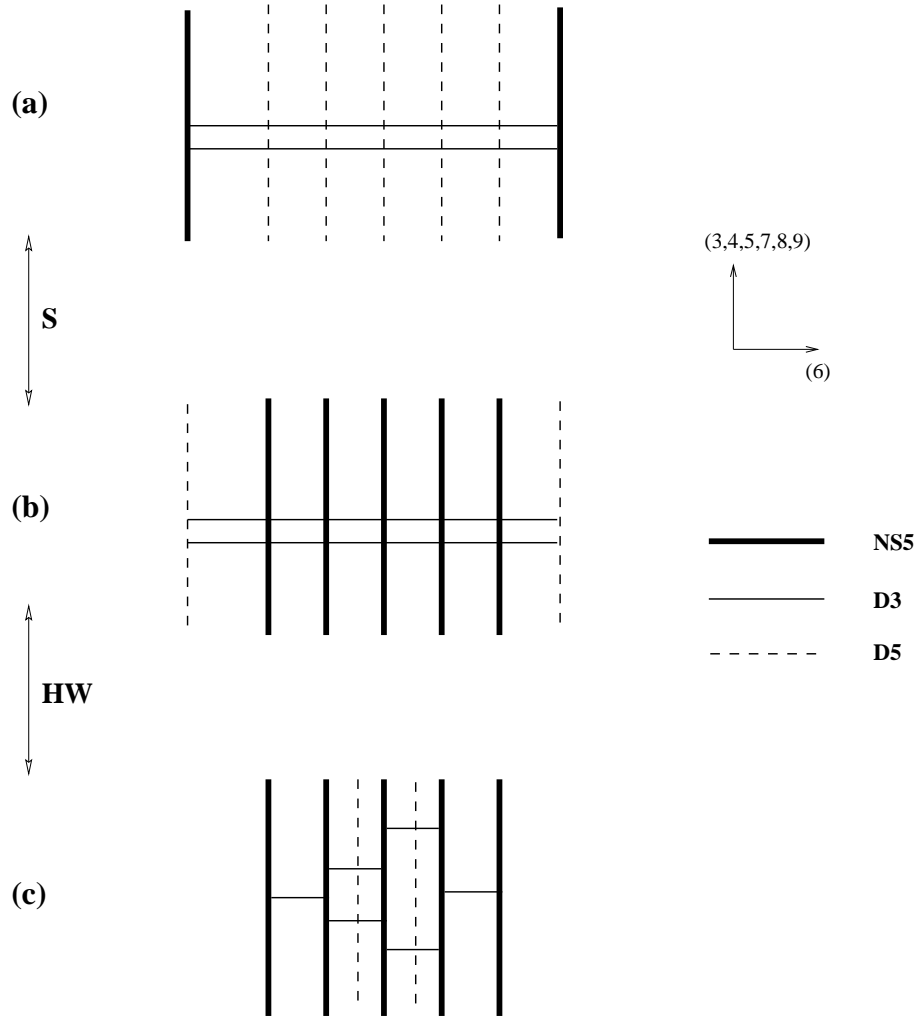}}
\vspace*{1cm}
\caption{S-duality of IIB string
theory implies mirror symmetry
of the three dimensional $N=4$ SYM
on $D3$-branes.}
\label{fourtyfive}
\end{figure}

Under S-duality~\footnote{S-duality
here corresponds to inverting the
coupling and exchanging $(x^3,x^4,x^5)
\leftrightarrow(x^7,x^8,x^9)$.}, 
the $NS5$-branes are exchanged
with the $D5$-branes while the $D3$-branes
remain invariant. The original configuration
is replaced by one where
$\nc$ $D3$-branes are stretched between
two $D5$-branes with $\nf$ $NS5$-branes
located between the two $D5$-branes (see
Fig.~\ref{fourtyfive}).

This is a configuration that is by now familar.
To exhibit the gauge group we have to reconnect
threebranes stretched between the two $D5$-branes
into pieces connecting $D5$-branes
and $NS5$-branes, and other pieces connecting different
$NS5$-branes. In doing that one has to take into account
the s-rule, which implies that the $\nc$ threebranes
attached to say the left $D5$-brane have to end on
different $NS5$-branes. Thus if we break the first
threebrane on the leftmost $NS5$-brane we have to break
the second on the second leftmost, etc. A similar
constraint has to be taken into account on the
right $D5$-brane.

\smallskip

The maximal gauge symmetry one can obtain depends on
$\nf$, $\nc$. The analysis is simplest for $\nf\geq2\nc$
and we will describe only this case here. The generalization
to $\nf<2\nc$ is simple.

Breaking the $\nc$ threebranes on the $NS5$-branes in
the most general way consistent with the s-rule leads
in this case to a magnetic gauge theory with
$G_m=U(1)\times U(2)\times \cdots U(\nc-1)\times U(\nc)^{\nf-2\nc+1}
\times U(\nc-1)\times \cdots U(2)\times U(1)$.
To see the hypermultiplets, it is convenient to
move the left $D5$-brane past the leftmost
$N_c$ $NS5$-branes (to which it is connected)
and similarly for the right $D5$-brane.

The hypermultiplets can now be read
off the brane configuration 
(Fig.~\ref{fourtyfive}(c)).
They transform under $G_m$ as:
$(1,2)\oplus(2,\bar3)\oplus\cdots\oplus(k-1,\bar k)
\oplus k \oplus (k,\bar k)\oplus\cdots\oplus (k,\bar k)
\oplus k \oplus (k, \overline{k-1})\oplus\cdots\oplus (3,2)
\oplus (2,1)$.

The original electric brane configuration at a certain
$\gs$ must, by S-duality, describe identical physics
to the magnetic one at $\tilde\gs=1/\gs$ (but the same
value of $l_{10}=l_s\gs^{1/4}$; see after eq. (\ref{BST30})).
In the low energy limit
$E\ll 1/l_{10}$ the electric configuration reduces to the
electric gauge theory with gauge group $G_e$, while
the magnetic one reduces to the magnetic gauge theory
with gauge group $G_m$ (and the specified matter).
Thus the two theories are clearly closely related.

However, as before, to go from one to the other, one
has to tune a parameter describing
the brane configuration to rather different
values. In the electric
theory the energy scale
we want to hold fixed as we take $l_{10}\to 0$ is set
by the three dimensional gauge coupling (\ref{gc3d}).
To ignore Kaluza-Klein excitations on the threebranes,
we must require $\gs/L_6\ll 1/L_6$, \ie\ $\gs\ll 1$.
Similarly, in the magnetic theory we must have $\tilde
\gs\ll 1$ to be able to ignore Kaluza-Klein excitations.

When $\gs$ is small, there exists an energy
scale for which all the complications of string theory
can be neglected and the running gauge coupling
of the electric gauge theory is still very small,
so that we are in the vicinity of
the UV fixed point of the gauge theory. The physics of
the brane configuration below this energy is well described
by gauge theory.
Similarly, the magnetic
gauge theory provides a good description of the low energy
behavior of the brane configuration
for large $\gs$ (or small $\tilde \gs$).

To relate the two gauge theories we must go to strong coupling
$\gs\simeq 1$. In this regime the brane configuration is still
described in the infrared by the same fixed point, but there is
no longer an energy range in which it is well approximated
by the full RG trajectory of either the electric or magnetic gauge
theories. The KK excitations of the threebranes modify
the RG flow at energies above $1/L_6\simeq g^2$, and one
would expect the correspondence between the two gauge theories
to break down.

In effect, the brane construction provides
a deformation of the RG trajectories of both the electric
and magnetic gauge theories that flow to the
same IR fixed point, but with different UV behavior. In particular,
the three dimensional dynamics is embedded in a four dimensional
setting; the fourth (compact) dimension decouples in the
extreme infrared but cannot be ignored at finite energies
or for large Higgs expectation values. Thus, the brane
construction  shows that the low energy
behavior of the electric
and magnetic theories is identical
in the strong coupling limit
$g\to\infty$; equivalently, it shows that the
infrared limits of the two models coincide for
Higgs expectation values $\langle\phi\rangle\ll g^2$.

In gauge theory, mirror symmetry
maps the Coulomb branch of the electric theory to the Higgs
branch of the magnetic one and vice-versa. It also exchanges
mass perturbations with FI D-terms. All this
is manifest in the brane construction.
As should be familiar by now, the Coulomb branch is
described by motions of threebranes
suspended between $NS5$-branes,
while the Higgs branch corresponds to motions of threebranes
stretched between $D5$-branes. Since under S-duality
$NS5$-branes are exchanged with $D5$-branes, the Coulomb
branch is exchanged with the Higgs branch. In the
example discussed in detail above, it is not
difficult to check that the
(complex) dimensions of the
electric Coulomb and Higgs branches are $\nc$ and
$2\nc(\nf-\nc)$, respectively, while in the magnetic
theory they are reversed.

Similarly, since masses correspond in the
brane language to relative displacements
of $D5$-branes
and FI D-terms  are described by relative
displacements of $NS5$-branes, S-duality
permutes the two.

\subsection{$N=2$ SUSY}
\label{D3N2}

In this subsection we will study three
dimensional $N=2$ SQCD.
We start with a summary of
field theoretic results followed by
the brane description.

\subsubsection{Field Theory}
\label{FTTD}

Consider $N=2$ SQCD with gauge group $G=U(\nc)$
and $\nf$ flavors of chiral multiplets
$Q^i$, $\tilde Q_i$ ($i=1,\cdots, \nf$)
in the fundamental representation of $G$.
This theory can be obtained from $N=1$
SQCD in four dimensions by dropping the
dependence of all fields on $x^3$. The vector
multiplet (\ref{RFTR1}) gives rise upon
reduction to three dimensions to a gauge field,
a real scalar field in the adjoint representation
of $G$, $X\equiv A_3$, and fermions. The chiral multiplets
$(Q,\tilde Q)$ reduce in an obvious way. The
four dimensional gauge interaction (\ref{RFTR7})
leads in three dimensions to a potential
for the bosonic components of $Q$, $\tilde Q$
\beq
V\sim \sum_i|XQ^i|^2+|X\tilde Q_i|^2
\label{potnt}
\eeq
More generally, one can compactify $x^3$ on a circle
of radius $R$ and interpolate smoothly between
four dimensional $(R\to\infty)$ and three dimensional
$(R\to0)$ physics. The three and four dimensional
gauge couplings are related (classically) by
$1/g_3^2=R/g_4^2$. Below, we describe the vacuum
structure of the theory as a function of $R$.

The classical theory has an $\nc$ complex dimensional
Coulomb branch.
At generic points in the classical Coulomb branch
the light degrees of freedom are the $\nc$
photons and scalars in the Cartan subalgebra
of $U(\nc)$, $A_\mu^{ii}$ and $X^{ii}$ $(i=1,\cdots,
\nc)$. Dualizing the photons
\beq
\partial_\mu\gamma^{ii}=\epsilon_{\mu\nu}^\lambda
\partial^\nu A^{ii}_\lambda
\label{dualph}
\eeq
gives rise to a second set of scalar fields
$\gamma^{ii}$ which together with $X^{ii}$
form $\nc$ complex chiral superfields whose
bosonic components are
\beq
\Phi^j=X^{jj}+i\gamma^{jj}
\label{pxg}
\eeq
The expectation values of $\Phi^j$
parametrize the classical Coulomb branch.

In the three dimensional limit $R=0$ the
scalars $\Phi^j$ live on a cylinder $R\times S^1$.
$X^{jj}$ are non-compact, while
$\gamma^{jj}$ live on a circle
of radius $g_3^2$. For finite $R$, $\Phi^j$
live on a torus since then ${\rm Re}\Phi^j$ also
live on a circle of radius $1/R$. In the
four dimensional limit $R\to\infty$, holding
the four dimensional gauge coupling fixed,
the torus shrinks to zero size and the Coulomb
branch disappears.
The quarks are generically massive on the Coulomb
branch (\ref{potnt}).

For $N_f \geq\nc$ the theory has a $2\nc\nf-\nc^2$
dimensional Higgs branch with completely broken
gauge symmetry (whose structure is the same
as in four dimensions and, in particular, independent
of $R$). There are also mixed Higgs-Coulomb
branches corresponding to partially broken gauge
symmetry.

In addition to
the complex mass terms, described by a quadratic
superpotential $W=m\tilde Q Q$, upon compactification
to three dimensions  one can write a
``real mass'' term for the quarks
\beq
\int d^4\theta Q^\dagger e^{m_r\theta\bar\theta}Q
\label{rmterm}
\eeq
We have encountered these real mass
terms before: in the previous section, where we
saw that the mass parameters in brane configurations
describing three dimensional $N=4$ SUSY gauge theories
have three components, and in (\ref{potnt}), which
describes a real mass term for the quarks proportional
to $\langle X\rangle$.

Quantum mechanically, the gauge coupling is a relevant
(=super-renormalizable) perturbation
and thus the theory is strongly
coupled in the infrared. Most or all of the Coulomb branch
and in some cases part of the Higgs branch
are typically lifted by strong coupling quantum effects.
We next turn to a brief description of these
effects as a function of $\nf$. A more detailed
discussion appears in~\cite{AHW,AHISS,BHO}.

\medskip
\noindent
{\bf 1.} ${\bf N_f=0}$

The dynamics of $U(1)\subset U(\nc)$ is trivial
in this case since there are no fields charged
under it. It gives a decoupled factor $R\times S^1$
in the quantum moduli space corresponding to
$(1/\nc)\sum\Phi^j$ (\ref{pxg}). The $SU(\nc)$
dynamics is non-trivial. A
non-perturbative superpotential is generated
by instantons, which in three dimensions are
the familiar monopoles of broken $SU(\nc)$
gauge theory.
By using the symmetries of the gauge theory
and the results of~\cite{VY,ADS} (for a review
see~\cite{IS95} and references therein) for
$\nc=2$ one can compute this superpotential
exactly.

Any point in the Coulomb branch can be
mapped by a Weyl transformation to the
Weyl chamber
$X^{11}\ge X^{22}\ge \cdots\ge X^{\nc\nc}$.
In this wedge the natural variables
are~\footnote{More precisely, the relation below
is valid far from the edges of the wedge
and for $R=0$; in general there are
corrections to the relation between
$Y_j$ and $\Phi^j$.}
\beq
Y_j=\exp\left({\Phi^j-\Phi^{j+1}\over
g_3^2}\right);\;\;\;j=1,\cdots, \nc-1
\label{yjg}
\eeq
and one can show that the exact superpotential
is
\beq
W=\sum_{j=1}^{\nc-1}{1\over Y_j}
\label{wyj}
\eeq
This theory has no stable vacuum. The superpotential
(\ref{wyj}) tends to push the moduli $\Phi^i$
away from each other to infinity.

When the radius of compactification
from four to three dimensions $R$ is
non-zero the analysis is modified.
The exact superpotential for finite
$R$ is
\beq
W=\sum_{j=1}^{\nc-1}{1\over Y_j}+
\eta\prod_{j=1}^{\nc-1}Y_j
\label{wstable}
\eeq
where $\eta$ is related to the four
dimensional QCD scale $\Lambda_4$:
\beq
\eta\sim \exp\left(-{1\over R g_3^2}\right)
\sim \exp\left(-{1\over g_4^2}\right)
\sim\Lambda_4^{3\nc-\nf}
\label{thfr}
\eeq
As $R\to0$ at fixed $g_3$, $\eta\to0$, while
in the four dimensional limit ($R\to\infty$)
$\eta$ turns into an appropriate power of
the QCD scale (\ref{thfr}).

The superpotential (\ref{wstable}) is stable.
Vacua satisfy $\partial_j W=0$ which leads to:
\beq
Z^{\nc}\eta^{\nc-1}=1;\;\;\;Z\equiv\prod_{i=1}^{\nc-1}
Y_i
\label{znc}
\eeq
Thus, for all $R\not=0$ there are $\nc$ vacua
corresponding to different phases of $Z$. As
$R\to0$, the vacua (\ref{znc}) recede to infinity.
Since $\eta$ remains finite as $R\to\infty$,
the $\nc$ solutions persist in the four dimensional
limit.

As we add light fundamentals $Q$, $\tilde Q$,
the vacuum structure becomes more intricate due to
the appearance of Higgs branches and additional
parameters such as real and complex masses, and
FI D-terms. As in four dimensions,
already classically there is a
difference between $\nf\ge \nc$ and $\nf<\nc$
massless fundamentals --
in the former case the gauge group can be broken
completely, while in the latter the maximal breaking
is $U(\nc)\to U(\nc-\nf)$. We next turn to the
quantum structure in the two cases.

\medskip
\noindent
{\bf 2.} ${\bf N_f\le \nc}$

The theory with $\nf=\nc$ and vanishing
real masses at finite $R$ is described
at low energies by a sigma model for
$N_c^2+2$ chiral superfields $V_\pm$,
$M^i_{\tilde i}$, with the superpotential
\beq
W=V_+V_-(\det M+\eta)
\label{wvpvm}
\eeq
$M$ should be thought of as representing
the meson field $M^i_{\tilde i}=Q^i
\tilde Q_{\tilde i}$, $V_\pm$ parametrize
the Coulomb branch and $\eta$ is given by
(\ref{thfr}). Note that most of the classical
$\nc$ complex dimensional
Coulomb branch is lifted in the quantum
theory; its only remnants are
$V_\pm$. The description (\ref{wvpvm})
is arrived at by a combination of
holomorphicity arguments, analysis of
low $\nc$ and inspired guesswork which we
will not review here (see~\cite{AHISS}).

Varying (\ref{wvpvm}) w.r.t. the fields $V_\pm$, $M$
gives rise to the equations of motion
\beq
V_\pm(\det M+\eta)=0; \;\; V_+V_-
(\det M) (M^{-1})_i^{\tilde i}=0
\label{eqq}
\eeq
Consider first the three dimensional case
($R=\eta=0$). There are three branches of
moduli space:
\begin{enumerate}
\item $V_+=V_-=0$; $M$ arbitrary.
\item $V_+V_-=0$; $M$ has rank
at most $N_c-1$.
\item $V_+$, $V_-$ arbitrary; $M$ has
rank at most $N_c-2$.
\end{enumerate}

\noindent
The first branch can be thought of as a Higgs
branch, while the last two are mixed Higgs-Coulomb
branches. The three branches meet on a complex
hyperplane on which the rank of $M$ is $N_c-2$
and $V_+=V_-=0$.

The understanding of the theory with $N_f=N_c$
allows us to study models with any $N_f\le N_c$
by adding masses to some of the flavors and
integrating them out.
Adding a complex quark mass term $W=-m M $ to (\ref{wvpvm}),
the following structure emerges. If the rank of
$m$ is one, one finds in the IR a theory
with $\nf-1=\nc-1$ massless flavors.
Integrating out the massive flavor
one finds a moduli space of vacua with
\beq
V_+V_-\det M=1
\label{ncmone}
\eeq
where $M$ is the $(N_f-1)\times
(N_f-1)$ matrix of classically massless mesons.
Equation (\ref{ncmone}) implies that
the classically separate Coulomb and Higgs branches
merge quantum mechanically into one smooth moduli space.
If the rank of $m$ is larger than one, one finds
a superpotential with a runaway behavior.
For example, if we add two non-vanishing masses:
\beq
W=V_+V_-\det M-m_1^1M^1_1-m_2^2M^2_2
\label{wmas}
\eeq
we find, after integrating out the massive mesons
$M_1^i$, $M_j^1$, $M_2^i$, $M_j^2$:
\beq
W=-{m_1^1m_2^2\over V_+V_-\det M}
\label{wintout}
\eeq
where, again, $M$ represents the $(N_f-2)^2$
classically massless mesons. Clearly, the
superpotential (\ref{wintout}) does not have a minimum
at finite values of the fields; there is no stable
vacuum.

For a mass matrix $m$ of rank $\nf$
\beq
W=V_+V_-\det M -m_i^{\tilde i} M^i_{\tilde i}
\label{mnf}
\eeq
we make contact with the case $\nf=0$. Integrating
out the massive meson fields $M$ gives rise
to the superpotential
\beq
W=-(\nf-1)\left({\det m\over V_+V_-}\right)^{1\over \nf-1}
\label{wpuregauge}
\eeq
This superpotential can be obtained from (\ref{wyj})
by integrating out the $Y_j$ keeping $Z$ (\ref{znc}),
and identifying it with $Z=V_+V_-$.

When the radius of the circle is not strictly zero
($\eta\not=0$), the analysis of (\ref{eqq}) changes somewhat.
There are now only two branches:
\begin{enumerate}
\item  $V_+=V_-=0$; $M$ arbitrary.
\item  $V_+V_-=0$; $\det M=-\eta$.
\end{enumerate}
\noindent
In particular, there is no analog of the
third branch of the three dimensional
problem. The two branches meet on a complex
hyperplane on which $\det M=-\eta$ and
$V_+=V_-=0$. The structure for all
$\eta\not=0$ agrees with the four dimensional
analysis of section \ref{D4N1}.

If we add to (\ref{wvpvm}) a complex mass term
$W=-mM$ with a mass matrix $m$ whose rank is
smaller than $N_f$, the vacuum is destabilized
(including the case of a mass matrix of rank
one where previously there was a stable vacuum).
If the rank of $m$ is $N_f$, so that
the low energy theory is pure $U(N_c)$
SYM, there are
$N_c(=N_f)$ isolated vacua which run off to
infinity as the radius of the circle $R$ goes
to zero (there is also a decoupled moduli space
for the $U(1)$ piece of the gauge group).
All this can be seen by adding to (\ref{wpuregauge})
the term proportional to $Z$,
$\eta V_+V_-=\eta Z$, and looking for extrema
of the superpotential
\beq
W=-(\nf-1) Z^{-{1\over \nf-1}}+\eta Z
\label{extyy}
\eeq

We next turn to the dependence of long distance
physics  on the real masses of the quarks. As we
saw before, real mass terms are described by D-terms
(\ref{rmterm}); therefore, the effective low energy
superpotential (\ref{wvpvm}) is independent of these
terms.

The effect of the real masses is to make some of
the low energy degrees of freedom in (\ref{wvpvm})
massive. To see that, consider weakly gauging the
(vector) $SU(\nf)$ flavor symmetry of (\ref{wvpvm}).
The real mass matrix $m_i^{\tilde i}$ corresponds
to the expectation values of the scalars in the $SU(\nf)$
vectormultiplet. A term analogous to (\ref{potnt})
in the Lagrangian of the $SU(\nf)$ theory will make some
of the components of $M$ massive. For a diagonal mass
matrix
\beq
\left( m_r\right)={\rm diag} \left(m_1,m_2,\cdots, m_{\nf}
\right)
\label{diagm}
\eeq
the off-diagonal components $M^i_{\tilde i}$
get a mass proportional to $|m_i-m_{\tilde i}|$.
When all the real masses $m_i$ are different, the low
energy limit is described by a sigma model for the
$\nf+2$ fields $V_+, V_-, M_1^1, M_2^2,\cdots,
M_{\nf}^{\nf}$ with the superpotential (compare
to (\ref{wvpvm})),
\beq
W=V_+V_-\left(M_1^1M_2^2\cdots M_{\nf}^{\nf}+\eta\right)
\label{wrealm}
\eeq
More generally, if
\beq
\left(m\right)={\rm diag} \left(m_1^{n_1},m_2^{n_2},
\cdots, m_k^{n_k}\right)
\label{mndeg}
\eeq
where $\{n_i\}$ are the degeneracies of
$m_i$ and $\sum_i n_i=\nf$, the low energy
limit includes $V_\pm$ and $k$ matrices
$M_i$ whose size is $n_i\times n_i$ $(i=1,
\cdots, k)$. The corresponding superpotential is
\beq
W=V_+V_-\left(\det M_1\det M_2\cdots\det M_k+\eta\right)
\label{wgenm}
\eeq
The moduli space corresponding to (\ref{wgenm})
is rather complicated in general. We have discussed
the case of equal real masses $k=1$ before. We will next
describe the other extreme case, $k=\nf$, leaving the
general analysis to the reader.

In the three dimensional limit $\eta\to0$,
(\ref{wrealm}) describes $\nf+2\choose2$
branches in each of which two of the $\nf+2$ fields
$\{V_\pm,M_i^i\}$ vanish. For non-zero $R$ (or $\eta$),
there are three branches:
\begin{enumerate}
\item $V_\pm=0$; $M^i_i$ arbitrary.
\item $V_+=0$, $V_-\not=0$; $\prod_{i=1}^{\nf}M_i^i=\eta$.
\item $V_-=0$, $V_+\not=0$; $\prod_{i=1}^{\nf}M_i^i=\eta$.
\end{enumerate}

\medskip
\noindent
{\bf 3.} ${\bf N_f>\nc}$

In this case, there is no (known) description of the
low energy physics in terms of a sigma model without
gauge fields. For vanishing real masses,
instanton corrections again lift all
but a two dimensional subspace of the Coulomb branch,
which can be parametrized, as before, by two chiral
superfields $V_\pm$. The Higgs branch is similar to
that of the four dimensional theory; it is parametrized
by the meson fields $M^i_{\tilde i}=Q^i\tilde Q_{\tilde i}$
subject to the classical
compositness constraints (such that only
$2\nf\nc-\nc^2$ of the $\nf^2$ components of $M$
are independent). An attempt to write a superpotential
for $V_+, V_-$ and $M^i_{\tilde i}$ using
holomorphicity and global symmetries leads in this case
to
\beq
W=\left( V_+V_-\det M\right)^{1\over \nf-\nc+1}
\label{wsing}
\eeq
which is singular at the origin, clearly indicating
that additional degrees of freedom that have been ignored
become massless there.

For non-vanishing real masses the phase structure
becomes quite intricate, and has not been analyzed
using gauge theory methods. We will see later using
brane techniques that when all the real masses
are different there are
$2\nf-\nc+2\choose \nc$, $\nc$-dimensional mixed
Higgs-Coulomb branches intersecting on lower
dimensional manifolds.

There are at least two other theories that have the same
infrared limit as $N=2$ SQCD. One is the ``mirror,''
which just like for $N=4$ SYM is easiest to
describe using branes~\cite{EGK,BHOY}; we will do this later.
The other is the 
``Seiberg dual''~\cite{Kar,Aha} which we will also
describe using branes below.
This is a gauge theory with $G_m=U(\nf-\nc)$,
$\nf$ flavors of magnetic quarks $q_i$, $\tilde q^{\tilde i}$,
and singlet fields $M^i_{\tilde i}$, $V_\pm$ which couple
to the magnetic gauge theory via the superpotential
\beq
W=M^i_{\tilde i}q_i\tilde q^{\tilde i} +V_+\tilde V_-
+V_-\tilde V_+
\label{dualsup}
\eeq
where $\tilde V_\pm$ are the effective fields parametrizing
the unlifted quantum Coulomb branch of the magnetic gauge
theory.

It should be emphasized that the three dimensional
``Seiberg duality'' is different from its four
dimensional analog in at least two respects. The
first is that it is not really a strong-weak coupling
duality. In three dimensions both the electric and magnetic
descriptions are strongly coupled (with the exception
of the case $\nf=\nc>1$ where the superpotential (\ref{wvpvm})
is dangerously irrelevant and, therefore, the sigma model is
weakly coupled in the IR, at least at the origin of moduli
space). Thus it is less useful as a tool to study
strong coupling dynamics.

The second is that the magnetic theory is not well
formulated throughout its RG trajectory. In particular,
the fields $\tilde V_\pm$ are effective low energy
degrees of freedom that emerge after taking into
account non-perturbative gauge dynamics. They are
ill defined in the high energy limit in which the
magnetic theory is (asymptotically) free.

Of course, the equivalence of the electric and magnetic
theories is expected to hold only in the IR, so this
is not necessarily a problem for the duality hypothesis.
However, it does seem to suggest that the
three dimensional Seiberg duality is a low
energy manifestation of a relation between
different theories that reduce to
low energy SYM in the IR but have quite different
high energy properties. We will next argue that the
relevant theories are theories on branes.

\subsubsection{Brane Theory}
\label{BDTD}

To study four dimensional $N=1$ SQCD
compactified on a circle of radius
$R$ using branes~\cite{EGKRS} we can simply
compactify the corresponding type IIA
configuration (\ie\ take $x^3\sim
x^3+2\pi R$ in Fig.~\ref{twentyfourb}). At large
$R$ we recover the results of section \ref{D4N1}.
For small $R$ it is convenient to perform a T-duality
on the $x^3$ circle; this transforms type IIA
to IIB and turns $D4$-branes wrapped around
$x^3$ into $D3$-branes at points on the dual
circle of radius
\beq
R_3={l_s^2\over R}
\label{rinv}
\eeq
The $NS5$ and \nsp-branes transform
to themselves, while the $D6$-branes
turn into $D5$-branes at points in
$(x^3,x^4,x^5,x^6)$. We will mostly
use the IIB language to describe the
physics.

\begin{figure}
\centerline{\epsfxsize=120mm\epsfbox{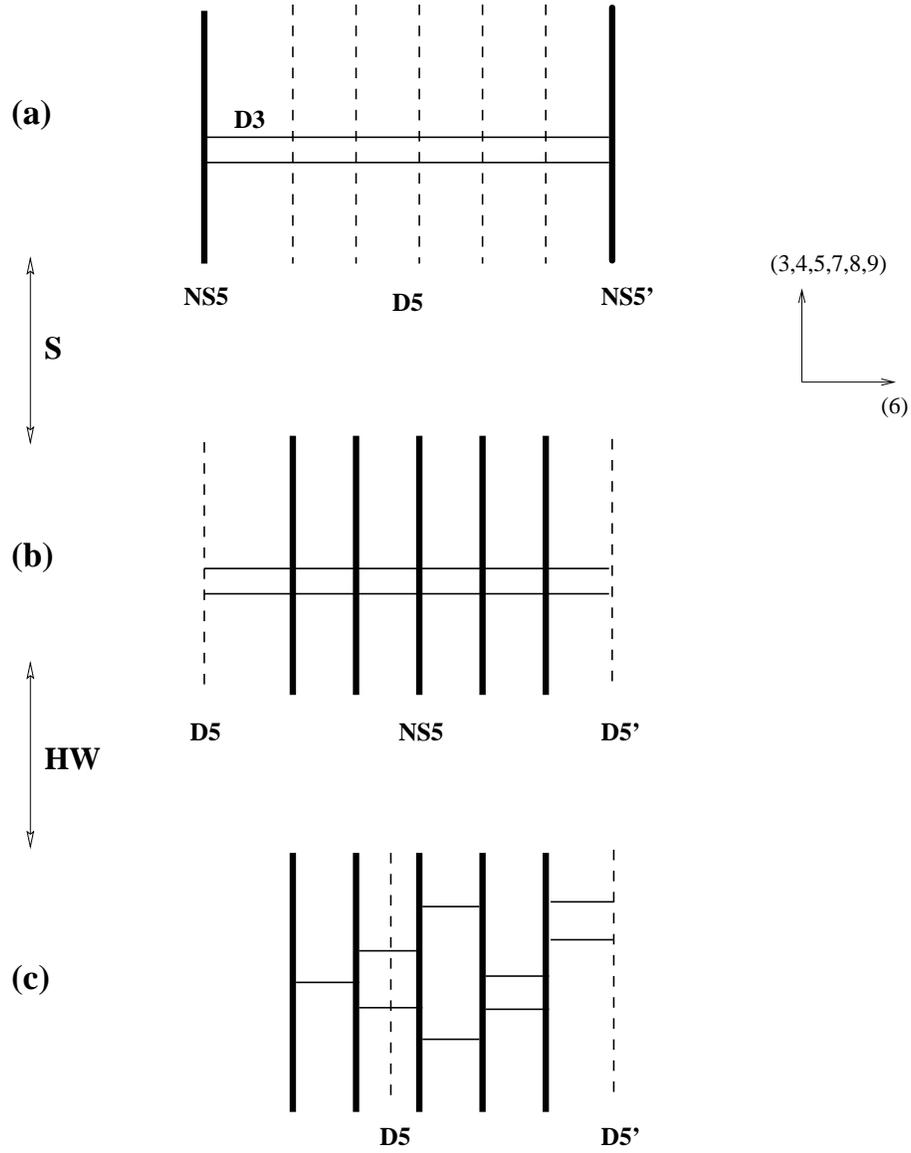}}
\vspace*{1cm}
\caption{The brane realization of mirror symmetry
in three dimensional $N=2$ SYM.}
\label{fourtysix}
\end{figure}
\smallskip

The IIB brane configuration corresponding
to three dimensional $N=2$ SQCD is depicted
in Fig.~\ref{fourtysix}. 
The classical analysis of deformations
and moduli mirrors closely the discussion of
section \ref{D4N1}. Compactification to three
dimensions gives rise to a new branch of moduli
space -- the Coulomb branch, and new parameters
in the Lagrangian -- the real masses. The former
correspond in the brane language to locations
in $x^3$ of $D3$-branes stretched between the $NS5$
and \nsp-branes. The latter are given by the positions
in $x^3$ of the $\nf$ $D5$-branes.

Note that due to (\ref{rinv}), as $R_3\to\infty$
we recover the three dimensional $N=2$ SQCD theory
with $\eta=0$ which was discussed in the previous
subsection, while the four dimensional limit
corresponds to $R_3\to0$. The three dimensional gauge
coupling is given in type IIB language by
(\ref{gc3d}), $1/g_3^2=L_6/\gs$. The four dimensional
gauge coupling is related to it by $1/g_3^2=R/g_4^2$
or, using (\ref{rinv}),
\beq
{1\over g_4^2}={L_6R_3\over \gs l_s^2}
\label{fourdcop}
\eeq
The gauge theory limit corresponds to
(\ref{limlsg}). To get a three dimensional
theory further requires $R_3\to\infty$ with $L_6/
\gs$ held fixed; the four dimensional limit is
$R_3\to 0$ with $g_4$ (\ref{fourdcop}) held fixed.
The instanton effects which give rise to the
term proportional to $\exp(-1/g_4^2)$ in
(\ref{wstable}) arise from Euclidean D-strings
which are stretched between the $NS5$ and \nsp-branes
and are wrapped around the $x^3$ circle.

As discussed in the previous subsection,
the infrared dynamics of the
gauge theory in question has at least
two alternative descriptions, the mirror
and the Seiberg dual. Both are easy to
understand using branes. To construct
the mirror, we apply an S-duality transformation
to the electric configuration; the result
is described in Fig.~\ref{fourtysix}(b).
The $NS5$ and \nsp-branes are exchanged with
$D5$ and $D5^\prime$ branes, while $D3$-branes
are invariant. The mirror brane configuration
is very similar to that found for $N=4$
SQCD in the section \ref{D3N4}. The only
difference is that one of the two $D5$-branes
has been rotated into a $D5^\prime$-brane.
$N=2$ mirror symmetry was suggested in~\cite{EGK}
and further investigated in~\cite{BHOY}.

To find the gauge symmetry of the mirror theory
we break the threebranes on the $NS5$-branes
in the most general way (Fig.~\ref{fourtysix}(b)). 
This leads to the gauge group
$U(1)\times U(2)\times \cdots\times U(\nc-1)\times
U(\nc)^{\nf-\nc}$.
There is still matter in bifundamental representations
of adjacent factors of the gauge group, and
since $D3$-branes stretched between $NS5$-branes
actually preserve $N=4$ SUSY, there are also
chiral multiplets transforming in the adjoint
of each factor. The $D3$-branes stretched between
the $D5^\prime$ brane and the closest $NS5$-brane
give rise as usual to $\nc$ scalars $M_\alpha$
which couple via a cubic superpotential to
$\nc$ fundamentals of the ``last'' $U(\nc)$
factor. The analysis of the magnetic
theory involves no new elements; details
appear in~\cite{BHOY}.

The Seiberg dual is obtained as usual
by exchanging fivebranes in $x^6$.
Since the two NS fivebranes are not
parallel, we expect the transition
to be smooth and the resulting
theory to be equivalent in the
infrared to the original one.
The magnetic brane configuration
one is led to is in fact very similar
to that obtained in the four dimensional
case, with $D4$ and $D6$-branes replaced
by $D3$ and $D5$-branes.
In particular, classically it seems
to correspond to a $G_m=U(\nf-\nc)$
gauge theory with magnetic quarks
and $\nf^2$ singlet mesons $M$
with the superpotential $W=Mq\tilde q$.
Comparing to the gauge theory result
(\ref{dualsup}), we seem to be missing
the two fields $V_\pm$ and their couplings to
the gauge degrees of freedom.

What saves the day is the fact that
the equivalence between the electric
and magnetic theories is expected
to be a quantum feature,
while our analysis of the magnetic brane
configuration so far was purely classical.
Thus our next task is to study the
quantum vacuum structure corresponding to
the electric and magnetic brane
configurations. We will first describe
the structure for the electric theory
and, in particular, reproduce the
gauge theory results of the previous
subsection. We will then turn to the
magnetic theory and show that in fact
the fields $V_\pm$ are secretly
present in the three dimensional  
analog of Fig.~\ref{twentyeight}
(but are not geometrical,
like the adjoint field with a polynomial
superpotential discussed in
section \ref{D4N1}). We will also
see evidence of the superpotential
(\ref{dualsup}).

The tool we will use
to analyze the vacuum structure is
the quantum brane interaction rules
described in section \ref{QEN13}.
As explained there, this allows one
to analyze the moduli space for
widely separated branes. The behavior
for branes that are close to each
other has to be addressed by other means.
Unfortunately, the M-theory analysis
is inapplicable for type IIB configurations
and there are at present no known alternatives.

Consider the electric configuration
of Fig.~\ref{fourtysix}(a)
with $\nf=0$. The $\nc$
threebranes stretched between the NS fivebranes
repel each other; therefore, the
classical $\nc$ dimensional
Coulomb branch is lifted.
The repulsive potential between
pairs of adjacent threebranes
can be thought of
in this case as due to Euclidean
D-strings stretched between the
$NS5$ and \nsp-branes and between
the $D3$-branes (as in Fig.~\ref{thirtyfour}).
They correspond to instantons in the
low energy three dimensional gauge theory.
Since there are two fermionic zero modes
in the presence of these instantons,
they lead to a superpotential on the classical
Coulomb branch.

In the three dimensional theory (with
$R_3=\infty$, or $\eta=0$ (\ref{thfr})),
the long range repulsion between threebranes
leads to runaway behavior, since there
is no stable vacuum with the threebranes
at finite distances; this is in agreement
with the gauge theory analysis of the
superpotential (\ref{wyj}).
For finite $R_3$ (or $\eta$) the threebranes
arrange around the $x^3$ circle at equal spacings,
maximizing the distances between them and
leading to an isolated vacuum. The
fact that there are $\nc$ vacua (\ref{znc})
has to do with the dual of the three dimensional
gauge field (\ref{dualph}, \ref{pxg}), and
is not expected to be seen geometrically
in the current setup. As $R_3\to\infty$ the
vacua run off to infinity, and we recover
the previous results.

In the presence of massless quarks, in the brane
description there are $D5$-branes  in the system
that can ``screen'' the interactions between the
threebranes. This screening can be seen directly
by studying Euclidean D-strings stretched between
$D3$-branes. If the worldsheet of such a D-string
intersects a $D5$-brane, two additional zero
modes appear and the contribution to the superprotential
vanishes.

For $1\le\nf\le\nc-2$ massless flavors
we saw before that the gauge theory
is unstable and exhibits a runaway
superpotential (given by (\ref{wintout}) for
$\nf=\nc-2$). In the brane picture we have
$N_c$ threebranes stretched
between $NS5$ and \nsp-branes, and $N_c-2$
$D5$-branes located at the same value of $x^3$
(we are restricting to the case of vanishing
real masses for now) between the $NS5$ and
\nsp-branes.

Due to the repulsion between unscreened threebranes
stretched between $NS5$ and \nsp-branes, $N_c-2$
of the $N_c$ threebranes must break on different
$D5$-branes. The s-rule implies that once this has
occurred, no additional threebranes attached to the
$NS5$-brane can break on these $D5$-branes.
We are left with two unbroken threebranes, one
on each side of the $D5$-branes (in $x^3$).
These threebranes repel each other, as well as
the pieces of the broken threebranes stretched
between the \nsp-brane and the $D5$-brane
closest to it. There is no screening in this
situation since all $N_c-2$ $D5$-branes are
connected to the $NS5$-brane; hence they can
be removed by moving them past the $NS5$-brane
in $x^6$, using the HW transition. The system is unstable,
and some or all of the threebranes mentioned above
must run away to infinity. 

This is in agreement with the gauge theory analysis of the
superpotential (\ref{wintout}). One can think of $V_\pm$ as the
positions in $x^3$ of the two threebranes stretched
between $NS5$ and \nsp-branes mentioned above (as usual,
together with the dual of the three dimensional gauge field).
The potential obtained from  (\ref{wintout}) indeed suggests a
repulsion between the different threebranes.

It is clear that the arguments above continue to hold
when the radius of the circle on which the threebranes
live is finite. While the two threebranes stretched between
the $NS5$ and \nsp-branes can no longer run away to
infinity in the $x^3$ direction, those connecting the
\nsp-brane to a $D5$-brane (representing components
of $M$) can, and there is still no stable vacuum.
This is in agreement with gauge theory; adding the term
$W=\eta V_+V_-$ to (\ref{wintout}) and integrating out $V_\pm$
leads to a superpotential of the form $W\sim (\det M)^{-1/2}$.

The above discussion can be repeated with the same
conclusions for all $1\le N_f\le N_c-2$.

For $N_f=N_c-1$ the gauge theory answer is different;
there is still no vacuum in the four dimensional case
$\eta\not=0$, while in three dimensions there is a
quantum moduli space of vacua with $V_+V_-\det M=1$.
In brane theory there are now $N_c-1$ $D5$-branes,
and the interaction
between the D threebranes stretched between $NS5$ and
\nsp-branes can be screened. Indeed, consider a situation
where $N_c-2$ of the $N_c$ threebranes stretched between
$NS5$ and \nsp-branes break on $D5$-branes.
This leaves two threebranes and one $D5$-brane that is
not connected to the $NS5$-brane. If $R_3=\infty$
(\ie\ $\eta=0$), the single $D5$-brane
can screen the repulsion between the two threebranes.
If the threebrane is at $x^3=0$, then using the rules
of subsection \ref{QEN13} we deduce that
any configuration where one of the threebranes is at
$x^3>0$ while the other is at $x^3<0$ is stable.
The locations in $x^3$ of the two threebranes give
the two moduli $V_\pm$.
Thus, the brane picture predicts correctly the existence
of the quantum moduli space and its dimension. The
precise shape of the moduli space (the relation
$V_+V_-\det M=1$) is a
feature of nearby branes
and thus is expected to be more difficult
to see; nevertheless, it is clear
that due to the repulsion there is no vacuum when
either $V_+$ or $V_-$ vanish. 

If the radius of the fourth dimension $R$ is not zero,
there is a qualitative change in the
physics. Since $R_3$ is now finite, the two threebranes
stretched between $NS5$ and \nsp-branes are no longer
screened by the $D5$-brane -- they interact through the
other side of the circle. Thus one of them has to
break on the remaining $D5$-brane, and one remains
unbroken because of the s-rule. The repulsion between
that threebrane and the threebranes stretched between
the \nsp-brane and a $D5$-brane which is no longer
screened leads to vacuum destabilization, in agreement
with the gauge theory analysis. 

For $N_f=N_c$ (and vanishing real masses) the
brane theory analysis is similar to
the previous cases, and the conclusions are again in
agreement with gauge theory. For $R_3=\infty$ one finds
three phases corresponding to a pure Higgs phase in which
there are no threebranes stretched between $NS5$ and
\nsp-branes, and two mixed Higgs-Coulomb phases in which
there are one or two threebranes stretched between the $NS5$
and \nsp-branes; the locations of the threebranes in
$x^3$ are parametrized by $V_\pm$. When there are two
unbroken threebranes, they must be separated in $x^3$ by
the $D5$-branes, which provide the necessary screening.

For finite $R_3$, the structure is similar, except
for the absence of the branch with two unbroken
threebranes, which is lifted by the same mechanism
to that described in the case $N_f=N_c-1$ above.
$\nf>\nc$ works in the same way as in the four
dimensional case described earlier.

So far we discussed the electric theory with
vanishing (more generally equal) real masses
for the quarks. Turning on real masses gives rise
to a rich phase structure of mixed Higgs-Coulomb
branches. We have seen an example in the theory
with $\nf=\nc$. The three dimensional
theory with $\eta=0$ and vanishing
real masses, which is equivalent in the infrared
to the sigma model
(\ref{wvpvm}), has three branches described
after equation (\ref{eqq}). When all the real
masses are different, the corresponding
sigma model (\ref{wrealm}) has $(\nf+1)
(\nf+2)/2$ $\nf$-dimensional branches
intersecting on lower dimensional manifolds.
For $\nf>\nc$ the problem has not been analyzed
in gauge theory.

Branes provide a simple way of studying
the phase structure. As an example we will
describe it for the case of arbitrary $\nf\ge
\nc$ with all $\nf$ real masses different.
The generalization to cases where some of
the real masses coincide is straightforward.

The brane configuration includes in this
case $\nf$ $D5$-branes at different values
of $x^3$. The $\nc$ $D3$-branes can either
stretch between the $NS5$ and \nsp-branes
or split into two components on $D5$-branes.
Different branches of moduli space
correspond to different ways of
distributing $D3$-branes between the
two options, in such a way that there are
no unscreened interactions. 

Stability implies
that any two $D3$-branes stretched between
the $NS5$ and \nsp-branes must be separated
(in $x^3$) by a $D5$-brane that screens the
repulsive interaction between them.
Similarly, a $D3$-brane stretched between
the $NS5$ and \nsp-branes and a second
one that is broken into two components
on a $D5$-brane must be separated by
such a $D5$-brane.

To describe
the different branches of the
quantum moduli space of vacua
we have to place
the $\nc$ $D3$-branes 
in such a way that
there are no repulsive interactions.
Each threebrane can either be placed
between two $D5$-branes, or on top of
one. There are therefore
$\nf+(\nf+1)=2\nf+1$ possible locations
for $D3$-branes corresponding to the
$\nf$ $D5$-branes and the $\nf+1$
spaces between and around them.
The quantum interactions between branes
mean that one cannot place two threebranes
at adjacent locations. The number of branches
is, therefore, the number of ways
to distribute $\nc$ identical objects between
$2\nf+1$ slots, with at most one object per slot
and no two objects sitting in adjacent slots.
It is not difficult to show that this number
is
\beq
{n\choose k};\;\;n\equiv2\nf-\nc+2,\; k\equiv\nc
\label{numbran}
\eeq

We next turn to the magnetic brane
configuration. Naively
it describes a $U(\bar \nc)$ gauge
theory with $\nf$ fundamentals
$q$, $\tilde q$,
a magnetic meson $M$ and the
standard superpotential
\beq
W=M\tilde q q
\label{standsup}
\eeq
To see whether this is in fact correct
we will study the resulting theory
for the special case $\bar\nc=\nf$
and compare the vacuum structure
of the gauge theory to that of the
brane configuration, which is described
by a three dimensional analog of
Fig.~\ref{twentyeight}.
We will furthermore discuss only the
three dimensional limit $\eta=0$.
Classically, the two definitely
agree. The theory has an $\nf^2+\nf$
dimensional moduli space of vacua;
in the brane language it corresponds to
independent motions of the color
branes along the NS fivebranes and
to breaking of flavor threebranes
on different $D5$-branes. In gauge
theory it is parametrized by expectation
values of $M$ and the $\Phi^j$
(\ref{pxg}). Quantum mechanically,
one finds a discrepancy, which we describe
next.

Before turning on the Yukawa
superpotential (\ref{standsup}),
the low energy dynamics of the gauge theory in
question is described by the sigma model
(\ref{wvpvm}) for the fields $\tilde M\equiv
\tilde q q$ and $\tilde V_\pm$ which parametrize
the potentially unlifted Coulomb branch of the
theory. Coupling the ``magnetic quarks''
$q$, $\tilde q$ to the singlet meson $M$
leads to a low energy sigma model
with the superpotential
\beq
W=\tilde V_+\tilde V_-\det\tilde M+
M\tilde M
\label{effss}
\eeq
Varying with respect to
$M$ sets $\tilde M=0$. Therefore,
we conclude that the quantum gauge
theory in question has a two complex
dimensional moduli space of vacua
parametrized by arbitrary expectation
values of the fields $\tilde V_\pm$.

On the other hand the brane configuration
has a unique vacuum at the origin where
all $\bar\nc=\nf$ threebranes are aligned
and can be thought of as stretching between
the $NS5$-brane and the $\nf$ $D5$-branes.
This is the only stable configuration,
taking into account the repulsive
interactions between color threebranes
and the attractive interactions between
color and flavor threebranes.

We conclude that the gauge theory
leading to the low energy sigma model
(\ref{effss}) cannot provide a full
description of the physics of the
brane configuration of Fig.~\ref{twentyeight}. 
In~\cite{EGKRS} it is proposed that the
magnetic brane configuration is in fact
described by the above gauge theory, but
there are two more fields $V_\pm$ that
are singlets under the $U(\bar\nc)$
gauge group and contribute the term
\beq
W_V=V_+\tilde V_-+V_-\tilde V_+
\label{wvvv}
\eeq
to the low energy superpotential.
Combining (\ref{effss}) with
(\ref{wvvv}) clearly gives the
right quantum vacuum structure
for $\bar\nc=\nf$, and also more generally.
Note that the term (\ref{wvvv}) is just
what has been seen to be needed in gauge
theory to generalize Seiberg's duality
to three dimensions (\ref{dualsup})!
In particular, it can be used to make
sense of the dual theory, which as
we discussed in the previous subsection
is not really well defined as a local
quantum field theory.
We see that the high energy theory
that underlies (\ref{dualsup}) is
best thought of as the theory on
the web of branes described by 
the three dimensional Fig.~\ref{twentyeight}. 

The fields $V_\pm$ and their interactions
(\ref{wvvv}) are not seen geometrically
in the brane configuration. 
This makes it more difficult in general
to compare the vacuum structure of three
dimensional Seiberg duals.
Nevertheless, in all cases that have
been checked, no disagreement has
been found, supporting the proposed
duality. Some tests of the equivalence
of the theories with vanishing real masses
appear in~\cite{Aha}.
We have further checked the
other extreme case of $\nf$
{\em different} real masses
in a few examples and find
agreement. For $\nf=\nc$
the magnetic theory can be shown
to reduce to the sigma model
(\ref{wvpvm}), or when all the
real masses are different,
(\ref{wrealm}). As we  have seen
above, the magnetic moduli space has
in this case $(\nf+2)(\nf+1)/2$
branches, in agreement with the
electric theory
(\ref{numbran}).

For $\nf=\nc+1$, the magnetic theory
reduces in the infrared to a $U(1)$
gauge theory with $\nf$ flavors and
the superpotential (\ref{dualsup}).
By using the results of~\cite{AHISS}
one can check that the phase structure
of the magnetic theory is again
in agreement with (\ref{numbran}).
It would be interesting to check
agreement for arbitrary $\nf>\nc$.

\section{Two Dimensional Theories}
\label{D2}

\subsection{Field Theory Results}
\label{D2FTR}

Two dimensional gauge theories with $N=(4,4)$ supersymmetry
can be obtained by the dimensional reduction of six dimensional
$N=(1,0)$ theories (or four dimensional $N=2$ theories).
The $1+5$ dimensional Lorentz symmetry is
broken to $SO(1,1)\times Spin(4)$ and the latter combines
with the $R$-symmetry to a $Spin(4)\times SU(2)_R$ global
symmetry group.

As in four dimensional 
$N=2$ theories, two dimensional $(4,4)$ gauge
theories have two massless representations: a
hypermultiplet and a vectormultiplet (also called
a twisted multiplet).
In terms of an $N=(2,2)$ superalgebra the
hypermultiplets decompose into two chiral multiplets
(see for example~\cite{W93} 
for a review of two dimensional 
$N=(2,2)$ theories).
The scalars in these multiplets parametrize a
``Higgs branch~\footnote{One should keep in mind that there
is no moduli space in two dimensions and we thus work
in the Born-Oppenheimer approximation.}'' which is a
Hyper-K\"ahler manifold.
The vectormultiplet decomposes into a chiral multiplet
and a twisted chiral multiplet. The scalars in these
superfields parametrize the ``Coulomb branch,'' which is
characterized by a generalized K\"ahler potential determining the
metric and torsion on target space~\cite{GHR}.

Next we consider $U(1)$ gauge theories with $N_f$
``electron'' hypermultiplets.
The Coulomb branch is parametrized
by the expectation values
${\vec\phi}\in R^4$ of the four scalars in the
twisted multiplet. Classically, the metric on the
Coulomb branch is flat. Quantum mechanically, the metric
receives a contribution
whose form is fixed by hyper-K\"ahler geometry,
and whose normalization
can be determined by an explicit one-loop
computation. In the massless case the metric is~\cite{RSS}
\beq
ds^2=\left({1\over g_2^2}+{N_f\over \vec\phi^2}\right)d{\vec\phi}^2
\label{D211}
\eeq
where $g_2$ is the $2d$ gauge coupling.
The coefficient in front of $d{\vec\phi}^2$ is the
effective gauge coupling.

Turning on bare masses $\vec m_i$, $i=1,...,N_f$, to the
hypermultiplets the metric becomes
\beq
ds^2=\left({1\over g_2^2}+
\sum_{i=1}^{N_f}{1\over |\vec\phi - \vec m_i|^2}\right)d{\vec\phi}^2
\label{D212}
\eeq
One notes that this is precisely the form of the metric
of a $2d$ CFT describing the propagation of a string near
$N_f$ parallel NS fivebranes (\ref{BST13}).
We shall see in the next section that this is not
an accident. Moreover, the torsion $H=dB$ on the Coulomb branch
is also given by equation (\ref{BST13}).

As in previous sections, it 
will also be interesting to consider
a compactification, in this case
from three to two dimensions on a circle 
of radius $R$. The 
Coulomb branch is four dimensional
and is parametrized
by the expectation values
of the three scalars in the vector
multiplet,
$\vec\rho\in R^3$, and the scalar
$\sigma$ dual to the  
$3d$ gauge field; $\sigma$
lives on a circle, $\sigma\sim \sigma+1/R$.
The metric on the Coulomb branch now takes the form~\cite{DS}
\beq
ds^2=2\pi R\left({1\over g_3^2}+
\sum_{i=1}^{N_f}{1\over |\vec\rho - \vec m_i|}
\Big\{{1\over 2}
+\sum_{n=1}^{\infty}e^{-2\pi Rn|\vec\rho - \vec m_i|}
\cos[2\pi Rn(\sigma-\sigma_i)] \Big\}\right)(d{\vec\rho}^2+d\sigma^2)
\label{D213}
\eeq
The coefficient in front of $d{\vec\rho}^2+d\sigma^2$ is
the effective gauge coupling of a $3d$ theory
compactified to $2d$ on $S^1_R$ and
$g_3$ is the three dimensional coupling constant,
which is related to the two dimensional coupling constant
by standard dimensional reduction
\beq
{1\over g_2^2}={2\pi R\over g_3^2}
\label{D2131}
\eeq

When $R\to 0$ the metric (\ref{D213}) approaches (\ref{D212})
with $\vec\phi\equiv (\vec\rho,\sigma)$.
For large compactification radius, $R\gg 1/|\vec\rho|$,
the effective gauge coupling becomes
\beq
{1\over g_2^2}+\pi R
\sum_{i=1}^{N_f}{1\over |\vec\rho - \vec m_i|}
\label{D214}
\eeq
This is similar to the metric (\ref{KK2}) on
an ALE space with a resolved $A_{N_f-1}$ singularity
which appeared when we discussed the
metric felt by a string in the
presence of $N_f$ parallel KK monopoles.
Again, as we shall discuss later, 
this is not an accident.

For $SU(2)\simeq Sp(1)$ gauge group 
with $N_f$ ``quark'' hypermultiplets,
the metric on the Coulomb branch of 
the two dimensional $(4,4)$
theory is~\cite{DS}
\beq
ds^2=\left({1\over g_2^2}
+\sum_{i=1}^{N_f}\Big\{{1\over |\vec\phi - \vec m_i|^2}
+{1\over |\vec\phi + \vec m_i|^2}\Big\}
-{2\over \vec\phi^2}\right)d{\vec\phi}^2
\label{D215}
\eeq
This metric is related to an ALE space 
with a resolved $D_{N_f}$
singularity for reasons that we shall point out later.

Next we turn to $N=(2,2)$ supersymmetric gauge theories
in two dimensions. $(2,2)$ SCFTs in $2d$ were studied,
in particular, in the context of standard perturbative
string compactifications since they 
lead to spacetime supersymmetric vacua.
Here we shall only
touch upon a small class of $(2,2)$ theories.

Two dimensional $N=(2,2)$ theories
can be obtained by dimensional reduction
of four dimensional $N=1$ supersymmetric theories.
Since anomaly constraints are milder in $2d$,
generic {\em chiral}, anomalous
$4d$ gauge theories typically lead to
consistent $2d$ theories. 
Therefore, we may consider gauged linear sigma models like
a $U(N_c)$ gauge theory with $n$ quarks
$Q$ in the fundamental representation ${\bf N}_c$ and
$\tilde n$ anti-quarks $\tilde Q$ in the anti-fundamental
${\bf \bar{N}}_c$, where $n$ is {\em not}
necessarily equal to $\tilde n$ 
(for a recent review and further references
on such
theories see~\cite{HH}).
When $\tilde n=0$ and $N_c=1$ this is the $CP^{n-1}$ model.
When $\tilde n=0$ and $N_c>1$ this theory is called the
``Grassmanian model.''
The space of its classical vacua is the complex Grassmanian
manifold $G(N_c,n)$.
The dynamics of vacua of the sigma model
with target space $G(N_c,n)$ is described by the
$U(N_c)/U(N_c)$ gauged WZW model 
with the level $k$ of $SU(N_c)$
being
\beq
k=n-N_c
\label{D2151}
\eeq
In this case there is a ``level-rank duality'' (see~\cite{NT}
and references therein) which exchanges:
\beq
N_c\leftrightarrow k
\label{D2152}
\eeq
This duality is the stetement
that the space of conformal blocks of 
an $SU(N_c)$ WZW model
at level $k$ is identical 
to the one of $SU(k)$ at level $N_c$ and, therefore,
the topological theory $U(N_c)/U(N_c)$ at level $k$
is equivalent to the theory $U(k)/U(k)$ at level $N_c$.

\subsection{Brane Theory I: $(4,4)$ Theories}
\label{D244}
Two dimensional unitary gauge theories with $(4,4)$ supersymmetry
appear on D-strings near $D5$-branes. In particular, the
low energy theory on a
$D1$-brane stretched in $(x^0,x^1)$
near $N_f$ parallel $D5$-branes
stretched in $(x^0,x^1,x^2,x^3,x^4,x^5)$ is
a $U(1)$ gauge theory with $N_f$ flavors.
The metric on the Coulomb branch of the theory -- parametrized
by the location of the $D1$-brane in the four directions
transverse to the $D5$-branes $l_s^2 \vec\phi =(x^6,x^7,x^8,x^9)$ --
should be equal to the background
metric of a D-string in the presence of $N_f$ parallel $D5$-branes
located at $l_s^2\vec m_i$, $i=1,...,N_f$. This type IIB system is
$S$-dual to a fundamental string in the presence of
$N_f$ parallel NS fivebranes. 
This explains the relation 
between the metric (\ref{D212})
(and torsion) on the Coulomb branch 
and those of a string propagating in 
the background of
solitonic fivebranes (\ref{BST13}).

Three dimensional gauge theories with eight supercharges
compactified to two dimensions on $S^1_R$ can be studied on
$D2$-branes stretched in $(x^0,x^1,x^6)$) 
near $N_f$ $D6$-branes stretched in
$(x^0,x^1,x^2,x^3,x^4,x^5,x^6)$, both 
wrapping a circle of
radius $R$ in the $x^6$ direction.
Consider a single $D2$-brane.
$T_6$-duality (\ref{BST27}) maps it
to a $D1$-brane near
$N_f$ $D5$-branes at points on a transverse
circle of radius
\beq
R_6=l_s^2/R
\label{D216}
\eeq
The background metric of a fivebrane transverse to
a circle in the $x^6$ direction and located, say,
at $(x^6,x^7,x^8,x^9)=0$
can be obtained by considering an infinite array of fivebranes
separated by a distance $2\pi R_6$. It gives rise
to an H-monopole background with metric and torsion
given by~\cite{GHL}:
\bea
G_{IJ}&=&e^{2(\Phi-\Phi_0)}\delta_{IJ}; \qquad I,J,K,M=6,7,8,9
\nonumber\\
H_{IJK}&=&-\epsilon_{IJKM}\partial^M\Phi \nonumber\\
e^{2(\Phi-\Phi_0)}-1
&=&\sum_{n=-\infty}^{\infty}{l_s^2\over {\vec x}^2+(x^6-2\pi R_6n)^2}
\nonumber\\
&=&{l_s^2\over 2R_6x}\, {\sinh(x/R_6)\over \cosh(x/R_6)-\cos(x^6/R_6)}
\nonumber\\
&=&{l_s^2\over 2R_6 x}
\Big\{1+\sum_{n=1}^{\infty}e^{-nx/R_6}\cos(nx^6/R_6)\Big\}
\label{D2161}
\eea
where $x=|\vec x|$, $\vec x=(x^7,x^8,x^9)$.
{}From (\ref{D216}, \ref{D2161}) we see that the
metric on the Coulomb branch (\ref{D213}) is precisely
the metric of $N_f$ H-monopoles located at
$(x^6_i,\vec x_i)=2\pi l_s^2(\sigma_i,\vec m_i)$,
$i=1,...,N_f$.
We thus see again how the geometry is probed by D-branes.

In the limit $R_6\to\infty$ we obtain a system
of $N_f$ fivebranes in non-compact space
which was discussed above.
This is compatible with the fact that in this limit
(\ref{D2161}) behaves like $1/{\vec\phi}^2$, where
$\vec\phi=(x^6,\vec x)/2\pi l_s^2$.

On the other hand, in the limit $R\to\infty$ we get on the
$D2$-brane a $1+2$
dimensional $N=2$ SUSY $U(1)$ gauge theory with $N_f$ flavors.
The background metric of the $D6$-branes should be
related to the metric on the Coulomb branch of that
gauge theory.
As discussed in section \ref{BST}, $D6$-branes are KK
monopoles in M-theory, and they are described
by the same metric
as KK monopoles in 
type IIA string theory.
A KK monopole with charge $R/l_s$
(Taub-NUT)
is related by T-duality (in an appropriate sense)
to an H-monopole (\ref{D2161}) 
for any value of $R$~\cite{GHM}.
In particular,
when $R\to\infty$ (\ref{D2161}) behaves like
$R/|\vec\rho|$, where $\vec\rho=\vec x/2\pi l_s^2$,
which is compatible with a Coulomb branch with
effective coupling (\ref{D214}).
Metrics similar to these showed up already
in other (related) situations in this review,
such as in sections \ref{metr} and \ref{D3N4}.

A $(4,4)$ SUSY gauge theory in two dimensions
with gauge group $Sp(1)\simeq SU(2)$ can be obtained
on a $D1$-brane (and its mirror image) near
an orientifold fiveplane parallel to $N_f$ $D5$-branes (and
their $N_f$ mirror images). On a transverse circle
of radius $R_6$ it describes a compactification
from three to two dimensions 
on a circle of radius $R$ (\ref{D216}).
T-duality in this transverse direction gives instead
an $O6$-plane parallel to $2N_f$ $D6$-branes, whose
background metric is related
to an ALE space with resolved $D_{N_f}$ 
singularity~\cite{Sei9606,SW9607,Sen9707}.
T-dualizing back to the original system (and
taking $R_6\to\infty$) gives rise to the
metric (\ref{D215}) in agreement with $2d$
field theory.

An alternative way to study $(4,4)$ $2d$ theories on branes
is to allow branes to end on branes, as in previous sections.
A typical configuration involves $N_c$ $D2$-branes
stretched between two $NS5$-branes, 
with $N_f$ $D4$-branes located between
them (or, equivalently by an 
HW transition, $N_f$ $D4$-branes to the
left (right) of the left (right) NS-brane,
each connected to the NS-brane by a single $D2$-brane).
This configuration is T-dual to configurations preserving
eight supercharges which were studied in
previous sections (Figs.~\ref{eleven},\ref{fourteen}).
The branes involved here have worldvolumes
\beq
\begin{array}{ll}
\mbox{$NS5:$}&  \mbox{$(x^0, x^1, x^2, x^3, x^4, x^5)$}\\
\mbox{$D2:$}&  \mbox{$(x^0, x^1, x^6)$}\\
\mbox{$D4:$}&  \mbox{$(x^0, x^1, x^7, x^8, x^9)$}
\end{array}
\label{D217}
\eeq

The low energy theory on the $D2$-branes is a
$U(N_c)$ gauge theory with $N_f$ quark flavors
and classical gauge coupling
\beq
{1\over g_2^2}={L_6l_s\over g_s}
\label{D218}
\eeq
where $L_6$ is the distance between the two NS-branes in $x^6$
(as before, we consider the limit $g_s,l_s\to 0$ such that $g_2$
is held fixed). The locations
of the $D2$-branes along the NS-branes
$\vec r_a=(x^2_a,x^3_a,x^4_a,x^5_a)$, $a=1,...,N_c$,
parametrize the Coulomb branch of the theory.
The locations of the $D4$-branes 
$\vec r_i=(x^2_i,x^3_i,x^4_i,x^5_i)$,
$i=1,...,N_f$, are the bare masses of quark hypermultiplets.
Higgsing correspond to breaking $D2$-branes
on $D4$-branes, and the relative motion
of the two NS-branes in $(x^7,x^8,x^9)$
corresponds to a FI D-term.

When $\vec r_a=\vec r_i=0$, the brane configuration
is invariant under rotations in $(x^2,x^3,x^4,x^5)$ and
$(x^7,x^8,x^9)$. These $Spin(4)_{2345}$ and $SU(2)_{789}$
rotations, respectively,  are associated with the global
$R$-symmetries of the $(4,4)$ gauge theory.

The interpretation of the torsion on the 
Coulomb branch in the brane picture is the
following~\cite{Bro}. A $D2$-brane ending on an NS fivebrane
looks like a string in the $(2,0)$ six dimensional theory on
the fivebrane. Strings in six dimensions couple to the
self-dual two-form $B$, which is identified with the $2d$
$B$-field. Each fundamental hypermultiplet corresponds
to a $D2$-brane ending on an NS-brane and contributes
to the torsion.

Quantum mechanically, the NS-branes bend due to
Coulomb-like interactions in four dimensions (\ref{BST40}).
For simplicity, we consider the $U(1)$ theory: a single
$D2$-brane located at $\vec r$.
As in previous sections, 
the resulting effective gauge coupling
is given by the distance between the
NS-branes in $x^6$ as a function of $\vec r, \vec r_i$:
\beq
{x^6l_s\over g_s}={1\over g_{eff}^2}={1\over g_2^2}+\sum_{i=1}^{N_f}
{l_s^4\over |\vec r - \vec r_i|^2}
\label{D219}
\eeq
This is indeed the exact effective coupling in field theory
(\ref{D212}) with $\vec\phi\equiv \vec r/l_s^2$,
$\vec m_i\equiv \vec r_i/l_s^2$.

As usual, the type IIA configuration at finite $g_s$ is
equivalent to M-theory
on a compact circle of radius $R_{10}=g_sl_s$.
The relative location of the ``NS fivebranes'' in $x^{10}$
correspond to a ``$\theta$ angle.'' This $\theta$ parameter
together with the FI D-term -- the relative position
of the fivebranes in $(x^7,x^8,x^9,x^{10})$ -- combine into
a ``quaternionic K\"ahler form.''

In M-theory, the $SU(2)_R$ symmetry is enhanced to
a $Spin(4)_{7,8,9,10}$. Indeed,~\cite{W9707}
argues that this should happen in field theory.
It thus seems from the brane picture that quantum mechanically
there is a ``mirror symmetry'' interchanging masses with
FI D-term and theta parameters, and the Coulomb branch
with the Higgs branch. 
For more details we refer the reader to~\cite{Bro}.

Brane configurations giving 
rise to three dimensional gauge theories
compactified to two dimensions 
on a circle of radius $R$
can be studied using the above 
configurations by compactifying
$x^2$ on a circle of radius $R_2=l_s^2/R$
(or their $T_2$-dual versions).
In particular, the NS fivebranes now bend due to Coulomb-like
forces in $R^3\times S^1$. The solution to Laplace equation
in this case gives rise to a distance in $x^6$ which
is compatible with the field theory effective gauge coupling
given in (\ref{D213}).

Finally, we may add to the configurations above an orientifold
twoplane (fourplane) parallel to the $D2$-branes ($D4$-branes)
and obtain symplectic or orthogonal gauge groups in two dimensions.
For example, considering two $D2$-branes
stretched between the $NS5$-branes in the presence of an
$O2$-plane, together with $2N_f$ $D4$-branes, gives rise to either
an $SO(2)\simeq U(1)$ or $Sp(1)\simeq SU(2)$ gauge theory,
depending on the sign of the orientifold charge.
Taking into account the sign flip of the orientifold charge,
and the Coulomb-like interactions associated with
$D2$-branes, their mirror images and the $O2$-plane,
gives rise to a bending of
the $NS5$-branes which is in agreement with the field theory
results (\ref{D212}) and (\ref{D215}):
\beq
{x^6l_s\over g_s}={1\over g_{eff}^2}={1\over g_2^2}+\sum_{i=1}^{N_f}
\Big\{ {l_s^4\over |\vec r - \vec r_i|^2}
      +{l_s^4\over |\vec r + \vec r_i|^2} \Big\}
      -{(1+1)(1+1)l_s^4\over |2\vec r|^2}
\pm {(1/2+1/2)l_s^4\over |\vec r|^2}
\label{D220}
\eeq
The second term on the right hand side of (\ref{D220})
is due to the $N_f$ flavors and their $N_f$ mirror images,
the third term is due to a $D2$-brane at $\vec r$ having
its mirror image at $-\vec r$, and the last term is due to
the $O2$-plane -- the ``$\pm$'' corresponding to orthogonal
or symplectic projections, respectively.
Obviously, this discussion can be generalized to other
dimensions and to compactifications from high
to lower dimensions performing the analysis with an
orientifold charge and Coulomb-like interactions
in the appropriate dimension~\footnote{This was done explicitly
in an unpublished work with M. Ro\v cek.}.

Brane configurations corresponding to two dimensional
$(4,4)$ gauge theories were also considered
in~\cite{Ali,IM}.

\subsection{Brane Theory II: $(2,2)$ Theories}
\label{D222}

We may now rotate branes in the configurations
of the previous section and get at low energy
two dimensional $N=(2,2)$ supersymmetric
gauge theories on the $D2$-branes.
As an example, we shall examine a configuration of
an $NS5$-brane connected to an \nsp-brane by
$N_c$ $D2$-branes in the presence of $N_f$ $D4$-branes.
The worldvolumes of the various objects are given in
(\ref{D217}, \ref{D4N11}).

\begin{figure}
\centerline{\epsfxsize=90mm\epsfbox{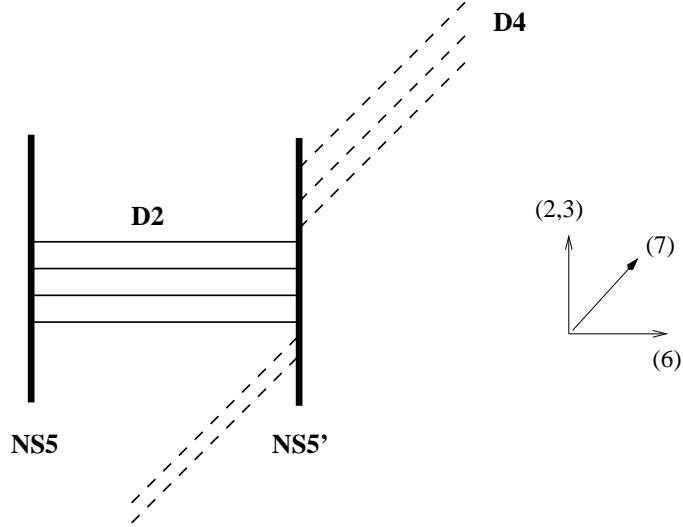}}
\vspace*{1cm}
\caption{$D4$-branes oriented as explained in the
text can split into two disconnected components
and separate along an NS fivebrane in $(x^2, x^3)$.}
\label{fourtyseven}
\end{figure}
\smallskip

A new ingredient which appears in such a brane configuration
is the possibility to put a $D4$-brane at the same
$x^6$ location of the \nsp-brane, then break it
and separate the two semi-infinite pieces along the
\nsp-brane in the $\sigma$ direction:
\beq
\sigma\equiv x^2+ix^3
\label{D221}
\eeq
We may break all the $N_f$ $D4$-branes on the \nsp-brane
and take part of the semi-infinite $D4$-branes to infinity.
One then obtains a configuration where, say, $n$ semi-infinite
$D4$-branes -- stretched in $x^7>0$ -- are located at
$\sigma_i$, $i=1,...,n$, along the \nsp-brane, and
$\tilde n$ semi-infinite
$D4$-branes -- stretched in $x^7<0$ -- are located at
$\sigma_{\tilde i}$, ${\tilde i}=1,...,{\tilde n}$ 
(see Fig~\ref{fourtyseven}).

The low energy theory on the $D2$-branes 
is a two dimensional $(2,2)$, $U(N_c)$ gauge theory with
$n$ quarks and $\tilde n$ anti-quarks.
There is a manifest chiral flavor symmetry
$U(n)\times U(\tilde n)$ which is broken for generic
values of $\sigma_i$, $\sigma_{\tilde i}$.
One can check that the classical moduli space of vacua and
deformations of the brane configuration (almost) agree with
a field theory analysis~\cite{HH}.

In M-theory, the $NS5$-brane and $D4$-branes ending on
the \nsp-brane turn into two
disconnected $M5$-branes. The type IIA $D2$-branes
turn into $M2$-branes connecting these $M5$-branes.
The dynamics of such open membranes stretched between
two fivebranes is not completely understood; nevertheless,
chiral features of the quantum $(2,2)$
theories can be studied in this way. We refer the reader
to~\cite{HH} for details.

We should remark that the Coulomb-like interactions
associated with the $D4$-branes ending on the
\nsp-brane give rise to terms which are logarithmic in $\sigma$
and which contribute to the 
quantum low energy $2d$ effective
superpotential. Logarithmic effective
superpotentials are indeed familiar in
such two dimensional $(2,2)$ theories (namely, in gauged
linear sigma models). Other relations between
the parameter space of $(2,2)$ theories in two dimensions
and the moduli space of $N=2$ four dimensional theories --
associated with the $D4$-branes ending on the \nsp-brane --
are discussed in~\cite{HH}.

Finally, let us consider a duality trajectory interchanging
the $NS5$ and \nsp-brane in the $x^6$ direction.
The details of this process can be worked out along
the lines of previous sections (up to an ambiguity which is
resolved quantum mechanically in M-theory~\cite{HH}).
Here we shall only state the
result in the case $\tilde n=0$, namely, for a $G(N_c,n)$ model
(see section \ref{D2FTR}). The duality trajectory takes
an electric $U(N_c)$ theory with $n$ quarks to
a magnetic $U(n-N_c)$ theory with $n$ quarks, \ie
\beq
G(N_c,n)\leftrightarrow G(n-N_c,n)
\label{D2211}
\eeq
providing a brane realization of the level-rank duality
(\ref{D2152}, \ref{D2151}) discussed in section \ref{D2FTR}.

\section{Five And Six Dimensional Theories}
\label{D5}

\subsection{Five Dimensional Field Theory Results}
\label{D5FTR}

$N=1$ supersymmetric gauge theories in five dimensions have
eight supercharges and an $SU(2)_R$ global symmetry.
The two possible multiplets in the theory
are the vectormultiplet in the adjoint of the gauge
group $G$ -- containing a vector field,
a real scalar $\phi$ and fermions -- and the
hypermultiplet in a representation $R_f$ of $G$ --
containing four real scalars and fermions. The Coulomb branch
is parametrized by the scalar components of the vector
multiplet $\phi^i$, $i=1,...,{\rm rank}G$,
in the Cartan sub-algebra of $G$.

The low energy theory is determined by the prepotential
$\FF(\phi)$, which is required to be at
most cubic due to $5d$ gauge invariance~\cite{Sei9608}.
The exact quantum prepotential is given by~\cite{IMS}
\beq
\FF ={1\over 2g_0^2}\phi^i\phi_i
     +{c_{cl}\over 6}d_{ijk}\phi^i\phi^j\phi^k
+{1\over 12}\left(\sum_{\alpha}|\alpha_i\phi^i|^3
-\sum_f\sum_{w\in R_f}|w_i\phi^i+m_f|^3\right)
\label{D51}
\eeq
Here $g_0$ is the bare coupling of the gauge theory
and $d_{ijk}$ is the third rank symmetric tensor:
$d_{abc}={\rm Tr}(T_a\{T_b,T_c\})/2$. The first sum
in (\ref{D51}) is over the roots of $G$ and the second sum is over
the weights of the representation $R_f$ of $G$;
$m_f$ are the (real) masses of the hypermultiplets in $R_f$.
$c_{cl}$ is a quantized parameter of the theory, related to
a $5d$ Chern-Simons term.
In terms of $\FF$ the effective gauge coupling is
\beq
\left({1\over g^2}\right)_{ij}=
{\partial^2\FF\over \partial\phi^i\partial\phi^j}
\label{D52}
\eeq

{}From now on we discuss simple groups $G$ with $N_f$
hypermultiplets in the fundamental representation of $G$:

\begin{itemize}

\item
$G=SU(N_c)$ ($N_c>2$):
The Coulomb branch
of the moduli space is given by $\phi={\rm diag}(a_1,...,a_{N_c})$
with $\sum_{i=1}^{N_c}a_i=0$. The prepotential in this case is
\beq
\FF ={1\over 2g_0^2}\sum_{i=1}^{N_c}a_i^2
     +{1\over 12}\left(2\sum_{i<j}^{N_c}|a_i-a_j|^3
     +2c_{cl}\sum_{i=1}^{N_c}a_i^3
     -\sum_{f=1}^{N_f}\sum_{i=1}^{N_c}|a_i+m_f|^3\right)
\label{D56}
\eeq
The conditions on $N_c$, $N_f$ and $c_{cl}$ in (\ref{D51}) are
\beq
c_{cl}+{1\over 2}N_f \in Z
\label{D53}
\eeq
\beq
SU(N_c): \qquad N_f+2|c_{cl}| \leq 2N_c
\label{D54}
\eeq

\item
$G=SU(2)$:
This case is somewhat special. There are two
pure gauge theories labeled by a $Z_2$ valued theta angle, since
$\pi_4(SU(2))=Z_2$. $c_{cl}$ is irrelevant since $d_{ijk}=0$
in (\ref{D51}), and the number of flavors allowed is $N_f\leq 7$.

\item
$G=SO(N_c)$ $(Sp(N_c/2))$: In this case $c_{cl}=0$ and
\beq
SO(N_c)\,\, (Sp(N_c/2)): \qquad N_f\leq N_c-4 \,\, (N_f\leq N_c+4)
\label{D55}
\eeq
The inequalities in (\ref{D54}, \ref{D55}) are necessary
conditions to have non-trivial fixed points which one can
use to define the $5d$ gauge theory.

\end{itemize}

In five dimensions there are no instanton
corrections to the metric and, therefore, the exact results
considered above are obtained already at one loop.
However, compactifying the theory gives
rise to non-perturbative corrections.
Supersymmetric $5d$ gauge theories compactified to
four dimensions on a circle were studied 
in~\cite{GA,Nek,GMS,NLa}.
The perturbative contributions to $\FF$ from KK modes were
found, and were shown to obtain the correct behavior in the
five and four dimensional limits. Non-perturbative
corrections to $\FF$ are conjectured to be related
to spectral curves of relativistic Toda systems.

\subsection{Webs Of Fivebranes And Five Dimensional Theories}
\label{D5FEF}

We begin by considering a $D5$-brane ending (classically)
on an $NS5$-brane~\cite{AH}. Recall (section \ref{DP}) that
type IIB fivebranes sit in a $(p,q)$ multiplet of the
$SL(2,Z)$ S-duality group, where the $NS5$-brane is
a $(0,1)$ fivebrane while the $D5$-brane is a $(1,0)$
fibebrane. We choose the worldvolumes of these fivebranes
to be
\beq
\begin{array}{ll}
\mbox{$NS5\,\, (0,1):$}&  \mbox{$(x^0, x^1, x^2, x^3, x^4, x^5)$}\\
\mbox{$D5\,\, (1,0):$}&  \mbox{$(x^0, x^1, x^2, x^3, x^4, x^6)$}
\end{array}
\label{D511}
\eeq
Classically, we may let the $D5$-brane end on the NS fivebrane,
say from the left in $x^6$. Such a configuration is allowed --
as discussed in section \ref{BEB} -- and
it is T-dual to situations where $D4$ or $D3$-branes
are ending on NS fivebranes as in previous sections.
Therefore, this configuration preserves
eight supercharges.

Quantum mechanically the NS fivebrane bends. Its bending is due
to the fact that the $D5$-brane ending on the NS fivebrane
looks like an electric charge in one dimension.
This causes a linear Coulomb-like interaction
(see (\ref{BST40})) which leads to the bending of the
$NS5$-brane in the $(x^5,x^6)$ plane into the location
\beq
x^6={g_s\over 2}(|x^5|+x^5)
\label{D512}
\eeq
Here and below we set $a=0$ in the complex type IIB
string coupling $\tau$ (\ref{BST205}), for simplicity.
Moreover, without lose of generality we chose
the intersection of the fivebranes to be located at the
origin $(x^5,x^6)=0$, and the extension of the $NS5$-brane
to be as in (\ref{D511}) when $x^5<0$.

\begin{figure}
\centerline{\epsfxsize=100mm\epsfbox{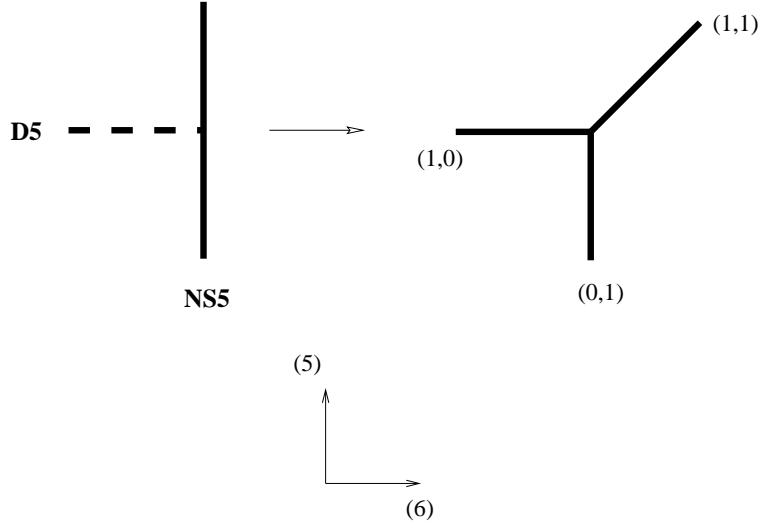}}
\vspace*{1cm}
\caption{The classical configuration of a $D5$-brane
ending on an NS fivebrane is replaced for finite
$\gs$ by a vertex in which $(1,0)$ and $(0,1)$
fivebranes merge into a $(1,1)$ fivebrane.}
\label{fourtyeight}
\end{figure}

While we started classically from a semi-infinite
fivebrane ending on an infinite fivebrane, what we
have obtained instead -- quantum mechanically -- is an intersection
of three semi-infinite fivebranes 
(Fig~\ref{fourtyeight}).
All fivebranes share the same
$1+4$ directions $(x^0,x^1,x^2,x^3,x^4)$ and are real straight
lines in the $(x^5,x^6)$ plane.
In the example above we have three semi-infinite fivebranes
meeting at the origin. The semi-infinite fivebrane
located at $x^6=0$ and stretched along $x^5<0$ is
the $NS5$-brane.
The semi-infinite fivebrane located at $x^5=0$ and stretched
along $x^6<0$ is the $D5$-brane.
The third semi-infinite fivebrane is located (\ref{D512})
at $x^6=g_s x^5$, $x^5,x^6>0$.

\smallskip

Which is the fivebrane located at $x^6=g_sx^5$? Clearly,
it is a $(1,1)$ fivebrane! Charge conservation does not
really allow a $D5$-brane to end on an $NS5$-brane. Instead,
at the intersection point the $(1,0)$ and $(0,1)$ fivebranes
merge together to form a $(1,1)$ fivebrane.
In order for this ``new'' $(1,1)$ fivebrane not to break
supersymmetry any further, it must merge from the intersection
point at an angle, as described above (\ref{D512}).

In the same way general vertices of $(p,q)$ fivebranes
are permitted provided that $(p,q)$ charge is preserved.
To write down a charge conservation condition we have to
pick up an orientation for the fivebranes. If we fix the
orientation of all $n$ fivebranes in the direction towards
the vertex the charge conservation reads
\beq
\sum_{i=1}^n p_i=\sum_{i=1}^n q_i =0
\label{D513}
\eeq
Moreover, requiring the vertex to preserve eight supercharges
implies that the $(p,q)$ fivebrane is stretched along the
semi-infinite line in the $(x^5,x^6)$ plane located at
\beq
q x^6=g_s p x^5
\label{D514}
\eeq
This condition is equivalent to the zero Coulomb-like
force condition required for the stability of the vertex.

We can easily extend the discussion above to
a situation where $n_L$ $D5$-branes end on a
fivebrane from the left and $n_R$ $D5$-branes
end on the fivebrane from the right.
Let $a_i$, $i=1,...,n_L$, be the $x^5$ locations
of the $D5$-branes from the left and
$b_j$, $j=1,...,n_R$, the locations
of the $D5$-branes from the right. The
bending of the fivebrane -- generalizing (\ref{D512}) -- is
\beq
x^6={g_s\over 2}\left(\sum_{i=1}^{n_L}|x^5-a_i|-
                      \sum_{j=1}^{n_R}|x^5-b_j|+
                      (n_L-n_R)x^5\right)
\label{D512a}
\eeq
This equation has the interpretation of a fivebrane
which is an $NS5$-brane at large negative $x^5$, and
it changes its charge and angle in $(x^5,x^6)$ in places
where a $D5$-brane ends on it. The change of charge and
angles is dictated by the conditions (\ref{D513}, \ref{D514}).

The presence of such $(p,q)$ fivebranes breaks the
$1+9$ dimensional Lorentz group of the
ten dimensional type IIB string to
\beq
SO(1,9)\to SO(1,4)\times SO(3)_{789}
\label{D515}
\eeq
The $SO(1,4)$ is the five dimensional Lorentz symmetry
preserved in the $(x^0,x^1,x^2,x^3,x^4)$ directions -- common
to all fivebranes -- while $SO(3)_{789}$ is the
three dimensional rotation symmetry preserved in the
$(x^7,x^8,x^9)$ directions -- transverse to all fivebranes.
The double cover of this group will be identified with the
five dimensional $R$-symmetry: $SU(2)_{789}\equiv SU(2)_R$.

To study five dimensional gauge theories on
type IIB fivebranes we need to describe webs of
$(p,q)$ fivebranes where some of the branes
are finite in one direction, say in $x^6$.
A web of fivebranes includes vertices (where fivebranes intersect)
legs (the segments of fivebranes) and faces.
In each vertex charge conservation is obtained and
the zero force condition (\ref{D514}) is applied
to fix the appropriate angles. In what follows we
shall not specify the orientation choices for
the legs which should be understood from the charge
assignments given in each case.

For example, we study webs describing an $SU(N_c)$ gauge
theory. Consider $N_c$ parallel $D5$-branes -- with
worldvolume as in (\ref{D511}) -- stretched
between other two fivebranes separated in the $x^6$ direction,
which we choose to be $(p_L,q_L)$ and $(p_R,q_R)$
fivebranes for large negative values of $x^5$.
The left and right fivebranes are broken into segments
between $x^5=-\infty$ and the lower (in $x^5$) $D5$-brane, between
$D5$-branes and between the upper $D5$-brane and $x^5=\infty$.

In other words, we consider a web with $N_c$
horizontal internal legs (stretched in the $x^6$ direction),
which are connected to each other
by $N_c-1$ internal legs on the left (in $x^6$)
and $N_c-1$ internal legs connecting them on the right.
In addition, there are four external legs, two from
above (in $x^5$) and two from below. The lower left and right
legs have charges $(p_L,q_L)$ and $(p_R,q_R)$,
respectively.

Charge conservation (\ref{D513}) implies that the left and
right internal legs between the $a$'th and $a+1$'st
$D5$-branes have charges $(p_L-a,q_L)$ and $(p_R+a,q_R)$,
respectively.
This means that for large positive values of $x^5$ the
left and right fivebranes will have charges
$(p_L-N_c,q_L)$ and $(p_R+N_c,q_R)$, respectively.
The different left and right fivebrane segments are oriented
in different directions in the $(x^5,x^6)$ plane,
in accordance with the zero force conditions in
each vertex (\ref{D514}), separately. The precise bending
of the left and right fivebranes can be obtained
by using an appropriate version of (\ref{D512a}).

As we saw in the four and three dimensional cases,
to study the gauge physics using branes we need to
consider a limit in which gravity and massive string
modes decouple. The relevant limit in this case
is $L_{max},l_s,g_s\to 0$,
where $L_{max}$ is the largest length of an internal leg.
If the gauge coupling at some scale $L$ satisfying
$L\gg l_s \gg L_{max}$ is finite, at larger distances
gravity decouples, massive KK modes can be integrated out,
and the dynamics on the brane configuration is governed
by gauge theory.

At low energy the theory on the $D5$-branes is
a pure $N=1$ supersymmetric $SU(N_c)$ gauge theory
in $1+4$ dimensions. Deformations of the web that do not change
the asymptotic locations of the external legs correspond
to moduli in the field theory.
Such locations $a_i$, $i=1,...,N_c$, of the $D5$-branes along
the $x^5$ direction parametrize the Coulomb branch of
the theory.

When $g_s\to 0$ the configuration
tends to $N_c$ parallel $D5$-branes stretched between
two parallel fivebranes, and the classical gauge
group is $U(N_c)$ with gauge coupling
\beq
{1\over g_0^2}={L_6\over g_sl_s^2}
\label{D516}
\eeq
Here $L_6$ is the distance between the left and right fivebranes.
To keep $g_0$ finite we need to take $L_6\to 0$
such that the ratio $L_6/g_sl_s^2$ is finite.
The $N_c$ values of $a_i$ are independent
and parametrize the Coulomb branch of $U(N_c)$.

For finite $g_s$ quantum effects
cause the fivebranes to bend -- as described above --
and ``freeze'' 
the $U(1)$ factor, as in the four
dimensional theories considered in sections
\ref{D4N2}, \ref{D4N1}. One of the $N_c$ independent
classical motions of the $N_c$ $D5$-branes is indeed frozen
-- once the left and right fivebranes bend
$\sum_{i=1}^{N_c} a_i=0$ is required to keep the asymptotic
locations of external legs fixed  --
leaving a total of $N_c-1$ real motions parametrizing the
Coulomb branch of $SU(N_c)$, as in (\ref{D56}).

\begin{figure}
\centerline{\epsfxsize=80mm\epsfbox{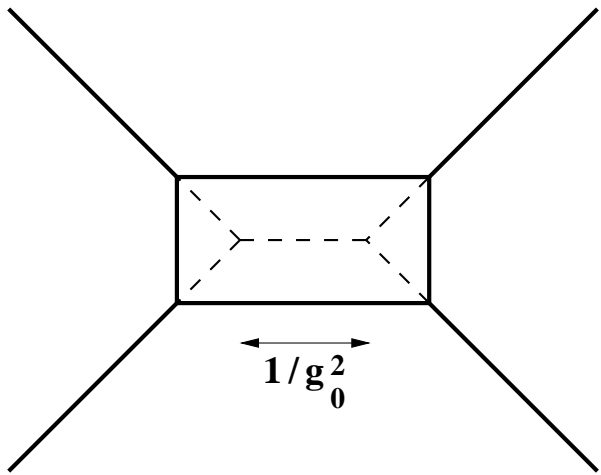}}
\vspace*{1cm}
\caption{A five dimensional $SU(2)$ gauge theory
described using fivebranes.}
\label{fourtynine}
\end{figure}

The asymptotic positions of the external legs are associated with
the gauge coupling. The classical gauge coupling $1/g_0^2$
can be obtained geometrically as follows. We set $a_i=0$, namely,
we deform the $D5$-branes to a position where they are
coincident without changing the locations of the external legs.
Then the length of the $D5$-branes $L_6$ is related to $g_0$
by equation (\ref{D516}) -- now with non-zero $g_s$ and $L_6$.
For general $a_i$, $L_6$ is still the distance between the point
where the (``imaginary'') continuation of the left external
legs meet and the point where the continuation of
right external legs meet (see
Fig~\ref{fourtynine}).

\smallskip

The semiclassical $SU(N_c)$ gauge coupling (which in this
case is exact) is related to the the ``size'' of
the brane configuration in $x^6$ as
in the four and three dimensional cases
(see sections \ref{D4N2}, \ref{D4N1}, \ref{D3}).
Indeed, it is  linear in $a_i$ -- as predicted by the
bending (\ref{D512a}, \ref{D514}) -- and in agreement with the
field theory result obtained from (\ref{D56}, \ref{D52})
(with $N_f=0$).

Not every charge assignment is allowed to be given
to the external legs while still describing an $SU(N_c)$
gauge theory on the $D5$-branes. In this respect, two questions
are interesting:

\begin{enumerate}
\item What is the gauge theory meaning of the charges
$(p_L,q_L)$ and $(p_R,q_R)$ on the external legs?
\item Which values of $(p_L,q_L)$ and $(p_R,q_R)$
are permitted?
\end{enumerate}

The answer to 2 is clear. The permitted values
$(p_L,q_L)$ and $(p_R,q_R)$ are such that the
external legs do not cross each other.
If the external legs do cross each other,
the brane configuration has more crossings
of fivebranes than those required to describe the Coulomb branch
of $N=1$, $SU(N_c)$ gauge theory in five dimensions.

To find the independent ``legal'' $SU(N_c)$ webs --
obeying all the above conditions -- and their moduli,
it is convenient
to describe a web by its dual grid diagram. The grid has
points, lines and polygons, which are dual to the faces, legs
and vertices of the web.
One can show~\cite{AHK} that for $N_c>2$ there are
$2N_c+1$ inequivalent webs.
Indeed, this is precisely the number of allowed
values of $c_{cl}$ as obtained from the conditions
(\ref{D53}, \ref{D54}) for $N_f=0$:
$c_{cl}=-N_c,-N_c+1,...,N_c-1,N_c$.
This answers question 1: the $2N_c+1$ different legal
webs are in one to one correspondence with the different
allowed values of $c_{cl}$.

Each different allowed $(p_L,q_L)$, $(p_R,q_R)$
corresponds to a different allowed $c_{cl}$.
The web corresponding to $-c_{cl}$ is obtained from
the web corresponding to $c_{cl}$ by the use of
$SL(2,Z)$ S-duality together with a rotation
-- a $Z_2$ reflection.
Indeed, in field theory there is a $Z_2$ symmetry
of the spectrum under the reflection $c_{cl}\to -c_{cl}$.
In particular, the configuration corresponding to $c_{cl}=0$ is
the web invariant under reflection, while configurations
corresponding to $|c_{cl}|=N_c$ are the two webs with
parallel external legs. In the latter case an equality
holds in the field theory constraint
(\ref{D54}).~\footnote{Webs with parallel external legs
seem to be inconsistent as $5d$ theories and
perhaps should not be considered; when two external legs
are parallel a string corresponding to either a gauge boson or
an instanton (see below) can ``leak'' out of the web~\cite{AHK}.}

\begin{figure}
\centerline{\epsfxsize=60mm\epsfbox{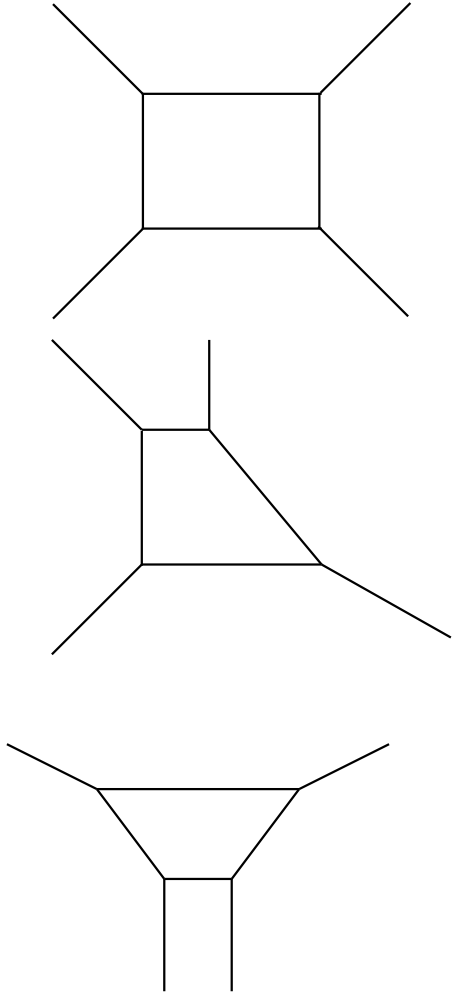}}
\vspace*{1cm}
\caption{The three possible configurations
corresponding to $SU(2)$ gauge theories. Only
the first two appear to give rise to non-trivial
fixed points.}
\label{fiftyone}
\end{figure}

The $N_c=2$ case is special. 
Here one finds three independent,
apparently legal, webs (Fig~\ref{fiftyone}).
Each web has (generically) four
vertices, four external legs, and four internal legs forming a single
face. One of the webs has two parallel external legs.
It is claimed, however, that in this case parallel external
legs do not correspond to a web describing an $SU(2)$
gauge theory. The remaining two webs -- one with a
rectangular face and the other with a right angle trapeze --
correspond to the two $SU(2)$ gauge theories found in field
theory, as discussed in subsection \ref{D5FTR}.

\smallskip

We see that the webs and grids are useful in
classifying five dimensional $N=1$, $SU(N_c)$ gauge theories.
We refer the reader to~\cite{AHK} for a detailed description
of the classification of $5d$ theories using grids.

\begin{figure}
\centerline{\epsfxsize=120mm\epsfbox{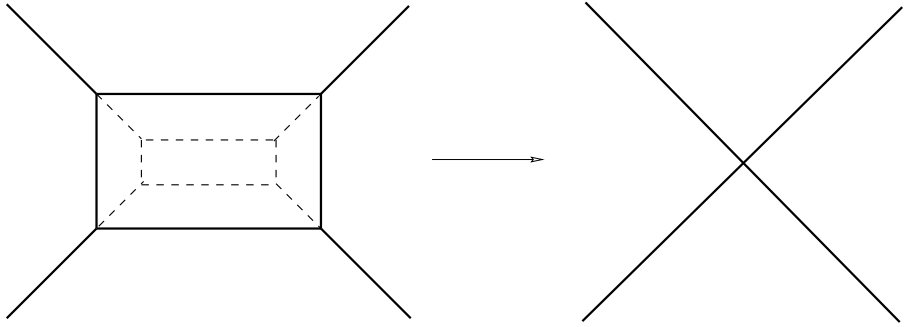}}
\vspace*{1cm}
\caption{Non-trivial fixed points are described
by vertices with more than three external legs.}
\label{fifty}
\end{figure}

In the webs above each vertex had three intersecting legs.
However, displacing the $D5$-branes in $x^5$ and/or changing
the locations of external legs may lead to
situations where faces shrink to zero size, thus forming
vertices with more than three intersecting legs.
For instance, the $SU(2)$ diagrams can be deformed to
a single vertex with four legs (see Fig~\ref{fifty}).
In particular, the gauge coupling
in such configurations tends to infinity, and we describe
a strong coupling fixed point of the theory.

\smallskip

The webs of fivebranes may thus be useful in classifying
five dimensional $N=1$ superconformal fixed points.
Such attempts were initiated in~\cite{AH,AHK}.
Indeed, many configurations corresponding to known
$5d$ superconformal field theories (SCFT) were identified,
as well as webs which lead to new SCFTs.

To add $N_f$ fundamental flavors we allow the inclusion
of a total of $N_f$ semi-infinite $D5$-branes, $N_L$ ($N_R$)
of which are connected to the left (right) of the left (right)
fivebrane ($N_L+N_R=N_f$).~\footnote{Alternatively,
we could add $D7$-branes stretched in
$(x^0,x^1,x^2,x^3,x^4,x^7,x^8,x^9)$. But sevenbranes
affect the asymptotic behavior of space-time and we shall
not consider them here. Without $D7$-branes
we will not be able to see the complete structure of the
Higgs branches of the theory geometrically.
Some Higgsing can be obtained, however, by deforming a sub-web
in the $(x^7,x^8,x^9)$ directions corresponding, classically,
to a FI D-term. This is possible when a configuration
is reducible, namely, when it can be considered to consist of
two independent webs. It may happen at the roots of the Higgs branches.}
The locations $m_f$, $f=1,...,N_f$,
of the semi-infinite D fivebranes in the $x^5$ directions
parametrize the $N_f$ real masses of quarks.
Of course, these additional $D5$-branes affect the
bending of the left and right fivebranes (see (\ref{D512a})).
In particular, for large positive values of $x^5$ the
left and right fivebranes will have charges $(p_L+N_L-N_c,q_L)$
and $(p_R+N_c-N_R,q_R)$, respectively.

As before, $SU(N_c)$ legal configurations are those
where the left and right fivebranes do not cross each other.
As a result, the number of inequivalent webs
describing $SU(N_c)$ with $N_f$ flavors and
obeying all the required conditions is indeed the
one indicated by the field theory conditions
(\ref{D53}, \ref{D54}). Allowed values of
$c_{cl}$ are in one to one correspondence with
such inequivalent legal webs.

Again, the gauge coupling related to the brane configuration
-- along the lines of the four and three dimensional discussion --
is in agreement with the field theory result
obtained from (\ref{D56}, \ref{D52}) for general
$N_c$, $N_f$, $m_f$ and $c_{cl}$. This can be seen by
using the relation between the bending (\ref{D512a})
and the gauge coupling as discussed in sections
\ref{D4N2}, \ref{D4N1}, \ref{D3}.

In addition to classifications and the study of the
structure of moduli space,
more aspects of five dimensional gauge theories can
also be considered in the brane configurations.
In particular, there are BPS saturated monopole
strings, which arise from $D3$-branes stretched along
faces in the brane configuration, and instantons --
BPS saturated particles in five dimensions -- corresponding
to D-strings parallel to (or inside) the $D5$-branes and ending on
the left and right fivebranes. Moreover, one can stretch
$(p,q)$ string webs ending on the
fivebranes~\footnote{Webs of $(p,q)$ strings can also
be stretched between $D3$-branes; in the context of section \ref{D4N4}
they describe 1/4 BPS states in $4$-$d$, $N=4$
SYM theory~\cite{Ber}.}.
These and other BPS states can be studied in the brane configurations
considered above; we refer the reader to~\cite{AHK,KR}
for details.

\subsection{Compactifying From Five To Four Dimensions}
\label{D5CFFD}

In this subsection we compactify the five dimensional gauge theory
with space-time
$(x^0,x^1,x^2,x^3,x^4)$ to four dimensions $(x^0,x^1,x^2,x^3)$
on a circle of radius $R$.
In the type IIB brane configuration of the previous
subsection this is also obtained by compactifying $x^4$
on a radius $R$ circle.

The semiclassical results of the previous subsection
are no longer exact for $R<\infty$. To study the exact
non-perturbative corrections in the brane configuration
we lift the web into an M-theory curve
(an $M5$-brane)~\cite{Kol,BISTY,AHK}.
For that purpose, we need to identify the type IIB parameters
$g_s,l_s$ and $R$ in terms of the M-theory parameters
$l_p$ and $R_i$, $i=1,...,10$.

To do that we first perform a T-duality in the $x^4$
direction $T_4$ which takes the compactified type IIB string to
a type IIA string compactified on a circle of radius $R_4$
(see section \ref{DP})
\beq
T_4: \quad R\to R_4={l_s^2\over R}
\label{D531}
\eeq
and string coupling
\beq
g_A={g_sl_s\over R}
\label{D532}
\eeq
Moreover, as explained in section \ref{MI}, the $D5$
and $NS5$-branes with worldvolumes as in (\ref{D511})
transform under $T_4$ to the type IIA $D4$ and $NS5$-branes,
respectively, with worldvolumes
\beq
\begin{array}{ll}
\mbox{$NS5:$}&  \mbox{$(x^0, x^1, x^2, x^3, x^4, x^5)$}\\
\mbox{$D4:$}&  \mbox{$(x^0, x^1, x^2, x^3, x^6)$}
\end{array}
\label{D533}
\eeq
Therefore, a web describing $SU(N_c)$ with $N_f$ flavors, namely,
$N_c$ finite and $N_f$ semi-infinite $D5$-branes --
as considered in the end of the previous subsection --
transforms into a type IIA
configuration where $N_c$ fourbranes are stretched between two
NS fivebranes while $N_L$ ($N_R$) semi-infinite fourbranes
are connected to the left (right) of the left (right) fivebrane,
$N_L+N_R=N_f$. In the following we will usually take $N_L=0$,
$N_R=N_f$.

Equivalently, for finite $g_A$ (\ref{D532}) we have M-theory
compactified on a rectangular two-torus in the $x^4$ and $x^{10}$
directions with sizes $R_4$ and $R_{10}$, respectively, where
\beq
R_{10}=l_sg_A
\label{D534}
\eeq
and with an eleven dimensional Planck scale
\beq
l_p^3=R_{10}l_s^2
\label{D535}
\eeq
As explained in section \ref{MI} (and can be re-derived from
(\ref{D531}, \ref{D532}, \ref{D534}, \ref{D535})),
the type IIB compactification radius
and string coupling are related to the M-theory parameters via
\beq
R={l_p^3\over R_4R_{10}}
\label{D536}
\eeq
and
\beq
g_s={R_{10}\over R_4}
\label{D537}
\eeq
The ten dimensional type IIB limit is obtained by taking
$R_4R_{10}\to 0$ while keeping $g_s$ (\ref{D537}) fixed.
Indeed, (\ref{D536}) implies that in this limit $R\to\infty$
and we recover the five dimensional field theory
configurations of the previous subsection.

As before, in M-theory the type IIA brane configuration is an
$M5$-brane with worldvolume $R^{1,3}\times \Sigma$, where
$R^{1,3}$ is the $1+3$ space-time $(x^0,x^1,x^2,x^3)$
and $\Sigma$ is a two dimensional surface embedded in the
four dimensional space $Q=S^1\times R^2\times S^1$ in the
$(x^4,x^5,x^6,x^{10})$ directions. Since both $x^4$ and $x^{10}$
are compact, to find the curve $\Sigma$ it is
convenient to parametrize $Q$ by the single valued
coordinates $t$ and $u$:
\bea
t=e^{-s/R_{10}}, &\quad s=x^6+ix^{10}\nonumber\\
u=e^{-iv/R_4}, &\quad   v=x^4+ix^5
\label{D538}
\eea
and describe the curve by the algebraic equation
\beq
F(t,u)=0
\label{D539}
\eeq

As in sections \ref{QEFF}, \ref{QEFFII},
the form of the curve should be
\beq
F(t,u)=A(u)t^2+B(u)t+C(u)=0
\label{D540}
\eeq
and we may set
\beq
A(u)=1
\label{D541}
\eeq
corresponding to all semi-infinite fourbranes being to the
right of the right fivebrane ($N_L=0$, $N_R=N_f$, see above).
Since both $u=\infty$ (namely, $x^5=\infty$)
and $u=0$ ($x^5=-\infty$) correspond to the asymptotic
region the multiplicity of the the zero roots of the
polynomials $B(u)$ and $C(u)$ is relevant.
Analyzing~\cite{BISTY} the asymptotic behavior
-- as in section \ref{QEFF} -- one finds that
curves describing consistent $SU(N_c)$ configurations with
$N_f$ fundamental flavors have
\bea
&B(u)=b\prod_{i=1}^{N_c}(u-A_i) \nonumber\\
&C(u)=cu^{N_c-N_f/2-c_{cl}}\prod_{f=1}^{N_f}(u-M_f)
\label{D542}
\eea
where $a,b,A_i,M_f,c_{cl}$ are constant parameters and $c_{cl}$
must obey the conditions (\ref{D53}, \ref{D54}).
Therefore, $c_{cl}$ in (\ref{D542}) corresponds precisely to $c_{cl}$
in field theory. In M-theory $c_{cl}$ must obey the condition (\ref{D53})
because otherwise the curve is not holomorphic, and
it must obey the condition (\ref{D54}) because otherwise
the $M5$-brane describes a type IIB configuration where
the external fivebranes cross each other.

Each monomial $u^nt^m$ in the curve is associated with the
point $(n,m)$ in the grid diagram dual to the web, namely,
to a face in the web where it is dominant. The curve is just
the sum of these monomials~\cite{Kol,AHK} with the coefficients constrained
to obey some consistency conditions.

The four dimensional field theory limit is obtained at $R\to 0$.
To consider this limit in M-theory, it is convenient to
rewrite the algebraic equation (\ref{D539}-\ref{D542})
in terms of $v$ (instead of $u$ (\ref{D538})).
By an appropriate choice of the constants $b$ and $c$
in (\ref{D542}) one finds the curve
\beq
t^2+te^{-iN_cv/2R_4}
\prod_{i=1}^{N_c}R_4\sin\Big({v-a_i\over 2R_4}\Big)+
e^{-iv(N_c-c_{cl})/R_4}
\prod_{f=1}^{N_f}R_4\sin\Big({v-m_f\over 2R_4}\Big)=0
\label{D543}
\eeq
where the parameters $A_i$ and $M_f$ in (\ref{D542})
are related to $a_i$ and $m_f$ in (\ref{D543}), respectively, by
\beq
A_i=e^{-a_i/R_4}, \qquad M_f=e^{-im_f/R_4}
\label{D544}
\eeq
The $R\to 0$ limit means $R_4\to\infty$ (see (\ref{D531})) and,
therefore, in the four dimensional limit the curve (\ref{D543}) becomes
\beq
t^2+t\prod_{i=1}^{N_c}(v-a_i)+\prod_{f=1}^{N_f}(v-m_f)=0
\label{D545}
\eeq
This is precisely the curve of $N=2$ supersymmetric $SU(N_c)$
gauge theory with $N_f$ flavors in four dimensions (\ref{sqcdnf}).

\subsection{Some Generalizations}
\label{D5SG}

To study $N=1$ supersymmetric symplectic (or orthogonal)
gauge theories in five dimensions we need to present
an orientifold fiveplane.
For instance, let us introduce an $O5_{+2}$ plane parallel to
the $D5$-branes (\ref{D511}) in the type IIB webs of
subsection \ref{D5FEF}. It gives rise to an $Sp(N_c/2)$
gauge theory on the $D5$-branes (because the orientifold has
a positive charge thus imposing a symplectic projection
on the parallel D fivebranes). The brane configuration
is necessarily invariant under the orientifold reflection and,
therefore, it is more constrained than the $SU(N_c)$
configurations. In particular, given $N_c$ and $N_f$
there is a unique possibility
(modulo equivalence transformations)
for the orientation of the external legs -- the
one invariant under the mirror reflection.
This single consistent configuration -- describing an
$Sp(N_c/2)$ gauge theory with $N_f$ fundamental hypermultiplets --
corresponds to the unique field theory obeying the
condition $c_{cl}=0$ (see subsection \ref{D5FTR}).
Moreover, as in the unitary case,
the field theory condition (\ref{D55}) translates in the
brane construction into
the requirement that the external legs do not cross each other.

Gauge theories with product gauge groups can also be
considered in the brane picture.
For example, webs corresponding to a product of unitary gauge groups
$SU(N_1)\times SU(N_2)\times \cdots \times SU(N_k)$
have -- in the $g_s\to 0$ limit -- $k+1$ parallel NS
fivebranes separated in $x^6$, and $N_i$ D fivebranes
connecting the $i$'th NS fivebrane (from the left in $x^6$)
to the $i+1$'st NS fivebrane, $i=1,...,k$.
For $g_s\neq 0$ the fivebranes bend -- according to the rules
discussed in this review -- describing the exact quantum corrections
to the five dimensional theories.
Again, the condition that external legs do not cross each other
must correspond to appropriate field theory constraints.

Webs of fivebranes can also be used to obtain new $N=2$
$3$-$d$ SCFTs from $5d$ fixed points. This is done by
considering two identical webs separated, say in the $x^7$
direction, and stretching between them $D3$-branes with
worldvolume in $(x^0,x^1,x^2,x^7)$. For more details we
refer the reader to~\cite{AH}.

Finally, we should remark that some deformations of consistent webs
may lead to theories with no gauge theory interpretation.

\subsection{Six Dimensional Theories}
\label{D5SDT}

In this subsection we discuss brane configurations
in the type IIA string with sixbranes ending on fivebranes
which describe at low energy six dimensional $(0,1)$
supersymmetric gauge theories~\cite{BKa}.

We consider NS fivebranes, D sixbranes and orientifold sixplanes
in the type IIA string with worldvolume
\beq
\begin{array}{ll}
\mbox{$NS5:$}&  \mbox{$(x^0, x^1, x^2, x^3, x^4, x^5)$}\\
\mbox{$D6/O6:$}&  \mbox{$(x^0, x^1, x^2, x^3, x^4, x^5, x^6)$}
\end{array}
\label{D551}
\eeq
With these objects we can construct several stable
configurations leading to consistent $6d$ supersymmetric
gauge theories with eight supercharges, a few examples of
which are being presented below:

\begin{itemize}

\item {\em $SU(N_c)$ with $N_f=2N_c$:}
Let us stretch $N_c$ $D6$-branes between two $NS5$-branes
which are separated in $x^6$. To the left (right) of the
left (right) fivebrane we place $N_L$ ($N_R$) semi-infinite
$D6$-branes. Since sixbranes ending on a fivebrane
behave like electric charges in zero dimensions, stability
implies that the total charge must vanish.
This zero charge condition on the fivebranes
implies that
\beq
N_L=N_R=N_c \Rightarrow N_f\equiv N_L+N_R=2N_c
\label{D552}
\eeq

The low energy theory on the $D6$-branes is, therefore,
an $SU(N_c)$ gauge theory with $N_f=2N_c$ fundamental hypermultiplets.
In field theory, the condition (\ref{D552}) is precisely the one
required for anomaly cancelation in $(0,1)$ SUSY
six dimensional theories. Again, we find that the
brane configuration is stable if and only if the
gauge theory is anomaly free.

We may compactify the theory on a three-torus
in the $(x^3,x^4,x^5)$ directions and perform
T-duality in these directions.
The brane configuration considered above is
T-dual to configurations describing $SU(N_c)$
gauge theories with eight supercharges and
$N_f=2N_c$ which were discussed in previous sections.
T-duality (followed by decompactifications)
in the $x^5$ direction $T_5$ takes
it to a web of fivebranes describing a particular
$N=1$ five dimensional gauge theory. $T_{45}$ leads to
an $N=2$ four dimensional configuration, while
$T_{345}$ gives an $N=4$ three dimensional case.

\item {\em $SO(N_c)$ with $N_f=N_c-8$:}
To get an orthogonal gauge group we add an $O6$-plane
and stuck the two NS fivebranes separated in $x^6$
on top of the orientifold.
As we have learned, there is a sign flip in the
RR charge of the orientifold
on the two sides of the fivebrane.

Consider the case where an $O6_{-4}$-plane
(the orientifold sixplane with
charge $-4$) is stretched between the two NS fivebranes.
Therefore, to the left (right) of the left (right) fivebranes
we must have semi-infinite $O6_{+4}$-planes.
Moreover, between the NS fivebranes we stretch
$N_c$ D sixbranes and to the left (right) of them we
place $N_L$ ($N_R$) semi-infinite $D6$-branes.
The zero force condition on the fivebranes now implies
that
\beq
N_L=N_R=N_c-8\Rightarrow N_f\equiv (N_L+N_R)/2=N_c-8
\label{D553}
\eeq
To obtain the consistency condition (\ref{D553})
we had to take into account the sign flip of the orientifold.

The theory on the sixbranes is a $(0,1)$ supersymmetric
$SO(N_c)$ gauge theory (because the orientifold segment
parallel to the $N_c$ finite sixbranes is an $O6_{-4}$
thus imposing an orthogonal projection) with $N_f$ hypermultiplets
in the vector representation. The requirement (\ref{D553})
is precisely the anomaly free condition in such a gauge
theory.

\item {\em $Sp(N_c/2)$ with $N_f=N_c+8$:}
To get a symplectic gauge group all we need to
do is to change the sign of the orientifold in the
previous example. The projection of the $O6_{+4}$-plane
stretched between the fivebranes on the parallel $N_c$
sixbranes is the symplectic one, leading to an $Sp(N_c/2)$
supersymmetric gauge theory. Moreover, the zero force
condition implies now that the theory has $N_c+8$
fundamental hypermultiplets:
\beq
N_L=N_R=N_c+8 \Rightarrow N_f\equiv (N_L+N_R)/2=N_c+8
\label{D554}
\eeq
which is precisely the anomaly free condition in
gauge theory.

\item {\em Product Groups:}
Configurations describing an alternating product
of $k$ orthogonal and symplectic gauge groups can be studied
by considering $k+1$ NS fivebranes separated in $x^6$
on top of an $O6$-plane. The zero force condition
implies the correct relations of colors and flavors
required for anomaly cancellation in field theory.

\end{itemize}

\noindent
Very recently, new works discussing branes and six dimensional
theories appeared~\cite{BKb,HZ9712}. In these works
configurations including also eightbranes and orientifolds
were considered, leading to classes of $6d$
models with non-trivial fixed points at strong coupling,
some of which were studied previously in field theory~\cite{Sei9609}
and using branes at orbifold singularities~\cite{Int}.

\section{Discussion}
\label{DISC}

\subsection{Summary}

The worldvolume physics of branes
in string theory provides a remarkably
efficient tool for studying many
aspects of the vacuum structure and
properties of BPS states in
supersymmetric gauge theories.
By embedding it
in a much richer dynamical
structure, brane dynamics provides a
new perspective on gauge theory and
in many cases explains phenomena that
are known to occur in field theory but
are rather mysterious there. Some examples 
of results which can be better understood using
branes that were described in this review are:

\begin{enumerate}

\item Montonen and Olive's
electric-magnetic duality in four
dimensional $N=4$ SUSY gauge theory as well
as Intriligator and Seiberg's mirror symmetry
in three dimensional $N=4$ SUSY gauge theory
are consequences of the non-perturbative
S-duality symmetry of type IIB string theory
\cite{Tse,GG96,HW}.

\item Nahm's construction of the moduli
space of magnetic monopoles can be derived
by using the description of monopoles
as D-strings
stretched between parallel $D3$-branes
in type IIB string theory~\cite{Dia}.
A similar description leads
to a relation between the moduli space
of monopoles in one field theory and
the quantum Coulomb branch of another
\cite{HW}.

\item The auxiliary Riemann surface
whose complex structure was proven by
Seiberg and Witten to determine the low
energy coupling matrix of four dimensional
$N=2$ SUSY gauge theory is naturally interpreted
as part of the worldvolume of a fivebrane
\cite{KLMVW,W9703}. Hence it is physical
in string theory; moreover, this geometrical
interpretation is very useful for studying
BPS states in $N=2$ SYM.

\item Seiberg's infrared equivalence between
different four dimensional $N=1$ supersymmetric
gauge theories is manifest in string
theory \cite{EGK,EGKRS}. The electric
and magnetic theories provide different
parametrizations of the same quantum moduli
space of vacua. They are related
by smoothly exchanging fivebranes in an
appropriate brane configuration. Many additional
features of the vacuum structure of $N=1$ SUSY
gauge theories can be reproduced by studying the
fivebrane configuration \cite{HOO,W9706,BIKSY}.
In particular, the QCD string of confining
$N=1$ SYM theory appears to be a membrane
ending on the fivebrane \cite{W9706}.

\item The vacuum structure
of $N=2$ supersymmetric gauge theories
in three dimensions and, in particular,
the generalization of Seiberg's duality
to such systems, can be understood
using branes~\cite{EGKRS}. An interesting feature
of Seiberg's duality in three dimensions
is that it relates two theories one of
which is a conventional field theory, while
the other does not seem to have a local field
theoretic formulation (but it does have a
brane description).

\item Webs of branes provide
a useful description of non-trivial fixed
points of the RG in five and six 
dimensions~\cite{AH,AHK,BKb,HZ9712}.

\end{enumerate}

\noindent
In fact, one could argue that all the
results regarding the vacuum structure
of strongly coupled supersymmetric
gauge theories
obtained in the last four years should
be thought of as
low energy manifestations of string
theory.

The improved understanding of the vacuum
structure obtained by embedding gauge
theory in the larger context of string
or brane theory is very interesting, but it
would be even more important to go beyond
the vacuum/BPS sector and obtain
new results on non-vacuum low energy
properties, \eg\ the masses and interactions
of low lying non-BPS states.
In field
theory not much is known about this
subject, but there are reasons to
believe that progress can be made
using branes.

The role of branes
in describing low energy gauge theory
so far is somewhat analogous to
that of Landau-Ginzburg theory
in critical phenomena. It provides
a remarkably accurate description
of the space of vacua of the theory
as a function of the parameters in
the Lagrangian,
including aspects that are quite
well hidden in the standard variables,
such as strong-weak coupling relations
between different theories.
As in critical phenomena, to
compute critical exponents or, more
generally, study the detailed structure
of the infrared CFT, one will have to
go beyond the analysis of the vacuum.
However, if the analogy to statistical
mechanics is a good guide, the brane
description -- which clearly captures correctly the
order parameters and symmetries of the theory --
should prove to be a very useful starting point
for such a study.

\subsection{Open Problems}

In the course of the discussion
we have encountered a few issues
that deserve better understanding.
Some examples are:

\subsubsection{$SU(\nc)$ Versus $U(\nc)$}

We have seen in sections \ref{D4N2},
\ref{D4N1} that brane configurations
describing four dimensional gauge theory
with a unitary gauge group seem to
have the peculiar property that
while classically the gauge group
is $U(\nc)$, quantum mechanically
it is $SU(\nc)$, with the gauge coupling
of the $U(1)$ factor vanishing logarithmically
as we turn on quantum effects. At the same
time, an interpretation of the physics
in terms of an $SU(\nc)$ gauge theory
seems to be in contradiction with
certain supersymmetric
deformations of the brane configuration
which appear to be
parameters in the Lagrangian rather
than moduli~\cite{GP}. It would be interesting
to resolve this apparent paradox,
especially because it is closely related
to other issues that one would like to
understand better. In particular, as
we have seen, some of the features
of the infrared physics are not visible
geometrically in brane theory. For example,
in four dimensional magnetic $N=1$ SQCD with
$\nf=\nc$, the mesons $M$ (\ref{FT6})
are clearly
visible, while the baryons $B$, $\tilde B$
(\ref{FT9}) are more difficult to see. Similarly,
in three dimensional $N=2$ SQCD,
the fields $V_\pm$ (\ref{dualsup})
are a form of dark matter,
visible only through their effect
on the quantum moduli space of vacua.

\subsubsection{Non-Trivial Fixed Points,
Intersecting
Fivebranes And Phase Transitions}

Brane configurations provide very useful 
descriptions of the classical and quantum
moduli spaces of vacua of different
gauge theories, but so far it proved
difficult to use them to study other
features of the long distance behavior.
The corresponding M-theory fivebrane 
becomes singular as one approaches a
non-trivial IR fixed point, and
thus it is not well described 
by eleven dimensional supergravity.
Only aspects of the fixed point that
can be studied by perturbing away
from it and continuing to unphysical
values of $L_6$, $R_{10}$, 
such as the superpotential,
dimensions of
chiral operators and global symmetries,
can be usefully studied using low energy
M-theory.
 
An important tool for studying
the low energy dynamics of $N=1$ SUSY
gauge theories using branes is
$N=1$ duality.
We have seen that the theory
on fourbranes stretched between
non-parallel fivebranes
changes smoothly when the
fivebranes meet in space
and exchange places.
In the case of parallel fivebranes,
this process corresponds to a phase
transition. It would be very interesting
to understand this phenomenon in more detail
by studying the theory on parallel versus
non-parallel fivebranes.

Specifically, we have seen
using branes that the quantum moduli
spaces of vacua and
quantum chiral rings
of the electric and
magnetic SQCD theories coincide.
This leaves open the question whether
Seiberg's duality extends to an
equivalence of the full infrared
theory, since in general the
chiral ring does not fully
specify the IR CFT. It is believed
that in gauge theory the answer is yes,
and to prove it in brane theory will
require an understanding of the smoothness
of the transition when fivebranes cross.

It is important to emphasize that the
question cannot be addressed using any currently
available tools. The M-theory approach 
fails since the characteristic size of the
fivebrane becomes small, and the brane
interactions relevant for this situation
are unknown. As we saw,
the fact that when parallel NS fivebranes
cross the theory on fourbranes stretched
between them undergoes a phase transition,
is related to the fact that when the fivebranes
coincide they describe a non-trivial six dimensional
CFT. It is a hallmark of non-trivial fixed
points that the physics seen when one is
approaching them from different
directions is different.
To show smoothness for non-parallel
branes one has to understand the theory
on intersecting non-parallel NS fivebranes.
At present such theories
are not understood.

\subsubsection{Orientifolds}

Brane configurations involving
orientifold fourplanes such as those
of 
Figures~\ref{nineteen},\ref{twenty},\ref{fourtytwo},
discussed in sections \ref{D4N2}, \ref{D4N1},
are still puzzling. It appears that  
when a D-brane
intersects an orientifold and divides
it into two disconnected components,
the charge of the orientifold flips
sign as one crosses the D-brane.
Also, upon compactification of
such brane configurations on a
longitudinal circle and T-duality
along the circle, one finds brane 
configurations
in type IIB string theory in which
the analysis is severely constrained
by S-duality. It is not clear how the
corresponding analysis is related to
the process of compacifying the low energy
gauge theory on the fourbranes from four
to three dimensions.

\subsubsection{Future Work}

Clearly, there is much that remains to be
done. The two main avenues for possible
progress at the moment seem to be the following:

\medskip
\noindent
{\bf 1. More Models}

One would like to find the sort of description
of the vacuum structure and low energy
physics that we presented for additional
models. Specific examples include models
with exceptional gauge groups, and 
more general matter representations of
the classical groups, such as $SO$ groups
with spinors. For example, 
if one believes that Seiberg's
duality is a string theory
phenomenon, it should be possible to 
find an embedding in string theory of
the set of Seiberg dual pairs studied 
by~\cite{Pou,BCKS} and others.

One way to proceed in the case
of four dimensional $N=2$ SUSY
is to study configurations
in which an $M5$-brane wraps the Seiberg-Witten
surface relevant for the particular
gauge theory, as is done in~\cite{KLMVW}.
It would be interesting to understand
the relation of these constructions
to the sort of configurations studied
here.
It is worth stressing that
one is looking for a brane construction
that does not only share with SYM its
vacuum structure. Rather, we want
to reproduce the whole RG trajectory
corresponding to the particular gauge
theory in some limit of string/brane 
dynamics. This means that there
has to be a weakly coupled description
of the brane configuration, suitable
for studying the vicinity of the UV fixed
point of SYM.

For $N=1$ supersymmetric theories, it would
be interesting to construct large classes
of chiral gauge theories that break supersymmetry
and study them using branes. This may
clarify the general requirements for
SUSY breaking and hopefully provide
the same kind of conceptual unification
of SUSY breaking that was achieved for
Seiberg-Witten theory and $N=1$ duality.

In this review we have mostly discussed
the worldvolume physics on branes which
have finite extent in one non-compact
direction. 
An interesting generalization corresponds 
to configurations containing branes
that are finite or semi-infinite
in more than one non-compact direction. 
The simplest case to examine is that of
branes that are finite in {\em two}
directions.

We have seen that such configurations
are necessary to describe Euclidean 
field configurations that give rise
to different non-perturbative effects
(see \eg\ 
Figs.~\ref{eight},\ref{thirtyfour}).
Similarly, in section \ref{D4N2} configurations
of $D2$-branes stretched between two
$NS5$-branes and two $D4$-branes were
used to describe magnetic monopoles
in four dimensional $N=2$ SYM.
Using U-duality, such configurations
can be mapped to other interesting
configurations. For example, one
can study $Dp$-branes (with worldvolume, say, in
$(x^0,..,x^{p-2},x^4,x^8)$, $p=2,3,4,5$)
stretched between a pair of $NS5$-branes separated 
by a distance $L_8$ in $x^8$
and a pair of \nsp-branes 
separated by a distance $L_4$
in $x^4$ (see (\ref{BSB1},\ref{D4N11})
for the conventions), or $D3$-branes
stretched between two $NS5$-branes
separated in $x^6$ and two $D5$-branes
(\ref{D511}) separated in $x^5$. 

\begin{figure}
\centerline{\epsfxsize=90mm\epsfbox{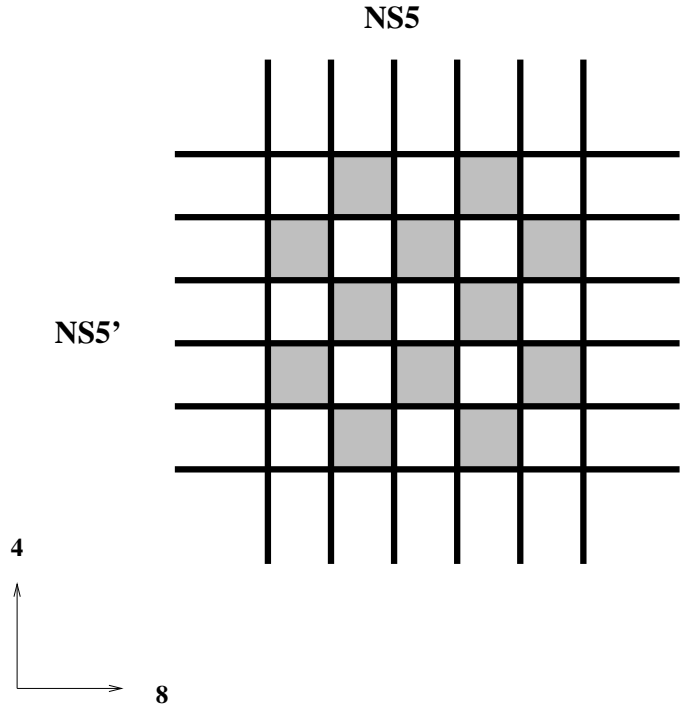}}
\vspace*{1cm}
\caption{Does chess 
play a role in string theory?}
\label{fiftytwo}
\end{figure}
\smallskip

In the latter case, it is easy to
check that the two dimensional
low energy theory on the threebranes
has $(4,0)$ SUSY and is thus
chiral. Therefore, these configurations
and their generalizations provide
a useful laboratory for the study
of chiral field theories. A large 
class of generalizations can be obtained
by studying ``chess board'' configurations
in which branes finite in two directions
stretch like rugs between different
segments of a two dimensional network
of intersecting branes (Fig.~\ref{fiftytwo}). 
Particular
configurations of this 
sort~\footnote{See~\cite{HHZZ} for a recent
discussion.}  are dual
to some of the chiral models discussed
in subsection \ref{CHIRAL}, using the
duality relating
an ALE space with a $Z_n$ orbifold
singularity to a vacuum containing
$n-1$ solitonic fivebranes on $R^4/Z_n$.

Clearly, it would be interesting to
study configurations of this type 
with more branes and/or orientifolds,
as well as consider branes that are finite
in more dimensions, which should lead at
low energies to many new models
and hopefully also some new understanding.

\medskip
\noindent
{\bf 2. The Dynamics Of Fivebranes}

It is clearly important to develop tools to
study the dynamics of fivebranes in string
theory. Four dimensional $N=4$ SYM can be thought
of as the six dimensional CFT on $\nc$
fivebranes compactified on a two torus
whose modulus $\tau$ is related to the
four dimensional SYM coupling~\cite{W9507}.
$N=1,2$ SYM can be thought of as
compactifications of the $(2,0)$ CFT
from six down to four dimensions
on the Seiberg-Witten Riemann surface
$\Sigma$~\cite{W9706}. 

Recently, the $(2,0)$ theory on $R^{5,1}$
as well as the compactified theory on
$R^{3,1}\times T^2$ were studied using
matrix theory~\cite{ABKSS,ABS,GS}. 
These attempts are still
at an early stage and it is not clear
whether they will eventually provide
efficient techniques for studying
these theories. In any case,
matrix descriptions of theories
like four dimensional $N=4$ SYM
are useful also as a testing ground
for matrix theory in general, as
the theory that one is trying to
describe is in this case
well defined and understood
(at least in certain corners of
parameter space),
unlike eleven dimensional
M-theory for which matrix theory was
originally proposed.

Another promising direction is
to understand the
theory of the QCD string.
At large $\nc$ the string coupling
of the QCD string is expected to be
small~\cite{thooft} and one may hope
that the theory can be described by
a more or less conventional worldsheet
formalism~\cite{polbook}. 
What kind of theory does
one expect to find? The brane construction
suggests a theory that
lives in six dimensions,
but is Lorentz invariant
only in four of these. There is
a non-trivial metric in the remaining
two directions, $\Phi$, which suppresses
fluctuations of the string in
these directions. The resulting
picture is very reminiscent of
non-critical superstrings that
were constructed in~\cite{DN1}
and of the recent work of~\cite{Polyakov}.
It would of course be very interesting
to make this more precise.

A long standing puzzle in the theory of QCD
strings is related to the work of~\cite{DN2},
who pointed out that in fundamental string
theory IR stability of the vacuum (absence
of tachyons) and unitarity imply asymptotic
supersymmetry of the spectrum.
Confining large $\nc$ gauge theories are traditionally
expected to have a string description even in
the absence of
supersymmetry~\cite{thooft,polbook}. The new ideas
on QCD string theory and, in particular, the relation
of the QCD string to the fundamental string,
might help resolve the puzzle. Perhaps a description
of QCD in terms of continuous  worldsheets requires
asymptotic SUSY. This may be
related to recent speculations
that SUSY appears to play a deep role in string
dynamics~\cite{DKPS,BFSS}.  For example,
there are indications that locality
in string theory is a consequence of asymptotic
SUSY.

Eventually, one would like to use branes
to study the infrared dynamics of non-supersymmetric
theories like QCD. At present, brane
constructions shed no light on strongly
coupled non-supersymmetric gauge theory.
Thus, if SUSY is 
dynamically broken for a particular
brane configuration, one can generally
say very little about the physics of the
non-supersymmetric ground state. It seems
quite likely that progress on one of the fronts
mentioned above will also allow one to
study non-supersymmetric gauge theories.

\medskip

\noindent

\acknowledgments
We thank O. Aharony, A. Hanany,
J. Harvey, B. Kol, E. Martinec,
Y. Oz, O. Pelc, R. Plesser, E. Rabinovici,
M. Ro\v cek, A. Schwimmer, N. Seiberg,
and especially S. Elitzur for discussions.
We are grateful to M. Dine for his support
and comments on the manuscript.
This work is supported in
part by the Israel Academy of Sciences and
Humanities -- Centers of Excellence Program.
The work of A. G. is supported in
part by BSF -- American-Israel Bi-National
Science Foundation. A. G. thanks
the EFI at the University of Chicago,
where part of this work was done, for its warm hospitality,
and the Einstein Center at the Weizmann Institute for partial
support. D. K. is supported in part by a DOE OJI grant.

\end{document}